\documentclass[11pt,a4paper,oneside]{article}
\pdfoutput=1

\usepackage[affil-it]{authblk}
\usepackage[a4paper,left=2.5cm,right=2.5cm,top=2.8cm,bottom=3.7cm]{geometry}

\usepackage[]{amsmath}
	\numberwithin{equation}{section}
\usepackage[]{amssymb}
\usepackage{amsfonts}
\usepackage{mathrsfs}

\DeclareMathOperator{\R}{\mathbb{R}}
\DeclareMathOperator{\C}{\mathbb{C}}
\DeclareMathOperator{\N}{\mathbb{N}}
\DeclareMathOperator{\Z}{\mathbb{Z}}
\DeclareMathOperator{\cs}{\mathbb{S}}

\newcommand{\mz}{\mathcal{Z}}
\newcommand{\mf}{\mathcal{F}}
\newcommand{\mN}{\mathcal{N}}
\newcommand{\sQ}{\mathsf{Q}}
\newcommand{\Th}{\mathscr{T}}
\newcommand{\surho}{\mathscr{S}}
\newcommand{\cutrho}{\mathscr{C}}

\newcommand{\dd}{\mathrm{d}}

\newcommand{\tr}{\mathrm{Tr}}
\newcommand{\Dslash}{\not{\!\!D}}

\DeclareMathOperator{\vol}{vol}

\usepackage[utf8]{inputenc}
\usepackage[english]{babel}

\usepackage[pdftex]{graphicx}
\usepackage[dvipsnames]{xcolor}

\definecolor{myb}{rgb}{.1,0,.72}
\definecolor{myr}{rgb}{.78,0,.1}
\usepackage[linktocpage=true,colorlinks=true,citecolor=myb,linkcolor=myr,urlcolor=myb]{hyperref}

\usepackage[toc,page]{appendix}
\usepackage[nottoc]{tocbibind}
\usepackage{caption}
\usepackage{enumerate}
\usepackage{ytableau}

\usepackage{tikz}
		\usetikzlibrary{er,positioning,arrows.meta}

\usepackage[numbers,compress,square,comma]{natbib}
	\numberwithin{figure}{section}

\begin{document}
\bibliographystyle{myJHEP}
\captionsetup[figure]{labelfont={bf,small},labelformat={default},labelsep=period,font=small}
\captionsetup[table]{labelfont={bf,small},labelformat={default},labelsep=period,font=small}

{\pagenumbering{roman} 

		\renewcommand*{\thefootnote}{\fnsymbol{footnote}}
	\title{\huge\textbf{Large \texorpdfstring{$N$}{N} limits\protect\\ of supersymmetric quantum field theories:\protect\\ \vspace{0.4cm} A pedagogical overview}}

\author{\vspace{0.4cm}{\Large Leonardo Santilli}\footnote{santilli@tsinghua.edu.cn}}
\affil{Yau Mathematical Sciences Center,\protect\\ Tsinghua University, Beijing, 100084, China.}

	\date{  }

	\maketitle
	\thispagestyle{empty}

	\begin{abstract}
		The different large $N$ limits of supersymmetric quantum field theories in three, four, and five dimensions are reviewed. We distinguish between the planar limit of SQCD theories, the M-theory limit suited in three and five dimensions, and the long quiver limit. The method to solve exactly the sphere partition functions in each type of limit is spelled out in a pedagogical way. 
		After a comprehensive general treatment of the saddle point approximation in the large $N$ limit, we present an extensive list of examples and detail the calculations. 
		The scope of this overview is to provide an entry-level, computation-oriented understanding of the techniques featured in the field theory side of the AdS/CFT correspondence.
	\end{abstract}

	\clearpage
	\tableofcontents
	\thispagestyle{empty}
}

	\clearpage
	\pagenumbering{arabic}
	\setcounter{page}{1}
		\renewcommand*{\thefootnote}{\arabic{footnote}}
		\setcounter{footnote}{0}

	\section{Introduction}
		Quantum field theory (QFT) is the formalism that, to date, yields the most accurate descriptions and predictions of particle physics. However, explicit computations remain mostly restricted to the perturbative regime. Furthermore, conformal QFTs (CFTs) in the large $N$ limit give us a glimpse of quantum gravity, through the AdS/CFT correspondence \cite{Aharony:1999ti}. Adding supersymmetry to the story gives us a handle to perform exact computations in strongly coupled dynamics.\par
		Supersymmetric QFTs in the large $N$ limit therefore provide an especially insightful area of research, thanks to the possibility of explicit and detailed computations. There are two approaches one may adopt:
		\begin{itemize}
			\item Take AdS/CFT for granted, and use large $N$ supersymmetric field theory techniques to gain insight into gravity for asymptotically-AdS spacetimes;
			\item Compute certain quantities independently using large $N$ supersymmetric field theory techniques and semiclassical gravity in AdS, and match them to enhance the AdS/CFT dictionary.
		\end{itemize}
		The second philosophy is especially instructive. Indeed, it is often the case that the task of matching quantities across the holographic duality entails shedding new light on either side. Successful examples of this interplay include compactifications on Riemann surfaces \cite{Maldacena:2000mw,Gaiotto:2009gz}, the volume minimization prescription \cite{Martelli:2005tp,Martelli:2006yb} and the counting of black holes microstates in AdS$_5$ \cite{Benini:2015eyy,Cabo-Bizet:2018ehj,Choi:2018hmj,Benini:2018ywd}, to name a few.\par
		\medskip
		Needless to say, the interest in the study of large $N$ limits of gauge theories (with or without supersymmetry) spans far beyond the AdS/CFT correspondence, and represents a fruitful field of research on its own. Large $N$ techniques \cite{tHooft:1973alw,Brezin:1977sv} have famously been instrumental in areas including the baryon spectrum of QCD \cite{Witten:1979kh}, lattice QCD \cite{Billo:1996pu}, and critical phenomena \cite{ZinnJustin:2007zz}.\par
		Yet another way in which supersymmetry comes in handy is to permit explicit computations that, combined with the large $N$ approximation, provide information about the gapless spectrum of a gauge theory. Indeed, the methods reviewed herein allow to study correlation functions of extended defects that preserve a fraction of supersymmetry. These defects are charged under a higher form symmetry and therefore, evaluating their expectation value in closed form, will tell if the symmetry is preserved or spontaneously broken, giving rise to higher spin Goldstone bosons.\par
		To give one concrete example, it was argued in \cite{BenettiGenolini:2020doj} that five-dimensional supersymmetric Yang--Mills theory with a Higgs field in the symmetric or adjoint representation exhibits spontaneous 1-form symmetry breaking. This statement was substantiated by a large $N$ computation of the Wilson loop expectation value in \cite{Santilli:2021qyt}, showing a second order phase transition accompanied by spontaneous symmetry breaking. Another recent example is provided by certain surface defects in maximally supersymmetric Yang--Mills theory in four dimensions, that undergo a deconfining phase transition \cite{Chen:2023lzq} (see also \cite{Amariti:2024bsr}). For higher form symmetries in four-dimensional supersymmetric field theories and their holographic dual, we refer to \cite{Hofman:2017vwr}.\par
		\medskip
		This overview will illustrate how to compute explicitly physical quantities in supersymmetric QFT, with emphasis on applications to the AdS$_{d+1}$/CFT$_d$ correspondence. The focus is on field theories with eight supercharges in spacetime dimensions $d=3,4,5$.\par
		The techniques to study the large $N$ limit of supersymmetric gauge theories are presented in a pedagogical way. Except for a few minor aspects, none of the results presented here is new. Nonetheless, we tried to give a uniform treatment of the many distinct derivations in the literature.\par 
		We wish to convey the intuition that, despite being presented in a superficially different fashion, all the results in the large $N$ limit and AdS/CFT are incarnations of the same underlying mechanism.\par
		
		\subsubsection*{Scope and style}
		This overview is not meant to represent an exhaustive review of the subject. Their goal, instead, is to provide an accessible, easily readable introduction `for pedestrians'. A minimal amount of specific background on supersymmetric field theories or AdS/CFT is required, if at all.\par 
		Our hope is that, after studying these notes, every reader will be able to perform the same computations and to adapt the general strategy to the case at hand.\par
		We therefore opt to review the general technique and provide an extensive list of examples. For pedagogical reasons, we decided to give many details of the computations and discuss several steps explicitly. We refrain from presenting a comprehensive but necessarily less detailed list of results, as well as from showing more technical or advanced computations, which are very close in spirit to but more obscure than what we present here.\par
		\begin{itemize}
			\item The ideal target audience for these lecture notes consists of graduate or advanced undergraduate students, as well as researchers without deep knowledge of large $N$ limits.\par
				AdS/CFT is an exciting subject that provides many hands-on problems. The philosophy of the present notes is that the reader is encouraged to learn the subject by doing. To comply with this scope we provide thorough details on how to perform the explicit calculations, without requiring any knowledge of AdS/CFT.
			\item The more experienced reader may find these notes a useful compendium of known results and techniques, presented in a unified framework. Besides, we enrich the text with observations and analysis of certain subtleties that are seldom discussed explicitly in the literature, and which may be relevant or interesting for the experts. 
		\end{itemize}

	\subsection{Overview of the topics}
		
		Large $N$ limits of vector and matrix models are an active area of research for which several excellent reviews and lecture notes have already been written. However, the method more often systematized and detailed in most reviews is, in many cases, not the best suited to study the large $N$ limit of theories relevant to AdS/CFT. This difficulty has been overcome in the papers that aim at matching AdS and SCFT computations but, due to their inherently technical nature, the methods of these papers may not be easily grasped by the beginner. These notes grew out of an effort to fill this gap.\par
		We will thoroughly discuss different large $N$ limits for different theories in three, four and five dimensions.
		
		\subsubsection*{Localization}
		Our starting point will be supersymmetric localization \cite{Pestun:2007rz}. Moreover, we focus on theories with eight supercharges: this is the minimal amount of supersymmetry for which localization on the round sphere $\cs^d$ is known in dimension $d \ge 4$, and captures a rich class of three-dimensional theories.\par
		There are various reviews entirely devoted to supersymmetric localization, and we refer to the existing literature on the subject --- listed in Appendix \ref{app:reading} --- for a deeper understanding of localization. For completeness, a lightning overview of localization is given in Section \ref{sec:ReviewLoc}, which is by no means exhaustive.\par 
		These notes can be read without knowing what `localization' means.\par

		\subsubsection*{Matrix models from localization}
		Taking the partition function of a supersymmetric QFT in the form of an $N$-dimensional integral, we will study its large $N$ limit. We will first explain the various ingredients that appear in the integral, and then describe their behavior at large $N$ in full generality. We will see how the saddle point argument allows us to obtain a universal answer, whose derivation does not depend on the details of the supersymmetric QFT considered.\par
		We will then apply the general formalism to a wealth of explicit examples, where many aspects of the derivation will become manifest, such as the competition among various terms in the integral and the cancellations at large $N$.\par
		\medskip
		The aim of these notes is to gather together in one place information that appears scattered in the literature, and to present it in a pedagogical manner. Many earlier reviews exist, each of them overlapping with parts of these notes. A fundamental reference for large $N$ limits in matrix models is \cite{DiFrancesco:1993cyw}. A thorough discussion of Chern--Simons-matter theories can be found in M. Mari\~{n}o's reviews \cite{Marino:2004eq,Marino:2011nm}. An overview of matrix models and five-dimensional supersymmetric gauge theories is \cite{Minahan:2016xwk}. For a more in-depth list of references, consult Appendix \ref{app:reading}.
		
		\subsubsection*{Large \texorpdfstring{$N$}{N} limits}
		The large $N$ limit can be taken in inequivalent ways. Which one is the correct one for any specific theory is ultimately a question about the embedding into string or M-theory, and about the supergravity dual.\par 
		In these lecture notes, we will remain agnostic about AdS/CFT or string theory embeddings, and we will show how the correct way to take the large $N$ limit is dictated by the form of the integral representation of the partition functions. The details of the supersymmetric QFT will be such that, at large $N$, the integrand will demand a certain scaling, leading us to consider one concrete way of taking the large $N$ limit.\par
		We will discuss different large $N$ limits, applied to different theories, in the forthcoming sections. The most famous large $N$ limit is 't Hooft's planar limit \cite{tHooft:1973alw} of gauge theories, reviewed for instance in S. Coleman's classic \cite[Ch.8]{Coleman:1985rnk}. This limit is suited to QCD and its supersymmetric generalizations, albeit other supersymmetric theories with more involved matter content will require different ways of taking the limit. We will define three ways of taking the limit in Section \ref{sec:LargeNtypes}, and treat them in full detail in the rest of the work.
		
		\subsubsection*{Saddle point and free energy}
		The ultimate goal of this overview is to compute the free energy of any supersymmetric QFT that admits a consistent large $N$ limit. This is achieved in two steps: after writing down the partition function as an integral and identifying the appropriate large $N$ regime, we will:
		\begin{enumerate}[(i)]
			\item\label{s1item1} Apply a saddle point analysis, valid in the large $N$ limit, to deduce the saddle point configuration that dominates the integral;
			\item Evaluate the free energy onto the saddle point configuration, thereby obtaining its large $N$ expression.
		\end{enumerate}
		The aim of these notes is to clarify why this is indeed the right procedure, and to perform it explicitly in a variety of distinct examples. For concreteness, we focus on the free energy, although, once the saddle point configuration is obtained, it is straightforward to apply the same strategy to compute other supersymmetric observables, such as half-BPS Wilson loops, and compare them with supergravity.\par
		\medskip
		As a word of caution, throughout these notes, only the leading order in the large $N$ limit is calculated. In particular, other saddle points giving sub-leading contributions compared to those in (\ref{s1item1}) will be neglected.

	\subsection{Organization}
		The content of this overview is structured as follows.\par
		We begin with two preliminary sections: Sections \ref{sec:SPEprelim} and \ref{sec:ReviewLoc} (indicated as warm-up sections in the table of contents) are inserted for completeness and to set the stage. The reader solely interested in acquiring the computational tools for large $N$ supersymmetric gauge theories may safely skip them. Section \ref{sec:SPEprelim} is a brief introduction to the large $N$ limit in simpler examples, while Section \ref{sec:ReviewLoc} quickly walks the reader through the basics of supersymmetric Lagrangians and the concept of supersymmetric localization.\par
		The core of the present overview starts with Section \ref{sec:SphereZ}, where we present the setup for the large $N$ limit in full generality. Three types of large $N$ limits, referred to as the planar limit, the M-theory limit, and the long quiver limit, are presented in Section \ref{sec:LargeNtypes}, while the general procedure for the solution is outlined in Sections \ref{sec:GeneralSPE}-\ref{sec:GeneralF}.\par
		From there, we explore the three approaches to the large $N$ limit, relevant in different supersymmetric field theories: 't Hooft's planar limit in Section \ref{sec:LargeNtHooft}, the M-theory limit in Section \ref{sec:LargeNMtheory}, and the long quiver limit in Section \ref{sec:LargeNLongQ}. These three sections need not be read in sequential order, albeit the long quiver limit is better understood after the M-theory limit.\par
\begin{figure}[tb]
		\centering
			\begin{tikzpicture}
				\node[align=center,rectangle, rounded corners, draw=black, fill=gray, fill opacity=0.3, text=black, text opacity=1] (c) at (0,0) {Partition functions\\ and\\ large $N$ limits};
				\node[align=center,rectangle, rounded corners, draw] (sp) at (-4,3.5) {(Warm-up)\\ Saddle point\\ approximation};
				\node[align=center,rectangle, rounded corners, draw] (lo) at (-4,-3.5) {(Warm-up)\\ Localization};
				
				\path[->] (sp) edge (c);
				\path[->,dotted] (lo) edge (c);
				\path[->,dotted] (sp) edge (lo);
				
				\node[align=center,rectangle, rounded corners, draw] (s2) at (4.5,0) {M-theory\\ limit};
				\node[align=center,rectangle, rounded corners, draw] (s1) at (4.5,3.5) {'t Hooft\\ limit};
				\node[align=center,rectangle, rounded corners, draw] (s3) at (4.5,-3.5) {Long quiver\\ limit};

				\node[anchor=south] at (c.north) {Sec. \ref{sec:SphereZ}};
				\node[anchor=west] at (s1.east) {Sec. \ref{sec:LargeNtHooft}};
				\node[anchor=west] at (s2.east) {Sec. \ref{sec:LargeNMtheory}};
				\node[anchor=west] at (s3.east) {Sec. \ref{sec:LargeNLongQ}};
				\node[anchor=east] at (lo.west) {Sec. \ref{sec:ReviewLoc}};
				\node[anchor=east] at (sp.west) {Sec. \ref{sec:SPEprelim}};

				\path[->] (c) edge (s1);
				\path[->,dotted] (c) edge (s2);
				\path[->,dotted] (c) edge (s3);
				\path[->] (s1) edge (s2);
				\path[->] (s2) edge (s3);

				\node[align=center,rectangle, rounded corners, draw] (aB) at (7.75,1.75) {Planar\\ matrix models};				
				\node[anchor=north west] at (aB.south) {App. \ref{app:MMHU}};
				\path[->,dotted] (s1) edge (aB);
				\path[->,dotted] (aB) edge (s2);
				
			\end{tikzpicture}
		\caption{Organization of the contents. The central node, shaded, contains the core concepts. The warm up sections can be safely skipped by the reader familiar with basic concepts of supersymmetry and saddle point approximation. Solid arrows indicate the suggested reading pattern.}
		\label{fig:organization}
		\end{figure}
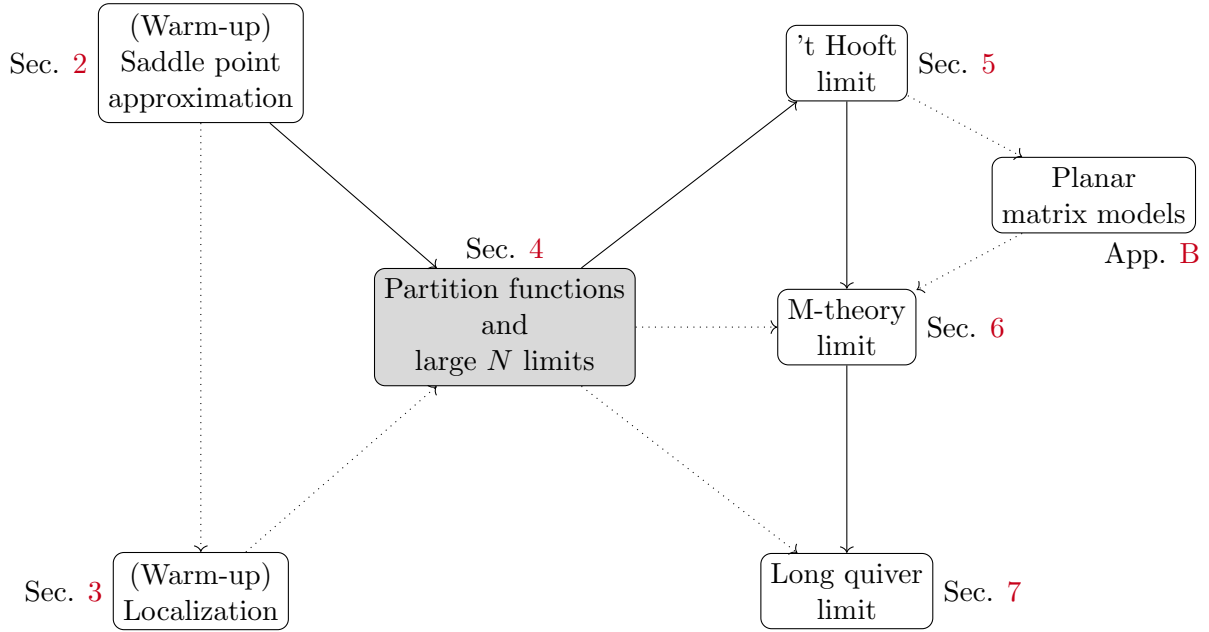

		The sections and relations among them are illustrated in Figure \ref{fig:organization}.\par
		\par
		\medskip
		The main text is complemented with two appendices. Appendix \ref{app:reading} contains an extensive list of references to topics not covered in these notes. Appendix \ref{app:MMHU} contains further material on the planar limit of matrix models. In particular, Appendix \ref{app:UMM} is devoted to a detailed pedagogical overview of the 't Hooft large $N$ limit of unitary matrix models. These matrix models appear in a variety of subjects and play a role in the partition functions on $\cs^{d-1} \times \cs^1$.\par
		Subsections indicated as aside discuss slightly more advanced topics, and may be safely skipped by the beginner. Along the way, we leave as an exercise to fill in some technical details that we omit for readability.

		\subsubsection*{Acknowledgements}
	I am indebted to Carlos Nu\~{n}ez, Miguel Tierz, and Christoph Uhlemann for the many insightful discussions, that enhanced and sharpened my understanding of this subject. I also thank Carlos Nu\~{n}ez and Lorenzo Ruggeri for helpful comments on the draft.\\
	Parts of this work grew out of lectures given at Swansea University and Tsinghua University. I gratefully acknowledge the Department of Physics of Swansea University for warm hospitality, and for the encouragement to write these notes. I also acknowledge the Shanghai Institute for Mathematics and Interdisciplinary Sciences, the Department of Physics of the University of Milan, and the Department of Mathematics of the University of Turin for hospitality during the completion of this work.\\
	This work was supported by the Shuimu Scholars program of Tsinghua University, by the NSFC grant W2433005 ``String theory, supersymmetry, and applications to quantum geometry'', and in part by the Beijing NSF grant IS23008 ``Exact results in algebraic geometry from supersymmetric field theory''.

\section[(Warm-up) Evaluating integrals by saddle point approximation]{Evaluating integrals by saddle point approximation}
\label{sec:SPEprelim}

This review is ultimately about learning how to evaluate integrals in the large $N$ limit. To this aim, we begin analyzing the asymptotic behaviour of simpler integrals, to understand the techniques and gain insight.\par
This section is meant as an exercise for those unfamiliar with studying integrals by saddle point approximations and is not related to the main topic of these notes, which is supersymmetric field theory.

\subsection{Saddle point approximation in a nutshell}
Throughout this review, we will extensively study integrals over $N$ variables $(\phi_1, \dots, \phi_N) \in \R^N$ of the form 
\begin{equation*}
	\mz = \frac{1}{N!}\int_{\R^N} \dd \phi ~e^{- S_{\mathrm{cl}}(\phi)} Z_{\text{1-loop}} (\phi) ,
\end{equation*}
with $\dd \phi$ being the Lebesgue measure on $\R^N$. Trivially, these integrals are rewritten as  
\begin{equation*}
	\mz = \frac{1}{N!}\int_{\R^N} \dd \phi ~\exp \left\{ - \underbrace{\left[ S_{\mathrm{cl}}(\phi) - \log  \left(Z_{\text{1-loop}} (\phi)\right) \right]}_{=: S_{\mathrm{eff}} (\phi)} \right\} ,
\end{equation*}
where we collectively refer to the term in square brackets as the effective action $S_{\mathrm{eff}}$. Due to traces of $\phi$ appearing in $S_{\mathrm{eff}}$, it typically grows proportionally to a positive power of $N$ in the large $N$ limit.\par
To start understanding how to deal with these expressions, we consider a simpler type of integral in this section. Namely, we assume there is only one integration variable, and $N$ is an external parameter such that 
\begin{equation}
\label{eq:Seff1dscaleN}
	S_{\mathrm{eff}} \propto N^{\chi} \qquad \text{when } N \to \infty ,
\end{equation}
for some $\chi >0$.\par
The integral we wish to compute is 
\begin{equation*}
	\mz = \int_{- \infty}^{+\infty} \dd \phi ~e^{- S_{\mathrm{eff}} (\phi)} .
\end{equation*}
By convergence, $S_{\mathrm{eff}} (\phi) $ must be such that the integrand is damped at $\phi \to \pm \infty$. In particular, it certainly cannot become arbitrarily negative in the regions $\phi \to \pm \infty$.\par 
By the hypothesis \eqref{eq:Seff1dscaleN}, the effective action becomes large in the large $N$ limit, thus, since it appears in the exponential with a negative sign, the integrand is sharply peaked around the minima of $S_{\mathrm{eff}} (\phi) $. Therefore, the leading contribution to the integral $\mz$ comes from a small neighborhood of the saddle points $\phi_{\ast}$ of $S_{\mathrm{eff}} (\phi) $. These are the points that solve the equation 
\begin{equation}
\label{eq:SPE1dint}
	\left. \frac{ \partial S_{\mathrm{eff}} (\phi) }{\partial \phi } \right\rvert_{\phi=\phi_{\ast}} = 0 .
\end{equation}\par
Let us assume that we have solved explicitly this saddle point equation and identified the absolute minimum $\phi_{\ast}$ of $S_{\mathrm{eff}} (\phi)$. It is then convenient to make a change of variables
\begin{equation*}
	\phi = \phi_{\ast} + \frac{\sigma}{N^{\chi/2}} .
\end{equation*}
We expand in Taylor series
\begin{equation*}
	S_{\mathrm{eff}} (\phi) = \underbrace{S_{\mathrm{eff}} (\phi_{\ast})}_{\propto N^{\chi}} + \underbrace{ \frac{\sigma}{N^{\chi/2}} S_{\mathrm{eff}}^{\prime} (\phi_{\ast})}_{\text{$0$ by \eqref{eq:SPE1dint}}} + \underbrace{\frac{1}{2} \frac{\sigma^2}{N^{\chi}} S_{\mathrm{eff}}^{\prime\prime} (\phi_{\ast}) }_{\text{indep. of $N$}} + \cdots 
\end{equation*}
and note that, by the scaling assumption \eqref{eq:Seff1dscaleN}, the terms beyond the second derivative are suppressed by inverse powers of $N$ in the large $N$ limit. In this way, we compute the free energy 
\begin{align*}
	\mf &= -\log \mz \\
		&\stackrel{\text{ large $N$ }}{=}  - \log \left[  \int_{- \infty}^{+\infty}\frac{ \dd \sigma}{N^{\chi/2}} ~\exp \left\{ -\left( S_{\mathrm{eff}} (\phi_{\ast}) + \frac{\sigma^2}{N^{\chi}} S_{\mathrm{eff}}^{\prime\prime} (\phi_{\ast}) \right) \right\} \right] \\
		&= S_{\mathrm{eff}} (\phi_{\ast})  - \log \left[ \int_{- \infty}^{+\infty} \frac{ \dd \sigma}{N^{\chi/2}} ~ e^{- \frac{\sigma^2}{2} \cdot \frac{S_{\mathrm{eff}}^{\prime\prime} (\phi_{\ast})}{N^{\chi}} }  \right] \\
		&= S_{\mathrm{eff}} (\phi_{\ast})  - \log  \left( \sqrt{\frac{2 \pi}{S_{\mathrm{eff}}^{\prime\prime} (\phi_{\ast})}} \right) \\
		&=\underbrace{S_{\mathrm{eff}} (\phi_{\ast})}_{\propto N^{\chi}} + \underbrace{\frac{1}{2}\log \left( \frac{S_{\mathrm{eff}}^{\prime\prime} (\phi_{\ast})}{2 \pi} \right) }_{\text{sub-leading}} .
\end{align*}
To pass from the third to the fourth line we have used the fact that $\phi_{\ast}$ is a minimum, thus $S_{\mathrm{eff}}^{\prime\prime} (\phi_{\ast}) >0$ and we can perform the Gaussian integral.\par
We conclude that integrals showing the behaviour \eqref{eq:Seff1dscaleN} are approximated at leading order in the large $N$ limit by the saddle point value of their integrand.

\subsection{Asymptotic behaviour of the Gamma function}
As a warm-up example to understand how the saddle point approximation works, we will derive the asymptotic behaviour of the Gamma function $\Gamma (N)$.\par
For integer argument $N \in \N$, the Gamma function satisfies $\Gamma (N)=(N-1)!$. For complex argument $N \in \C$ with $\Re (N) >0$, it admits the integral representation 
\begin{equation}
\label{eq:Gammaint}
	\Gamma (N) = \int_0 ^{\infty} \phi^{N-1} e^{- \phi} \dd \phi .
\end{equation}
When $\Re (N) \to \infty$, $\Gamma (N)$ is approximated by Stirling's formula: 
\begin{equation}
\label{eq:StirlingGamma}
	\log \Gamma (N) = N \log (N) -N - \frac{1}{2} \log \left( \frac{N}{2\pi} \right) .
\end{equation}
We now proceed to show how to obtain this large $\Re (N)$ limit from the integral \eqref{eq:Gammaint}. To simplify the discussion, we assume $N \in \R_{>0}$ from now on.\par

\subsubsection*{Saddle points of the Gamma function}
\begin{figure}
			\centering
			\includegraphics[width=0.5\textwidth]{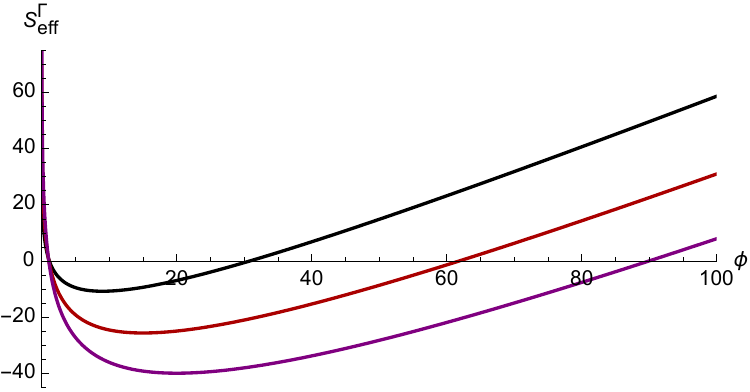}
			\caption{The function $S_{\mathrm{eff}}^{\Gamma} (\phi)$ defined in \eqref{eq:GammaPot}, shown as a function of $\phi$ for $N=10$ (black), $N=16$ (red) and $N=21$ (purple).}
			\label{fig:SeffGamma}
\end{figure}\par
As a preliminary step, we rewrite \eqref{eq:Gammaint} in the form
\begin{equation*}
	\Gamma (N) = \int_0 ^{\infty} \dd \phi ~\exp \left\{ \underbrace{(N-1)\log(\phi) -\phi}_{=: - S_{\mathrm{eff}}^{\Gamma} (\phi) } \right\}.
\end{equation*}
When $\Re (N)$ is very large, the argument of the exponential has a very large absolute value. In particular, the function 
\begin{equation}
\label{eq:GammaPot}
	S_{\mathrm{eff}}^{\Gamma} (\phi) =  \phi - (N-1)\log(\phi) 
\end{equation}
controls the integrand. At very large $\Re (N)$, the exponential is very small when \eqref{eq:GammaPot} is positive and is very large when \eqref{eq:GammaPot} is negative. For values of $\phi$ that minimize $S_{\mathrm{eff}}^{\Gamma} (\phi)$, the integrand is much larger than for values of $\phi$ far away from the absolute minimum of \eqref{eq:GammaPot}.\par
Therefore, the integral representation of the Gamma function in the large $\Re (N)$ limit is dominated by a neighborhood of the saddle points of \eqref{eq:GammaPot}. The function $S_{\mathrm{eff}}^{\Gamma} $ is shown in Figure \ref{fig:SeffGamma} for various values of $N \in \R_{>0}$: it possesses a unique minimum at a positive value of $\phi$. The saddle point value of $\phi_{\ast}$ satisfies the equation 
\begin{equation*}
	\left. \frac{ \dd S_{\mathrm{eff}}^{\Gamma} (\phi) }{\dd \phi } \right\rvert_{\phi=\phi_{\ast}} = 0 ,
\end{equation*}
that is,
\begin{equation*}
	1- \frac{N-1}{\phi_{\ast}} =0 .
\end{equation*}
The saddle point is thus $\phi_{\ast} = N-1$. However, we are allowed to focus on a neighborhood of the saddle point $\phi_{\ast}$ only when $\Re (N) \gg 1$. We thus ought to approximate $S_{\mathrm{eff}}^{\Gamma} (\phi)$ by $\phi -N \log (\phi) $, whereby the saddle point is approximated by $\phi_{\ast} = N $ in the large $N$ limit.\par

\subsubsection*{Free energy}
Having observed that the integrand is sharply peaked around $\phi_{\ast}$ in the large $N$ limit, it is convenient to change variables 
\begin{equation*}
	\phi = \phi_{\ast} + \sigma \stackrel{\text{ large $N$ }}{=} N+ \sigma .
\end{equation*}
We then write 
\begin{align*}
	\Gamma (N) &=\int_{-N} ^{\infty} \dd \sigma ~ \exp \left\{ (N-1)\log (N+\sigma) - (N+\sigma)\right\}  \\
		&= \exp \left( N \log (N) -N - \log (N) \right) \int_{-N} ^{\infty} \dd \sigma~ \exp \left\{ (N-1)\log \left( 1+ \frac{\sigma}{N}\right) - \sigma  \right\} \\
		&\stackrel{\text{ large $N$ }}{=} \exp \left( N \log (N) -N - \log (N) \right) \int_{-\infty} ^{\infty} \dd \sigma~ e^{- \frac{1}{2N} \left( \sigma^2 + 2 \sigma \right) } .
\end{align*}
In the last line we have approximated the argument in the exponential at leading non-trivial order in the large $N$ limit, 
\begin{align*}
	- \sigma + (N-1)\log \left( 1+ \frac{\sigma}{N}\right)  & = - \sigma + N \left( \frac{\sigma}{N} - \frac{\sigma^2}{2N^2}  + \cdots \right) - \left( \frac{\sigma}{N} - \frac{\sigma^2}{2N^2} + \cdots \right) \\
	 & =  - \sigma + \left( \sigma- \frac{\sigma^2}{2N} + \cdots \right) + \left( - \frac{\sigma}{N} + \frac{\sigma^2}{2N^2}  - \cdots \right) \\
	 &= - \frac{\sigma^2}{2N}  - \frac{\sigma}{N} + \mathcal{O} (N^{-2}) .
\end{align*}
We have also approximated the lower bound $-N$ of the integral by $-\infty$. Taking the logarithm on both sides we write 
\begin{equation*}
	\log \Gamma (N) = N \log (N) -N - \log (N)  + \log \left[ \int_{-\infty} ^{\infty} \dd \sigma~ e^{- \frac{1}{2N} \left( \sigma^2 + 2 \sigma \right) } \right] .
\end{equation*}
Shifting again variables $\sigma^{\prime}=\sigma+1$ we obtain 
\begin{align*}
	\log \Gamma (N) & =  N \log (N) -N - \log (N) + \underbrace{ \frac{1}{2N}}_{\text{sub-leading}} + \log \left[ \int_{-\infty} ^{\infty} \dd \sigma^{\prime}~ e^{- \frac{(\sigma^{\prime})^2}{2N}} \right] \\
	&=  N \log (N) -N - \log (N) + \log \left[ \sqrt{2 \pi N} \right] \\
	&= N \log (N) -N - \frac{1}{2} \log \left( \frac{N}{2\pi} \right) .
\end{align*}
The $\frac{1}{2N}$ term in the first line, originating from the change of variables, is a sub-leading correction which must be discarded as it is of the same order as the other corrections that we neglect in the leading large $N$ limit. In the second line, we have discarded the sub-leading term and solved the Gaussian integral. Note that, had we kept the integration domain for $\sigma$ to be $(-N, \infty)$, the integration domain for $\sigma^{\prime}$ would be $(-N+1, \infty)$ and finiteness of the lower bound would give a correction of order $1/N$, to be discarded at this stage. In the last line, we have simplified the expression, obtaining the claimed result \eqref{eq:StirlingGamma}.

\subsection{An example from the \texorpdfstring{$O(N)$}{O(N)} model}
For our next warm-up example we study the zero-dimensional $O(N)$ model. The presentation in this subsection follows the review \cite{Klebanov:2018fzb}.\par
We consider a real scalar field with $N$ components, $\phi= (\phi_1, \dots, \phi_N)$, and an interaction Lagrangian of the form $(\tr \phi^2)^2$. Explicitly, we wish to study the integral 
\begin{equation*}
	\mz [O(N)] = \int_{\R^N} \dd \phi ~\exp \left\{ - \frac{g_2}{2} \sum_{a=1}^{N} \phi_a ^2 - \frac{g_4}{4} \left(  \sum_{a=1}^{N} \phi_a ^2 \right)^2 \right\}
\end{equation*}
in the limit $N \to \infty$. Here $\dd \phi$ is the Lebesgue measure on the Euclidean space $\R^N$. The coefficient $g_2$ can be reabsorbed in a change of variables, 
\begin{equation}
\label{eq:ZONphi4}
	\mz [O(N)] = g_2 ^{-N/2} \int_{\R^N} \dd \phi ~\exp \left\{ - \frac{1}{2} \sum_{a=1}^{N} \phi_a ^2 - \frac{g_4}{4 g_2^2} \left(  \sum_{a=1}^{N} \phi_a ^2 \right)^2 \right\} ,
\end{equation}
from where we see that the dependence on $g_2$ is extremely simple, while the parameter that controls the physical system is 
\begin{equation*}
	g_4^{\prime} =\frac{g_4}{g_2^2}. 
\end{equation*}

\subsubsection*{Integral transformation}
It is convenient to rewrite this integral using a so-called Hubbard--Stratonovich transformation: 
\begin{equation*}
	\mz [O(N)] = g_2 ^{-N/2} \int_{\R^N} \dd \phi \sqrt{\frac{1}{2\pi}\frac{2}{ g_4^{\prime}}} \int_{-\infty}^{+\infty} \dd \sigma ~\exp \left\{ - \frac{1}{2} \sum_{a=1}^{N} \phi_a ^2 - i \sigma \left( \sum_{a=1}^{N} \phi_a ^2 \right) - \frac{2}{g_4^{\prime}} \frac{\sigma^2}{2} \right\} .
\end{equation*}
This is a standard trick to solve integrals with double-trace (or multi-trace) interactions. Performing the Gaussian integral over the auxiliary scalar $\sigma$ gives back \eqref{eq:ZONphi4}. On the other hand, this trick has left us with $N$ Gaussian integrals over decoupled fields $\phi_a$:
\begin{equation*}
	\mz [O(N)] = \frac{1}{\sqrt{g_2 ^{N}} \sqrt{\pi  g_4^{\prime}}} \int_{-\infty}^{+\infty} \dd \sigma~e^{- \frac{\sigma^2}{ g_4^{\prime}}} ~\prod_{a=1}^{N} \left[ \int_{\R} \dd \phi_a   ~\exp \left\{ - \frac{1}{2} \left( 1+ 2i \sigma \right)\phi_a ^2 \right\} \right] .
\end{equation*}
Integrating out these $N$ free fields we get 
\begin{align}
	\mz [O(N)] & = \left(\sqrt{\frac{2\pi}{g_2}} \right)^{N}  \frac{1}{\sqrt{\pi g_4^{\prime}}} \int_{-\infty}^{+\infty} \dd \sigma ~   \frac{e^{- \frac{\sigma^2}{g_4^{\prime}}}}{\left( \sqrt{1+2 i \sigma} \right)^N}  \notag \\
	& = \left(\sqrt{\frac{2\pi}{g_2}} \right)^{N}  \frac{1}{\sqrt{\pi g_4^{\prime}}}  \int_{-\infty}^{+\infty} \dd \sigma ~ \exp \left\{ - \frac{\sigma^2}{g_4^{\prime}} - \frac{N}{2} \log (1+2 i \sigma) \right\} .  \label{eq:ZONsigmalog}
\end{align}
In this way, we have cast the problem of an integral in $N$ variables $(\phi_1, \dots, \phi_N)$ with a double-trace interaction into an ordinary integral of a single variable $\sigma$. Therefore, the study in this example, despite originating from an $O(N)$-symmetric theory, reduces in practice to the saddle point evaluation of a one-dimensional integral.

\subsubsection*{Large \texorpdfstring{$N$}{N} limit}
We observe that the second term in the exponential in \eqref{eq:ZONsigmalog} grows linearly with $N$. Therefore, this term will dominate the integrand unless we scale 
\begin{equation*}
	\frac{1}{g_4^{\prime}} = \frac{N}{\lambda} 
\end{equation*}
for a parameter $\lambda$ that stays finite in the large $N$ limit. With this scaling, we write 
\begin{equation*}
	\mz [O(N)] = \left(\sqrt{\frac{2\pi}{g_2}} \right)^{N}  \frac{1}{\sqrt{\pi g_4^{\prime}}}  \int_{-\infty}^{+\infty} \dd \sigma ~ \exp \left\{ - N \left[ \frac{\sigma^2}{\lambda} + \frac{1}{2} \log (1+2 i \sigma) \right] \right\} .
\end{equation*}
We now send $N \to \infty$. Expressing the integrand in the form $\exp \left( - S_{\mathrm{eff}}^{\mathrm{HS}} (\sigma) \right)$ with 
\begin{equation*}
	 S_{\mathrm{eff}}^{\mathrm{HS}} (\sigma) = N \left[ \frac{\sigma^2}{\lambda} + \frac{1}{2} \log (1+2 i \sigma) \right] ,
\end{equation*}
we see that it is suppressed by the damping of the exponential away from the saddle points of $ S_{\mathrm{eff}}^{\mathrm{HS}}$. In other words, the integrand is sharply peaked around the saddle points of $S_{\mathrm{eff}}^{\mathrm{HS}}$, which thus yield the leading large $N$ contribution to the integral. The saddle point equation is:
\begin{align*}
	0 & = \left. \frac{\partial S_{\mathrm{eff}}^{\mathrm{HS}} (\sigma) }{\partial \sigma} \right\rvert_{\sigma = \sigma_{\ast}} \\
	& = \frac{2}{\lambda} \sigma_{\ast} + \frac{i}{1 + 2 i \sigma_{\ast}} .
\end{align*}
Multiplying both sides by $-i$, we have to solve 
\begin{equation*}
	 \frac{2}{\lambda} \tilde{\sigma} = \frac{1}{1+2 \tilde{\sigma}}
\end{equation*}
where $\tilde{\sigma}=i \sigma_{\ast}$. The two solutions are 
\begin{equation}
\label{eq:sigmatildeHSsol}
	\tilde{\sigma} = \frac{1}{4} \left[ -1 \pm \sqrt{1 + 4 \lambda} \right] .
\end{equation}

\subsubsection*{Identify the dominant saddle}
To get the correct saddle point approximation, we ought to understand which of the two branches of the square root in \eqref{eq:sigmatildeHSsol} gives the solution that dominates the large $N$ limit. To this aim, let us step back to \eqref{eq:ZONsigmalog} and change variables $\sigma^{\prime}= \sqrt{\frac{g_4 ^{\prime}}{2}} \sigma$. We get
\begin{equation*}
	\mz [O(N)] = \left(\sqrt{\frac{2\pi}{g_2}} \right)^{N}  \int_{-\infty}^{+\infty} \frac{\dd \sigma^{\prime}}{\sqrt{2\pi}} ~ \exp \left\{ - \frac{ (\sigma^{\prime})^2}{2} - \frac{N}{2} \log \left( 1+ i \sqrt{ 2 g_4 ^{\prime}} \sigma^{\prime} \right) \right\} .
\end{equation*}
Sending $g_4^{\prime} \to 0$ we have 
\begin{equation*}
	\frac{N}{2} \log \left( 1+ i \sqrt{ 2 g_4 ^{\prime}} \sigma^{\prime} \right)  = i \sigma^{\prime} N \sqrt{\frac{g_4 ^{\prime}}{2}} + \cdots = i \sigma^{\prime} \sqrt{\frac{N \lambda}{2}} + \cdots .
\end{equation*}
Thus, if we send $\lambda \to 0$ fast enough so that $\sqrt{N \lambda} \to 0$, the logarithmic dependence on $\sigma^{\prime}$ trivializes, the integral over $\sigma^{\prime}$ is Gaussian, and we must recover 
\begin{equation*}
	\lim_{\lambda \to 0^{+}}  \mz [O(N)] = \left(\sqrt{\frac{2\pi}{g_2}} \right)^{N}  .
\end{equation*}
This is indeed the correct behaviour because, in this limit, the $O(N)$-symmetric integral we have started with reduces to $N$ Gaussian integrals over the initial fields $\phi_a$.\par
\begin{figure}[tb]
			\centering
			\includegraphics[width=0.5\textwidth]{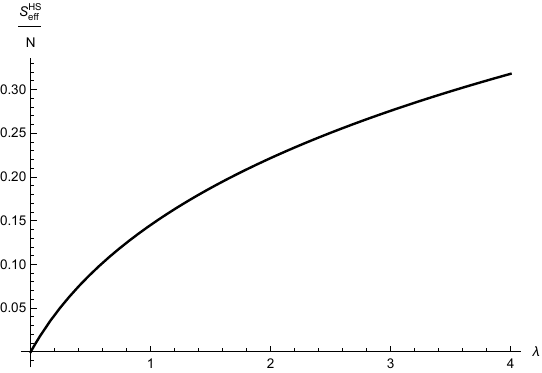}
			\caption{The function $\frac{1}{N}S_{\mathrm{eff}}^{\mathrm{HS}}(\sigma_\ast)$ evaluated on the saddle point.}
			\label{fig:SeffHSON}
\end{figure}

We conclude that the branch of the solution to be kept is the one for which $S_{\mathrm{eff}}^{\mathrm{HS}} (\sigma_\ast)$ behaves smoothly in the limit $\lambda \to 0^{+}$, and in particular 
\begin{equation*}
	\lim_{N \lambda \to 0^{+}} S_{\mathrm{eff}}^{\mathrm{HS}} (\sigma_\ast) = 0 .
\end{equation*}
This reasoning fixes the positive branch of the square root in $\tilde{\sigma}$. The plot of $ S_{\mathrm{eff}}^{\mathrm{HS}} (\sigma_\ast)$ taken with the positive branch of the square root is shown in Figure \ref{fig:SeffHSON} as a function of $\lambda$.\par
\par
\medskip
{\small\textbf{Exercise:} Evaluate $S_{\mathrm{eff}}^{\mathrm{HS}}$ on the two saddle points. Plot the two expressions as functions of $\lambda$.}\par
\medskip

\subsubsection*{Free energy}
Defining the free energy 
\begin{equation*}
	\mf [O(N)]= - \log \mz [O(N)] ,
\end{equation*}
in the large $N$ limit we get 
\begin{equation*}
	\mf [O(N)]= \frac{N}{2} \log \left( \frac{g_2}{2\pi}\right) + S_{\mathrm{eff}}^{\mathrm{HS}}  (\sigma_{\ast}) .
\end{equation*}
Evaluating the effective action on the saddle point we obtain 
\begin{align*}
	\mf [O(N)] &= N \left[ \frac{1}{2} \log \left( \frac{g_2}{2\pi}\right) + \frac{-2 \lambda +\sqrt{4 \lambda +1}+4 \lambda  \log \left(\frac{1}{2} \left(\sqrt{4 \lambda +1}+1\right)\right)-1}{8 \lambda } \right] \\
   	&= N \left[ \frac{1}{2} \log \left( \frac{g_2}{2\pi}\right) + \frac{\lambda}{4} -\frac{\lambda ^2}{4} +\frac{5 \lambda ^3}{12}  -\frac{7 \lambda ^4}{8} + \cdots \right]
\end{align*}
where in the second line we have shown the small $\lambda$ expansion.\par

\subsubsection*{(Aside) Caveat on the steepest descent method}
Before concluding, we should emphasize that the initial integration domain for $\sigma$ was $\R$, but we have found an imaginary saddle point $\sigma_{\ast}$.\par
For the large $N$ analysis to be valid, the integration contour should be carefully deformed in a neighborhood of $\sigma=0$ to pass through the saddle point on the imaginary axis, in such a way that the contour crosses the saddle point along the steepest descent path. This is an important part of the analysis which however will not be instrumental for the rest of these notes, thus we skip it.\par

\section[(Warm-up) Overview of supersymmetry and localization]{Overview of supersymmetry and localization}
\label{sec:ReviewLoc}

This section contains a very brief and schematic introduction to supersymmetric gauge theories and supersymmetric localization. It serves to help the reader with no previous experience to familiarize with a rough idea of the setup. For this reason, in this section we aim at qualitatively conveying the basic concepts rather than pursuing an accurate exposition.\par
\medskip
Every reader who is inclined to accept the fact that partition functions of supersymmetric gauge theories reduce from a path integral to an ordinary integral, is encouraged to move directly to Section \ref{sec:SphereZ}.

\subsection{Supermultiplets and Lagrangians in a nutshell}
\label{sec:supermultiplet}

Before diving into localization arguments and the discussion of partition functions, we describe the field content of supersymmetric quantum field theories with eight supercharges. This amount of supersymmetry is indicated as:
\begin{equation*}
	d=3,\ \mN=4  \qquad \vert \qquad d=4,\ \mN=2 \qquad \vert \qquad d=5,\ \mN=1 .
\end{equation*}
In this notation, $\mN$ counts the number of Killing spinors, and the number of supercharges equals $\mN$ times the number of minimal spinor components in dimension $d$.\par
We are interested in supersymmetric gauge theories $\Th$ with eight supercharges, which contain two types of building blocks:
\begin{itemize}
	\item The vector multiplet, that forms the supersymmetric extension of the gauge fields;
	\item The hypermultiplet, which describes the supersymmetric matter content.
\end{itemize}

\subsubsection*{Vector multiplet}
Consider a compact, connected gauge group $G$. For theories with eight or more supercharges, the vector multiplet is a supermultiplet with components $(\phi, A, \lambda, \mathrm{D})$, all taking values in the Lie algebra $\mathfrak{g}$. The field $A$ is the usual gauge connection, while $\lambda$ is a fermion called the gaugino. The scalar $\mathrm{D}$ is an auxiliary field, in the sense that the action will not contain a kinetic term for it. The field $\phi$, which consists of the bottom component of the vector multiplet in the superfield formalism, is a scalar which, by the amount of supersymmetry provided by the eight supercharges, is real in $d=5$, complex in $d=4$, and forms a triplet of an $SU(2)$ R-symmetry in $d=3$.\par
To place these theories on a sphere $\cs^d$ in a way that preserves supersymmetry, one considers a Lagrangian of the schematic form 
\begin{equation*}
	\frac{1}{2g_{\mathrm{YM}}^2} \tr \left[ F \wedge \ast F + \text{ supersymmetric completion } + \text{ curvature couplings } \right] ,
\end{equation*}
where $F= \dd A + A \wedge A$ is the curvature of $A$ and $g_{\mathrm{YM}}$ the gauge coupling.\par 
The first term is just the usual Yang--Mills action, which must be supplemented by terms in $\phi, \lambda, \mathrm{D}$ such that the total action is invariant under supersymmetry transformations generated by each of the eight supercharges. The supersymmetric completion contains, in particular, the gaugino kinetic term $-2\lambda \Dslash_A \lambda$, where the covariant derivative is twisted by both the spin and the gauge connections.\par
Moreover, being interested in theories on curved manifolds, we add curvature couplings of the form 
\begin{equation*}
	\frac{\phi \mathrm{D}}{r} \quad \text{ and } \quad \frac{(d-2)}{2} \frac{\phi^2}{r^2} ,
\end{equation*}
together their supersymmetric completion, where $r$ here is the radius of $\cs^d$. The complete supersymmetric Lagrangian can be consulted in the references listed in Appendix \ref{app:reading}.

\subsubsection*{Hypermultiplet}
Let $\mathfrak{R}$ be a (symplectic) representation of $G$, which in general is reducible. The hypermultiplet is a supermultiplet with components $(q, \tilde{q}, \psi, \mathrm{F})$, all taking values in $\mathfrak{R}$. Here $q$ is a complex scalar, and $\tilde{q}$ is the scalar in Euclidean signature that descends from the complex conjugate $\bar{q}$ in Lorentizian signature. Besides, $\psi$ is the fermion superpartner and $\mathrm{F}$ is an auxiliary field. The supersymmetric Lagrangian on $\cs^d$ is schematically 
\begin{equation*}
	\tr\left[ \dd_A \tilde{q} \wedge \ast \dd_A q + \text{ supersymmetric completion } + \text{ curvature couplings } \right] ,
\end{equation*}
where $\dd_A= \dd +A \wedge $ is the standard covariant derivative involving the gauge connection $A$. The first entry is therefore the usual kinetic term, with flavor indices in the representation $\mathfrak{R}$ contracted.\par 
The supersymmetric completion will include covariant derivatives for the fermion $\psi$ as well as a coupling $\tr \left( \tilde{q} \phi^2 q \right)$, with $\phi$ the scalar in the vector multiplet.  The curvature coupling is proportional to $\tr ( \tilde{q} q /r^2)$. These terms are quadratic in the hypermultiplet scalar and give it a mass. The term proportional to $\tr (\tilde{q} \phi^2 q)$ must be accompanied by the Yukawa coupling $\tr (\psi^{\dagger} \phi \psi )$ for the supersymmetric partner $\psi$ of $q$. For the full supersymmetric Lagrangian, consult the references listed in Appendix \ref{app:reading}.\par

\subsubsection*{(Aside) Superpotential}

Another important ingredient in constructing a supersymmetric field theory is the superpotential. However, thanks to the amount of supersymmetry, it is constrained; moreover, it does not show up explicitly in the partition functions we are going to consider (albeit it affects them through fixing the R-charges). The superpotential can thus be ignored for all practical purposes treated here.\par

\subsubsection*{Quivers}

\begin{figure}[tbh]
		\centering
		\begin{tikzpicture}[auto,square/.style={regular polygon,regular polygon sides=4}]
			\node[circle,draw] (gauge1) at (1,0) {$N_1$};
			\node[circle,draw] (gauge2) at (-1,0) {$N_2$};
			\node at (-2,0) {$\cdots$};
			\node at (2,0) {$\cdots$};
			
			\node[square,draw] (fl1) at (1,-1.5) { \hspace{6pt} };
			\node[draw=none] (aux1) at (1,-1.5) {$\mathsf{F}$};
			
			\draw[-] (fl1) -- (gauge1);
			\draw (gauge1)--(gauge2);
		\end{tikzpicture}
		\caption{Portion of a quiver representing a supersymmetric gauge theory. The two round nodes represent the gauge groups $U(N_1)$ and $U(N_2)$. The edge between the two round nodes represents a bifundamental hypermultiplet of $U(N_1)\times U(N_2)$, and the edge between the square node and the round node represents $\mathsf{F}$ additional hypermultiplets in the fundamental representation of $U(N_1)$.}
		\label{fig:quivex}
		\end{figure}
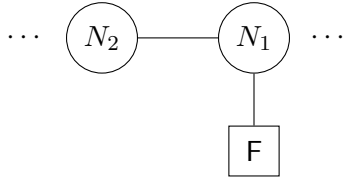\par
		
A convenient graphical way to encode the field content of a supersymmetric gauge theory is through a quiver. A quiver is a graph, which we decorate by specifying the gauge and flavour groups.\par
It is not necessary for the sake of these notes to know anything about quivers, and it suffices to think of them as graphs with round or square vertices, each of which carries a decoration by an integer. Round nodes represent simple factors of the gauge group, with the integer specifying the rank. Square nodes represent flavour symmetry groups. Finally, edges indicate hypermultiplets.\par 
For instance, the portion of quiver shown in Figure \ref{fig:quivex} describes a theory with gauge group containing $U(N_1) \times U(N_2)$, hypermultiplets in the bifundamental representation of $U(N_1) \times U(N_2)$, a flavour symmetry $SU(\mathsf{F})$, and hypermultiplets in the bifundamental of the gauge-flavour group $U(N_1) \times SU(\mathsf{F})$. Note that in the common parlance this latter set of hypermultiplets is usually referred to as having $\mathsf{F}$ hypermultiplets in the fundamental representation of $U(N_1)$. Hence, in the example of Figure \ref{fig:quivex}, the representation $\mathfrak{R}$ is reducible and contains at least the direct sum of the bifundamental of $U(N_1) \times U(N_2)$ with $\mathsf{F}$ copies of the fundamental of $U(N_1)$.\par

\subsubsection*{(Aside) More on quivers}
According to the Merriam-Webster dictionary, a quiver is a case for carrying arrows. However, throughout this lecture notes, there will be no arrows in our quivers. Instead, they will be used for carrying unoriented edges.\par 
This is due to the amount of supersymmetry and a common convention for drawing. Theories with eight supercharges are non-chiral, hence for every arrow in the quiver there should be another arrow with opposite orientation. It is then customary to replace the pair of oppositely oriented arrows, which graphically encode chiral multiplets of opposite gauge charge and equal R-charge, by a single unoriented edge, which graphically encodes a hypermultiplet.\par
More precisely, for gauge theories with eight supercharges we use doubled, framed quivers. Moreover, we ought to mention that what is often referred to as `a quiver' in the physicists's parlance, is in fact the representation of a quiver. The quiver is the graph itself, whereas a quiver representation is the assignment of integer labels to the vertices, and of homomorphisms to the arrows.\par 
These aspects will play no role throughout, and we will simply use quivers as a depiction of the field content of a supersymmetric gauge theory with eight supercharges.

\subsection{Localization in a nutshell}
This subsection contains a sketchy overview of the idea behind supersymmetric localization. It is meant to set the stage and to familiarize the reader with the type of partition functions considered in the next sections. For more details we refer to exhaustive topical reviews, some of which we list at the beginning of Appendix \ref{app:reading}. An accessible account of all the necessary features for the beginner is \cite{Cremonesi:2013twh}.\par

\subsubsection*{A simplified example}
To start with an analogy, consider the following scenario: let $\mathscr{M}$ be a manifold, and $\omega$ a top form on $\mathscr{M}$, i.e. a $\dim (\mathscr{M})$-dimensional differential form. For integration to be well defined, we may take $\mathscr{M}$ to be compact, $\partial \mathscr{M} = \emptyset$, or we restrict our attention to forms $\omega$ that decay exponentially fast at the boundaries $\partial \mathscr{M}$. By dimensional reasons, $\omega = \tilde{\omega} \vol_{\mathscr{M}}$, where $\tilde{\omega}$ is a scalar section and $\vol_{\mathscr{M}}$ is the volume form on $\mathscr{M}$. In practice, this simply means that we can choose local coordinates $(x_1, \dots, x_{\dim (\mathscr{M})})$ such that 
\begin{equation*}
	\omega = \tilde{\omega} (x_1, \dots, x_{\dim (\mathscr{M})}) \dd x_1 \wedge \cdots \wedge \dd x_{\dim (\mathscr{M})} .
\end{equation*}\par
Moreover, let $\sQ$ be a differential on $\mathscr{M}$, or more generally on some ambient space in which $\mathscr{M}$ is embedded. Thus, the input data includes $\sQ$ and $\omega$ such that, by hypothesis,
\begin{equation}
\label{eq:Qdiffcoho}
	\sQ^2 =0, \qquad \sQ \omega =0. 
\end{equation}
We are interested in evaluating 
\begin{equation}
\label{eq:integralomega}
	\int_{\mathscr{M}} \omega .
\end{equation}\par
$\mathscr{M}$ may be an untamed manifold,\footnote{In fact, in QFT $\mathscr{M}$ is typically \emph{not} a manifold. In tractable cases it will at best be a stack. The integration over $\mathscr{M}$ should then be replaced by an appropriate notion of (virtual) fundamental homology class. These technicalities play no role in this review and will not be discussed further.} and evaluating \eqref{eq:integralomega} explicitly may be complicated in practice. It would thus be convenient to leverage the additional differential structure $\sQ$ on $\mathscr{M}$ to simplify the problem.\par
We consider a well-behaved function $V: \mathscr{M} \to \R$ such that $\sQ V$ is positive (semi-)definite with finitely many minima, and study 
\begin{equation*}
	\mz (t) := \int_{\mathscr{M}} \omega ~e^{-t \sQ V} .
\end{equation*}
Clearly the integral \eqref{eq:integralomega} we wish to obtain is $\mz (0)$. We have 
\begin{align*}
	- \frac{ \dd \ }{\dd t} \mz (t) & = \int_{\mathscr{M}} \omega ~e^{-t \sQ V}  (\sQ V)\\
		&\underbrace{=}_{\text{by parts}}  \int_{\mathscr{M}} \left[ \sQ (\omega V e^{-t \sQ V} ) -  V (\sQ \omega e^{-t \sQ V} ) \right] \\
		&\underbrace{=}_{\text{chain rule}}  \int_{\mathscr{M}} \left[ \underbrace{\sQ (\omega V e^{-t \sQ V} )}_{\text{total derivative}} -  V \underbrace{(\sQ \omega)}_{0\text{ by \eqref{eq:Qdiffcoho}}}~ e^{-t \sQ V}  -  V  \omega~ \underbrace{(\sQ e^{-t \sQ V} )}_{0\text{ by \eqref{eq:Qdiffcoho}}} \right] \\
		&=0 .
\end{align*}
The first equality is well-posed thanks to our assumptions on $\mathscr{M}$ in this simplified example. The second equality follows from integration by parts. The three terms in the last expression vanish as a consequence of \eqref{eq:Qdiffcoho}. In more detail: (i) the first term vanishes because either $\partial \mathscr{M}= \emptyset$ or $e^{-t \sQ V} \omega$ decays (at least) exponentially at the boundaries of $\mathscr{M}$; (ii) the second term is zero because $\omega$ is $\sQ$-closed by assumption; (iii) the third term vanishes because $\sQ^2 V=0$.\par
The vanishing of the derivative means that $\mz (t)$ is a constant function of $t$, 
\begin{equation*}
	\mz (0) = \mz (t) \qquad \forall t >0.
\end{equation*}
Therefore, instead of computing \eqref{eq:integralomega} directly, we can compute $\mz (t)$ at any value of $t$ which we find convenient. In particular, we can evaluate $\mz (0)$ by deforming it by $e^{- t \sQ V}$ first and then sending $t \to + \infty$. In this way, the exponential factor damps the integrand everywhere except the minima of $\sQ V$. Up to rescaling by $1/\sqrt{t}$, the Gaussian integration around the minima of $\sQ$ produces a factor $\left[ \det \mathrm{Hess} ( \sQ V) \right]^{-1/2}$, where $\mathrm{Hess} $ is the Hessian matrix, which contains second derivatives with respect to the local coordinates on $\mathscr{M}$. Its determinant is positive by assumption on $V$.\par
The outcome is that \eqref{eq:integralomega} equals
\begin{equation*}
	\mz (0) = \sum_{p \in \min (\sQ V)} \left. \frac{\tilde{\omega}}{\sqrt{ \det \mathrm{Hess} ( \sQ V)} } \right\rvert_{p} ,
\end{equation*}
with sum over the minima $p\in \mathscr{M}$ of the deforming function $\sQ V$.\par
We emphasize that, despite the appearances, the final result is independent of the choice of $V$ among those well-behaved functions compatible with the necessary assumptions.

\subsubsection*{Equivariant localization}
A version of the above reasoning that extends to quantum field theories is equivariant localization. Here we roughly sketch the idea, and refer to the review \cite{Pestun:2016qko} (or the extensive \cite{Szabo:1996md}) for more details. The mathematical foundation of equivariant localization was initially formalized by J. Duistermaat and G. Heckman \cite{Duistermaat:1982vw}. Typically, the presentation suitable for QFT is modelled on the equivariant localization formula of \cite{Atiyah:1984px}.\par
Let us now consider a situation in which $\mathscr{M}$ has an isometry $\mathsf{T}$. The fixed locus of the $\mathsf{T}$-action on $\mathscr{M}$ is denoted $\mathscr{M}^{\mathsf{T}}$, and we have the natural inclusion $\iota : \mathscr{M}^{\mathsf{T}} \hookrightarrow \mathscr{M}$. We further assume $\iota \left( \mathscr{M}^{\mathsf{T}}\right) \cap \partial \mathscr{M} = \emptyset$, that is to say, the boundary contains no $\mathsf{T}$-fixed points.\par
Morally, one would like to leverage the invariance of integrals under the isometry to reduce them to the quotient space $\mathscr{M}/\mathsf{T}$. However, the latter is, generically, not a manifold, thus we cannot define its de Rham cohomology. Rather, the correct procedure is to work with the $\mathsf{T}$-equivariant cohomology of $\mathscr{M}$.\par 
For ease of exposition let us restrict to the simple case in which $\mathsf{T} \cong U(1)$. Let $\xi$ denote the vector field on $\mathscr{M}$ generating the isometry $\mathsf{T}$. The first step to work in $\mathsf{T}$-equivariant cohomology consists in replacing the above differential $\mathsf{Q}$, which satisfies $\mathsf{Q}^2=0$, by 
\begin{equation*}
	\widehat{\sQ} = \sQ + \xi \llcorner .
\end{equation*}
The first term is a genuine differential, and the second part is the contraction with the vector field $\xi$. Observe that $\widehat{\sQ}$ mixes forms of different degrees: the action of $\mathsf{Q}$, which is an analogue of the de Rham differential, increases the form degree by 1, whereas acting with $\xi \llcorner$ decreases the form degree by 1.\par
This new differential squares to $\mathscr{L}_{\xi}$, the Lie derivative along the integral curves of $\xi$: 
\begin{equation*}
\begin{aligned}
	\widehat{\sQ}^2 &= \underbrace{\sQ^2}_{0} + \sQ \circ \xi \llcorner + \xi \llcorner  \circ \sQ + \underbrace{\xi \llcorner \xi \llcorner }_{0} \\
		&= \left\{ \sQ, \xi \llcorner  \right\} = \mathscr{L}_{\xi} ,
\end{aligned}
\end{equation*}
where the last term in the first line vanishes when acting on differential forms, because it is symmetric and contracts anti-symmetric indices. The last line is the definition of Lie derivative.\par
In the equivariant setting, the hypotheses \eqref{eq:Qdiffcoho} are replaced by 
\begin{equation}
\label{eq:Qequiv}
	\widehat{\sQ}^2 = \mathscr{L}_{\xi} , \qquad \widehat{\sQ} \omega = 0.
\end{equation}
Notice that now the second requirement, namely that we have an equivariantly closed form $\omega$, is not automatically satisfied by choosing a top form, differently from the previous example.\par
\medskip
As before, we wish to compute 
\begin{equation*}
	\int_{\mathscr{M}} \omega ,
\end{equation*}
subject to \eqref{eq:Qequiv}. Running the analogous argument as above, one considers instead 
\begin{equation*}
	\mz (t) := \int_{\mathscr{M}} \omega ~e^{-t \widehat{\sQ} V}  
\end{equation*}
and shows that, for suitably chosen functions $V$ on $\mathscr{M}$, $\frac{\dd \mz }{\dd t }=0$. In the previous example, $\sQ^2 V =0$ was used to prove the independence of $\mz (t)$ on $t$, which held true for any $V$ due to \eqref{eq:Qdiffcoho}. Here, however, $\widehat{\sQ}^2 V =\mathscr{L}_{\xi} V$, thus the function $V$ must be chosen such that 
\begin{equation*}
	\mathscr{L}_{\xi} V=0 ,
\end{equation*}
namely, it is equivariantly constant with respecto to the action of $\mathsf{T}$ on $\mathscr{M}$. With this assumption, the computation above goes through. For example, writing in local coordinates $\xi = \sum_{\mu=1}^{\dim (\mathscr{M})}\xi^{\mu} \partial_{\mu}$, it is possible to choose $V$ such that $\partial_{\mu} V = \xi_{\mu}$. In global form, this means that we require $\dd V$ to be the 1-form dual to the vector field $\xi$.\par
Therefore, to obtain $\mz (0)$, one computes $\lim_{t \to \infty} \mz (t)$, which reduces to 
\begin{equation*}
	\int_{\min (\widehat{\sQ} V)} \frac{\omega }{\sqrt{ \det \mathrm{Hess} ( \widehat{\sQ} V)} } .
\end{equation*}
Here we have not assumed that the minima are isolated points. Choosing $\dd V$ to be dual to $\xi$, one has that 
\begin{equation*}
\begin{aligned}
	\exp \left(  -t \widehat{\sQ} V \right) &= \exp \left(  -t \dd V -t \xi^{\mu} \partial_{\mu} V \right) \\
		&= e^{- t \| \xi \|^2 } \left( 1- t \dd V + \cdots \right) .
\end{aligned}
\end{equation*}
Hence, the $t \to \infty$ limit is controlled by the exponential damping $e^{-t \| \xi \|^2}$, minimized at points $p \in \mathscr{M}$ where $\xi \vert_p =0$. In other words, the locus of minima of the equivariantly closed term $\widehat{\sQ} V$ can be chosen to be the locus in $\mathscr{M}$ where $\xi$ vanishes. Being $\xi$ the generator of $\mathsf{T}$, such locus is precisely the fixed locus $\mathscr{M}^{\mathsf{T}}$. We conclude that
\begin{equation*}
	\mz (0) = \int_{\mathscr{M}^{\mathsf{T}}} \frac{\iota^{\ast}\omega}{\mathrm{Pf} (\mathscr{L}_{\xi})} .
\end{equation*}
Here $\iota^{\ast}\omega$ is the pullback to $\mathscr{M}^{\mathsf{T}}$ of the form $\omega$, via the embedding map $\iota : \mathscr{M}^{\mathsf{T}} \hookrightarrow \mathscr{M}$; besides, we have identified the square root of the determinant of the Hessian matrix with the Pfaffian of the Lie derivative.\par 
The meaning of the equivariant localization formula can be summarized as: 
\begin{itemize}
	\item $\mathsf{T}$-equivariant integrals localize to the fixed locus of the $\mathsf{T}$-action;
	\item Splitting the integral between the normal and tangent bundles to the fixed locus, the normal directions are integrated out by a Gaussian integral and contribute a one-loop determinant.
\end{itemize}
See the review \cite{Pestun:2016qko} or the seminal work \cite{Atiyah:1984px} for the mathematical details behind this qualitative statement.

\subsubsection*{Interpretation}
Physically, we are interested in the analogue of this reasoning when $\mathscr{M}$ is a certain space of fields. In this scenario, $\mathscr{M}$ is typically infinite-dimensional and the integration operation is a path integral (swiping all the mathematical difficulties under the rug).\par
\medskip
The schematic interpretation of the procedure exemplified above is that, if the path integral possesses a powerful enough symmetry, it localizes around some simple configurations, and one only needs to worry about the one-loop determinants of the fields around their fixed configuration. More precisely:
\begin{itemize}
	\item The simple configuration of fields that dominates the path integral is called localization locus, or `BPS locus', and consists of the fixed points of the above-mentioned symmetry. Ideally, one wishes to reduce the path integral over $\mathscr{M}$ to an ordinary integral over a finite-dimensional BPS locus.
	\item Then, one has to account for the one-loop determinants coming from field modes along the normal directions to the BPS locus inside the huge field space $\mathscr{M}$. These fluctuations around the BPS locus are Gaussian integrals, and are the analogue of the determinant of the Hessian.\par
\end{itemize}\par
\medskip
The actual situation is more nuanced than that, because typically the path integral admits several saddle points. One is the semi-classical one, which gives the perturbative partition function, while the other field configurations are non-perturbative (such as instantons, vortices, and so on) and are typically hard to compute explicitly. As explained below, for the large $N$ limits to be considered in the rest of these notes, non-perturbative contributions can be safely neglected. Hence, we shall not discuss such technical challenges further.

\subsubsection*{Supersymmetric localization}

To study supersymmetric localization, we wish to incorporate fermions in the above discussion. We thus replace the infinite-dimensional manifold $\mathscr{M}$ with a supermanifold and the one-loop determinants become super-determinants. As usual, integrating out fermions produces $\det^{1/2}$ while integrating out bosons produces $\det^{-1/2}$.\par
\medskip
In a supersymmetric field theory $\Th$, $\sQ$ is a supercharge and 
\begin{equation*}
	\omega = e^{-S} O_{\text{BPS}} ~\mathscr{D} [\text{fields}], 
\end{equation*}
where $S$ is the classical action and $O_{\text{BPS}}$ is a $\sQ$-closed operator, not necessarily a local one. The measure factor $\mathscr{D} [\text{fields}]$ stands for the path integral volume form on the space $\mathscr{M}_{\text{susy}}$ of all quantum fields in $\Th$.\par
Taking $O_{\text{BPS}}=1$ for simplicity of exposition, the goal is to evaluate the partition function of the supersymmetric field theory:
\begin{equation*}
	\mz [\Th] = \int_{\mathscr{M}_{\text{susy}}} \mathscr{D} [\text{fields}] ~e^{-S} .
\end{equation*}
By definition, $\mz [\Th]$ is the path integral over the space $\mathscr{M}_{\text{susy}}$ of all fields in $\Th$.\par 
Locally, $\mathscr{M}_{\text{susy}}$ can be written as the product $\mathscr{M} [0] \times \mathscr{M} [1]$ of an even part $\mathscr{M} [0]$ and an odd part $\mathscr{M} [1]$. In physical terms, $\mathscr{M} [0]$ is the space of all field configurations of bosonic fields, and $\mathscr{M} [1]$ is the space of all field configurations of fermionic fields. Acting with the supercharge $\sQ$ on $\mathscr{M} [0]$ lands in $\mathscr{M} [1]$ and vice versa.\par 
The action $S$ is, by hypothesis, invariant under supersymmetry transformations generated by a suitable subset of the eight supercharges. For instance, one such suitable subset consists of a pair of (linear combinations of) supercharges that square to an isometry of the spacetime with non-empty fixed locus. The prototypical example to bear in mind is supersymmetric localization on $\cs^3$, with respect to supercharges that square to a rotation of the Hopf fibre.\par
We therefore choose a supercharge $\sQ$ and a suitable function $V$ of the fields, and consider 
\begin{equation*}
	\mz [\Th] (t) = \int_{\mathscr{M} [0] \times \mathscr{M} [1]} \mathscr{D} [\text{fields}] ~e^{-S - t \sQ V} .
\end{equation*}
Sending $t \to \infty$ and using the previous argument for the invariance under $\sQ$-exact deformations, we find:
\begin{equation*}
	\mz [\Th] = \int_{\text{BPS locus}} \vol_{\text{BPS}} \left. e^{-S} \right\vert_{\text{BPS locus}} ~ \sqrt{ \frac{\det_{\mathscr{M} [1]} (\text{1-loop}) }{\det_{\mathscr{M} [0]} (\text{1-loop})} } .
\end{equation*}
The ratio of one-loop determinants is obtained integrating out massive fluctuations of fermionic (in the numerator) and bosonic (in the denominator) fields around their BPS configuration. Here the localization locus has been denoted BPS locus to emphasize its supersymmetry invariance, and $\vol_{\text{BPS}}$ schematically denotes a volume form on the BPS locus.\par
\medskip
For the supersymmetric gauge theories introduced in Section \ref{sec:supermultiplet}, the ratio of determinants factorizes into the product of a vector multiplet part and a hypermultiplet part. Typically, the BPS locus consists of a constant configuration for the real component of the scalar $\phi$ in the vector multiplet, further restricted to a Cartan subalgebra of the Lie algebra $\mathfrak{g}$, with all other fields set to zero. Instanton configurations for the gauge field give rise to non-perturbative corrections to the partition function. On the other hand, hypermultiplets are essentially always set to vanish at the localization locus.\par
Thus, in all practical cases of interest to us, the measure $\mathrm{vol}_{\text{BPS}}$ is actually given by the normalized Haar measure on the gauge algebra $\mathfrak{g}$ and 
\begin{equation*}
	\mz [\Th] = \int_{\mathcal{C}_{\mathfrak{g}}} \dd \phi \prod_{\alpha \in \triangle} (\alpha, \phi) ~ \sqrt{(\text{1-loop})}
\end{equation*}
with $\mathcal{C}_{\mathfrak{g}}$ the principal Weyl chamber of the Cartan subalgebra of $\mathfrak{g}$ and $\dd \phi$ the Lebesgue measure on the Weyl chamber. The product is over all roots $\alpha \in \triangle$ of $\mathfrak{g}$. When $\mathfrak{g}=\mathfrak{u}(N)$, the product over roots gives the square of the Vandermonde determinant. The rest of the integrand is the leftover after cancellations between of fermionic and bosonic 1-loop determinants.\par
\medskip
An important feature of supersymmetric field theories is that their field content is not just a pair of independent collections of bosons and fermions. Rather, bosons and fermions come paired, and for every bosonic mode there is fermionic mode, except for the ground state. This gives rise to huge cancellations in the ratio of one-loop determinants between fermions and bosons, leaving eventually a tractable determinant.

\section{Sphere partition functions}
\label{sec:SphereZ}

	We discuss supersymmetric quantum field theories on round spheres $\cs^d$, and analyze their partition function starting from the matrix model formulation derived via localization. The large $N$ limit of superconformal field theories (SCFTs) on $\cs^d$ is predicted by holography to match with the supergravity computations in $(d+1)$-dimensional global AdS.\par
	
	\subsubsection*{Notation}
	Let us consider a supersymmetric QFT $\Th$, which we assume to be a quiver gauge theory with gauge group $G$ and hypermultiplets in the (in general, reducible) representation $\mathfrak{R}$ of $G$. Let $\mathrm{rk} (G)$ denote the rank of $G$, $\lvert \mathrm{W} (G) \rvert $ the order of the Weyl group of $G$, $\triangle $ the simple roots of $G$ and $\Lambda_{\mathfrak{R}}$ the weight system of the representation $\mathfrak{R}$ of $G$.

	\subsection{Partition functions}
		The sphere partition function of the theory $\Th$ takes the form 
		\begin{equation}
		\label{eq:genericZ}
			\mz_{\cs^d} [\Th] = \frac{\varpi}{ \lvert \mathrm{W} (G) \rvert } \int_{\mathbb{R}^{\mathrm{rk} (G)}} \dd \phi ~ e^{- S_{\mathrm{cl}} (\phi) } Z_{\mathrm{vec}} (\phi) Z_{\mathrm{hyp}} (\phi ) ~Z_{\text{non-pert.}}  .
		\end{equation}
		We now elucidate all the ingredients in this expression.
		\begin{itemize}
		\item The overall complex number $\varpi $ in \eqref{eq:genericZ} is a phase, of which we will not keep track. In odd dimensions, it has the physical meaning of a background Chern--Simons term \cite{Closset:2012vp}.
		\item The integral is over constant configurations for the lowest component of the vector multiplet, conjugated into a Cartan subalgebra of the gauge algebra, with all other fields set to zero. The scalar $\phi $ is adimensional and expressed in units of the radius of the sphere.\par
			The action of the Weyl subgroup of the gauge group shuffles the zero-mode eigenvalues, and hence configurations $\phi \in \R^{N}$ which only differ by reordering the components are gauge-equivalent. The overall factor $1/ \lvert \mathrm{W} (G) \rvert $ divides by the order of the Weyl group, canceling the overcounting of these equivalent configurations.\par
		\item In \eqref{eq:genericZ}, $S_{\mathrm{cl}} (\phi) $ is the classical action evaluated on this field configuration, 
		\begin{equation}
			S_{\mathrm{cl}} (\phi) = \tr \left(  V_{d} (\phi) \right) .
		\end{equation}
		The polynomial 
		\begin{equation}
		\label{eq:Vpotential}
			V_{d} (\phi) = \begin{cases} - i \pi k \phi^2  & d=3 , \\ g \phi^2 & d=4 , \\ \frac{\pi}{3} k \phi^3  + \pi g \phi^2  & d=5 , \end{cases}
		\end{equation}
		henceforth referred to as potential, encodes the Chern--Simons couplings $k$ and the inverse Yang--Mills couplings $g= \frac{8 \pi^2}{g_{\text{\tiny YM}}^2 }$. This parameter $g$ is an exactly marginal coupling in $d=4$ and is a mass parameter in $d=5$ (expressed here in units of the sphere radius). When the gauge group is $U(N)$, we may also include a Fayet--Iliopoulos (FI) term, which corresponds to a linear term $i 2 \pi \xi \phi$ in $V_d(\phi)$. For quiver theories with several gauge nodes, distinct $\left\{ k_j, g_{j} , \xi_j \right\}$ are allowed, for $j$ running over the gauge nodes.
		\item Furthermore, in \eqref{eq:genericZ},
		\begin{align*}
			Z_{\mathrm{vec}} (\phi ) & = \prod_{\alpha \in \triangle} e^{ - v_d \left( ( \alpha, \phi ) \right) } \\
			Z_{\mathrm{hyp}} (\phi ) & = \prod_{w \in \Lambda_{\mathfrak{R}} } e^{ - h_d \left(  w ( \phi ) \right) } 
		\end{align*}
		are the one-loop determinants of the vector multiplet\footnote{To lighten the formulae, we have combined the Vandermonde factor coming from diagonalizing the $\phi$ modes with $Z_{\text{vec}} (\phi)$.} and hypermultiplet respectively. The explicit forms of the functions $v_d (\phi)$ and $h_d (\phi)$ depend on $d$ and on the amount of supersymmetry. For eight supercharges on the round sphere, they are:
		\begin{subequations}
		\begin{align}
			v_{d} (\phi) & =  \begin{cases} - \log 2 \sinh (\pi \phi)  & d=3 , \\ - \log (\pi \phi )	- \log H(i \phi) & d=4 ,   \\ 	- \frac{1}{2} f (i \phi) - \log 2 \sinh (\pi \phi) 	 & d=5 ,  \end{cases}   \\
			h_{d} (\phi) & =  \begin{cases}   \log 2 \cosh (\pi \phi)  & d=3 , \\ 				 	  \log H(i \phi)	& d=4 ,   \\ 	  \frac{1}{4} f \left( \frac{1}{2} +  i \phi \right) +  \frac{1}{4}  f \left( \frac{1}{2} +  i \phi \right) - \log 2 \cosh (\pi \phi) 	& d=5 ,  \end{cases}   
		\end{align}
		\label{eq:defvdhd}
		\end{subequations}
		where the functions $H (z)$, in $d=4$, and $f(z)$ in $d=5$, have been defined respectively in \cite{Pestun:2007rz} and \cite{Kallen:2012cs}, and read 
		\begin{align*}
			H (z) & = G(1+z) G(1-z) \\
				& = \prod_{n=1} ^{\infty} \left( 1 - \frac{z^2}{n^2}   \right)^n e^{\frac{z^2}{n}} , \\
			f (z) & = - \frac{i \pi}{3} z^3 +z^2 \log \left( 1-e^{ i 2 \pi z } \right) - \frac{i z}{\pi} \mathrm{Li}_2 \left( e^{ i  2 \pi z } \right) + \frac{1}{2 \pi^2} \mathrm{Li}_3 \left( e^{i  2 \pi z } \right) - \frac{\zeta (3)}{2 \pi^2} \\
				& \overbrace{=}^{\zeta \text{-reg.}} \sum_{n=1} ^{\infty} n^2 \log \left( 1 - \frac{z^2}{n^2} \right)  .
		\end{align*}
		In the last equality, the  symbol $\overbrace{=}^{\zeta \text{-reg.}}$ means that the left-hand side comes from the $\zeta$-function regularization of the divergent right-hand side. The latter expression has the advantage of showing that $f(z)$ is manifestly even.\par
		\item Finally,
		\begin{equation*}
			Z_{\text{non-pert.}} = \begin{cases} 1 &  d=3 , \\ 1 + \cdots & d \ge 4 , \end{cases}
		\end{equation*}
		includes instantonic contributions, which we will neglect in what follows, since they are suppressed in the large $N$ limit.\par
		\end{itemize}

		\subsubsection*{(Aside) Mass deformations}
			Turning on a mass deformation for a hypermultiplet in the representation $\mathfrak{R}$ of the gauge group and in the representation $\widetilde{\mathfrak{R}}$ of the flavour group, corresponds to shift $w (\phi) \mapsto w (\phi) + \widetilde{w} (m)$, with $\widetilde{w} \in \Lambda_{\widetilde{\mathfrak{R}}}$, and $m$ indicating the lowest component of the background vector multiplet for the flavour symmetry, in unit of the sphere radius. $m$ is usually simply referred to as the mass.\par

		\subsubsection*{Linear quivers with $U(N)$ gauge nodes}
			Let us consider for concreteness a quiver gauge theory $\Th$ described by a linear quiver with $L$ nodes, with gauge group 
			\begin{equation*}
				U(N_1) \times U(N_2) \times \cdots \times U(N_L) .
			\end{equation*}\par
			The reducible representation $\mathfrak{R}$, encoding the hypermultiplets, decomposes into the bifundamental representation of $U(N_{j}) \times U(N_{j+1}) $, $\forall j=1, \dots, L-1$, as well as $\mathsf{F}_j$ copies of the fundamental representation and $\mathsf{A}_j$ copies of the adjoint representation of $U(N_j)$, $\forall j=1, \dots , L$. The partition function \eqref{eq:genericZ} specialized to this case can be written as 
			\begin{equation}
			\label{eq:ZlinearQSeff}
				\mz_{\cs^d} = \frac{\varpi}{ \lvert \mathrm{W} (G) \rvert } \int_{\mathbb{R}^{\mathrm{rk} (G)}} \dd \phi ~ e^{- S_{\mathrm{eff}} (\phi) }
			\end{equation}
			with effective action 
			\begin{equation}
			\begin{aligned}
				S_{\mathrm{eff}} (\phi)  = \sum_{j=1} ^{L} \sum_{a=1} ^{N_j} & \left\{ \underbrace{V_d \left( \phi_{j,a} \right) }_{\text{classical}} + \sum_{\alpha =1} ^{\mathsf{F}_j} \underbrace{ h_{d} \left( \phi_{j,a} + m^{\mathsf{F}}_{j,\alpha}  \right) }_{\text{fund. hyper.}}  + \sum_{\beta =1} ^{\mathsf{A}_j} \sum_{b=1}^{N_j} \underbrace{ h_{d} \left( \phi_{j,a} - \phi_{j,b} + m_{j,\beta} ^{\mathsf{A}}\right) }_{\text{adjoint hyper.}}  \right.  \\
					&  \left. \ + \sum_{b=1} ^{N_{j+1}} \underbrace{ h_{d} \left( \phi_{j,a} - \phi_{j+1,b} \right) }_{\text{bifund. hyper.}}  + \sum_{b=1} ^{N_{j}}  \underbrace{ v_{d} \left( \phi_{j,a} - \phi_{j,b} \right) }_{\text{vec.}}  \right\} ,
			\end{aligned}
			\label{eq:SeffFiniteN}
			\end{equation}
			where we have turned on generic masses for the fundamental and adjoint hypermultiplets.
			
		\subsubsection*{From $U(N)$ to $SU(N)$}
			If we are interested in theories in which a $U(N_j)$ gauge group is replaced by $SU(N_j)$, we can turn on the FI term 
			\begin{equation*}
				V_d \left( \phi_{j,a} \right) \mapsto V_d \left( \phi_{j,a} \right) + i 2 \pi \xi_j \phi_{j,a} ,
			\end{equation*}
			and then promote the parameter $\xi_j$ to the lowest component of a dynamical multiplet and integrate over it. The identity 
			\begin{equation*}
				\int_{- \infty} ^{+ \infty } \dd \xi_j ~ e^{i 2 \pi \xi_j \sum_{a=1} ^{N_j} \phi_{j,a} } [\text{integrand}] (\phi) = [\text{integrand}](\phi) \delta \left( \sum_{a=1} ^{N_j} \phi_{j,a} \right) .
			\end{equation*}
			shows that the integration effectively enforces the traceless condition. In $d=3$, this procedure corresponds to gauging the $U(1)$ topological symmetry associated to the $U(N)$, thus flowing to the $SU(N)$ theory in which such symmetry is absent.\par
			
		\subsubsection*{(Aside) Decoupling instantons}
			We have claimed above that, in the large $N$ limit, instantons are suppressed and drop out of the computation, thus we can safely set $Z_{\text{non-pert.}} $ to $1$. This assumption is based on the physical intuition that instantons acquire a large mass in the large $N$ limit and decouple.\par 
			This statement is subtle in $d>4$, where the SCFT is necessarily realized at infinite coupling. Therefore, instantons in $d > 4$ can become massless at the conformal point. However, by construction, we will always first flow to a gauge theory description, so that we can apply localization. We then take the large $N$ limit at fixed gauge coupling, in which instantons decouple, and only at the end we take the strong coupling limit in terms of the 't Hooft coupling.\par
			It is not a priori completely obvious which is the most appropriate order of limits, namely 
			\begin{enumerate}[(a)]
				\item\label{instantonitema} First large $N$, then strong coupling (as we do);
				\item First strong coupling, then large $N$.
			\end{enumerate}
			Only prescription (\hyperref[instantonitema]{a}) is tractable using localization, because instantons are untamed in a generic supersymmetric SCFT in $d \ge 4$, and it is the approach we adopt here. It turns out, based on evidence, that this is the correct order of limits in AdS/CFT, and the results obtained with prescription (\hyperref[instantonitema]{a}) match with the supergravity calculations.

		\subsection{Convergence of the localized partition function}
		\label{sec:exampleZconv}
			For the partition function \eqref{eq:genericZ} to be a well-defined quantity, the theory $\Th$ must satisfy additional conditions. These convergence conditions for the integral \eqref{eq:genericZ}, which could be formulated directly in the field theory language, are derived by looking at the asymptotic behaviour of the functions $v_d (\phi)$ and $h_d (\phi)$ from \eqref{eq:defvdhd}.\par
			The functions \eqref{eq:defvdhd} have the large argument asymptotic:
			\begin{subequations}
			\begin{align}
				v_d (\phi) & \xrightarrow{ \quad \lvert \phi \rvert \to \infty \quad } \begin{cases} - \pi \lvert \phi \rvert & d=3 , \\   \frac{1}{2} \phi^2 \log \phi^2 - \frac{3}{2} \phi^2 & d=4 , \\  \frac{\pi}{6} \lvert \phi \rvert^3 - \pi \lvert \phi \rvert & d= 5  , \end{cases} \\
				h_d (\phi) & \xrightarrow{ \quad \lvert \phi \rvert \to \infty \quad } \begin{cases}   \pi \lvert \phi \rvert & d=3 , \\ - \frac{1}{2} \phi^2 \log \phi^2 + \frac{3}{2} \phi^2 & d=4 , \\ -\frac{\pi}{6} \lvert \phi \rvert^3 - \frac{\pi}{8} \lvert \phi \rvert & d= 5 . \end{cases} 
			\end{align}
			\label{eq:asymptvdhdlargex}
			\end{subequations}
			\medskip
			{\small\textbf{Exercise:} Starting from \eqref{eq:defvdhd}, prove \eqref{eq:asymptvdhdlargex} for both $\phi \to \pm \infty$.}

			\subsubsection*{Example: Good theories in $d=3$}
			Let, for instance, $\Th$ be a $U(N)$ gauge theory in $d=3$ without Chern--Simons term, i.e. set $k =0$ in $V(x)$, and with $\mathsf{F}$ hypermultiplets in the fundamental representation. This theory is called three-dimensional SQCD, or SQCD$_3$ for short. The partition function is 
			\begin{equation*}
				\mz_{\cs^3} [\text{SQCD}_3] = \frac{1}{N!} \int_{\R^N} \dd \phi \frac{ \prod_{1 \le a \ne b \le N} 2 \sinh \pi (\phi_a - \phi_b)}{ \prod_{a=1} ^{N} \prod_{\alpha=1} ^{\mathsf{F}} 2 \cosh \pi (\phi_a + m_{\alpha}) }  .
			\end{equation*}
			From the large $\lvert \phi \rvert$ behaviour of the integrand, we deduce that the integral $\mz_{\cs^3} [\text{SQCD}_3] $ converges only if 
			\begin{equation*}
				- 2 N (N-1) + N \mathsf{F} >0 \quad \Longrightarrow \quad \mathsf{F} \ge 2 N -1 .
			\end{equation*}\par
			More generally we may consider $\Th$ in $d=3$ by connecting a linear chain of $L$ such theories with hypermultiplets in the bifundamental representation of $U(N_j) \times U(N_{j+1})$, $\forall j=1, \dots, L-1$. Defining the quantity 
			\begin{equation*}
				\Delta_j = -2N_j + N_{j-1} + N_{j+1} , \quad \forall j=1, \dots , L
			\end{equation*}
			with $N_0 =0 = N_{L+1}$ understood, we derive the convergence condition 
			\begin{equation*}
				\Delta_j + \mathsf{F}_j \ge -1 , \quad \forall j=1, \dots , L  . 
			\end{equation*}
			Theories in $d=3$ that satisfy this condition with strict inequality are dubbed `good' theories in the Gaiotto--Witten classification \cite{Gaiotto:2008ak}, whilst those that saturate the inequality are called `ugly' theories \cite{Gaiotto:2008ak}.\footnote{Ugly theories flow to a good theory with a decoupled sector. With clever manipulations, one should always be able to rewrite $\mz_{\cs^3}$ of an ugly theory as the product of an overall factor, due to decoupled fields, times the partition function of a good theory.} Besides, good theories that satisfy the equality $\Delta_j + \mathsf{F}_j =0$ $\forall j=1, \dots, L$ are called balanced.\par

			\subsubsection*{Example: $\mathcal{N}=4$ super-Yang--Mills in $d=4$}
			Four dimensions is peculiar, compared to $d \ne 4$, because the functions $v_{d=4} (x)$ and $h_{d=4} (x)$ are almost the opposite of one another: their difference leaves behind $-\log\lvert \phi \rvert$ coming from the Vandermonde factor. This implies that if we consider a gauge theory with simple and simply connected gauge group $G$ coupled to one adjoint hypermultiplet, the vector multiplet and hypermultiplet contributions cancel exactly if the mass is turned off.\par 
			This description gives the partition function of the $G$-gauge $\mN=4$ (i.e. maximally supersymmetric) super-Yang--Mills theory on $\mathbb{S}^4$ \cite{Pestun:2007rz}:
			\begin{equation*}
				\mz_{\cs^4} [\mN=4 \text{ SYM}] = \frac{1}{\lvert \mathrm{W} (G) \rvert} \int_{\R^{\mathrm{rk} (G)}} \dd \phi ~ \prod_{\alpha \in \triangle } (\alpha, \phi) ~ e^{- g \tr \phi^2 } ,
			\end{equation*}
			which is the celebrated Gaussian ensemble over the Lie algebra $\text{Lie} (G) =\mathfrak{g} $. It converges for every $g >0$.

			\subsubsection*{Example: UV complete theories $d=5$}
			As a further example, consider a theory with $U(N)$ gauge group in $d=5$ coupled to $\mathsf{F}$ hypermultiplets in the fundamental representation. This theory is called SQCD$_5$.\par
			Reading the corresponding line in \eqref{eq:asymptvdhdlargex} and accounting for the cubic piece in $V(x)$, we see that $\mz_{\cs^5} [\text{SQCD}_5]$ converges if 
			\begin{equation*}
				\frac{\pi}{6} 2 N(N-1)  - \frac{\pi}{6} N \mathsf{F} - \frac{\pi}{3} N \lvert k \rvert >0 \quad \Longrightarrow \quad  \frac{\mathsf{F}}{2} + \lvert k \rvert \le N .
			\end{equation*}

			\subsubsection*{Convergence conditions}
			Combining the asymptotic expressions \eqref{eq:asymptvdhdlargex} with the classical potential \eqref{eq:Vpotential} we derive the following
			\begin{equation*}
				\text{convergence cond. :} \quad \begin{cases}  k_j \ne 0 \text{ or } \mathsf{A}_j > 0  \text{ or } \Delta_j + \mathsf{F}_j \ge 0  & d= 3 , \\ \Delta_j + \mathsf{F}_j + N_j \mathsf{A}_j + \lvert 2 k_j \rvert \le 0  & d=5 . \end{cases}
			\end{equation*}
			These conditions must be satisfied for every simple factor of the gauge group, labeled by the index $j$. It is of course possible to include other representations besides fundamental and adjoint, such as symmetric or rank-two antisymmetric, and derive similar constraints.\par
			We stress that, in $d=5$, the convergence is an upper bound on the number of matter fields, which constrains the number of well-posed models. In contrast, the condition in $d=3$ is a lower bound, thus it is always allowed to add flavours to a convergent model.\par
			\medskip
			{\small\textbf{Exercise:} Derive the convergence conditions.}

	\subsection{Large \texorpdfstring{$N$}{N} limits}
	\label{sec:LargeNtypes}
		
		In defining the large $N$ limit, we fix $N \in \mathbb{N}$ and write the ranks of the gauge groups as 
		\begin{equation*}
			N_j = N \nu_j
		\end{equation*}
		for some rational numbers $\left\{  \nu_j \right\}_{j=1, \dots, L}$. We then take $N \to \infty $ keeping $\nu_j = \mathcal{O}(1)$, i.e. independent of $N$.\par
		In this limit, the integer indices $a=1, \dots , N_j$ are replaced by the continuous variables $\mathsf{a} = \frac{a}{N}$, $\mathsf{a} \in [0, \nu_j]$ and the integration variables $\phi_{j,a}$ become functions of such continuous indices $\phi_j (\mathsf{a})$. We then make the replacement 
		\begin{equation*}
			\frac{1}{N} \sum_{a=1} ^{N_j}  F \left( \phi_{j,a} \right) \quad \mapsto \quad \int_{0} ^{\nu_j} \dd \mathsf{a}  F \left( \phi_{j} (\mathsf{a}) \right)
		\end{equation*}
		in \eqref{eq:SeffFiniteN}. Besides, we introduce the densities of eigenvalues 
		\begin{equation}
		\label{eq:defrhodelta}
			\rho_j ( \phi ) = \frac{1}{N} \sum_{a=1}^{N_j} \delta \left( \phi - \phi_{j,a} \right) ,
		\end{equation}
		which are normalized by definition, 
		\begin{equation}
		\label{eq:normrhoj}
			\int \dd \phi \rho_j ( \phi ) = \nu_j  .
		\end{equation}
		Notice that $\phi \in \R$ is an auxiliary, one-dimensional integration variable, related to but distinct from the original scalar field with zero-mode components $ \phi_{j,a} $. We also stress that the eigenvalue density is a well-posed quantity at any $N$. At large $N$, we can use \eqref{eq:defrhodelta} to further replace 
		\begin{equation*}
			 \int_{0} ^{\nu_j} \dd \mathsf{a}   F \left( \phi_{j} (\mathsf{a}) \right)  \quad \mapsto \quad  \int \dd \phi \rho_j (\phi)  F \left( \phi   \right)
		\end{equation*}
		which is tantamount to the relation $ \rho (\phi) \dd \phi = \dd \mathsf{a} $ among measures. It also follows from \eqref{eq:defrhodelta} that $\rho_j$ is compactly supported at large $N$ (assuming the equilibrium configuration exists).\par
		\medskip
		The problem of solving a large $N$ limit corresponds to finding the collection $\left\{ \rho_j (\phi) \right\} $, from which all physical observables that have a matrix model description can be computed. Among all such quantities, we will focus on the sphere free energy 
		\begin{equation*}
			\mathcal{F}_{\cs^d} = - \log \left\lvert \mz_{\cs^d} \right\rvert .
		\end{equation*}
		We warn the reader that the overall sign convention in the definition of the free energy may vary. In particular, our sign convention differs from the standard F-theorem convention when $d=5$.\par
		\medskip
		A given QFT $\Th$ may admit several large $N$ limits (see e.g. \cite{Klebanov:2018fzb}). In the following we introduce three of them:
		\begin{enumerate}[(i)]
			\item 't Hooft large $N$ limit;
			\item M-theory limit;
			\item Long quiver limit.
		\end{enumerate}
		They will be explicitly analyzed in Sections \ref{sec:LargeNtHooft}, \ref{sec:LargeNMtheory} and \ref{sec:LargeNLongQ} respectively. We list the main examples of each limit in Table \ref{tab:limitsex}.\par
		
\begin{table}[th]
\centering
\begin{tabular}{|c |c |c |c|}
	\hline 
	Regime & \ &  Milestone example & \ \\
	\hline
	't Hooft limit & Section \ref{sec:LargeNtHooft} & Four-dimensional planar $\mN=4$ SYM & Section \ref{sec:PlanarN4SYM}\\
	M-theory limit & Section \ref{sec:LargeNMtheory} & Five-dimensional higher rank $E_n$ & Section \ref{sec:MJafferisPufu} \\
	 \ & \ & Five-dimensional MSYM & Section \ref{sec:MthLimit5d} \\
	Long quiver limit & Section \ref{sec:LargeNLongQ} & Three-dimensional long quivers & Section \ref{sec:3dLQ} \\
\hline
\end{tabular}
\caption{Different large $N$ regimes. The reader is encouraged to work out the milestone examples of each limit.}
\label{tab:limitsex}
\end{table}\par

		\subsubsection*{'t Hooft limit}
			The most well-known large $N$ limit is the 't Hooft limit \cite{tHooft:1973alw}. It consists in taking $N\to \infty $ while scaling the gauge couplings linearly with $1/N$. This limit goes under the name of planar limit, because the perturbation theory of the QFT in this limit is dominated by planar Feynman diagrams, that is, those which can be drawn on a plane.\par
			More precisely, in considering the 't Hooft planar limit, we adopt a slightly more general definition, in which all the parameters of the theory are scaled with $N$ is such a way that all terms in \eqref{eq:SeffFiniteN} contribute to leading order \cite{Brezin:1977sv}.\par
			\medskip
			We can assume that the potential is analytic and admits a Maclaurin expansion
			\begin{equation}
			\label{eq:VtHooftscaleW}
				V (x) = N  W(x) \quad \text{ where } \quad W(x) = \sum_{p \ge 1} \frac{c_p}{p} x^p .
			\end{equation}
			Then, all the coefficients $\left\{ c_p \right\}_{p \ge 1}$ are kept fixed as $N \to \infty$ \cite{Brezin:1977sv}. We will denote these parameters as 't Hooft couplings. To make contact with the standard notation, when considering the physical classical potential \eqref{eq:Vpotential} we will write 
			\begin{equation*}
				\frac{k}{N} = \frac{1}{t} , \quad \frac{g}{N} = \frac{1}{\lambda} ,
			\end{equation*}
			which implies that the potential \eqref{eq:Vpotential} in the 't Hooft limit has 
			\begin{equation*}
				c_{p} = \begin{cases}  \frac{i  }{2 t} \delta_{p,2} , & d=3 , \\ \frac{1}{2 \lambda} \delta_{p,2} & d=4 , \\  \frac{\pi}{2 \lambda} \delta_{p,2} + \frac{\pi}{3t} \delta_{p,3}  & d=5 . \end{cases}
			\end{equation*}
			Besides, we will keep the number of adjoint hypermultiplets fixed, while $\mathsf{F}_j \to \infty $ keeping 
			\begin{equation*}
				\frac{\mathsf{F}_j}{N_j} = \zeta_j 
			\end{equation*}
			fixed. This choice of scaling, in which the number of flavours is taken proportional to $N$, is called Veneziano limit. We emphasize, though, that with our definition of 't Hooft planar limit \cite{Brezin:1977sv} and from the point of view of the matrix model, the Veneziano limit is naturally included in the 't Hooft large $N$ limit.\par

		\subsubsection*{M-theory limit}
			The M-theory limit is relevant for $d=3$ $\mN \ge 2$ and $d=5$ $\mN =1 $ Chern--Simons-matter theory that admit a holographic dual in M-theory. A famous example is ABJM theory \cite{ABJM}.\par
			In this limit, the ranks of the gauge groups are sent to infinity, $N \to \infty$, with the parameters kept fixed and independent of $N$. The eigenvalues $\phi_{j,a}$ develop a non-trivial dependence on $N$, 
			\begin{equation*}
				\phi_{j,a} = N^{\gamma} x_{j,a} , \quad x_{j,a} = \mathcal{O} (1) 
			\end{equation*}
			for some power $\gamma >0$. Here we are implicitly assuming that the theory has enough symmetries to guarantee that $\gamma$ is independent of $j$, while it is necessarily independent of the label $a$ due to the permutation symmetry of the integrand.\par
			In this limit, one typically encounters simplifications of the equilibrium equation using the asymptotic expansion \eqref{eq:asymptvdhdlargex}. Then, the scaling power $\gamma$ is determined by requiring that at least two terms in the effective action compete to determine a non-trivial equilibrium configuration.

		\subsubsection*{Long quiver limit}
			A limitation of the 't Hooft limit is its increasing complexity with increasing length of the quiver. Indeed, one has to solve a system of coupled equilibrium equations to find the collection $\left\{ \rho_j \right\}_{j=1, \dots , L}$ of eigenvalue densities at each gauge node.\par
			However, in the limit $L \to \infty$, the problem simplifies, very much in the same way as the problem of solving $N$ coupled saddle point equations simplifies at large $N$ into the problem of determining a density of eigenvalues. We have two options: either impose additional scaling of the 't Hooft couplings with $L$, or allow for a non-trivial dependence of the eigenvalues on $L$, akin to the M-theory limit:
			\begin{equation*}
				\phi = L^{\alpha} x ,\quad x = \mathcal{O} (1)  .
			\end{equation*}
			Let us emphasize that this limit is taken only after a large $N$ limit. For this reason, the scaling law is imposed on the integration variable $\phi$, and not on the scalar field zero-modes $\phi_{j,a}$.\par
			Again in analogy with the M-theory limit, the power $\alpha >0$ is determined requiring that at least two terms in the effective action compete at leading order in $L$. Again, using the scaling with the large number $L$ and \eqref{eq:asymptvdhdlargex}, the resulting expressions are hugely simplified.\par
			When $L \to \infty$, we can replace the index $j$ running over the gauge nodes with a continuous variable 
			\begin{equation*}
				z = \frac{j}{L+1} , \quad 0 < z < 1 .
			\end{equation*}
			In circular quivers, we add one node $j=0$, whence $z \in [0,1)$, and impose the periodic identification $z+1 \sim z$. The rational numbers $ \left\{ \nu_j \right\}_{j=1, \dots, L}$ are then collected into a rank function $\nu (z) \ge 0$ and the eigenvalue densities  $ \left\{ \rho_j (\phi) \right\}_{j=1, \dots, L}$ become a unique function of two variables $\varrho (z, x)$, where 
			\begin{equation*}
				\varrho (z,x) \dd x = \rho_{zL} (\phi) \vert_{\phi =L^{\alpha} x} \dd \phi  .
			\end{equation*}
			It satisfies the normalization condition
			\begin{equation*}
				\int \dd x \varrho (z,x)  = \nu (z) , \quad \forall 0 \le z \le 1 .
			\end{equation*}

			\subsubsection*{(Aside) Double-scaling limit}
			As reviewed above, the natural definition of the planar limit \`a la 't Hooft requires scaling all the quantities so that the effective action retains as many terms as possible at large $N$ \cite{Brezin:1977sv}. This includes the Veneziano limit among others.\par
			In the physics literature, scaling multiple parameters in the 't Hooft limit is sometimes referred to as double-scaling limit. This nomenclature can be misleading, because the double-scaling limit as originally defined in \cite{Gross:1989vs,Brezin:1990rb,Douglas:1989ve} is a different procedure. The latter is also the way the double-scaling limit is defined in the mathematical literature on random matrix theory.\par
\begin{figure}[htb]
			\centering
			\includegraphics[width=0.6\textwidth]{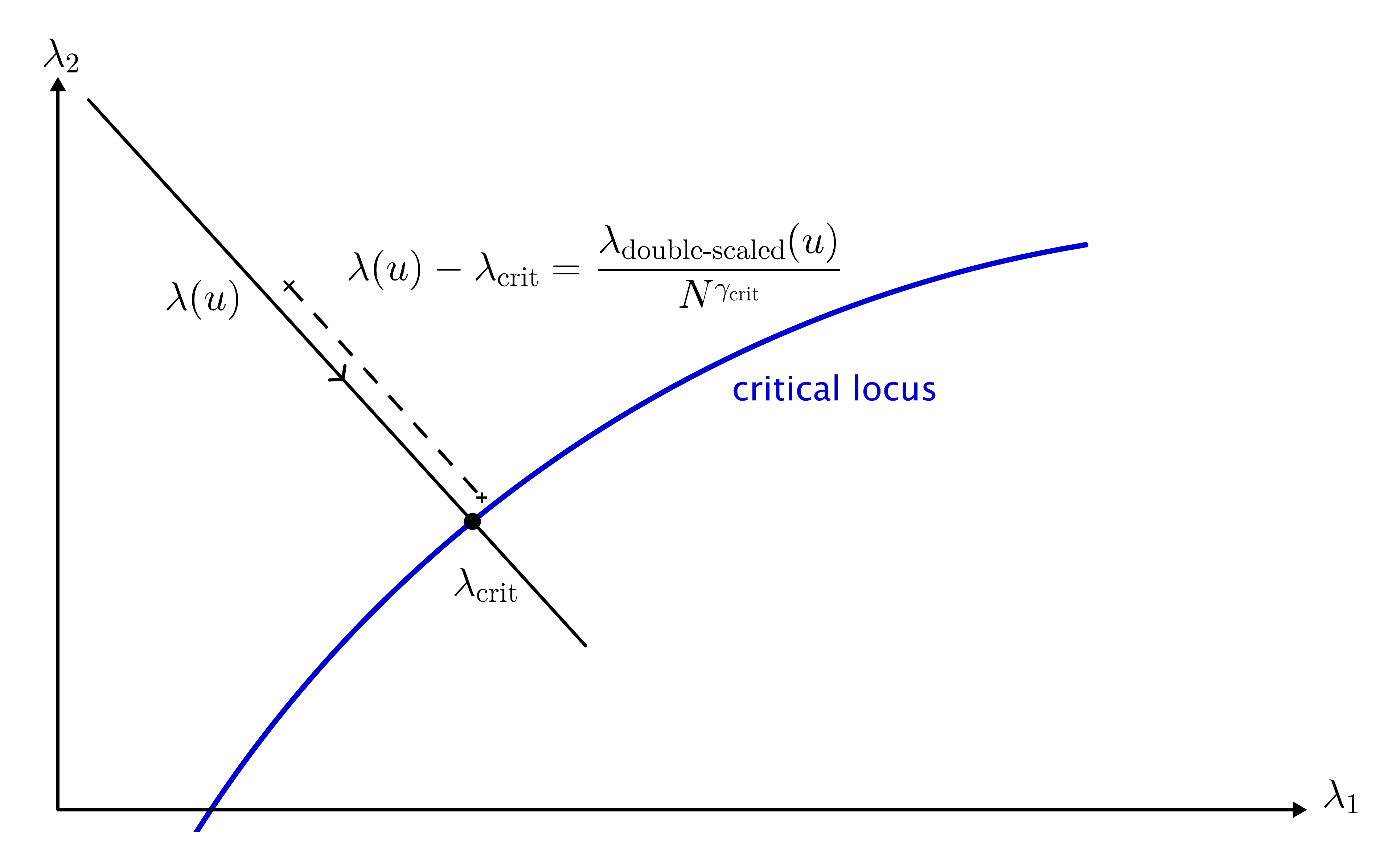}
			\caption{Sketch of the double-scaling limit. In this picture, the parameter space of 't Hooft couplings is the $(\lambda_1,\lambda_2)$-plane. The critical locus is shown in blue. In black, a curve $u \mapsto \lambda (u)$ in this space that intersects the critical locus at a single point $\lambda_{\mathrm{crit}}$. The double-scaling limit consists in sending $N \to \infty$ and $\lvert \lambda (u) - \lambda_{\mathrm{crit}}\rvert \to 0$ such that $ N^{\gamma_{\mathrm{crit}}} [\lambda (u) - \lambda_{\mathrm{crit}}]$ stays finite, for a power $\gamma_{\mathrm{crit}} >0$ that depends only on the universality class of the model. The scaled variable $\lambda_{\text{double-scaled}} (u)$ becomes the control parameter of the problem in the double-scaling limit.}
			\label{fig:doublescaling}
			\end{figure}

			The double-scaling limit refers to taking the large $N$ 't Hooft limit while at the same time a parameter is tuned toward its critical value \cite{Gross:1989vs,Brezin:1990rb,Douglas:1989ve}. More generally, it is possible to move along a curve in the space of parameters and approach a codimension-one critical locus at a rate that depends on $N$. This concept is schematically illustrated in Figure \ref{fig:doublescaling}.\par
			In practice, to identify the critical value it is typically needed to first solve the 't Hooft limit. The double-scaling thus comes as a second step, tailored to zoom-in close to the critical behaviour of a matrix model.\par
			
			\subsubsection*{(Aside) \texorpdfstring{$U(N)$}{U(N)} versus \texorpdfstring{$SU(N)$}{SU(N)}}
			Let us pause for a quick digression on the large $N$ limit of $SU(N)$ gauge theories. The difference between $\mathfrak{u}(N)$ and $\mathfrak{su}(N)$ is the constraint on the trace which, by power counting, is an $\mathcal{O}(1/N)$ correction. Thus, we expect that the two Lie algebras coincide in the large $N$ limit.\par
			The particle theorist's rephrasing of the above statement is that the only difference between the Feynman diagrams of $\mathfrak{su}(N)$ and $\mathfrak{u}(N)$ gauge theories with otherwise equal content consists of a meson propagator. This propagator can be attached to matter propagators but not to gluons and therefore, by power counting, it only appears in diagrams that are suppressed by $\mathcal{O}(1/N)$. The difference between $\mathfrak{su}(N)$ and $\mathfrak{u}(N)$ is thus expected to disappear at leading order in the large $N$ limit.\par
			\medskip
			Nevertheless, it is possible to consider a different definition of the large $N$ limit of $\mathfrak{su}(N)$. The sequence  
			\begin{equation*}
				\cdots \subset \mathfrak{su} (N-1) \subset \mathfrak{su} (N) \subset \mathfrak{su} (N+1) \subset \cdots 
			\end{equation*}
			suggests a definition of $\mathfrak{su} (\infty)$ in which one first takes the direct limit Lie algebra, and then imposes the traceless constraint. In practice, this amounts to define the large $N$ limit of $ \mathfrak{su} (N) $ as the traceless subspace of the direct limit Lie algebra $\mathfrak{u}(\infty)$. This is a valid mathematical definition which differs from the one taken in most works.\par
			It is often the case that the theory $\Th$ possesses a symmetry that forces the large $N$ solution of the partition function to be traceless, 
			\begin{equation*}
				\int_{-\infty}^{+\infty} \phi \rho_j (\phi) \dd \phi = 0.
			\end{equation*}
			This happens, for example, when the integrand of $\mz_{\cs^d} [\Th]$ is invariant under $\phi_a \mapsto - \phi_a$. Therefore, in many applications, the two different definitions of $\mathfrak{su}(\infty)$ yield the same answer. However, in general, the two definitions are inequivalent.\par
			\medskip
			When dealing with a more subtle example in which the saddle point configuration is not automatically traceless, it is important to decide which definition of $\mathfrak{su}(\infty)$ is to be used. If one follows the physical intuition that there is no such a thing as $\mathfrak{su}(\infty)$, but only $\mathfrak{u} (\infty)$, then one has to live with the fact that, in some instances, the observables of a $\mathfrak{su}(N)$ gauge theory are evaluated in the large $N$ limit on a non-traceless field configuration.

		\subsection{Saddle point approximation}
		\label{sec:GeneralSPE}
			Let us return to expression \eqref{eq:ZlinearQSeff}, 
			\begin{equation*}
				\mz_{\cs^d} \propto  \int_{\mathbb{R}^{\mathrm{rk} (G)}} \dd \phi ~ e^{- S_{\mathrm{eff}} (\phi) } ,
			\end{equation*}
			with effective action $S_{\mathrm{eff}}$ given in \eqref{eq:SeffFiniteN}. We now discuss its large $N$ behaviour in general terms. Concrete instances are detailed in the subsequent sections.\par

			\subsubsection*{Effective action}
				As we will see explicitly in the next sections, the effective action $S_{\mathrm{eff}} (\phi)$ becomes large as $N \to \infty$. This should be manifest by looking at its explicit form \eqref{eq:SeffFiniteN}, which includes traces and therefore will grow with some power of $N$. Concretely, at leading order in the large $N$ limit one finds
				\begin{equation}
				\label{eq:SeffscalingNchi}
					S_{\mathrm{eff}} (\phi)  = N^{\chi} S_{\mathrm{finite}} (\phi) , \qquad \chi >0 ,
				\end{equation}
				where $S_{\mathrm{finite}} (\phi)$ is a function of $\phi$ which stays finite in the large $N$ limit, and is independent of $N$ at leading order. The power $\chi >0$ is a rational number, and it is a characteristic of each class of theories. One important check in AdS/CFT is to match the power $\chi$ with the supergravity predictions. Typically, cancellations or enhancements that depend on the type of limit considered will adjust the value of $\chi$, compared to its naive power-counting prediction, to match with supergravity. This mechanism will be shown explicitly in the examples below.\par
				In the large $N$ limit, $S_{\mathrm{eff}}$, and thus $S_{\mathrm{finite}}$, become functionals of the eigenvalue densities $\left\{ \rho_j (\phi) \right\}$. We now restrict our attention to theories for which the effective action is of the form 
				\begin{subequations}
				\begin{equation}
				\label{eq:Sfinitesplit12a}
					S_{\mathrm{finite}} = S_1 - S_2 ,
				\end{equation}
				where 
				\begin{align}
					S_1 [\rho] &= \sum_{j} \int \dd \phi \rho_j (\phi) \mathcal{L}_{j}^{(1)} (\phi) , \label{eq:Sfinitesplit12b} \\
					S_2 [\rho] &= \sum_{(j,\tilde{\jmath})} \int \dd \phi \rho_j (\phi) \int \dd \sigma \rho_{\tilde{\jmath}} (\sigma) ~\mathcal{L}^{(2)} _{j,\tilde{\jmath}} (\lvert \phi - \sigma \rvert )  .  \label{eq:Sfinitesplit12c}
				\end{align}
				\label{eq:Sfinitesplit12}
				\end{subequations}
				In these expressions, $\mathcal{L}^{(i=1,2)}$ encode the contribution from every mode of the vector multiplet and hypermultiplets, the sum over $j$ is understood over the nodes of the quiver (that is, over the simple factors of the gauge group) and the sum over pairs $(j,\tilde{\jmath})$ includes the case $\tilde{\jmath}=j$ as well as pairs of neighboring nodes of the quiver.\par
				Let us clarify the meaning of this assumption.
				\begin{itemize} 
					\item In practice, \eqref{eq:Sfinitesplit12} means that the effective action is a quadratic functional of the array of densities $(\rho_1, \dots, \rho_L)$, and we isolate the quadratic and the linear pieces. 
					\item This is an extremely general assumption, encompassing a very vast class of theories. At this stage we are not restricting the gauge group or shape of the quiver, but only the field content, and only mildly. The quadratic dependence corresponds to restricting to fields that transform in representations of the gauge group with at most two indices. These include: 
					\begin{itemize}
						\item[---] The vector multiplet, which must transform in the adjoint and thus contributes to $S_2$ with $\tilde{\jmath}=j$;
						\item[---] Hypermultiplets in the fundamental representation, which contribute to $S_1$;
						\item[---] The classical piece $V (\phi)$, which is a trace and thus enters in $S_1$;
						\item[---] Hypermultiplets in the bifundamental representation, which contribute to $S_2$ with $\tilde{\jmath}$ and $j$ indexing adjacent nodes;
						\item[---] Hypermultiplets in the adjoint, rank-2 symmetric and rank-2 anti-symmetric representations, all of which contribute to $S_2$ with $\tilde{\jmath}=j$.
					\end{itemize}
					\item The overall sign of \eqref{eq:Sfinitesplit12a} depends only on the spacetime dimension $d$. It can be traced back to the UV behaviour of the supersymmetric field theory, and it will turn out that $S_1, S_2 >0$ in $d=3$, while $S_1,S_2 <0$ in $d=5$.
					\item The relative sign on the right-hand side of \eqref{eq:Sfinitesplit12a} imposes the competition between the two terms, to find a non-trivial configuration in the large $N$ limit.
				\end{itemize}

			\subsubsection*{Saddle point equation}	
				The behavior \eqref{eq:SeffscalingNchi} implies that the integrand in $\mz_{\cs^d}$ is sharply peaked, and it is very suppressed away from the minima of the effective action due to the super-exponential behaviour $e^{-N^{\chi} (\cdots)}$.\par
				Therefore, in the large $N$ limit, the leading contribution to $\mz_{\cs^d}$ comes from the saddle points of the effective action. This leads us to consider the saddle point equation (SPE). We have already explained that, in the large $N$ limit, $S_{\mathrm{eff}}$ becomes a functional of the densities of eigenvalues $\left\{ \rho_j (\phi) \right\}$. We have therefore to extremize $S_{\mathrm{eff}}$ over the set of (appropriately normalized) eigenvalue densities. The system of SPEs is:
				\begin{equation}
				\label{eq:variationalSPEgeneral}
					\frac{ \delta S_{\mathrm{eff}}}{\delta \rho_{j_{\ast}} (\phi_{\ast})} =0 , 
				\end{equation}
				for all $j_{\ast} =1, \dots, L$ and for all $\phi_{\ast} \in \R$. Imposing the form \eqref{eq:Sfinitesplit12} and taking the functional derivative, we have:
				\begin{align*}
					\frac{ \delta S_{1}}{\delta \rho_{j_{\ast}} (\phi_{\ast})} &= \mathcal{L}_{j_{\ast}}^{(1)} (\phi_{\ast}) , \\
					\frac{ \delta S_{2}}{\delta \rho_{j_{\ast}} (\phi_{\ast})} &= \sum_{\tilde{\jmath}} \int \dd \sigma \rho_{\tilde{\jmath}} (\sigma) ~\mathcal{L}^{(2)} _{j_{\ast},\tilde{\jmath}} (\lvert \phi_{\ast} - \sigma \rvert ) + \sum_{\tilde{\jmath}} \int \dd \phi \rho_{j} (\phi) ~\mathcal{L}^{(2)} _{j,j_{\ast}} (\lvert \phi_{\ast} - \phi \rvert ) \\
						&= 2 \sum_{\tilde{\jmath}} \int \dd \sigma \rho_{\tilde{\jmath}} (\sigma) ~\mathcal{L}^{(2)} _{j_{\ast},\tilde{\jmath}} (\lvert \phi_{\ast} - \sigma \rvert ) .
				\end{align*}
				The derivative of $S_2$ includes two pieces: one from $j=j_{\ast}$ and $\phi=\phi_{\ast}$, and the other from $\tilde{\jmath}=j_{\ast}$ and $\sigma=\phi_{\ast}$. We have used the symmetry of the system to write their sum as twice the same quantity. In the last expression, the sum runs over the label $\tilde{\jmath}$ of neighboring nodes to $j_{\ast}$, including $j_{\ast}$ itself.\par
				The SPE in this general setup is therefore written in the form:
				\begin{equation}
				\label{eq:generalSPEnodiff}
					2 \sum_{\tilde{\jmath}} \int \dd \sigma \rho_{\tilde{\jmath}} (\sigma)  \mathcal{L}^{(2)} _{j_{\ast},\tilde{\jmath}} (\lvert \phi_{\ast} - \sigma \rvert ) =  \mathcal{L}^{(1)}_{j_{\ast}} (\phi_{\ast}) ,
				\end{equation}
				to hold for all $j_{\ast} =1, \dots, L$ and for all $\phi_{\ast} \in \R$.

			\subsubsection*{(Aside) Saddle point equation, improved}
				We have derived the SPE \eqref{eq:generalSPEnodiff} setting to zero the variation of the effective action with respect to each density $\rho_j (\phi)$. Note however that the collection of saddle point equations at finite $N$ is 
				\begin{equation*}
					\frac{\partial S_{\mathrm{eff}}}{\partial \phi_{j,a}} = 0 \qquad  \forall a=1, \dots, N_j , \ \forall j=1, \dots, L .
				\end{equation*}
				To reduce clutter, we have written $j_{\ast} \mapsto j$, $\phi_{\ast} \mapsto \phi$. Differentiating \eqref{eq:Sfinitesplit12c} and only then substituting the eigenvalue density, we arrive at 
				\begin{equation*}
					2\sum_{\tilde{\jmath}} \int \dd \sigma \rho_{\tilde{\jmath}} (\sigma) ~\partial_{\phi} \mathcal{L}^{(2)} _{j,\tilde{\jmath}} (\lvert \phi - \sigma \rvert ) = \partial_{\phi} \mathcal{L}^{(1)}_j (\phi) .
				\end{equation*}
				This is not quite \eqref{eq:generalSPEnodiff}, but rather its derivative with respect to $\phi$.\par
				This apparent discrepancy is rooted in the fact that the effective action can always be redefined by adding constant terms, which is tantamount to picking a different overall normalization for $\mz_{\cs^d}$. This normalization constant does not affect our argument, and it should be fixed once and for all at the beginning to avoid ambiguities.\par
				More precisely, note that it is always possible to shift $\mathcal{L}^{(1)}_j$ in \eqref{eq:Sfinitesplit12c} by a constant term $C_j$, independent of $\phi$. This shift corresponds to 
				\begin{equation*}
					S_1 \mapsto S_1 + \sum_{j}C_j \underbrace{\int \dd \phi \rho_j (\phi)}_{=\nu_j \text{ by \eqref{eq:normrhoj}}} = S_1 + \sum_{j}C_j \nu_j .
				\end{equation*}
				That is, redefining the effective action by $\phi$-independent shifts does not change the details of the theory under consideration, but only adds an overall factor in front of $\mz_{\cs^d}$. Since these shifts do not affect the dynamics, we should study the saddle point equations up to such shifts.\par 
				Schematically, the SPE \eqref{eq:variationalSPEgeneral} should be understood as 
				\begin{equation*}
					\frac{ \delta S_{\mathrm{eff}}}{\delta \rho_{j_{\ast}} (\phi_{\ast})} =0 \quad \text{modulo} \quad (\mathcal{L}^{(1)}_j (\phi) \ \mapsto \ \mathcal{L}^{(1)}_j (\phi) + C_j) .
				\end{equation*}
				To consider an equation modulo constant shifts really means to consider its derivative. We are finally led to the system of SPEs
				\begin{equation}
				\label{eq:generalSPE1}
					\frac{\partial \ }{\partial \phi_{\ast}} \frac{ \delta S_{\mathrm{eff}}}{\delta \rho_{j_{\ast}} (\phi_{\ast})} =0 .
				\end{equation}
				This is the equation of motion that governs the variational problem of extremizing $e^{-S_{\mathrm{eff}}}$.\par
				Imposing the form \eqref{eq:Sfinitesplit12} to $S_{\mathrm{eff}}$ and differentiating, we arrive at
				\begin{equation}
				\label{eq:mostgeneralSPE}
					2 \sum_{\tilde{\jmath}} \int \dd \sigma \rho_{\tilde{\jmath}} (\sigma) ~\partial_{\phi} \mathcal{L}^{(2)} _{j,\tilde{\jmath}} (\lvert \phi - \sigma \rvert ) = \partial_{\phi} \mathcal{L}^{(1)}_j (\phi) ,
				\end{equation}
				for all $j$, and it must hold at every point $\phi \in \R$. This expression agrees with exchanging the order: first take the saddle point and then send $N \to \infty$.\par
				The goal is to find the collection of eigenvalue densities $\left\{ \rho_j (\phi) \right\}$ that solves this system of equations.

			\subsubsection*{(Aside) Saddle point equation as an Euler--Lagrange equation}
				The saddle point equation in its most general form \eqref{eq:generalSPE1} is the Euler--Lagrange equation for the variational problem dictated by the effective action $S_{\mathrm{eff}}$. To simplify the discussion we present the argument for the case of a single gauge node, but the procedure goes thorough in the general case as well.\par
				Indeed, recall that at the beginning of Section \ref{sec:LargeNtypes} we have written $S_{\mathrm{eff}}$ by replacing the discrete index $a=1, \dots, N_j$ with a continuous index $\mathsf{a} \in [0, 1]$. This substitution trades the traces of the gauge indices for integrals over $\mathsf{a} \in [0, 1]$. This step was before introducing the eigenvalue density. At that stage, we have the effective action as a functional of the map
				\begin{align*}
					\phi \ : \ [0, 1 ]  \ & \to \ \R \\
						\mathsf{a} \ & \mapsto \ \phi (\mathsf{a}) .
				\end{align*}
				Formally inverting this map, one obtains $\mathsf{a}: \R \to [0, 1]$. It is then possible to define the eigenvalue density equivalently as
				\begin{equation}
				\label{eq:rhoisdadphi}
					\rho (\phi) = \frac{ \dd \mathsf{a} ( \phi)}{\dd \phi} .
				\end{equation}\par
				\medskip
				{\small\textbf{Exercise:} Show that the definition \eqref{eq:rhoisdadphi} is equivalent to \eqref{eq:defrhodelta} at large $N$.}\par
				\medskip
				Thus, at least formally, writing $S_{\mathrm{eff}}$ as a functional of $\rho$ corresponds to writing it as a functional of the derivative $\dot{\mathsf{a}}$ of the classical field $\mathsf{a} : \R \to [0,1]$.\par 
				The problem of extremizing $S_{\mathrm{eff}}$ has been reduced to a classical field theory problem for an action in terms of the field $\mathsf{a} : \R \to [0,1]$ and its derivatives, with ``time'' variable $\phi \in \R$. The solution to this variational problem is determined by the Euler--Lagrange equations
				\begin{equation*}
					\frac{\partial \ }{\partial \phi } \left( \frac{\delta S_{\mathrm{eff}}}{\delta \dot{\mathsf{a}} (\phi)} \right)  -  \frac{\delta S_{\mathrm{eff}}}{\delta \mathsf{a} (\phi)}  =0.
				\end{equation*}
				Remember that $\mathsf{a}$ is the continuous version of the gauge index $a$, acted upon by the Weyl group of the gauge group. Gauge invariance prevents the action to depend on $\mathsf{a}$ alone, thus the second term vanishes. Using \eqref{eq:rhoisdadphi} we have that $\dot{\mathsf{a}} (\phi)$ is nothing but $\rho (\phi)$. The Euler--Lagrange equation for this classical field theory problem is precisely \eqref{eq:generalSPE1}.

		\subsection{Free energy}
		\label{sec:GeneralF}
			Let us now assume we have obtained the saddle point eigenvalue densities $\rho_j (\phi)$ that solve the SPE \eqref{eq:mostgeneralSPE}. How to compute them in practice will be explained in the rest of these notes.\par
			Then, by the saddle point argument, the leading contribution to the free energy $\mf_{\cs^d} = - \log \lvert \mz_{\cs^{d}} \rvert $ comes from the on-shell value of the effective action:
			\begin{equation}
			\label{eq:FisSonshell}
				\mf_{\cs^d} = S_{\text{on-shell}} := \left. S_{\mathrm{eff}} [\rho] \right\rvert_{\rho_j \text{ that solves \eqref{eq:mostgeneralSPE}}} .
			\end{equation}
			This is because the integrand $e^{- S_{\mathrm{eff}}}$ in $\mz_{\cs^d}$ is sharply peaked around its saddle point value, thus the integral is dominated by a tiny neighborhood of the saddle point eigenvalue configuration. Taking the logarithm, this means that the free energy is dominated by the on-shell free energy.\par
			To see this, one may write 
			\begin{equation*}
				 S_{\mathrm{eff}}  = S_{\text{on-shell}} + S_{\text{fluctuations}} ,
			\end{equation*}
			where $S_{\text{fluctuations}}$ is suppressed as $N \to \infty$ compared to $S_{\text{on-shell}}$. That is to say, 
			\begin{equation}
			\label{eq:Sfluctuationsmall}
				\lim_{N \to \infty} \frac{ S_{\text{fluctuations}} }{ N^{\chi}} =0 ,
			\end{equation}
			with the power $N^{\chi}$ dictated by the large $N$ behaviour of $S_{\mathrm{eff}}$ in \eqref{eq:SeffscalingNchi}. Then, schematically 
			\begin{align*}
				 \mf_{\cs^d}  & = - \log \left[ \int e^{- S_{\mathrm{eff}}}  \right] \\
				 	& =   - \log \left[ \int e^{ - S_{\text{on-shell}} - S_{\text{fluctuations}} } \right] \\
				 	& = S_{\text{on-shell}}  - \log \left[ \int e^{- S_{\text{fluctuations}} }  \right] \\
				 	& = S_{\text{on-shell}}  - N^{\chi} \log \left[ \left( \int e^{-S_{\text{fluctuations}}} \right)^{1/{N^{\chi}} } \right] \\
				 	& = S_{\text{on-shell}} + o (N^{\chi}) ,
			\end{align*}
			where in the last line, which is a consequence of \eqref{eq:Sfluctuationsmall}, $o (N^{\chi})$ indicates terms that grow slower than $N^{\chi}$.\par
			Let us emphasize that the term `on-shell' here means the effective action at the saddle point value, but the effective action includes all quantum effects in the supersymmetric field theory.\par
			\medskip
			We claim that, if the effective action is of the form \eqref{eq:Sfinitesplit12}, then the free energy in the large $N$ limit is 
			\begin{equation}
			\label{eq:FishalfS1}
				\mf_{\cs^d} = \left. \frac{1}{2} N^{\chi} S_1 \right\rvert_{\rho_j \text{ that solves \eqref{eq:mostgeneralSPE}}}  .
			\end{equation}
			In other words, we claim that, when evaluated on the solution to \eqref{eq:mostgeneralSPE}, $S_2$ cancels half of $S_1$. We stress that this is a very general result, that does not rely on the details of the theory under consideration but only on the structure \eqref{eq:Sfinitesplit12} of its effective action. We have already noted how extensive the list of supersymmetric theories that fulfill that assumption is.\par
			\medskip
			{\small\textbf{Exercise:} Compute $S_{\text{on-shell}}$ explicitly in the models discussed in the next sections, by evaluating each term. Check that \eqref{eq:FishalfS1} is satisfied.}\par
			\medskip
			We now proceed to derive the claim \eqref{eq:FishalfS1} in two ways. The first way (also recommended in \cite{Marino:2004eq}) is the most appropriate. The second derivation, presented for pedagogical purposes, provides a useful handle in many examples, but it is heuristic because it ultimately relies on a normalization assumption that may not be true in many cases.\par

			\subsubsection*{Proof of formula \eqref{eq:FishalfS1}}
				Explicitly, \eqref{eq:FisSonshell} reads 
				\begin{equation}
				\label{eq:FSdgeneralonshell}
					\mf_{\cs^d} = \left. N^{\chi} \sum_j \int \dd \phi \rho_j (\phi) \left[ \mathcal{L}_j ^{(1)} (\phi) - \sum_{\tilde{\jmath}} \int \dd \sigma \rho_{\tilde{\jmath}} (\sigma) \mathcal{L}_{j,\tilde{\jmath}} ^{(2)} (\lvert \phi - \sigma \rvert ) \right] \right\rvert_{\rho_j \text{ that solves \eqref{eq:mostgeneralSPE}}}.
				\end{equation}
				At this point, we observe that the SPE \eqref{eq:generalSPEnodiff} tells us that the first piece in square brackets is exactly twice the second piece in square brackets. Substituting 
				\begin{equation*}
					  \sum_{\tilde{\jmath}} \int \dd \sigma \rho_{\tilde{\jmath}} (\sigma) \mathcal{L}_{j,\tilde{\jmath}} ^{(2)} (\lvert \phi - \sigma \rvert ) = \frac{1}{2} \mathcal{L}_j ^{(1)} (\phi)
				\end{equation*}
				in the square brackets in \eqref{eq:FSdgeneralonshell}, we get 
				\begin{align*}
					\mf_{\cs^d} &= \left. N^{\chi} \sum_j \int \dd \phi \rho_j (\phi) \left[ \mathcal{L}_j ^{(1)} (\phi) - \frac{1}{2}  \mathcal{L}_j ^{(1)} (\phi) \right] \right\rvert_{\rho_j \text{ that solves \eqref{eq:mostgeneralSPE}}} \\
					 &=  N^{\chi} \cdot \frac{1}{2} \cdot \underbrace{ \sum_j \int \dd \phi \rho_j (\phi)  \mathcal{L}_j ^{(1)} (\phi) }_{S_1 \text{ by \eqref{eq:Sfinitesplit12b}}} \Big\rvert_{\rho_j \text{ that solves \eqref{eq:mostgeneralSPE}}} \\ 
					 &= \left. \frac{1}{2}  N^{\chi} S_1 \right\rvert_{\rho_j \text{ that solves \eqref{eq:mostgeneralSPE}}} 
				\end{align*}
				as claimed in \eqref{eq:FishalfS1}.

			\subsubsection*{(Aside) Proof of formula \eqref{eq:FishalfS1}, improved}
				We have explained above that it should be more appropriate to consider the SPE \eqref{eq:mostgeneralSPE}, instead of its primitive \eqref{eq:generalSPEnodiff}. This SPE tells us that the first term in square bracket in \eqref{eq:FSdgeneralonshell} is exactly twice the second term, but only up to $\phi$-independent shifts $ \mathcal{L}_j ^{(1)} \mapsto  \mathcal{L}_j ^{(1)} + C_j$. As discussed along the derivation of \eqref{eq:mostgeneralSPE}, these shifts are nothing but a choice of overall normalization for the partition function. It is clear that $\mf_{\cs^d}$ is only well-defined once this normalization is fixed once and for all.\par 
				In summary, \eqref{eq:mostgeneralSPE} or \eqref{eq:variationalSPEgeneral} allow us to write 
				\begin{equation*}
					 \frac{1}{2} \mathcal{L}_j ^{(1)} (\phi) = \sum_{\tilde{\jmath}} \int \dd \sigma \rho_{\tilde{\jmath}} (\sigma) \mathcal{L}_{j,\tilde{\jmath}} ^{(2)} (\lvert \phi - \sigma \rvert ) \ + \ \text{arbitrary constant} ,
				\end{equation*}
				with the arbitrary constant uniquely fixed by the arbitrary choice of normalization. We henceforth assume without loss of generality that the constants $C_j=0$, because the overall normalization plays no role in our discussion and, being a spectator all along, it can always be adjusted at the end. We thus proceed as we did and arrive at \eqref{eq:FishalfS1}.

			\subsubsection*{(Aside) Alternative proof of formula \eqref{eq:FishalfS1}}
				A convenient way to derive \eqref{eq:FishalfS1} is based on the following trick. Let $\mz_{\cs^d}^{\Th}$ be the sphere partition function of the theory under consideration, and let $\mz^{\Th} (\epsilon)$ be the partition function of the theory, with replacement $S_1 \mapsto \epsilon S_1$, depending on an auxiliary parameter $\epsilon > 0$. Obviously 
				\begin{equation*}
					\mz_{\cs^d}^{\Th} = \mz^{\Th} (\epsilon=1).
				\end{equation*}
				We will denote $\mf^{\Th} (\epsilon) := - \log \lvert \mz^{\Th} (\epsilon) \rvert $. At $\epsilon \to 0^{+}$, two scenarios can happen.
				\begin{enumerate}[(i)]
					\item $S_2$ is a repulsive interaction among the eigenvalues, such that the partition function diverges at $\epsilon \to 0$, and hence $\lim_{\epsilon \to 0^{+}} \mf^{\Th} (\epsilon) = - \infty$. In this case, because at finite $\epsilon$ there is a balance between $\epsilon S_1$ and $S_2$, necessarily $S_1$ is a confining term that attracts the eigenvalues towards a minimum, and hence $\lim_{\epsilon \to \infty} \mf^{\Th} (\epsilon) =0$.
					\item $S_2$ is an attractive interaction among the eigenvalues, which tends to gather them together. In this case, $\lim_{\epsilon \to 0^{+}} \mf^{\Th} (\epsilon) \ge 0$ (possibly infinite). Necessarily, to balance the attractive term $S_2$, $S_1$ must correspond to a force that decreases (becomes more energetically convenient) as the eigenvalues attain larger values. This means that when $S_1$ dominates the integral, $\mz^{\Th} (\epsilon \to \infty)$ diverges and hence $\lim_{\epsilon \to \infty} \mf^{\Th} (\epsilon) =- \infty$.
				\end{enumerate}
				In either case, we have that $\mf^{\Th} (\epsilon)$ is a continuous function of $\epsilon$ that changes from negative to positive, or vice versa, as $\epsilon$ is varied from $0$ to $\infty$. By Bolzano's theorem, there exists $\epsilon_0$ such that $\mf^{\Th} (\epsilon_0) =0$. Throughout this alternative proof, we assume that it is possible to normalize the partition function such that $\lim_{\epsilon \to 0^{+}} \mz^{\Th} (\epsilon) =1$, thus $\epsilon_0=0$ and $\mf^{\Th} (0)=0$.\par
				We can therefore write 
				\begin{align*}
					\mf^{\Th}_{\cs^d} &= \mf^{\Th} (\epsilon=1) - \mf^{\Th} (\epsilon = 0) \\
						&= \int_{0}^{1} \frac{ \dd \ }{\dd \epsilon } \mf^{\Th} (\epsilon) \dd\epsilon \\
						&= - \int_{0}^{1} \frac{ \dd \ }{\dd \epsilon }  \log \mz^{\Th} (\epsilon) \dd\epsilon \\
						&= - \int_{0}^{1} \frac{1}{ \mz^{\Th} (\epsilon)} \frac{ \dd \ }{\dd \epsilon }  \mz^{\Th} (\epsilon) \dd\epsilon.
				\end{align*}
				At this stage, we observe that, in the large $N$ limit, 
				\begin{equation*}
					 \mz^{\Th} (\epsilon) = \int \dd \phi e^{- N^{\chi} \left[ \epsilon S_1 - S_2 \right] } 
				\end{equation*}
				and therefore 
				\begin{align*}
					- \frac{1}{ \mz^{\Th} (\epsilon)} \frac{ \dd \ }{\dd \epsilon }  \mz^{\Th} (\epsilon) & = N^{\chi}  \cdot \frac{1}{\mz^{\Th} (\epsilon)} \int \dd \phi e^{- N^{\chi} \left[ \epsilon S_1 - S_2 \right]} S_1 \\
						&= N^{\chi} \langle S_1 \rangle^{\epsilon} .
				\end{align*}
				Here we have recognized the expectation value of $S_1$, denoted $\langle S_1 \rangle^{\epsilon}$ to stress that it is being computed at arbitrary $\epsilon$. In the large $N$ limit, the expectation value of the single-trace part $S_1$ is obtained by evaluating it on shell,
				\begin{equation*}
					\langle S_1 \rangle^{\epsilon} = S_1 \vert_{\rho_j^{\epsilon} \text{ that solves SPE}} .
				\end{equation*}\par
				Combining these pieces we arrive at the relation 
				\begin{equation}
				\label{eq:FepsisS1eps}
					\mf^{\Th}_{\cs^d} = N^{\chi} \int_{0}^{1} \dd\epsilon  S_1 \vert_{\rho_j^{\epsilon} \text{ that solves SPE}} .
				\end{equation}
				Let us now derive the saddle point equation for $\mz^{\Th} (\epsilon)$. It is immediate to check that all the above procedure goes through, except for a factor $\epsilon$ in front of $S_1$. The saddle point equation is \eqref{eq:mostgeneralSPE} with the right-hand side multiplied by $\epsilon$, explicitly 
				\begin{equation}
				\label{eq:generalSPEepsilon}
					\sum_{\tilde{\jmath}} \int \dd \sigma \rho_{\tilde{\jmath}}^{\epsilon} (\sigma) ~\partial_{\phi} \mathcal{L}^{(2)} _{j,\tilde{\jmath}} (\lvert \phi - \sigma \rvert ) = \epsilon \partial_{\phi} \mathcal{L}^{(1)}_j (\phi) .
				\end{equation}
				Here we use the notation $\rho_j ^{\epsilon}$ to denote the eigenvalue density deformed by the arbitrary $\epsilon >0$. The saddle point equation of interest is recovered setting $\epsilon =1$.\par
				The crucial consequence of our assumption on $S_{\mathrm{eff}}$ is that the saddle point equation has a linear functional dependence on $\rho_j$. In practice, this simply means that if we have found a solution $\rho_j$ to \eqref{eq:mostgeneralSPE}, then clearly 
				\begin{equation*}
					\rho_j ^{\epsilon} (\phi) = \epsilon \rho_j (\phi)
				\end{equation*}
				is a solution to \eqref{eq:generalSPEepsilon}.\par
				Equipped with this solution, we compute 
				\begin{equation*}
					\langle S_1 \rangle^{\epsilon} = S_1 \vert_{\epsilon \rho_j} = \epsilon S_1 \vert_{\text{on-shell}} .
				\end{equation*}
				Here, by $S_1 \vert_{\text{on-shell}} $ we mean it evaluated on the solution to the authentic SPE \eqref{eq:mostgeneralSPE}. It does not depend on $\epsilon$. The second equality follows immediately from the definition \eqref{eq:Sfinitesplit12b} of $S_1$.\par
				Plugging this solution into \eqref{eq:FepsisS1eps} we finally obtain 
				\begin{align*}
					\mf^{\Th}_{\cs^d} &= N^{\chi} \int_{0}^{1} \epsilon \dd\epsilon  S_1 \vert_{\text{on-shell}} \\
					&=  N^{\chi} S_1 \vert_{\text{on-shell}} \left.\frac{\epsilon^2}{2} \right\rvert_{0}^{1} \\
					&=  N^{\chi} S_1 \vert_{\text{on-shell}} \cdot \frac{1}{2} ,
				\end{align*}
				giving \eqref{eq:FishalfS1}.\par 
				More generally, this procedure would yield $\mf^{\Th}_{\cs^d} = \frac{1}{2}  N^{\chi} S_1 \vert_{\text{on-shell}} - \mf^{\Th} (0)$, where the latter piece is just a choice of normalization whenever $\mz^{\Th} (\epsilon)$ is normalizable at $\epsilon \to 0$. The latter requirement on the finiteness of $\lim_{\epsilon \to 0^{+}} \mz^{\Th} (\epsilon)$, may in general fail, whereby the previous derivation of \eqref{eq:FishalfS1} is to be preferred.

	\section{'t Hooft limit in SQCD}
\label{sec:LargeNtHooft}

Let us now delve into the setup introduced in Section \ref{sec:LargeNtypes}. Throughout this section, we make the following assumptions:
\begin{itemize}
	\item[(tH1)] The theory consists of a single $U(N)$ gauge node;
	\item[(tH2)] The partition function is a random matrix ensemble.
\end{itemize}
These two constraints are restrictive and significantly circumscribe the set of allowed QFTs. In particular, the second point requires that we must be able to recast the interaction terms that depend on $( \phi_{a} - \phi_b )$ in the effective action in the form of the square of a Vandermonde determinant.\par
Despite the drawback of constraining the number of accessible theories, the assumptions above guarantee that the large $N$ 't Hooft limit can be solved by standard methods in random matrix theory.\par

\subsubsection*{Matrix model potential}
	We now discuss the large $N$ 't Hooft limit of these theories and solve it at leading order in $N$. This corresponds to scale all the coefficients in the potential $V(x)$ linearly with $N$, as in \eqref{eq:VtHooftscaleW}. That is, the partition function is 
	\begin{equation}
	\label{eq:ZMM}
		\mz = \frac{1}{N!} \int_{\R^{N}} \dd \phi ~ e^{- N \sum_{a=1}^{N} W (\phi_a) } ~ \prod_{1 \le a < b \le N} ( \phi_a - \phi_b)^2 .
	\end{equation}\par
	We will make a further technical assumption, which is not necessary but that simplifies the exposition:
	\begin{itemize}
		\item[(tH3)]\label{assump:tH3} The derivative $W^{\prime} (z) = \sum_{p \ge 0} c_{p+1} z^p$ of $W(z)$ is a meromorphic function of $z \in \mathbb{P}^1$ with real coefficients.
	\end{itemize}
	Notice that we allow infinitely many coefficients $\left\{ c_p \right\}_{p \ge 1}$, as long as they are real and satisfy this additional condition.\par
	The three hypothesis outlined so far lead us to restrict our attention to:
	\begin{itemize}
		\item $\mN=4$ $U(N)$ or $SU(N)$ super-Yang--Mills in $d=4$, and
		\item $\mN=4$ $U(N)$ SQCD$_3$.
	\end{itemize}
	However, the procedure we describe is a purely matrix model results and holds more in general, whenever the three assumptions are satisfied.

	\subsection{Planar matrix models}
	\label{sec:MMplanarlimit}
		Throughout this subsection we show the generic solution of matrix models at large $N$. Several excellent presentations exist reviewing this procedure. Here we follow \cite{DiFrancesco:1993cyw,Marino:2004eq}.\par
		We are interested in the leading order contribution to the free energy 
		\begin{equation*}
			\mf = - \log \mz
		\end{equation*}
		as $N \to \infty $, with the 't Hooft scaling as prescribed above \cite{Brezin:1977sv}. This limit is usually referred to as the planar limit because, in a diagrammatic expansion in perturbation theory, only Feynman diagrams that can be drawn on the plane or the sphere, referred to as planar diagrams, contribute at leading order in $N$ (but to all orders in the couplings).\par

		\subsubsection*{Effective action}
			The integrand in \eqref{eq:ZMM} is suppressed by a factor $e^{-N^2 ( \cdots )}$ as $N \to \infty$, while the number of variables is $N$. This makes \eqref{eq:ZMM} well suited to apply the steepest descent method.\par
			This in particular means that, in the planar limit, \eqref{eq:ZMM} is dominated by the saddle points of the effective action
			\begin{equation}
			\label{eq:SeffMMfiniteN}
				S_{\mathrm{eff}} (\phi) = N \sum_{a=1} ^{N} W(\phi_a) - \sum_{1 \le a \ne b \le N} \log \lvert \phi_a - \phi_b \rvert .
			\end{equation}
			Notice that we have used the simple identity 
			\begin{equation*}
				\sum_{1 \le a < b \le N} \log ( \phi_a - \phi_b)^2 =  \sum_{1 \le a \ne b \le N} \log \lvert \phi_a - \phi_b \rvert 
			\end{equation*}
			when passing the Vandermonde factor to the exponential.\par
			Introducing the eigenvalue density $\rho (\phi)$ from \eqref{eq:defrhodelta} and using the substitution 
			\begin{equation*}
				\frac{1}{N} \sum_{a=1} ^{N}  F( \phi_a) \mapsto \int \dd \phi \rho (\phi ) F(\phi)
			\end{equation*}
			explained in Section \ref{sec:LargeNtypes}, the effective action at large $N$ is written as 
			\begin{equation}
			\label{eq:SeffMMlargeN}
				S_{\mathrm{eff}} ( \phi ) = N^2 \int \dd \phi \rho (\phi) \left[  W(\phi)  - \int_{\sigma \ne \phi} \dd \sigma \rho (\sigma) \log \lvert \phi - \sigma \rvert  \right] .
			\end{equation}

		\subsubsection*{Saddle point equation}
			When $N \to \infty$, the leading contribution to $\mz$ comes from the saddle points of the effective action $S_{\mathrm{eff}}$. In the form \eqref{eq:SeffMMlargeN}, the saddle point problem is translated into the problem of finding a measure $\rho (\phi) \dd \phi $ that minimizes \eqref{eq:SeffMMlargeN}. The solution is called equilibrium measure.\par
			We observe that there are two contributions to \eqref{eq:SeffMMlargeN}:
			\begin{itemize}
				\item The confining potential $W(\phi)$, that tends to gather the eigenvalues at its minimum;
				\item The repulsive log-interaction, that tends to spread the eigenvalues far away from each other.
			\end{itemize}
			It is the competition between these two terms that determines the eigenvalue density $\rho (\phi)$.\par
			Functional differentiation of \eqref{eq:SeffMMlargeN} leads us to the saddle point equation (SPE)
			\begin{equation}
			\label{eq:SPEMM}
				\mathrm{P}\!\int \dd \sigma \frac{ \rho (\sigma) }{\phi - \sigma } = \frac{1}{2} W^{\prime} (\phi) ,
			\end{equation}
			where the symbol $\mathrm{P}\!\int $ stands for the principal value integral. This is a singular integral equation to be solved for $\rho (\phi)$. \par
			\medskip
			{\small\textbf{Exercise:} Observe that \eqref{eq:SPEMM} is $\frac{\partial \ }{\partial \phi } \left( \frac{ \delta S_{\mathrm{eff}}}{\delta \rho} \right) =0$. Check that it comes from the Euler--Lagrange equation for the variational problem of minimizing \eqref{eq:SeffMMlargeN}.}\par
			\medskip
			Alternatively, a perhaps simpler way to get \eqref{eq:SPEMM} is to look at the effective action \eqref{eq:SeffMMfiniteN} before taking the large $N$ limit. The system of saddle point equations is 
			\begin{equation*}
				\frac{\partial S_{\mathrm{eff}} }{\partial \phi_a } = 0 \qquad \forall~a=1, \dots, N.
			\end{equation*}
			Differentiating \eqref{eq:SeffMMfiniteN} and dividing by $N$ we find
			\begin{equation*}
				W^{\prime} (\phi_a) - \frac{2}{N} \sum_{\substack{b=1\\ b \ne a}}^{N} \frac{1}{\phi_a - \phi_b} =0. 
			\end{equation*}
			Notice the factor of $2$ coming from the derivative of the double sum. Explicitly, renaming the dummy variables summed over as $b,c$ and differentiating the double sum in \eqref{eq:SeffMMfiniteN} with respect to $\phi_a$, we get two contributions: from $a=c$ and from $a=b$, 
			\begin{align*}
				\frac{\partial }{\partial \phi_a } \sum_{b \ne c } \log \lvert \phi_b - \phi_c \rvert & = \frac{\partial }{\partial \phi_a } \sum_{b=1}^{N}\left[ \sum_{c : \ \phi_c < \phi_b} \log (\phi_b - \phi_c ) + \sum_{c : \ \phi_c > \phi_b} \log (\phi_c - \phi_b) \right] \\
					&= \delta_{ab} \underbrace{\left[ \sum_{\substack{ c \ne b \\ \phi_c <\phi_b } } \frac{1}{\phi_b - \phi_c } + \sum_{\substack{ c \ne b \\ \phi_c >\phi_b } } \frac{(-1)}{\phi_c - \phi_b }  \right] }_{= \sum_{c : c \ne b} \frac{1}{\phi_b - \phi_c }} + \delta_{ac} \underbrace{\sum_{b=1}^{N}\left[ \begin{cases} \frac{(-1)}{\phi_b - \phi_c }  & \text{\footnotesize if }\phi_c <\phi_b \\ \frac{1}{\phi_c - \phi_b } & \text{\footnotesize if } \phi_c >\phi_b  \end{cases}  \right] }_{= \sum_{b: b\ne c} \frac{1}{\phi_c - \phi_b }} \\
					&= \sum_{c \ne a} \frac{1}{\phi_a - \phi_c } + \sum_{b \ne a} \frac{1}{\phi_a - \phi_b } \\
					&= 2 \sum_{b \ne a} \frac{1}{\phi_a - \phi_b } .
			\end{align*}\par
			In the large $N$ limit, the system of $N$ coupled equations is gathered into the unique singular equation \eqref{eq:SPEMM}.\par
			\medskip
			{\small\textbf{Exercise:} Differentiate \eqref{eq:SeffMMfiniteN} without introducing the eigenvalue density. Convince yourself about the factor of 2. Check that, at large $N$, this system of saddle point equations is equivalent to \eqref{eq:SPEMM}.}\par
			\medskip
			The solution to \eqref{eq:SPEMM} is textbook material \cite{Muskhe:1977,Pipkin:1991} if hypothesis (\hyperref[assump:tH3]{tH3}) holds. The assumption can be relaxed to allow for branch cuts in $W^{\prime} (z)$, albeit not without technical complications. We stick to (\hyperref[assump:tH3]{tH3}) and refer to \cite{Marino:2004eq} for a detailed example in which this assumption fails. We now proceed to show the solution.

	\subsubsection*{Obtaining the eigenvalue density}
		We have already mentioned that the potential $W(\phi)$ tends to push the eigenvalues towards its minima, whereas the logarithmic interaction pushes them far apart from each other. Therefore, the first step in solving \eqref{eq:SPEMM} is to determine the number $\ell$ of minima of $W(\phi)$, for $\phi \in \R $.\par
		Once the number $\ell$ is known, we make the ansatz that the eigenvalues are spread on $\ell$ disjoint intervals, one around each minimum of $W (\phi)$. The ansatz corresponds to an eigenvalue density supported on 
		\begin{equation}
		\label{eq:branchcutmulticut}
			\surho : = \text{supp} \rho = \bigcup_{s=1} ^{\ell} \left[ A_{s} , B_{s} \right] .
		\end{equation}
		The parameters $\left\{ A_s, B_s \right\}_{s=1, \dots, \ell}$, with $B_{s} < A_{s+1}$, are determined from \eqref{eq:SPEMM}. Solutions of this type are known as $\ell$-cut solutions.\par
		The next ingredient needed is the planar resolvent (which we will henceforth simply denote as resolvent)
		\begin{equation}
		\label{eq:defresolvent}
			\omega (z) = \int_{ \surho } \dd \sigma \frac{ \rho (\sigma) }{z - \sigma } , \quad z \in \mathbb{P}^1 \setminus \surho .
		\end{equation}
		We therefore see where the nomenclature ``$\ell$-cut solution'' comes from: the $\ell$ intervals onto which $\rho (\phi)$ is supported are precisely the $\ell$ branch cuts in $\mathbb{P}^1$ for $\omega (z)$.\par
		It follows from the definition \eqref{eq:defresolvent} that, once we find the resolvent $\omega (z)$, we immediately recover $\rho (\phi)$ through the discontinuity equation 
		\begin{equation}
		\label{eq:disceqrhoomega}
			\lim_{\varepsilon \to 0^{+}} \left[ \omega (\phi + i \varepsilon ) - \omega (\phi - i \varepsilon )  \right] = - i 2 \pi \rho (\phi) , \quad \phi \in \surho .
		\end{equation}
		On the other hand, the SPE \eqref{eq:SPEMM} becomes an equation for $\omega (z)$:
		\begin{equation}
		\label{eq:jumpomegaplus}
			\lim_{\varepsilon \to 0^{+}} \left[ \omega (\phi + i \varepsilon ) + \omega (\phi - i \varepsilon )  \right] = W^{\prime} (\phi) , \quad \phi \in \surho .
		\end{equation}\par
\begin{figure}[ht]
		\centering
			\includegraphics[width=0.7\textwidth]{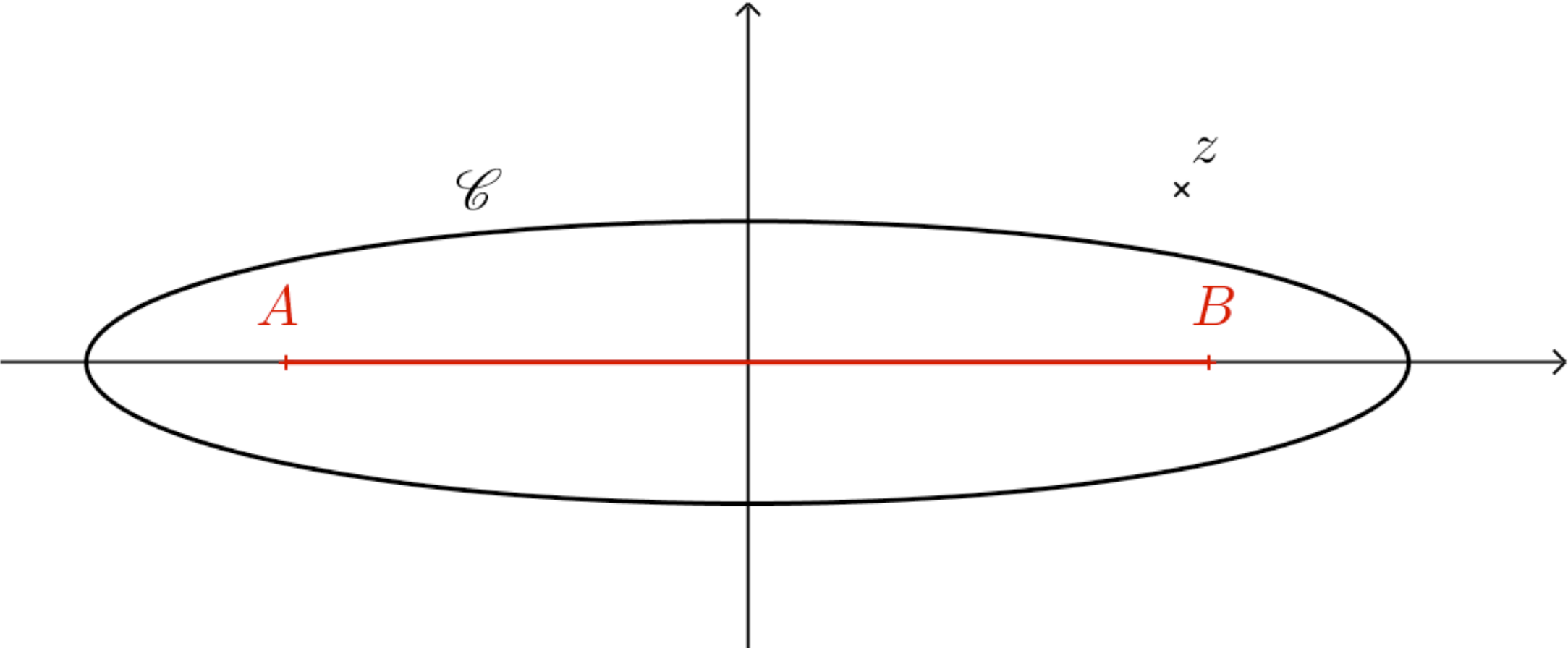}
			\caption{Integration contour $\mathscr{C}$ in the one-cut case. It encircles $\surho$ (red) leaving $z$ outside.}
			\label{fig:contourmmplanar}
		\end{figure}

		The function $\omega (z)$ satisfying this equation and with a branch cut along $\surho $ is given by the contour integral 
		\begin{equation}
		\label{eq:contouromega}
			\omega (z) = \sqrt{ \prod_{s=1} ^{\ell} (z-A_s)(z-B_s) } \oint_{\mathscr{C}} \frac{\dd u }{2 \pi i } \frac{\frac{1}{2} W^{\prime} (u)}{(z-u) \sqrt{ \prod_{s=1} ^{\ell} (u-A_s)(u-B_s) } } ,
		\end{equation}
		where the contour $\mathscr{C}$ encircles the branch cut \eqref{eq:branchcutmulticut} but leaves outside the point $z \in \C \setminus \surho $, as in Figure \ref{fig:contourmmplanar}. This expression is complemented by the normalization condition 
		\begin{equation*}
			\lim_{\lvert z \rvert \to \infty } z \omega (z) = 1
		\end{equation*}
		that follows directly from the normalization of $\rho (\phi)$ through \eqref{eq:defresolvent}:
		\begin{align*}
			z \omega (z) & = \int_{ \surho } \dd \sigma \frac{ \rho (\sigma) }{1- \frac{\sigma}{z} } \\
				& =  \int_{ \surho } \dd \sigma \rho (\sigma) \left[ 1 + \mathcal{O}\left( \frac{\sigma}{z} \right) \right] = \underbrace{ \int_{ \surho } \dd \sigma \rho (\sigma)}_{=1 \text{ by normaliz. of }\rho} +  \mathcal{O}\left( \frac{1}{z} \right) .
		\end{align*}
		Positive powers of $\frac{\sigma}{z}$ yield a vanishing contribution, because $\sigma \in \surho$ is bounded and $\lvert z \rvert \to \infty$.\par
		\par
		If $W^{\prime} (z)$ satisfies the hypothesis (\hyperref[assump:tH3]{tH3}), i.e. it has no branch cuts, it is possible to deform the integration contour in \eqref{eq:contouromega} to a circle at infinity, accounting for the residues of the poles picked up in the process. Very schematically, the idea is to write 
		\begin{equation*}
			\oint_{\mathscr{C}} \frac{\dd u }{2 \pi i } ~(\text{integrand}) = \oint_{\mathbb{S}^1 _{\infty}} \frac{\dd u }{2 \pi i } ~(\text{integrand}) - \sum_{\text{poles between $\mathscr{C}$ and $\mathbb{S}^1 _{\infty}$}} \mathrm{Res} (\text{integrand}) ,
		\end{equation*}
		where $\mathbb{S}^1 _{\infty}$ is a circle of arbitrarily large radius, in particular, larger than all poles of the integrand. We get 
		\begin{equation}
		\begin{aligned}
			\omega (z) & = \underbrace{ \frac{1}{2} W^{\prime} (z) }_{\text{pole at $u=z$}} - \underbrace{  \sum_{\left\{ z_{p} \right\} } \sqrt{  \prod_{s=1} ^{\ell} \frac{ (z-A_s)(z-B_s) }{ (z_{\ast}-A_s)(z_{\ast}-B_s) } } \frac{ \mathrm{Res}_{u=z_{\ast}} W^{\prime} (u) }{2 (z-z_{\ast})}  }_{\text{poles $z_{\ast}$ of $ W^{\prime}$}} \\ 
			& + \sqrt{ \prod_{s=1} ^{\ell} (z-A_s)(z-B_s) } \underbrace{ \oint_{\mathbb{S}^1 _{\infty} } \frac{\dd u }{2 \pi i } \frac{\frac{1}{2} W^{\prime} (u)}{(z-u) \sqrt{ \prod_{s=1} ^{\ell} (u-A_s)(u-B_s) } } }_{\text{contour integral at $\infty$}} ,
		\end{aligned}
		\label{eq:omegadefcontour}
		\end{equation}
		where the first term is the residue at $u=z$ (notice the sign), the second term is a sum over the poles $\left\{ z_{\ast} \right\}$ of $W^{\prime}$, and the last term is the residual integration along a large circle $\mathbb{S}^1 _{\infty}$. The number $\ell$ of cuts must be consistently chosen so that this piece converges.\par
		We have thus reduced the problem to a simple residue computation. Once $\omega (z)$ is obtained from \eqref{eq:omegadefcontour}, the eigenvalue density is recovered from \eqref{eq:disceqrhoomega}. Notice that the contribution to \eqref{eq:omegadefcontour} from the pole at $u=z$ gives the regular part of $\omega (z)$. The square root has a branch cut along $\surho$, thus yields opposite contributions when approaching $\phi \in \surho $ from above and below. We check that both \eqref{eq:jumpomegaplus} and \eqref{eq:disceqrhoomega} are satisfied.\par
		\medskip
		We do not enter into too many details in the main text, but provide concrete and explicit examples of how these computations work in Appendix \ref{app:HMM}.

	\subsubsection*{Support of the eigenvalue density}
		\begin{figure}[ht]
		\centering
			\includegraphics[width=0.35\textwidth]{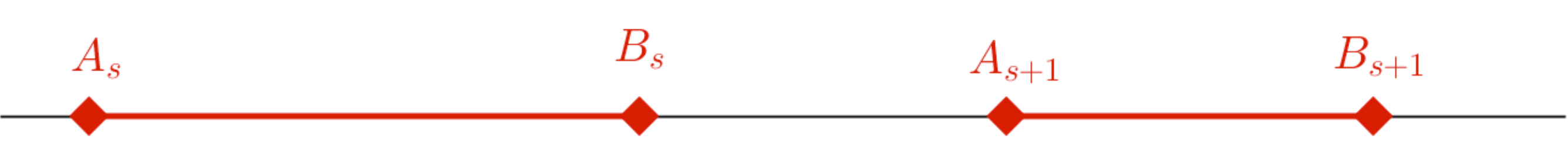}%
			\hspace{0.1\textwidth}\includegraphics[width=0.35\textwidth]{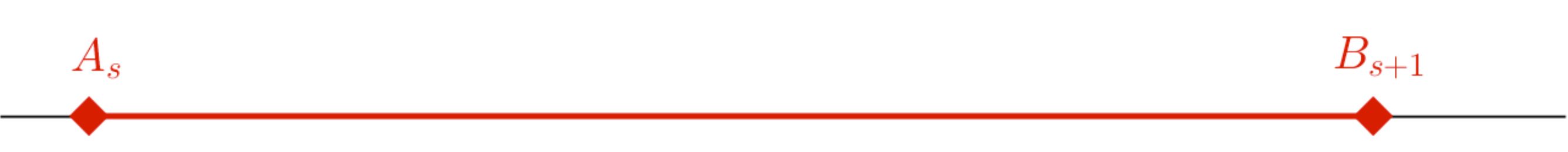}%
			\caption{The endpoints $B_s, A_{s+1}$ of two cuts of the $\ell$-cut solution (left) merge, developing a $(\ell-1)$-cut solution (right).}
			\label{fig:twocut-onecut}
		\end{figure}\par
	
		The support $\surho$ of the eigenvalue density $\rho$ is determined by the number $\ell$ of branch cuts of $\omega (z)$. Generically on the parameter space, $\ell$ equals the number of saddle points of $W$, which is the number of zeros of $W^{\prime}$. At positive-codimensional loci on the parameter space, certain zeros of $W^{\prime}$ coalesce, corresponding to the joining of two intervals 
		\begin{equation*}
			[A_s, B_s] \cup [A_{s+1}, B_{s+1} ] \longrightarrow  [A_s, B_{s+1} ]
		\end{equation*}
		when the endpoints $B_s$ and $A_{s+1}$ meet, as illustrated in Figure \ref{fig:twocut-onecut}.\par
		Let us write the large argument behaviour of $W^{\prime}$ as 
		\begin{equation*}
			W^{\prime} (u) = \frac{u^{\ell}}{\lambda_{\ell+1}} + \mathcal{O} (u^{\ell -1}) .
		\end{equation*}
		Note that when $W$ is a polynomial, this notation corresponds to the normalization $W (u)= \frac{ u^{\ell +1}}{(\ell+1)\lambda_{\ell+1}} + \mathcal{O} (u^{\ell})$. The contour integral along the circle at infinity $\mathbb{S}^1 _{\infty} $ thus becomes 
		\begin{align*}
			\lim_{\mathsf{r} \to \infty} \oint_{\lvert u \rvert = \mathsf{r}} \frac{\dd u }{2 \pi i } \frac{\frac{1}{2} W^{\prime} (u)}{(z-u) \sqrt{ \prod_{s=1} ^{\ell} (u-A_s)(u-B_s) } } &= \lim_{\mathsf{r} \to \infty} \left[ - \oint_{\lvert u \rvert = \mathsf{r}} \frac{\dd u }{2 \pi i u^{1 + \ell}} \cdot \frac{u^{\ell}}{2\lambda_{\ell +1}}+ \mathcal{O} \left( \frac{1}{\mathsf{r}}\right) \right] \\
			&= - \frac{1}{2\lambda_{\ell +1}} .
		\end{align*}
		Plugging this expression into \eqref{eq:omegadefcontour} gives $\omega (z)$:
		\begin{equation}
		\begin{aligned}
			\omega (z) & = \frac{1}{2} W^{\prime} (z) - \sum_{\left\{ z_{p} \right\} } \sqrt{  \prod_{s=1} ^{\ell} \frac{ (z-A_s)(z-B_s) }{ (z_{\ast}-A_s)(z_{\ast}-B_s) } } \frac{ \mathrm{Res}_{u=z_{\ast}} W^{\prime} (u) }{2 (z-z_{\ast})}  \\ 
			& - \frac{1}{2\lambda_{\ell +1}}\sqrt{ \prod_{s=1} ^{\ell} (z-A_s)(z-B_s) } .
		\end{aligned}
		\label{eq:omegaMMpoly}
		\end{equation}\par 
		It remains to determine the support. We expand $\omega (z)$ for large $z$, and impose that all the non-negative powers of $z$ vanish, while the coefficient of the order $1/z$ must be equal to 1. Both $W^{\prime} (z)$ and the last term start with $\frac{z^{\ell}}{2 \lambda_{\ell +1}}$ and they have opposite signs, so the $\mathcal{O}(z^{\ell})$ term cancels out. The terms with the poles of $W^{\prime}$ starts at $\mathcal{O} (z^{\ell-1})$.\par
		We are left with $(\ell+1)$ equations, derived from the normalization condition imposed to the large $z$ expansion of $\omega (z)$ from order $z^{\ell-1}$ to order $z^{-1}$. However, we have $2\ell$ constants to be determined, the $\ell$ pairs of endpoints $\left\{ (A_s, B_s) \right\}_{s=1, \dots, \ell}$.\par
		The solution is uniquely fixed specifying the filling fractions $\left\{ \nu_s \right\}_{s=1, \dots, \ell}$ (occupation numbers) of the intervals. That is, we fix fractions $0 \le \nu_s \le 1$, with $\sum_{s=1}^{\ell} \nu_s =1$, such that 
		\begin{equation}
		\label{eq:rhoplanarfilling}
			\int_{A_s}^{B_s} \dd \phi \rho (\phi) = \nu_s \qquad \forall s=1, \dots \ell .
		\end{equation}
		These equations provide exactly $(\ell-1)$ additional equations, yielding a total of $2\ell$ equations for the $2\ell$ endpoints $\left\{ (A_s, B_s) \right\}_{s=1, \dots, \ell}$. Note that one of the $\ell$ equations \eqref{eq:rhoplanarfilling} is trivially satisfied:
		\begin{align*}
			1= \sum_{s=1}^{\ell} \nu_s &= \sum_{s=1}^{\ell} \int_{A_s}^{B_s} \dd \phi \rho (\phi) \\
				&= \int_{\bigcup_{s=1} ^{\ell} \left[ A_{s} , B_{s} \right]} \dd \phi \rho (\phi) = 1 .
		\end{align*}
		In summary:
		\begin{itemize}
			\item[---] To determine the support in the $\ell$-cut phase, we have to solve for $2 \ell$ endpoints $\left\{ (A_s, B_s) \right\}_{s=1, \dots, \ell}$.
			\item[---] The  normalization of the eigenvalue density provides $(\ell+1)$ equations.
			\item[---] Fixing the filling fractions yields $\ell$ additional equations, but only $(\ell-1)$ of them are linearly independent. These provide the remaining $2\ell - (\ell +1)= (\ell -1)$ equations to completely solve the problem.
		\end{itemize}

	\subsection{Planar four-dimensional \texorpdfstring{$\mN=4$}{N=4} super-Yang--Mills}
	\label{sec:PlanarN4SYM}
		Let us apply the machinery to the celebrated example of $d=4$ $\mN=4$ super-Yang--Mills theory.\par
		We have seen in Section \ref{sec:exampleZconv} that the partition function of this theory is given by the Gaussian ensemble, which we recall here for $G=U(N)$:
		\begin{equation}
		\label{eq:ZSYM}
				\mz_{\cs^4} [\mN=4 \text{ SYM}] = \frac{1}{N!} \int_{\R^{N}} \prod_{a=1}^{N}\dd \phi_a ~ \prod_{1 \le a<b \le N} (\phi_a-\phi_b)^2  ~ e^{- g \sum_{a=1}^{N} \phi^2 } .
		\end{equation}
		In our conventions from Section \ref{sec:SphereZ}, we have defined $g= \frac{8 \pi^2}{g_{\text{\tiny YM}}^2 }$ and we set $\lambda = \frac{N}{g}$ the 't Hooft gauge coupling in our normalization. 
		
		\subsubsection*{Wigner's semicircle}
		The potential is 
		\begin{equation*}
			V_{\text{SYM}} (\phi) = \frac{N}{\lambda} \phi^2 \quad \Longrightarrow \quad W^{\prime} _{\text{SYM}} (\phi) = \frac{2}{\lambda} \phi .
		\end{equation*}
		In particular, $W (z)$ is a harmonic well, with no poles nor branch cuts, and admits a unique minimum at $\phi =0$. We then expect the eigenvalues to be gathered in a unique interval around $\phi=0$, leading us to consider a 1-cut ansatz for the eigenvalue density $\rho_{\text{SYM}} ( \phi )$, 
		\begin{equation*}
			\surho = \text{supp} \rho_{\text{SYM}}  = [A,B] .
		\end{equation*}\par
		\medskip
		{\small\textbf{Exercise:} Without any computation, write $A$ as a function of $B$ in this example.}\par
		\medskip
		The integral representation of the planar resolvent 
		\begin{equation*}
			\omega_{\text{SYM}}  (z) =  \sqrt{(A-z)(B-z)} \oint_{\cutrho } \frac{\dd u }{2 \pi i } \frac{ \frac{1}{2} W^{\prime} (u)}{(z-u) \sqrt{(A-u)(B-u)} } 
		\end{equation*}
		can be computed by the residue theorem, after deforming the contour $\cutrho$ to infinity. It only receives the contributions from the pole at $u=z$ and from the pole at $u = \infty $. We get 
		\begin{equation*}
			\omega_{\text{SYM}}  (z) =   \frac{1}{2} W^{\prime} (z) - \frac{1}{\lambda} \sqrt{(A-z)(B-z)} .
		\end{equation*}
		From the discontinuity equation \eqref{eq:disceqrhoomega} we immediately obtain 
		\begin{equation*}
			\rho_{\text{SYM}}  ( \phi ) = \frac{1}{\pi \lambda} \sqrt{(\phi - A)(B-\phi)}  .
		\end{equation*}
		Imposing the normalization condition 
		\begin{equation*}
			\lim_{\lvert z \rvert \to \infty} z \omega_{\text{SYM}}  (z) = 1 
		\end{equation*}
		and comparing with the large $z$ expansion 
		\begin{equation*}
			z \omega_{\text{SYM}}  (z) = \frac{A+B}{2 \lambda} z + \frac{ (A-B)^2}{8 \lambda } + \mathcal{O} (z^{-1}) ,
		\end{equation*}
		we infer the pair of equations 
		\begin{equation*}
			\begin{cases} A+B = 0 \\ \frac{ (A-B)^2}{8 \lambda } = 1 \end{cases}  \quad \Longrightarrow \quad \begin{cases} A = - \sqrt{ 2 \lambda }  \\ B = \sqrt{ 2 \lambda }  .\end{cases}
		\end{equation*}
		The equilibrium measure for $\mN=4$ super-Yang--Mills is 
		\begin{equation*}
			\rho_{\text{SYM}}  ( \phi ) \dd \phi  = \frac{1}{\pi \lambda} \sqrt{2 \lambda - \phi^2 } \dd \phi , \quad \phi \in \left[ - \sqrt{ 2 \lambda }  ,  \sqrt{ 2 \lambda }  \right]  ,
		\end{equation*}
		which goes under the name of Wigner semicircle distribution.

		\subsubsection*{The planar free energy}
			Now that we have obtained the eigenvalue density $\rho_{\text{SYM}}  ( \phi )$, we can use it to evaluate the sphere free energy in the planar limit. We observe that 
			\begin{align*}
				\frac{\partial \ }{\partial \lambda  } \mf_{\cs^4} [\text{SYM}] & = - \frac{1}{\lambda^2} \frac{\partial \ }{\partial ( \lambda^{-1})  } \mf_{\cs^4} [ \text{SYM} ] \\
					& =  \frac{N}{\lambda^2} \frac{\partial \ }{\partial ( N/\lambda ) } \log \mz_{\cs^4} [ \text{SYM} ] \\
					& =  \frac{N}{\lambda^2} \frac{1}{\mz_{\cs^4} [ \text{SYM} ] } \frac{\partial \ }{\partial g } \mz_{\cs^4}  [ \text{SYM} ] \\
					& = - N^2 \left\langle \frac{1}{N \lambda^2 } \tr \phi^2 \right\rangle_{\text{SYM}} ,
			\end{align*}
			where in the last expression we have denoted $\langle \cdot \rangle_{\text{SYM}} $ the average in the Gaussian ensemble that characterizes $\mN=4$ super-Yang--Mills theory. In the planar limit, we arrive at the relation 
			\begin{subequations}
			\begin{align}
				\frac{\partial \ }{\partial  \lambda } \mf_{\cs^4} [ \text{SYM} ]  & = - N^2 \int \frac{ \dd \phi }{\lambda^2} \rho_{\text{SYM}}  ( \phi ) \phi^2 \\
					& = - N^2 \int_{- \sqrt{ 2 \lambda} } ^{\sqrt{ 2 \lambda} } \frac{ \dd \phi }{ \pi \lambda^3 } \phi^2 \sqrt{2 \lambda - \phi^2 } = - \frac{N^2}{2\lambda} .
			\end{align}
			\label{eq:derivFN4SYM}
			\end{subequations}
			We now observe that 
			\begin{equation*}
				\lim_{g \to 0}  \mz_{\cs^4}  [ \text{SYM} ] = + \infty , \quad  \lim_{g \to + \infty }  \mz_{\cs^4}  [ \text{SYM} ] = 0 .
			\end{equation*}
			Combining this with the defining relations $g \propto 1/\lambda $ and $\mf_{\cs^4} [ \text{SYM} ] = - \log \mz_{\cs^4}  [ \text{SYM} ]  $, we can integrate \eqref{eq:derivFN4SYM} with the prescribed boundary condition and we obtain 
			\begin{equation}
			\label{eq:FreeEnergyN4SYM}
				\mf_{\cs^4} [\text{SYM}] = - \frac{N^2}{2} \log \lambda .
			\end{equation}
			We note that, when written in terms of the inverse gauge coupling $g$, \eqref{eq:FreeEnergyN4SYM} shows a $- \frac{1}{2}N^2 \log N$ leading term contribution. We will rederive it later in Section \ref{sec:N4SYMMthLimit} from the M-theory limit.

	\subsection{Planar three-dimensional SQCD}
	
	\begin{figure}[tbh]
		\centering
		\begin{tikzpicture}[auto,square/.style={regular polygon,regular polygon sides=4}]
			\node[circle,draw] (gauge1) at (1,0) {$N$};
			
			\node[square,draw] (fl1) at (2.5,0) { \hspace{6pt} };
			\node[draw=none] (aux1) at (2.5,0) {$\mathsf{F}$};
			
			\draw[-] (fl1) -- (gauge1);
		\end{tikzpicture}
		\caption{SQCD$_3$. The circular node represents the gauge group $U(N)$, the square node represents $\mathsf{F}$ fundamental hypermultiplets.}
		\label{fig:QCD3quiver}
		\end{figure}
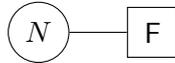\par
		We now focus on three-dimensional gauge theory with gauge group $U(N)$ and $\mathsf{F}$ hypermultiplets in the fundamental representation, known as SQCD$_3$. The quiver is shown in Figure \ref{fig:QCD3quiver}.
		We restrict our attention to the balanced case $\mathsf{F} =2N$ and set the masses of all hypermultiplets to zero for simplicity. More generic cases have been addressed in \cite[Ch.5]{Santilli:2022tjt} (also \cite{Shimizu:2018pnd}).\par
		The sphere partition function is 
		\begin{equation*}
			\mz_{\cs^3} [\text{SQCD}_3] = \frac{1}{N!} \int_{\R^N} \dd \phi \frac{ \prod_{1 \le a < b \le N} \left( 2 \sinh \pi (\phi_a - \phi_b) \right)^2 }{\prod_{a=1}^{N} \left( 2 \cosh \pi \phi_a \right)^{\mathsf{F}} } .
		\end{equation*}
		To recast this expression in the matrix model formalism, we use the change of variables $x_a = e^{2 \pi \phi_a}$ \cite{Tierz:2002jj}. By virtues of the identities 
		\begin{align*}
			 \prod_{1 \le a < b \le N} \left( 2 \sinh \pi (\phi_a - \phi_b) \right)^2 &=  \prod_{1 \le a \ne b \le N} e^{- \pi (\phi_a + \phi_b)} \prod_{1 \le a < b \le N}\left( e^{2 \pi \phi_a} - e^{2 \pi \phi_b} \right)^2 \\
			 &= \left( \prod_{a=1} ^{N} x_a ^{-N+1} \right) \prod_{1 \le a 1 b \le N} (x_a-x_b)^2 , \\
			 \prod_{a=1}^{N} \left( 2 \cosh \pi \phi_a \right)^{\mathsf{F}} &= \prod_{a=1}^{N} \left[e^{-\pi \phi_a} \left( 1+ e^{2 \pi \phi_a}\right) \right]^{\mathsf{F}} \\
			 &= \prod_{a=1}^{N} x_a ^{-\frac{\mathsf{F}}{2} } \left( 1 + x_a \right)^{\mathsf{F}} , \\
			 \prod_{a=1}^{N} \dd \phi_a &= \prod_{a=1}^{N} \frac{ \dd x_a }{2 \pi x_a} ,
		\end{align*}
		specialized to $\mathsf{F} = 2N$, we get 
		\begin{equation*}
			\mz_{\cs^3} [\text{SQCD}_3] = \frac{1}{(2 \pi)^N N!} \int_{\R^N} \dd x \frac{ \prod_{1 \le a < b \le N} (x_a-x_b)^2 }{ \prod_{a=1}^{N} \left( 1 + x_a \right)^{2N}  } .
		\end{equation*}
		In the notation above, the potential is 
		\begin{equation*}
			W(x) = 2 \log (1+x) ,
		\end{equation*}
		whose derivative 
		\begin{equation*}
			\frac{1}{2} W^{\prime} (x) = \frac{1}{1+x}
		\end{equation*}
		satisfies (\hyperref[assump:tH3]{tH3}).\par
		As opposed to the $d=4$ $\mN=4$ case, the potential in this matrix model is not confining. As a consequence, we expect the eigenvalues to spread all along $\phi \in \R$. It is convenient in practice to regularize the matrix model by introducing an auxiliary confining potential 
		\begin{equation}
		\label{eq:WauxSQCD3}
			W_{\text{aux}}(\phi) = \frac{\phi^2}{2\lambda}, 
		\end{equation}
		which corresponds to an imaginary Chern--Simons coupling. In terms of the variable $x$, 
		\begin{equation*}
			W_{\text{aux}}=\frac{1}{2\lambda} (\log x)^2 .
		\end{equation*}\par
		The large $N$ limit of this theory has been worked out in \cite{Barranco:2014tla}. Following the steps outlined above, the saddle point equation in this case is:
		\begin{equation*}
			\mathrm{P}\!\int \dd \tilde{x} \frac{ \hat{\rho} (\tilde{x}) }{x-\tilde{x} } = \frac{1}{2} W_{\mathrm{new}}^{\prime} (x) ,
		\end{equation*}
		where 
		\begin{equation*}
			\frac{1}{2} W_{\mathrm{new}}^{\prime} (x) = \frac{1}{\lambda}\left( \frac{\lambda}{1+x} + \frac{1}{x} \log(x) \right) .
		\end{equation*}
		We are denoting the eigenvalue density for the exponential variable $x$ as $\hat{\rho} (x)$, to distinguish it from $\rho (\phi)$. The measures are related through 
		\begin{equation}
		\label{eq:fromhatrhotorho}
			\hat{\rho}(x) \frac{\dd x}{2\pi} = \rho (\phi) \dd \phi .
		\end{equation}
		Introducing the auxiliary potential \eqref{eq:WauxSQCD3} we have obtained a potential that suppresses the tails of the distribution polynomially, instead of logarithmically, at the expense of getting a $W_{\mathrm{new}}^{\prime} $ which is not meromorphic (in the variable $x$). We are eventually interested in $\lambda \to \infty$.\par
		We start with the single-cut ansatz 
		\begin{equation*}
			\mathrm{supp} \hat{\rho} =[A,B] \subset (0, \infty) .
		\end{equation*}
		We introduce the resolvent $\hat{\omega} (z)$ for $\hat{\rho} (x)$, as explained in Section \ref{sec:MMplanarlimit}. Following the manipulations therein, in particular \eqref{eq:contouromega}, we write it split into three pieces:
		\begin{equation*}
			\hat{\omega} (z) = \frac{1}{2} W_{\mathrm{new}}^{\prime} (z) + \frac{1}{\lambda} \hat{\omega}_{\mathrm{aux}} (z) + \tilde{\omega} (z)  ,
		\end{equation*}
		with the first piece corresponding to picking the pole at $z$, the second piece comes from the contribution of $W_{\mathrm{aux}}^{\prime}$ to residual integration, and the third piece is 
		\begin{equation*}
			\tilde{\omega} (z) = \sqrt{ (z-A)(z-B)} \oint_{\tilde{\mathscr{C}}} \frac{ \dd u}{2\pi i} \frac{1}{(1+u)(z-u)\sqrt{(u-A)(u-B)}} .
		\end{equation*}
		In $\tilde{\omega} $, the integration contour $\tilde{\mathscr{C}}$ encircles $[A,B]$, has the point $z \in \mathbb{P}^1 \setminus [A,B]$ in its interior and the pole $u=-1$ in the exterior.\par
		The part $\hat{\omega}_{\mathrm{aux}} (z)$ includes the non-meromorphic dependence, but only involves the auxiliary part. It can be read off from \cite{Marino:2004eq}. The piece $\tilde{\omega} (z) $, instead, is obtained as explained in Section \ref{sec:MMplanarlimit}. We deform the contour $\tilde{\mathscr{C}}$ to infinity and pick the pole at $u=-1$ along the way. Noting that 
		\begin{equation*}
			\lim_{u \to \infty}\frac{u}{2(1+u)(z-u)\sqrt{(u-A)(u-B)}}  =0,
		\end{equation*}
		there is no contribution from the integration along the circle at infinity $\mathbb{S}^1 _{\infty}$. The only contribution to $\tilde{\omega}$ comes from the residue at the pole $u=-1$ of $W^{\prime}$ and reads 
		\begin{equation*}
			\tilde{\omega} (z) = \sqrt{ \frac{(z-A)(z-B)}{(A+1)(B+1)}} \frac{1}{(z+1)} .
		\end{equation*}
		We thus get 
		\begin{equation*}
			\hat{\omega} (z) = \underbrace{- \frac{1}{1+z} + \sqrt{ \frac{(z-A)(z-B)}{ (1+A)(1+B) } } \frac{1}{1+z} }_{\text{SQCD$_3$ part}} \ + \ \underbrace{\frac{1}{\lambda} \left( \frac{\log z}{z} + \hat{\omega}_{\mathrm{aux}} (z)  \right)}_{\text{aux part}} .
		\end{equation*}\par
		We now need to expand the result for large $\lvert z\rvert$ and obtain two equations to fix $A$ and $B$. In fact, we can use the symmetry $\phi_a \mapsto -\phi_a$ of the matrix model $\mz_{\cs^3} [\text{SQCD}_3]$, preserved by $W_{\text{aux}}$. This translates into a symmetry $x_a\mapsto x_a^{-1}$ for the exponential eigenvalues, and therefore into a symmetry $x \mapsto x^{-1}$ of $\hat{\rho}(x)$. We thus deduce 
		\begin{equation}
		\label{eq:AequalsBinSQCD3}
			A=B^{-1}.
		\end{equation}
		Then, we observe that 
		\begin{equation}
		\label{eq:expandhatomegaSQCD3}
			\hat{\omega} (z) = \frac{1}{2+A+B} + \frac{1}{\lambda} c_{\mathrm{aux}} (A,B) + \mathcal{O}(z^{-1}) ,
		\end{equation}
		where $c_{\mathrm{aux}} (A,B)$ encapsulates the contribution from the auxiliary term. For our purposes, it suffices to know that $c_{\mathrm{aux}}=\mathcal{O}(1)$ in $\lambda$, thus the term $\lambda^{-1}c_{\mathrm{aux}}$ drops out when we send $\lambda \to \infty$.\par
		We stress that the computation of the auxiliary part can be done completely explicitly, and goes exactly as in \cite{Marino:2004eq}. We refer to \cite{Barranco:2014tla} for the more complete expressions. Here we refrain from a thorough analysis of the auxiliary part for the sake of simplicity.\par
		\medskip
		Taking $\lambda \to \infty$ to remove the auxiliary part and plugging \eqref{eq:AequalsBinSQCD3}, we immediately see that \eqref{eq:expandhatomegaSQCD3} together with $\hat{\omega} (z) =\mathcal{O}(z^{-1}) $ implies 
		\begin{equation*}
			\lim_{\lambda \to \infty }B = \infty , \qquad \Longrightarrow \qquad \lim_{\lambda \to \infty }A = 0.
		\end{equation*}\par
		\medskip
		{\small\textbf{Exercise:} Based on the steps in Section \ref{sec:MMplanarlimit}, find $\hat{\omega}_{\mathrm{aux}} (z)$. Compare the result with \cite{Barranco:2014tla}. Expand the expression at large $\lvert z \rvert$ to find $c_{\mathrm{aux}} (A,B)$ explicitly. Write down the equations for $A$ and $B$ derived from \eqref{eq:expandhatomegaSQCD3}. Solve them (with the aid of Mathematica if necessary) and find $A,B$ as functions of the auxiliary coupling $\lambda$. Analyze them in the limit $\lambda \to \infty$.}\par
		\medskip
		With these expressions, we obtain $\hat{\rho}(x)$ in closed form from the jump of $\hat{\omega}$ along $0 < z < \infty$:
		\begin{equation*}
			\hat{\rho}(x)= \frac{1}{\pi} \frac{ \sqrt{(x-A)(B-x)}}{(x+1)\sqrt{(1+A)(B+1)}} + \text{ aux part} .
		\end{equation*}
		Sending $\lambda \to \infty$ to kill the contributions to the auxiliary term we are left with 
		\begin{equation*}
			\hat{\rho}(x)= \frac{\sqrt{x}}{\pi(x+1)} .
		\end{equation*}
		Undoing the change of variables to get $\rho (\phi)$, and taking into account the factor $2\pi$ from \eqref{eq:fromhatrhotorho} the final result is 
		\begin{equation*}
			\rho (\phi) = \frac{1}{\cosh (\pi \phi )} ,
		\end{equation*}
		which is normalized and supported on $\phi \in \R$.

		\subsubsection*{The planar free energy}
		
		Equipped with the density of eigenvalues $\rho(\phi)=(\cosh (\pi \phi))^{-1}$ we can compute the free energy 
		\begin{align*}
			\mf_{\cs^3}  [\text{SQCD}_3, \mathsf{F}=2N] & = 2 N^2 \int_{- \infty}^{+ \infty} \dd \phi \rho (\phi) \log 2 \cosh (\pi \phi ) \\
			&= 4N^2 \log (2) .
		\end{align*}

		\subsubsection*{Exact solution and planar limit}
\begin{figure}
			\centering
			\includegraphics[width=0.5\textwidth]{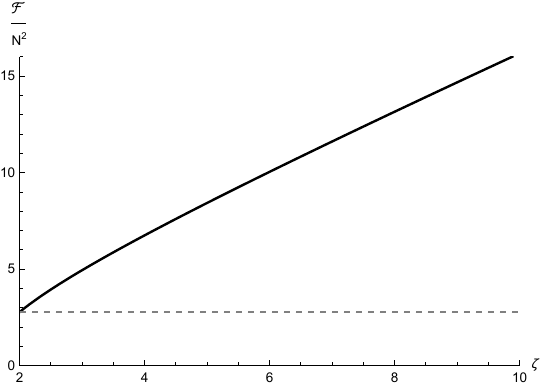}
			\caption{Free energy $\frac{1}{N^2} \mf_{\cs^3} [\text{SQCD}_3]$ as a function of the Veneziano parameter $\zeta = \frac{\mathsf{F}}{N}$ (black). The gray dashed line is the value $4 \log 2$, which intersects the black curve at $\zeta =2$.}
			\label{fig:FSQCD3}
			\end{figure}

			The partition function of SQCD$_3$ with arbitrary $\mathsf{F} \ge 2N $ flavours admits a closed form expression \cite{Tierz:2018fsn}. It follows from recognizing a Selberg integral after the change of variables $x_a = e^{2 \pi \phi_a}$, which can be evaluated to 
			\begin{equation}
			\label{eq:ZSQCD3exact2N}
				\mz_{\cs^3} [\text{SQCD}_3] = \frac{1}{(2 \pi)^N} \frac{ G \left( 1+ \frac{\mathsf{F}}{2}\right)^2 G(1+\mathsf{F} -N)}{  G \left( 1+ \frac{\mathsf{F}}{2} -N \right)^2 G(1+\mathsf{F})}
			\end{equation}
			in terms of the Barnes $G$-function. The asymptotic behaviour of $\log G(1+z)$ is well-known:
			\begin{equation*}
				\log G(1+z) = \frac{z^2}{2} \log (z) - \frac{3}{4} z^2 + \frac{\log (2 \pi)}{2} z - \frac{1}{12} \log (z) + \cdots 
			\end{equation*}
			from which we get \cite{Tierz:2018fsn}
			\begin{equation}
			\label{eq:FSQCD3exact}
				\mf_{\cs^3}  [\text{SQCD}_3] = \frac{\mathsf{F} ^2}{2} \log \left( 2 \frac{\mathsf{F}}{N} \right) + \frac{ (\mathsf{F}-2N)^2}{2} \log \left(  \frac{ \mathsf{F}-2N}{2N} \right) - (\mathsf{F} -N)^2 \log \left(  \frac{ \mathsf{F}-N}{N} \right) .
			\end{equation}
			It is a positive, monotonically increasing function of $\mathsf{F} \ge 2 N$, as required by the $\mf$-theorem. Indeed, it is always possible to reach theories with lower values of $\mathsf{F}$ by decoupling matter fields, and $\mf_{\cs^3}$ should decrease along these RG flows. The plot is shown in Figure \ref{fig:FSQCD3}.\par
			Specializing the right-hand side of \eqref{eq:FSQCD3exact} to the balanced case $\mathsf{F}=2N$ gives $4 N^2 \log (2)$, in agreement with the previous calculation.\par
			\par
			\medskip
			{\small\textbf{Exercise:} Substitute $\mathsf{F}=2N$ in \eqref{eq:ZSQCD3exact2N}. Using the large $z$ behaviour of $\log G(1+z)$, confirm that $\mf_{\cs^3}  [\text{SQCD}_3, \mathsf{F}=2N] = 4 N^2 \log (2)$.}\par

\section{M-theory limit}
\label{sec:LargeNMtheory}

	The idea behind the M-theory limit, at the QFT level, is to take $N \to \infty $ keeping all the couplings fixed. Then, in order to reach an equilibrium configuration, the eigenvalues develop a dependence on $N$ of the form 
	\begin{equation*}
		\phi = N^{\gamma} x , \quad x = \mathcal{O} (1)
	\end{equation*}
	for some $\gamma >0$ to be determined, and the scaled variable $x$ is kept independent of $N$.
	
	\subsubsection*{(Aside) M-theory versus Type IIA}
		The M-theoretic large $N$ limit originates from supersymmetric field theories with a holographic dual in M-theory. These theories typically live in odd spacetime dimensions and therefore, by conformal invariance, the gauge couplings are sent to infinity and drop out of the computation. In this case, the appropriate limit is $N \to \infty$, with the only remaining couplings being the Chern--Simons levels $k_j$, with $j$ labeling the gauge nodes.\par
		On the other hand, theories with a holographic dual in Type IIA string theory are expected to exhibit such duality in the 't Hooft limit, with $t_j = \frac{N}{k_j}$ fixed $\forall j$. In cases that admit both descriptions, it is expected that the strong coupling limit $t_j \to \infty$ will agree with the M-theoretic result. These matters have been discussed in \cite{Santamaria:2010dm,Azeyanagi:2012xj,Grassi:2014vwa}.\par
		\medskip
		The explicit check of the match between the M-theory limit and the strong coupling expansion of the 't Hooft limit in the overlapping regime of validity may be a hard task in general. Here we limit ourselves to a general remark.\par
		Assume that a theory $\Th$ admits both 't Hooft and M-theoretic large $N$ limits. The M-theory limit of $\Th$ yields a free energy 
		\begin{equation*}
			\lim_{\text{M-theory}} \mf [\Th] = N^{p_{\mathrm{M}}} F_{\mathrm{M}} ( \left\{k_j \right\}_j )
		\end{equation*}
		where $p_{\mathrm{M}} \in \mathbb{Q}_{>0}$ is a positive rational power and $F_{\mathrm{M}} $ is a function of the Chern--Simons levels. Let us write the Chern--Simons levels as $k_j = k \kappa_j$ for some rational numbers $\kappa_j$, fixed throughout the entire procedure. We will write the small $k$ expansion of $F_{\mathrm{M}} $ as 
		\begin{equation*}
			 F_{\mathrm{M}} ( \left\{k_j \right\}_j ) = k^{\eta_{\mathrm{M}}}  f^{(0)}_{\mathrm{M}} ( \left\{ \kappa_j \right\}_j ) + \cdots .
		\end{equation*}\par
		The 't Hooft limit of $\Th$ yields instead 
		\begin{equation*}
			\lim_{\text{'t Hooft}} \mf [\Th] = N^{p_{\mathrm{IIA}}} F_{\mathrm{IIA}} ( \left\{ t_j \right\}_j )
		\end{equation*}
		where again $p_{\mathrm{IIA}} \in \mathbb{Q}_{>0}$, generically different from $p_{\mathrm{M}}$, and $t_j=N/k_j$. In the previous notation, the latter expression is recast into a function of a single 't Hooft coupling  $t=N/k$ together with the rational numbers $\left\{ \kappa_j \right\}$. Let us generically write the strong coupling (i.e. large $t$) expansion of $F_{\mathrm{IIA}}$ as 
		\begin{equation*}
			 F_{\mathrm{IIA}} ( \left\{ t_j \right\}_j ) = t^{\eta_{\mathrm{IIA}}}  f^{(0)}_{\mathrm{IIA}} ( \left\{ \kappa_j \right\}_j ) + \cdots .
		\end{equation*}
		A necessary condition for the consistency of the two limits is 
		\begin{align*}
			\left. 	\lim_{\text{M-theory}} \mf [\Th] \right\rvert_{\text{small $k$}} & = \left. \lim_{\text{'t Hooft}} \mf [\Th] \right\rvert_{\text{large $t$}} \\
			 \Longrightarrow \qquad  N^{p_{\mathrm{M}}} k^{\eta_{\mathrm{M}}}  f^{(0)}_{\mathrm{M}} ( \left\{ \kappa_j \right\}_j ) & = \left. N^{p_{\mathrm{IIA}}} t^{\eta_{\mathrm{IIA}}}  f^{(0)}_{\mathrm{IIA}} ( \left\{ \kappa_j \right\}_j ) \right\rvert_{t=N/k} .
		\end{align*}
		Once we plug $t= N/k$ into $F_{\mathrm{IIA}}$, the necessary conditions become:
		\begin{align*}
			p_{\mathrm{IIA}} + \eta_{\mathrm{IIA}} = p_{\mathrm{M}} , \qquad \eta_{\mathrm{IIA}} = - \eta_{\mathrm{M}} , \qquad f^{(0)}_{\mathrm{IIA}} = f^{(0)}_{\mathrm{M}} .
		\end{align*}
		Checking that these identities are satisfied guarantees a match between the M-theory limit and 't Hooft limit results at leading order in the $1/k$ expansion. Refined tests beyond leading order have been carried out in \cite{Hong:2021bsb,Beccaria:2023hhi}.

	\subsection{M-theory limit of four-dimensional \texorpdfstring{$\mN=4$}{N=4} super-Yang--Mills}
	\label{sec:N4SYMMthLimit}
		As a warm-up example to begin with and to gain insight into how the machinery works, we derive in this subsection the large $N$ limit of $\mN=4$ super-Yang--Mills in four dimensions, by taking what we call the M-theory limit. The results are shown to agree with the 't Hooft limit analysis of Section \ref{sec:PlanarN4SYM}.\par
		Recall that the partition function of $U(N)$ $\mN=4$ super-Yang--Mills is 
		\begin{align*}
			 \mz_{\cs^4}  [ \text{SYM} ] & = \int_{\R^N} \dd \phi ~ e^{-S_{\mathrm{eff}} ^{\text{SYM}} (\phi) }   , \\
			S_{\mathrm{eff}} ^{\text{SYM}} (\phi) & = -  \sum_{1 \le a \ne b \le N} \log \lvert \phi_a - \phi_b \rvert  + g \sum_{a=1}^{N} \phi_a ^2  .
		\end{align*}
		In taking the large $N$ limit with $g$ fixed, we make the scaling ansatz 
		\begin{equation*}
			\phi = N^{\gamma} x 
		\end{equation*}
		for fixed $x$ and $\gamma >0$. Besides, we define the density of scaled eigenvalues $\varrho (x)$. In this way, the effective action becomes 
		\begin{equation*}
			S_{\mathrm{eff}} ^{\text{SYM}} (\phi) = - N^2 \log N^{\gamma} - N^2 \int \dd x \varrho (x) \int \dd y \varrho (y)  \log \lvert x-y\rvert + g N^{1 + 2 \gamma} \int \dd x \varrho (x) x^2 .
		\end{equation*}
		For the second and third terms to compete in determining a non-trivial equilibrium configuration, the corresponding powers of $N$ must match. This leads us to the condition 
		\begin{equation*}
			2 = 1 + 2 \gamma \quad \Longrightarrow \quad  \gamma = \frac{1}{2} .
		\end{equation*}
		From here, we already match the $g$-independent term of the free energy $- \frac{1}{2} N^2 \log N$ with the derivation in the 't Hooft limit \eqref{eq:FreeEnergyN4SYM}.\par
		The saddle point equation in this case is 
		\begin{equation*}
			\mathrm{P}\!\int \dd y \frac{\varrho (y)}{x-y} = g x .
		\end{equation*}
		At this point, the solution is identical to Section \ref{sec:PlanarN4SYM} with only two modifications: 
		\begin{itemize}
			\item Replace, from the SPE on, the variables $\phi,\sigma$ with $x,y$, and the measure $\rho (\phi) \dd \phi $ with $\varrho (x) \dd x$;
			\item Replace the 't Hooft coupling $\lambda$ with the fixed gauge coupling $1/g$.
		\end{itemize}
		
		\subsubsection*{(Aside) M-theory limit and $\mN=4$ super-Yang--Mills}
			The physical significance of this example is limited. We do not expect the M-theory limit to shed any light on the holographic dual to $\mN=4$ super-Yang--Mills in four dimensions, and the 't Hooft limit of Section \ref{sec:PlanarN4SYM} is the correct procedure.\par
			We have nevertheless decided to present the result here to show explicitly in a simple example how certain models admit more than one type of limit. In the overlapping regime of validity, the different approaches must produce equal answers.
		
		\subsubsection*{(Aside) Exact solution}
			The partition function \eqref{eq:ZSYM} is ultimately the normalization constant of the Gaussian unitary ensemble. Changing variables $\tilde{\phi}_a = \sqrt{g} \phi_a$ produces the changes:
			\begin{align*}
				\prod_{a=1}^{N} \dd \phi_a &= \left(\frac{1}{\sqrt{g}} \right)^N \prod_{a=1}^{N} \dd \tilde{\phi}_a , \\
				g \sum_{a=1}^{N} \phi_a^2 &=  \sum_{a=1}^{N} \tilde{\phi}_a^2 , \\
				\prod_{1 \le a<b \le N} (\phi_a-\phi_b)^2 &= \left(\frac{1}{\sqrt{g}} \right)^{N^2-N} \prod_{1 \le a<b \le N} (\tilde{\phi}_a-\tilde{\phi}_b)^2.
			\end{align*}
			Therefore \eqref{eq:ZSYM} equals (omitting the tilde to reduce clutter) 
			\begin{equation*}
				\mz_{\cs^4} [\mN=4 \text{ SYM}] = g^{-\frac{N^2}{2}} \left[ \frac{1}{N!} \int_{\R^{N}} \prod_{a=1}^{N}\dd \phi_a ~ \prod_{1 \le a<b \le N} (\phi_a-\phi_b)^2  ~ e^{- \sum_{a=1}^{N} \phi^2 } \right] ,
			\end{equation*}
			with the integral in square brackets independent of the gauge coupling. It is a Gaussian integral over the space of $N \times N$ Hermitian matrices, thus exactly solvable. It yields:
			\begin{equation*}
				\mz_{\cs^4} [\mN=4 \text{ SYM}] = g^{-\frac{N^2}{2}} \left[ 2^{-\frac{N^2}{2}} (\sqrt{2\pi})^{N} G(N+1) \right] = (2\pi)^{\frac{N}{2}} \frac{G(N+1)}{(2g)^{N^2/2} } ,
			\end{equation*}
			with $G(\cdot)$ Barnes's $G$-function. The large $N$ limit at fixed $g$ is dominated by the asymptotic behaviour $\log G(N+1) = \frac{N^2}{2} \log (N) + \mathcal{O} (N^2)$; in turn, substituting $g= \frac{N}{\lambda}$, the denominator grows with $N^{N^2/2}$ and competes with the $G$-functions, canceling the $N^2\log N$ piece. Thus,
			\begin{equation*}
				\mf_{\cs^4} [\text{SYM}] = \begin{cases} - \frac{N^2}{2} \log N & \text{large $N$, fixed $g$} \\ - \frac{N^2}{2} \left( \log \lambda + c_{\text{norm}} \right) & \text{large $N$, 't Hooft limit} \end{cases}
			\end{equation*}
			where $c_{\text{norm}}$ is a numerical constant that depends solely on the choice of normalization.

	\subsection{M-theory limit in three dimensions: ABJM theory}
		
		\begin{figure}[tbh]
		\centering
		\begin{tikzpicture}[auto]
			\node[circle,draw] (gauge1) at (1,0) {$N$};
			\node[circle,draw] (gauge2) at (-1,0) {$N$};

			\path (gauge1) edge [bend left] (gauge2);
			\path (gauge2) edge [bend left] (gauge1);
			
			\node[anchor=north] (k1) at (gauge1.south east) {${}_{-k}$};
			\node[anchor=north] (k2) at (gauge2.south east) {${}_k$};
		\end{tikzpicture}
		\caption{ABJM theory, drawn as a circular quiver with two nodes and Chern--Simons levels $\pm k$.}
		\label{fig:ABJM1}
		\end{figure}
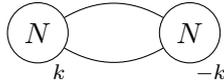\par
			We start the discussion in three dimensions computing the sphere free energy of ABJM theory \cite{ABJM} at large $N$ and fixed Chern--Simons coupling \cite{Herzog:2010hf}.\par
			ABJM theory is a $U(N)_k \times U(N)_{-k}$ Chern--Simons quiver with two bifundamental hypermultiplets. It is represented as a circular quiver with two nodes, shown in Figure \ref{fig:ABJM1}.\par
			The ABJM partition function is 
			\begin{equation*}
			\begin{aligned}
				\mz_{\cs^3} [\text{ABJM}] = \frac{1}{(N!)^2} & \int_{\R^{N}} \dd \phi \int_{\R^{N}} \dd \tilde{\phi} ~ e^{i \pi k \sum_{a=1}^{N} \left( \phi_a ^2 - \tilde{\phi}_a ^2 \right)} \\
				\times & \frac{\prod_{1 \le a \ne b \le N}  \left\lvert  2 \sinh \pi (\phi_a - \phi_b ) \right\rvert \cdot \left\lvert 2  \sinh \pi (\tilde{\phi}_a - \tilde{\phi}_b ) \right\rvert }{ \prod_{a=1}^{N} \prod_{b=1}^{N} \left[ 2 \cosh \pi (\phi_a - \tilde{\phi}_a) \right]^2 } .
			\end{aligned}
			\end{equation*}
			Notice the square in the denominator, that comes from the fact that two copies of the bifundamental representation are taken. We now proceed to review the derivation of C. Herzog, I. Klebanov, S. Pufu and T. Tesileanu \cite{Herzog:2010hf}, who pioneered the method for solving $\mN \ge 2$ Chern--Simons-matter theories at large $N$.\par
			As customary, we write
			\begin{equation*}
				\mz_{\cs^3} [\text{ABJM}] = \frac{1}{(N!)^2} \int_{\R^{N}} \dd \phi \int_{\R^{N}} \dd \tilde{\phi} ~ e^{-S_{\mathrm{eff}}^{\text{ABJM}} (\phi, \tilde{\phi})} ,
			\end{equation*}
			with 
			\begin{equation*}
			\begin{aligned}
				S_{\mathrm{eff}}^{\text{ABJM}} = & - \sum_{1 \le a \ne b \le N} \left[ \log \left( \left\lvert  2 \sinh \pi (\phi_a - \phi_b ) \right\rvert \right) + \log \left(\left\lvert 2  \sinh \pi (\tilde{\phi}_a - \tilde{\phi}_b ) \right\rvert \right) \right] \\
					&+ \sum_{a=1}^{N} i\pi k\left( \phi_a^2 - \tilde{\phi}_a^2 \right) + \sum_{a,b=1}^{N} 2 \log \left( 2 \cosh \pi (\phi_a - \tilde{\phi}_b)\right) .
			\end{aligned}
			\end{equation*}
			Next, we approximate the sums over discrete indices $a,b$ with integrals over the continuous variable $\mathsf{a}= \frac{a}{N}$. The eigenvalues are then replaced by functions, 
			\begin{equation*}
				\left\{ \phi_a , \ a=1, \dots, N \right\} \qquad  \mapsto \qquad \left\{ \phi (\mathsf{a}) , \ 0<\mathsf{a}\le 1 \right\} ,
			\end{equation*}
			and likewise for $\tilde{\phi} (\mathsf{a})$. We get 
			\begin{equation*}
			\begin{aligned}
				S_{\mathrm{eff}}^{\text{ABJM}} = & - N^2 \int_0^1 \dd \mathsf{a} \int_0^1 \dd \mathsf{b}\left[ \log \left( \left\lvert  2 \sinh \pi (\phi (\mathsf{a}) - \phi (\mathsf{b}) ) \right\rvert \right) + \log \left(\left\lvert 2  \sinh \pi (\tilde{\phi} (\mathsf{a}) - \tilde{\phi} (\mathsf{b}) ) \right\rvert \right) \right] \\
					&+ N^2 \int_0^1 \dd \mathsf{a} \left[ \frac{i\pi k}{N} \left( \phi (\mathsf{a})^2 - \tilde{\phi} (\mathsf{a})^2 \right) + \int_0 ^1 \dd \mathsf{b} 2 \log \left( 2 \cosh \pi (\phi (\mathsf{a}) - \tilde{\phi} (\mathsf{b}))\right) \right] .
			\end{aligned}
			\end{equation*}
			At this point we note that the contributions from the two gauge nodes only differ by $k \leftrightarrow -k$, and that this term is the only imaginary contribution to the effective action. The natural ansatz for the large $N$ limit is thus \cite{Herzog:2010hf} 
			\begin{equation}
			\label{eq:ABJMAnsatz}
				 2 \pi \phi (\mathsf{a}) = N^{\gamma} x(\mathsf{a}) + i y(\mathsf{a}) , \qquad 	2 \pi \tilde{\phi}(\mathsf{a}) = N^{\gamma} x(\mathsf{a}) - i y(\mathsf{a}) ,
			\end{equation}
			for some $\gamma >0$.\par
			The insight of \cite{Herzog:2010hf} is that, instead of introducing the two eigenvalue densities $\rho_1 (\phi), \rho_2 (\tilde{\phi})$, it is more convenient to use the rewriting \eqref{eq:ABJMAnsatz} and parametrize the solution with two functions $\varrho (x)$ and $y(x)$, with $\varrho (x) \dd x = \dd \mathsf{a} $ and $y(x)$ obtained from $y(\mathsf{a})$ after implicitly inverting the relation and writing $\mathsf{a}=\mathsf{a}(x)$. The factors of $2 \pi $ in the left-hand side of \eqref{eq:ABJMAnsatz} are included to make contact with the conventions in the literature, and to reduce clutter in the subsequent expressions.\par

		\subsubsection*{Scaled effective action of ABJM}
			In the large $N$ limit, the previous discussion leads us to look for a pair of functions: the scaled eigenvalue density $\varrho (x)$ and the distribution of the imaginary part $y(x)$.\par
			With the ansatz \eqref{eq:ABJMAnsatz}, the quadratic piece in $x$ from the Chern--Simons term cancels, 
			\begin{equation*}
			\begin{aligned}
				i\pi k N\int_0^1 \dd \mathsf{a} \left( \phi (\mathsf{a})^2 - \tilde{\phi} (\mathsf{a})^2 \right) \ \mapsto \ & i \frac{k}{4 \pi} N\int \dd x \varrho(x) \left[ (N^{\gamma} x +iy)^2 - (N^{\gamma} x -iy)^2 \right] \\
				= & i  \frac{k}{4 \pi} N^{1+\gamma} \int \dd x \varrho(x) (4 i x) y(x) \\
				= & - N^{1+\gamma} \frac{k}{\pi}  \int \dd x \varrho(x) y(x) x .
			\end{aligned}
			\end{equation*}\par
			\begin{figure}[htb]
			\centering
				\includegraphics[width=0.4\textwidth]{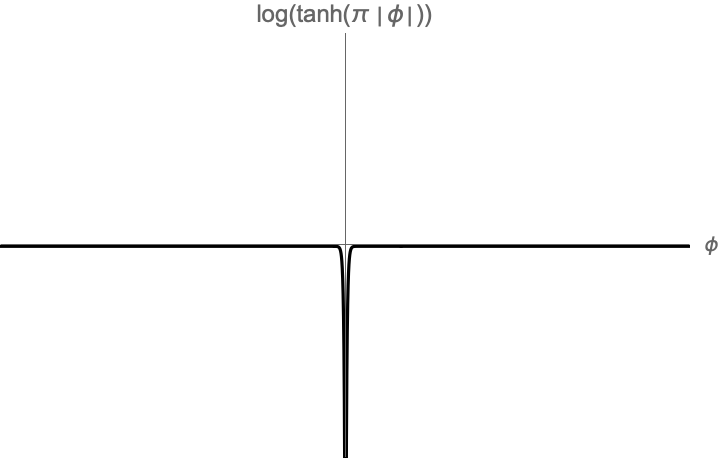}
				\caption{The contribution of the vector multiplets and bifundamental hypermultiplets combines into $\log \tanh \pi (\phi - \tilde{\phi} )$. For large argument, it vanishes everywhere except at $\tilde{\phi} - \phi =0$.}
			\label{fig:LogTanhCancel}
			\end{figure}\par
			The contribution from vector and bifundamental hypermultiplets is 
			\begin{align*}
				\int \dd x \varrho (x) & \left\{  \underbrace{- \int_{\tilde{x} \ne x}  \dd \tilde{x} \varrho (\tilde{x}) \log \left\lvert 2 \sinh \frac{1}{2} \left[ \left( N^{\gamma}  x +i y(x) \right) -  \left( N^{\gamma} \tilde{x} +i y(\tilde{x}) \right) \right] \right\rvert}_{\text{vec. $1^{\text{st}}$ node}} \right. \\
					& \ \underbrace{- \int_{\tilde{x} \ne x}  \dd \tilde{x} \varrho (\tilde{x}) \log \left\lvert 2 \sinh \frac{1}{2} \left[ \left( N^{\gamma}  x -i y(x) \right) -  \left( N^{\gamma} \tilde{x} -i y(\tilde{x}) \right) \right] \right\rvert}_{\text{vec. $2^{\text{nd}}$ node}} \\
					& \left. \underbrace{ \int \dd \tilde{x} \varrho (\tilde{x}) \log \left( 2 \cosh \frac{1}{2} \left[ \left( N^{\gamma}  x +i y(x) \right) -  \left( N^{\gamma}  \tilde{x} -i y(\tilde{x}) \right) \right] \right)^2 }_{\text{bifund. hyper}} \right\} .
			\end{align*}
			We now write 
			\begin{equation*}
			\begin{aligned}
				 \sinh \frac{1}{2} \left[ \left( N^{\gamma}  x +i y(x) \right) -  \left( N^{\gamma} \tilde{x} +i y(\tilde{x}) \right) \right] &= \cos \left( \frac{y(x)-y(\tilde{x})}{2} \right) \sinh \frac{N^{\gamma}}{2} (x-\tilde{x}) \\ & + i \sin \left( \frac{y(x)-y(\tilde{x})}{2} \right) \cosh \frac{N^{\gamma}}{2} (x-\tilde{x}) , \\
				 \sinh \frac{1}{2} \left[ \left( N^{\gamma}  x +i y(x) \right) -  \left( N^{\gamma} \tilde{x} +i y(\tilde{x}) \right) \right] &= \cos \left( \frac{y(x)-y(\tilde{x})}{2} \right) \sinh \frac{N^{\gamma}}{2} (x-\tilde{x}) \\ & - i \sin \left( \frac{y(x)-y(\tilde{x})}{2} \right) \cosh \frac{N^{\gamma}}{2} (x-\tilde{x}) , \\
				 \cosh \frac{1}{2} \left[ \left( N^{\gamma}  x +i y(x) \right) -  \left( N^{\gamma} \tilde{x} -i y(\tilde{x}) \right) \right] &= \cos \left( \frac{y(x)+y(\tilde{x})}{2} \right) \cosh \frac{N^{\gamma}}{2} (x-\tilde{x}) \\ & + i \sin \left( \frac{y(x)+y(\tilde{x})}{2} \right) \sinh \frac{N^{\gamma}}{2} (x-\tilde{x}) .
			\end{aligned}
			\end{equation*}
			For $x \ne \tilde{x}$, we can approximate at large $N$ and see that the three terms behave, respectively, as 
			\begin{equation*}
				e^{\frac{N^{\gamma}}{2} (x-\tilde{x}) + i  \frac{y(x)-y(\tilde{x})}{2} } , \quad e^{\frac{N^{\gamma}}{2} (x-\tilde{x}) - i  \frac{y(x)-y(\tilde{x})}{2} } , \quad e^{\frac{N^{\gamma}}{2} (x-\tilde{x}) + i  \frac{y(x)+y(\tilde{x})}{2} } .
			\end{equation*}\par
			\medskip
			{\small\textbf{Exercise:} Derive these expressions from the large $N$ behaviour of the functions above.}\par
			\medskip
			Taking the logarithms with the corresponding signs, the leading order contributions cancel everywhere if $x \ne \tilde{x}$, see Figure \ref{fig:LogTanhCancel}. This effect leaves behind a $\delta$-function $\delta (N^{\gamma}(x-\tilde{x})) = N^{-\gamma} \delta (x-\tilde{x})$. 
			The double integral then undergoes a simplification and reduces to a single integral, schematically: 
			\begin{equation*}
			\begin{aligned}
				N^2 \int \dd x \varrho  (x) \int_{\tilde{x} \ne x}  \dd \tilde{x} \varrho (\tilde{x}) \log \left[ \cdots \right] & \approx N^2 \int_{\tilde{x} \ne x}  \dd \tilde{x} \varrho (\tilde{x}) \delta \left( N^{\gamma}(x-\tilde{x}) \right) \log \left[ \cdots \right] \\
				& = N^{2-\gamma} \int \dd x \varrho  (x)^2 \log \left[ \cdots \right]_{\tilde{x}=x}.
			\end{aligned}
			\end{equation*}
			After manipulations and simplifications of the logarithmic piece, we get explicitly 
			\begin{equation*}
				N^{2-\gamma} \int \dd x \varrho  (x)^2 \log \left[ \cdots \right]_{\tilde{x}=x} = N^{2-\gamma} \int \dd x \varrho  (x)^2  \left[ \pi^2 - 4 y(x)^2 \right].
			\end{equation*}\par
			\medskip
			{\small\textbf{Exercise:} Derive the term $\pi^2-4 y(x)^2$. (Hint: compare with \cite[App. A]{Herzog:2010hf}).}\par
			\medskip
			Putting all the pieces together we obtain the effective action 
			\begin{equation}
			\label{eq:SeffABJMLargeN}
				S_{\mathrm{eff}} ^{\text{ABJM}} =  \frac{k}\pi N^{1+\gamma} \int \dd x \varrho(x) x y(x)  +  N^{2-\gamma} \int \dd x \varrho (x)^2 \left[ \pi^2 - 4 y(x)^2 \right] .
			\end{equation}
			For the two terms to compete and reach a non-trivial saddle point configuration we need 
			\begin{equation*}
				1+\gamma = 2-\gamma \quad \Longrightarrow \quad \gamma = \frac{1}{2} .
			\end{equation*}
			Plugging the value $\gamma= \frac{1}{2}$ in \eqref{eq:SeffABJMLargeN} we already expect the $N^{\frac{3}{2}}$ behaviour of the free energy, in agreement with the M-theory prediction.\par
			
		\subsubsection*{Saddle point equation}
			Following \cite{Herzog:2010hf} we enforce the normalization condition at the level of the effective action by introducing a Lagrange multiplier into \eqref{eq:SeffABJMLargeN}:
			\begin{equation*}
				S_{\mathrm{eff}} ^{\text{ABJM}} \ \mapsto \ S_{\mathrm{eff}} ^{\text{ABJM}} + N^{\frac{3}{2}} \frac{\xi}{2 \pi } \left[ \int \dd x \varrho (x)  - 1 \right] .
			\end{equation*}
			Notice that, in order to keep track of the normalization constraint at large $N$, we are working already with the scaled variable $x$ and consider an overall scaling of the Lagrange multiplier with a factor $N^{\frac{3}{2}}$, so that the constraint is enforced at leading order.\par
			Functional differentiation to minimize with respect to both $\varrho (x)$ and $y(x)$ yields 
			\begin{align*}
				\frac{k}{\pi} x y(x) + 2 \varrho (x) \left[ \pi^2 - 4 y(x)^2 \right]  + \frac{\xi}{2 \pi} & = 0 , \\
				\frac{k}{\pi} x \varrho (x) - 8  \varrho (x)^2 y(x) & = 0 .
			\end{align*}
			The system is solved by 
			\begin{equation}
			\label{eq:SolyandrhoABJM}
				y (x) = - \frac{\pi^2}{2 \xi} k x , \quad \varrho (x) = - \frac{\xi}{4 \pi^3 } .
			\end{equation}
			From the symmetries of the problem, inherited by the ansatz \eqref{eq:ABJMAnsatz}, $\varrho (x)$ has a symmetric support $[-B,B]$. Plugging the constant solution into the normalization condition, we express $B$ as a function of the Lagrange multiplier $\xi$:
			\begin{equation*}
				1 = \int_{-B} ^{B} \dd x \varrho (x) = \int_{-B} ^{B} \dd x \left(  - \frac{\xi}{4 \pi^3 } \right) \quad \Longrightarrow \quad B = - \frac{ 2 \pi^3}{\xi} .
			\end{equation*}
			
		\subsubsection*{Free energy}
			With these ingredients, we can evaluate the effective action \eqref{eq:SeffABJMLargeN} at the saddle point, expressed as a function of the Lagrange multiplier $\xi$. Plugging \eqref{eq:SolyandrhoABJM} into \eqref{eq:SeffABJMLargeN} and integrating, we find 
			\begin{align*}
				\mathcal{F}_{\cs^3} [\text{ABJM}] (\xi) &= N^{\frac{3}{2}}\int_{-B}^{B} \dd x \varrho (x) \left[  \underbrace{\frac{k}{\pi} x y(x) }_{\text{CS}} + \underbrace{ \varrho (x) (\pi^2 - y(x)^2) }_{\text{ vec. + hyp.}}   \right] \\
					&= N^{\frac{3}{2}}  \left(  - \frac{\xi}{4 \pi^3 } \right) \int_{-B}^{B} \dd x \left\{ \frac{k}{\pi} x \left( - \frac{\pi^2}{2 \xi} k x \right) +  \left(  - \frac{\xi}{4 \pi^3 } \right) \left[ \pi^2 - \left( - \frac{\pi^2}{2 \xi} k x \right)^2  \right] \right\} \\
					& = - N^{\frac{3}{2}}\left[  \frac{\xi}{4 \pi} + \frac{ \pi^{7} k^2}{3 \xi^3} \right] .
			\end{align*}
			The remaining step is to minimize this quantity with respect to $\xi$. The extremum is attained at $\xi_{\ast}$ determined by
			\begin{equation*}
				\frac{1}{4 \pi} - \frac{ \pi^{7} k^2}{\xi_{\ast}^4} =0 .
			\end{equation*}
			The solution is $\xi_{\ast} = - \pi^2 \sqrt{2  k} $, which yields 
			\begin{equation*}
				y(x) = \sqrt{\frac{k}{8}} x, \quad \varrho (x) = \frac{1}{\pi} \sqrt{\frac{k}{8}} , \quad x \in \left[ -  \pi  \sqrt{\frac{2}{k}} ,\pi  \sqrt{\frac{2}{k}}  \right] .
			\end{equation*}\par
			The free energy at leading order is 
			\begin{equation}
			\label{eq:ABJMFMth}
				\mathcal{F}_{\cs^3} [\text{ABJM}] = N^{\frac{3}{2}} \frac{\pi}{3} \sqrt{2k} .
			\end{equation}\par

		\subsubsection*{M-theory limit versus planar limit}
			The result \eqref{eq:ABJMFMth} matches the M-theory prediction, valid in the large $N$ limit and fixed $k$. The ABJM matrix model has also been solved in the 't Hooft limit, sending $N \to \infty$ with $t= \frac{N}{k}$ fixed \cite{Drukker:2010nc}. The 't Hooft limit successfully probes the type IIA regime. As a side remark, notice that replacing $k= N/t$ in \eqref{eq:ABJMFMth} we get the $N^2$ behaviour that characterizes the planar limit, a necessary condition for the agreement of the two procedures at strong coupling.\par
		
		\subsubsection*{(Aside) Matrix models and internal geometry}
			A relation between the eigenvalue densities in the M-theory limit and the geometry of the internal space in M-theory has been recently highlighted \cite{Boido:2023ojv}. The result applies to three-dimensional theories with $\mN \ge 2$ supersymmetry, which live on an M2-brane placed at the tip of a four-complex dimensional conical singularity. The holographic dual to this theory is AdS$_4 \times M_7$ where $M_7$ is a Sasaki--Einstein manifold. The work \cite{Boido:2023ojv} then relates the functions $\varrho (x)$ and $y(x)$ to the geometry of $M_7$ and its embedding in M-theory. This intriguing correspondence begs for further elucidation and a deeper understanding.

		\subsection{(Aside) M-theory limit in three dimensions: Circular Chern--Simons quivers}	
			The procedure used to solve the large $N$ limit of the ABJM free energy in the M-theory regime can be extended to study circular quiver Chern--Simons theories $\Th_{L, \vec{k}}$ \cite{Herzog:2010hf}. We take the gauge group to be 
			\begin{equation*}
				U(N)_{k_1} \times U(N)_{k_2} \times \cdots \times U(N)_{k_L} ,
			\end{equation*}
			with the further condition \cite{Imamura:2008nn,Jafferis:2008qz}
			\begin{equation*}
				\sum_{j=1} ^{L} k_j =0 .
			\end{equation*}
			Physically, this requirement guarantees that these theories have holographic duals in M-theory. The scenario $\sum_j k_j \ne 0$ would entail a holographic dual in Type IIA string theory, amenable to a 't Hooft limit. This case was considered explicitly in \cite{Jafferis:2011zi}.\par
			The hypermultiplet content consists of a bifundamental representation of $U(N)_{k_j} \times U(N)_{k_{j+1}}$, $\forall j=1, \dots , L$, with periodic identification $L+1 \sim 1$. Notice that the addition of a bifundamental between the last and first gauge nodes closes the quiver in a circle.\par
			The partition function of these theories is 
			\begin{equation*}
				\mz_{\cs^3} [\Th_{L, \vec{k}}] = \frac{1}{(N!)^L} \int_{\R^{NL}} \dd \phi ~ \prod_{j=1} ^{L} e^{i \pi  k_j \sum_{a=1}^{N} \phi_{j,a}^2 } \frac{ \prod_{1 \le a \ne b \le N} 2 \lvert \sinh \pi (\phi_{j,a} - \phi_{j,b} ) \rvert }{ \prod_{a,b=1}^{N} 2 \cosh \pi (\phi_{j,a} - \phi_{j+1,b}) } .
			\end{equation*}
			Reasoning as above, it is a simple exercise to find the effective action at large $N$:
			\begin{align*}
				S_{\mathrm{eff}} ^{\Th_{L,\vec{k}}} = \sum_{j=1} ^{L} & \left\{  \underbrace{ i \pi  k_j N \int \dd \phi \rho_j (\phi) \phi^2 }_{\text{CS term}} - \underbrace{ N^2 \int  \dd \phi \rho_j (\phi) \int_{\sigma \ne \phi}  \dd \sigma \rho_j (\sigma ) \log \lvert 2 \sinh \pi (\phi - \sigma ) \rvert }_{\text{vec.}} \right. \\
					& \left. \ +  \underbrace{ N^2 \int  \dd \phi \rho_j (\phi) \int  \dd \sigma \rho_{j+1} (\sigma ) \log  2 \cosh \pi (\phi - \sigma )  }_{\text{bifund. hyper.}} \right\} .
			\end{align*}
			This leads to a set of $L$ coupled SPEs 
			\begin{equation}
			\label{eq:SPECircQuiverCS3d}
				i \frac{2 \pi k_j}{N} \phi - 2 \int_{\sigma \ne \phi } \dd \sigma \rho_j (\sigma) \coth \pi (\phi- \sigma) + \int \dd \sigma \left[ \rho_{j-1} (\sigma) +  \rho_{j+1} (\sigma) \right]\tanh \pi (\phi- \sigma) = 0,
			\end{equation}
			$\forall j=1, \dots, L$. We again make an ansatz of the form 
			\begin{equation*}
				\frac{\phi}{2 \pi} = N^{\gamma} x + i y (x) . 
			\end{equation*}
			We thus look for the collection of pairs $\left\{ \varrho_j (x), y_j (x)\right\}_{j=1, \dots, L}$. By the assumption $\sum_{j=1} ^{L} k_j =0$, the term $k_j x^2$ cancels from the effective action, exactly as in ABJM.\par
			There is a remaining $\mathcal{O}(N^2)$ term in the effective action, which leads to the hyperbolic functions in \eqref{eq:SPECircQuiverCS3d}. For large argument, these functions are approximated by sign functions,
			\begin{equation*}
				\coth \pi N^{\gamma}(x -\tilde{x}) \approx \mathrm{sgn} (x -\tilde{x}) , \qquad \tanh \pi N^{\gamma}(x -\tilde{x}) \approx \mathrm{sgn} (x -\tilde{x}) .
			\end{equation*}
			At this point, the $\mathcal{O}(N^2)$ terms in effective action contribute to the SPEs \eqref{eq:SPECircQuiverCS3d} as 
			\begin{equation*}
				\int \dd \tilde{x} \left[ -2 \varrho_j (\tilde{x}) +  \varrho_{j-1} (\tilde{x}) + \varrho_{j+1} (\tilde{x}) \right] \mathrm{sgn} (x-\tilde{x}) = 0 , \quad \forall j=1, \dots, L ,
			\end{equation*}
			solved by identical eigenvalue densities $\varrho_j (x)= \varrho (x)$, $\forall j=1, \dots, L$.\par
			We now look at the rest of the contributions, as in ABJM. We use the $\mathcal{O} (N^2)$ constraint that all eigenvalue densities are equal, and write the effective action in the form 
			\begin{align*}
				S_{\mathrm{eff}} ^{\Th_{L,\vec{k}}} =  & \underbrace{ N^{1+\gamma} \int \dd x \varrho (x)\sum_{j=1} ^{L}  x y_j (x) \frac{k_j}{2\pi} }_{\text{CS term}}  \\
					& \ +  \underbrace{ \frac{N^{2-\gamma}}{2} \int  \dd x \varrho (x)^2  \sum_{j=1}^L \left[ \pi^2 - (y_j (x) - y_{j+1} (x))^2 \right] }_{\text{vec. + bifund. hyper.}} .
			\end{align*}
			Here we have used the same simplifications as in ABJM. Note that the last term involves the function $\pi^2 - (y_j (x) - y_{j+1} (x))^2$, and for ABJM we have $y_{2}(x)=-y_1(x)$ by symmetry.\par
			For the theory to admit a non-trivial saddle point configuration we have $\gamma = \frac{1}{2}$. This already predicts a behaviour $\mathcal{O}(N^{\frac{3}{2}})$, in agreement with the predicted behaviour for a three-dimensional theory with an M-theory dual.\par
			\medskip
			{\small\textbf{Exercise:} Check the derivation of the effective action and compare with ABJM.}\par
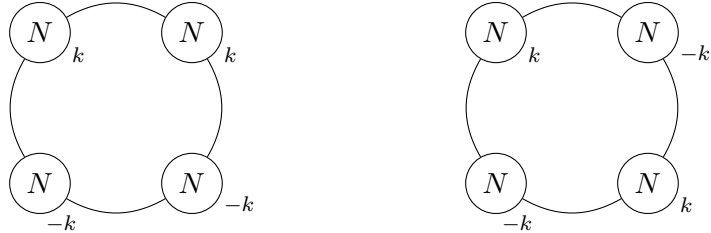
\begin{figure}[tbh]
			\centering
			\begin{tikzpicture}[auto]
			\node[circle,draw] (gauge1) at (-2,0) {$N$};
			\node[circle,draw] (gauge2) at (-4,0) {$N$};
			\node[circle,draw] (gauge4) at (-2,2) {$N$};
			\node[circle,draw] (gauge3) at (-4,2) {$N$};

			\path (gauge1) edge [bend left] (gauge2);
			\path (gauge2) edge [bend left] (gauge3);
			\path (gauge3) edge [bend left] (gauge4);
			\path (gauge4) edge [bend left] (gauge1);
			
			\node[anchor=west] (k1) at (gauge1.south east) {${}_{-k}$};
			\node[anchor=north] (k2) at (gauge2.south east) {${}_{-k}$};
			\node[anchor=west] (k3) at (gauge3.south east) {${}_{k}$};
			\node[anchor=west] (k4) at (gauge4.south east) {${}_k$};
			
			\node[circle,draw] (hauge1) at (2,0) {$N$};
			\node[circle,draw] (hauge2) at (4,0) {$N$};
			\node[circle,draw] (hauge4) at (2,2) {$N$};
			\node[circle,draw] (hauge3) at (4,2) {$N$};

			\path (hauge1) edge [bend right] (hauge2);
			\path (hauge2) edge [bend right] (hauge3);
			\path (hauge3) edge [bend right] (hauge4);
			\path (hauge4) edge [bend right] (hauge1);
			
			\node[anchor=north] (q1) at (hauge1.south east) {${}_{-k}$};
			\node[anchor=west] (q2) at (hauge2.south east) {${}_{k}$};
			\node[anchor=west] (q3) at (hauge3.south east) {${}_{-k}$};
			\node[anchor=west] (q4) at (hauge4.south east) {${}_k$};
		\end{tikzpicture}
		\caption{Left: Circular quiver with gauge group $U(N)_k \times U(N)_k \times U(N)_{-k} \times U(N)_{-k} $. Right: Circular quiver with gauge group $U(N)_k \times U(N)_{-k} \times U(N)_{k} \times U(N)_{-k} $.}
		\label{fig:circular4}
		\end{figure}

			{\small\textbf{Exercise:} Compute the free energy for the two circular quiver with four nodes and Chern--Simons levels $(k,k,-k-k)$ and $(k,-k,k,-k)$, respectively, shown in Figure \ref{fig:circular4}.}\par
			\medskip
			
			\par

		\subsection{M-theory limit in three dimensions: Flavoured ABJM theory}
		
		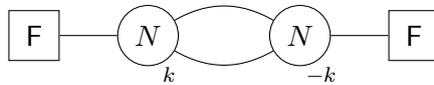
\begin{figure}[tbh]
		\centering
		\begin{tikzpicture}[auto,square/.style={regular polygon,regular polygon sides=4}]
			\node[circle,draw] (gauge1) at (1,0) {$N$};
			\node[circle,draw] (gauge2) at (-1,0) {$N$};

			\path (gauge1) edge [bend left] (gauge2);
			\path (gauge2) edge [bend left] (gauge1);
			
			\node[anchor=north] (k1) at (gauge1.south east) {${}_{-k}$};
			\node[anchor=north] (k2) at (gauge2.south east) {${}_k$};
			
			\node[square,draw] (fl1) at (2.5,0) { \hspace{6pt} };
			\node[draw=none] (aux1) at (2.5,0) {$\mathsf{F}$};
			\node[square,draw] (fl2) at (-2.5,0) { \hspace{6pt} };
			\node[draw=none] (aux2) at (-2.5,0) {$\mathsf{F}$};
			
			\draw[-] (fl1) -- (gauge1);
			\draw[-] (fl2) -- (gauge2);
		\end{tikzpicture}
		\caption{Flavoured ABJM theory. The two circular nodes represent the gauge groups $U(N)_k$ and $U(N)_{-k}$, the square nodes represent additional $\mathsf{F}$ fundamental hypermultiplets.}
		\label{fig:ABJM2}
		\end{figure}\par
			One additional example to consider in three dimensions is the flavoured ABJM theory, obtained by adding $\mathsf{F}$ fundamental hypermultiplets charged under $U(N)_k$ and $\mathsf{F}$ fundamental hypermultiplets charged under $U(N)_{-k}$ to the ABJM quiver. The corresponding matrix model has been analyzed using different techniques in \cite{Santamaria:2010dm} and in \cite{Gulotta:2011si} (here we follow \cite{Gulotta:2011si}).\par
			The partition function $\mz_{\cs^3} [\text{FlavABJM}]$ is a mild variation of the one of the original ABJM theory, with the insertion of $[2 \cosh \pi \phi_a]^{-\mathsf{F}}[2 \cosh \pi \tilde{\phi}_a]^{-\mathsf{F}}$ . Therefore, the effective action is:
			\begin{equation*}
			\begin{aligned}
				S_{\mathrm{eff}}^{\text{FlavABJM}} = & - \sum_{1 \le a \ne b \le N} \left[ \log \left( \left\lvert  2 \sinh \pi (\phi_a - \phi_b ) \right\rvert \right) + \log \left(\left\lvert 2  \sinh \pi (\tilde{\phi}_a - \tilde{\phi}_b ) \right\rvert \right) \right] \\
					&+ \sum_{a=1}^{N}  \left[ i\pi k\left( \phi_a^2 - \tilde{\phi}_a^2 \right)  + \mathsf{F} \left( \log \left(2 \cosh \pi \phi_a \right)  +  \log \left(2 \cosh \pi \tilde{\phi}_a \right) \right) \right] \\
					& +\sum_{a,b=1}^{N} 2 \log \left( 2 \cosh \pi (\phi_a - \tilde{\phi}_b)\right)  .
			\end{aligned}
			\end{equation*}
			
		\subsubsection*{Scaled effective action of flavoured ABJM}
			We have been careful in adding fundamental hypermultiplets in a way that preserves the $\mathbb{Z}_2$ symmetry of the two nodes. Furthermore, in the M-theory limit, we are keeping $\mathsf{F}$ fixed, and not scale it in the Veneziano way. We can therefore insert the same M-theory ansatz as in pure ABJM.\par
			For large argument we have 
			\begin{equation*}
			\begin{aligned}
				& \ \log \left[2 \cosh \left( \frac{ N^{\gamma} x + i y(x)}{2} \right) \right] + \log \left[2 \cosh  \left( \frac{ N^{\gamma} x - i y(x)}{2} \right)\right] \\
				& = \log \left[ 2 \cosh (N^{\gamma} x) + 2 \cosh (y (x)) \right] \\
				& =  \log \left[ 2 \cosh (N^{\gamma} x) \right] +  \log \left[ 1 + \frac{2 \cosh (y (x)) }{2 \cosh (N^{\gamma} x)}\right] \\
				& \approx N^{\gamma} \lvert x \rvert +  e^{- N^{\gamma} \lvert x \rvert} 2 \cosh (y (x)) 
			\end{aligned}
			\end{equation*}
			where we have used the exponential growth of $\cosh(N^{\gamma} x)$ for $\gamma >0$. We thus obtain that the contribution from the added flavours is $\mathcal{O}(N^{1+\gamma})$, exactly as the Chern--Simons couplings. The non-trivial saddle point again imposes $\gamma=\frac{1}{2}$.\par
			We arrive at the leading order effective action:
			\begin{equation}
			\label{eq:SeffABJMflav}
				S_{\mathrm{eff}} ^{\text{FlavABJM}} = N^{1+\gamma} \frac{k}\pi \int \dd x \varrho(x) x y(x) + N^{1+\gamma} \mathsf{F} \int \dd x \varrho(x) \lvert x \rvert +  N^{2-\gamma} \int \dd x \varrho (x)^2 \left[ \pi^2 - 4 y(x)^2 \right] ,
			\end{equation}
			which is a mild modification of the ABJM one.\par
			Proceeding as before, we introduce a Lagrange multiplier to enforce the normalization of $\varrho (x)$, 
			\begin{equation*}
				S_{\mathrm{eff}} ^{\text{ABJM}} \ \mapsto \ S_{\mathrm{eff}} ^{\text{ABJM}} + N^{1 + \gamma} \frac{\xi}{2 \pi } \left[ \int \dd x \varrho (x)  - 1 \right] ,
			\end{equation*}
			and solve as a function of $\xi$, to be extremized over $\xi$ at the end. Taking the functional saddle points of \eqref{eq:SeffABJMflav} with respect to both $\varrho (x)$ and $y(x)$ yields 
			\begin{align*}
				\frac{k}{\pi} x y(x) +\mathsf{F} \lvert x \rvert + 2 \varrho (x) \left[ \pi^2 - 4 y(x)^2 \right]  + \frac{\xi}{2 \pi} & = 0 , \\
				\frac{k}{\pi} x \varrho (x) - 8  \varrho (x)^2 y(x) & = 0 .
			\end{align*}
			The solution as a function of $\xi$ is 
			\begin{equation*}
				y(x) = -\frac{\pi ^2 k x}{2 (2 \pi \mathsf{F} \lvert x \rvert +\xi )}, \qquad  \varrho (x) = \frac{-2 \pi  \mathsf{F} \lvert x \rvert -\xi }{4 \pi ^3} .
			\end{equation*}
			The normalization condition imposes 
			\begin{equation*}
				1 = \int_{-B} ^{B} \dd x \varrho (x) \quad \Longrightarrow \quad B = \frac{-\xi + \sqrt{\xi ^2-8 \pi ^4 \mathsf{F}}}{2 \pi \mathsf{F}} .
			\end{equation*}\par
			\medskip
			{\small\textbf{Exercise:} (i) Use this eigenvalue density to calculate $\mf_{\cs^3}[\text{FlavABJM}]$ as a function of the Lagrange multiplier $\xi$. (ii) Approximate the result for large $\mathsf{F}\gg 1$ and extremize the free energy with respect to $\xi$. Observe that the result is a monotonically increasing function of $\mathsf{F}$.}\par

	\subsection{M-theory limit in five dimensions: Maximally supersymmetric Yang--Mills}
	\label{sec:MthLimit5d}
	
			A widely studied five-dimensional gauge theory that is connected to an SCFT via RG flow, is $\mN=2$ Yang--Mills. This is the maximal amount of supersymmetry in five dimensions, and is does not embed into a superconformal algebra. In fact, $d=5$ $\mN=2$ super-Yang--Mills descends from the maximally supersymmetric theory in $d=6$ compactified on a circle of radius 
			\begin{equation}
			\label{eq:rqadius6dgYM5d}
				\beta = \frac{g_{\text{\tiny YM}}^2}{8 \pi^2} = \frac{\pi}{g} .
			\end{equation}\par
			Because of its six-dimensional origin, $d=5$ $\mN=2$ Yang--Mills was predicted to follow a $N^3$ behaviour at large $N$. This was proven in \cite{Kallen:2012zn}, by working in the 't Hooft limit and then going to strong coupling. Notice that, from the relation \eqref{eq:rqadius6dgYM5d}, the six-dimensional nature of the SCFT arises precisely at strong coupling.\par
			Here we work directly in the M-theory limit, with $g$ (equivalently the radius $\beta$) fixed. The outcome agrees with the result of \cite{Kallen:2012zn}, as expected.\par
			
			\subsubsection*{Effective action}
			The field content of maximally supersymmetric Yang--Mills (MSYM) can be written in the eight supercharges notation as a vector multiplet for the simple gauge group $G$ coupled to a hypermultiplet in the adjoint representation. We assume the gauge group is $G=SU(N)$, the adaptation to the other cases being a simple exercise. In fact, we work with $U(N)$ gauge group and argue at the end that the distinction between $U(N)$ and $SU(N)$ is negligible at leading order in $N$.\par
			The partition function is 
			\begin{equation*}
				\mz_{\cs^5} [\text{MSYM}] = \int_{\R^N} \dd \phi ~ \exp - \left[  \underbrace{ g \pi \sum_{a=1}^{N} \phi_a ^2 }_{\text{class.}} + \underbrace{ \sum_{1 \le a \ne b \le N } v_{5} (\phi_a - \phi_b) }_{\text{vec.}} +  \underbrace{ \sum_{1 \le a \ne b \le N } h_{5} (\phi_a - \phi_b) }_{\text{adj. hyper.}}  \right] ,
			\end{equation*}
			where we recall that the functions $v_d (\phi)$ and $h_d (\phi)$ are given in \eqref{eq:defvdhd}. We consider the large $N$ limit in the M-theory regime, with $g$ fixed, and make the scaling ansatz 
			\begin{equation*}
				\phi = N^{\gamma} x 
			\end{equation*}
			for some $\gamma>0$ to be determined momentarily and $x=\mathcal{O}(1)$. We then need the asymptotic expressions \eqref{eq:asymptvdhdlargex}, that we report here:
			\begin{subequations}
			\begin{align}
				v_5 (N^{\gamma} x) \to & \frac{\pi}{6} N^{3 \gamma} \lvert x \rvert^{3} - \pi N^{\gamma} \lvert x \rvert , \\
				h_5 (N^{\gamma} x) \to & - \frac{\pi}{6} N^{3 \gamma} \lvert x \rvert^{3} - \frac{\pi}{8} N^{\gamma} \lvert x \rvert .
			\end{align}
			\label{eq:asymptv5dh5dMth}
			\end{subequations}
			From the replacement $\phi =N^{\gamma} x$, introducing the eigenvalue density $\varrho (x) $ for the reduced variable $x$, and exploiting these large argument expressions, we can write the effective action as
			\begin{align*}
				S_{\mathrm{eff}} ^{\text{MSYM}} = N \int \dd x \varrho (x) & \left\{  \underbrace{ g \pi N^{ 2 \gamma} x^2  }_{\text{class.}}  + \underbrace{N  \int_{\tilde{x} \ne x} \dd \tilde{x} \varrho (\tilde{x}) \left[ \frac{\pi}{6} N^{3 \gamma} \lvert x - \tilde{x} \rvert^{3} - \pi N^{\gamma} \lvert x - \tilde{x} \rvert  \right] }_{\text{vec.}} \right. \\ 
				& \left. \ + \underbrace{N  \int \dd \tilde{x} \varrho (\tilde{x}) \left[ - \frac{\pi}{6} N^{3 \gamma} \lvert x - \tilde{x} \rvert^{3} - \frac{\pi}{8} N^{\gamma} \lvert x- \tilde{x} \rvert  \right] }_{\text{adj. hyper.}} \right\} .
			\end{align*}
			The cubic contributions from vector and hypermultiplet cancel out:
			\begin{align}
				\underbrace{ \frac{\pi}{6} N^{3 \gamma}  \int_{\tilde{x} \ne x} \dd \tilde{x} \varrho (\tilde{x}) \lvert x - \tilde{x} \rvert^{3} }_{\text{vec.}}  \underbrace{ -  \frac{\pi}{6} N^{3 \gamma}  \int \dd \tilde{x} \varrho (\tilde{x})\lvert x - \tilde{x} \rvert^{3} }_{\text{adj. hyper.}}  & = -  \frac{\pi}{6} N^{3 \gamma}  \underbrace{\int \dd \tilde{x} \varrho (\tilde{x})\lvert x - \tilde{x} \rvert^{3}  \delta (x-\tilde{x}) }_{=0} , \label{eq:Mlimcancelx3}
			\end{align}
			while the linear pieces combine into a non-trivial term:
			\begin{align*}
				\underbrace{ - \pi N^{\gamma} \int_{\tilde{x} \ne x} \dd \tilde{x} \varrho (\tilde{x})  \lvert x - \tilde{x} \rvert }_{\text{vec.}}  \underbrace{ - \frac{\pi}{8} N^{\gamma} \int \dd \tilde{x} \varrho (\tilde{x}) \lvert x - \tilde{x} \rvert }_{\text{adj. hyper.}}  =& - \frac{9 \pi}{8} N^{\gamma} \int \dd \tilde{x} \varrho (\tilde{x}) \lvert x - \tilde{x} \rvert \\ 
				& - \underbrace{\left( - N^{\gamma} \int \dd \tilde{x} \varrho (\tilde{x}) \lvert x - \tilde{x} \rvert   \delta (x-\tilde{x}) \right) }_{=0} .
			\end{align*}
			After these straightforward simplifications, the effective action takes the form 
			\begin{equation*}
				S_{\mathrm{eff}} ^{\text{MSYM}} = \pi \int \dd x \varrho (x) \left\{  g  N^{ 1+ 2 \gamma} x^2   - \frac{9}{8} N^{2+\gamma}  \int \dd \tilde{x} \varrho (\tilde{x})  \lvert x - \tilde{x} \rvert  \right\}  .
			\end{equation*}
			
			\subsubsection*{Eigenvalue density}
			The leading contributions from the vector multiplet and adjoint hypermultiplet have canceled (as in \cite{Kallen:2012zn}). The two remaining contributions can lead to an equilibrium if 
			\begin{equation*}
				1 + 2 \gamma = 2 + \gamma \quad \Longrightarrow \quad \gamma = 1 ,
			\end{equation*}
			already manifesting the $N^3$ behaviour. The SPE for this effective action is 
			\begin{equation}
			\label{eq:MthMSYMSPE}
				\frac{9}{4} \int \dd \tilde{x} \varrho (\tilde{x}) \text{sgn} (x-\tilde{x}) = 2 g x .
			\end{equation}
			Moreover, the classical potential is quadratic, whence we look for a one-cut solution, $\surho = [A, B]$. The $\Z_2$ action reflecting the eigenvalues $\phi \mapsto -\phi$ is a symmetry of the problem, ensuring that $A=-B$.\par
			To find the eigenvalue density $\varrho (x)$, we split the integration domain $[-B,B]$ into two pieces, such that the sign function is constant on each piece. That is, we rewrite \eqref{eq:MthMSYMSPE} as 
			\begin{equation*}
				 \frac{9}{4} \left[  \int^{x}_{-B} \dd \tilde{x} \varrho (\tilde{x}) (+1) +   \int_{x}^{B} \dd \tilde{x} \varrho (\tilde{x}) (-1) \right] = 2 g x 
			\end{equation*}
			Differentiating both sides with respect to $x$ gives 
			\begin{equation*}
				\frac{9}{2} \varrho (x) = 2 g \quad \Longrightarrow \quad \varrho (x) = \frac{4g}{9} .
			\end{equation*}
			The normalization condition and the reflection symmetry of the classical potential imply 
			\begin{equation*}
				1 = \int_{-B} ^{B} \dd x \varrho (x) = \frac{8}{9} g B \quad \Longrightarrow \quad  B = \frac{9}{8g} .
			\end{equation*}\par
			We observe that the resulting eigenvalue density, computed in a $U(N)$ gauge theory, is automatically traceless:
			\begin{equation*}
				\int_{-B} ^{B} \dd x \varrho (x) x = 0 .
			\end{equation*}
			Therefore, the same result applies to the $SU(N)$ gauge theory, without the necessity of imposing any further constraint.\par
			
			\subsubsection*{Free energy}
			We can now use the constant eigenvalue density to compute the sphere free energy. It is easier to use the relation 
			\begin{align*}
				\frac{ \partial \ }{\partial g} \mf_{\cs^5} [\text{MSYM}] & = - \frac{ \partial \ }{\partial g} \log \mz_{\cs^5} [\text{MSYM}] \\
					& = - \frac{1}{\mz_{\cs^5} [\text{MSYM}] } \frac{ \partial \ }{\partial g} \mz_{\cs^5} [\text{MSYM}] \\
					& = + \pi \left\langle \sum_{a=1}^{N} \phi_a ^2 \right\rangle_{\text{MSYM}} \\
					& = \pi N^{1+2\gamma} \int_{B} ^{B} \dd x \varrho (x) x^2  \\
					& = N^3 g \pi \frac{4}{9}\int_{-\frac{9}{8h}} ^{\frac{9}{8h}} x^2  \dd x  = \left( \frac{3N}{4}\right)^3 \frac{\pi}{g^2} ,
			\end{align*}
			where in the last line we plugged the explicit expressions of $\varrho (x)$ and $B$. Integrating, we immediately obtain 
			\begin{equation*}
				\mf_{\cs^5} [\text{MSYM}] = - N^3 \frac{3^3 \pi}{2^6 g} = - N^3 \frac{3^3 }{2^9 \pi}  g_{\text{\tiny YM}} ^2 ,
			\end{equation*}
			up to a $g$-independent constant, negligible in the strong coupling regime of physical interest. The second equality follows from the definition $g= \frac{8 \pi^2}{g_{\text{\tiny YM}} ^2 }$ from Section \ref{sec:SphereZ}.\footnote{Recall that we are defining all the massive parameters in units of the radius of $\cs^5$. In this way, our $ g_{\text{\tiny YM}} ^2$ is the gauge coupling measured in units of the radius length $r_{\cs^{5}}$. To reintroduce the units, simply substitute $g \mapsto g r_{\cs^5}$.}\par
			We find that the derivation in the M-theory limit produces the same result as the derivation of \cite{Kallen:2012zn}, that is, first take the 't Hooft limit and then go to strong coupling.

		\subsection{M-theory limit in five dimensions: Circular quivers}
		\label{sec:circ5d}
			Let us now consider a circular quiver with gauge group $SU(N)^{L}$ and hypermultiplets in the bifundamental representation of two adjacent $SU(N)$ nodes. The large $N$ limit of this theory was first solved in \cite{Minahan:2013jwa,Minahan:2016xwk}, but here we provide a slightly different and complementary derivation.
			As in the previous example, we work directly with gauge groups $U(N)$, as the eigenvalue density will turn out to be the same as in the $SU(N)$ case.\par
			The effective action for this theory is:
			\begin{equation*}
				S_{\mathrm{eff}}  = \sum_{j=1} ^{L} \left[   \underbrace{ g_j \pi \sum_{a=1}^{N} \phi_{j,a} ^2 }_{\text{class.}}  + \underbrace{ \sum_{1 \le a \ne b \le N } v_{5} (\phi_{j,a} - \phi_{j,b} ) }_{\text{vec.}}  +  \underbrace{ \sum_{1 \le a \ne b \le N } h_{5} (\phi_{j,a} - \phi_{j+1,b} ) }_{\text{bifund. hyper.}}  \right]
			\end{equation*}
			with periodic identification $L+1 \sim 1$. 
		
		\subsubsection*{Effective action}
			We send $N \to \infty$ with the usual ansatz $\phi = N^{\gamma} x$, and again use the symmetry of the system to predict that $\gamma$ is the same at each node, i.e. does not depend on the index $j=1, \dots, L$. Using \eqref{eq:asymptv5dh5dMth} we get: 
			\begin{align*}
				S_{\mathrm{eff}} & = \sum_{j=1} ^{L} \pi N \int \dd x \varrho_j (x) \left\{ g_j N^{2 \gamma} x^2  + N  \int_{\tilde{x} \ne x} \dd \tilde{x} \varrho_j (\tilde{x}) \left[ \frac{N^{3 \gamma}}{6}  \lvert x - \tilde{x} \rvert^{3} -  N^{\gamma} \lvert x - \tilde{x} \rvert  \right]  \right. \\
					& \left. \ -  N  \int \dd \tilde{x} \varrho_{j+1} (\tilde{x}) \left[ \frac{N^{3 \gamma}}{6}  \lvert x - \tilde{x} \rvert^{3} + \frac{N^{\gamma}}{8}  \lvert x- \tilde{x} \rvert  \right] \right\} .
			\end{align*}
			The leading $\mathcal{O}(N^{2 + 3 \gamma})$ term yields 
			\begin{equation*}
				- \frac{\pi}{6} N^{2 + 3 \gamma} \int \dd x \varrho_j (x) \left[ \int \dd \tilde{x} \varrho_{j+1} (\tilde{x})  \lvert x - \tilde{x} \rvert^{3} - \int_{\tilde{x} \ne x} \dd \tilde{x} \varrho_{j} (\tilde{x})  \lvert x - \tilde{x} \rvert^{3} \right] ,
			\end{equation*}
			which is minimized by the choice $\varrho_{j} (x)= \varrho_{j+1} (x)$, $\forall j=1, \dots, L \mod L$. Therefore, we are left with a problem very similar to the theory with one adjoint hypermultiplet: in the M-theory limit, the difference between adjoint and bifundamental hypermultiplets becomes sub-leading in $N$.\par
			More explicitly, after imposing $\varrho_{j} = \varrho$ $\forall j=1, \dots, L$, we arrive at 
			\begin{align}
			\label{eq:Seff5dCircNLO}
				S_{\mathrm{eff}} & = \pi \int \dd x \varrho (x) \left\{ \left( \sum_{j=1} ^{L}  g_j \right) N^{1+2 \gamma} x^2  - \frac{9}{8} N^{2+\gamma} L \int \dd \tilde{x} \varrho (\tilde{x}) \lvert x - \tilde{x} \rvert \right\} .
			\end{align}
			Here we have used the obvious fact that $\lvert x-\tilde{x}\rvert$ vanishes at $\tilde{x}=x$ to combine the residual contribution from the bifundamental hypermultiplets with the one from the vector multiplets:
			\begin{equation*}
				\underbrace{- \int_{\tilde{x} \ne x} \dd \tilde{x} \varrho (\tilde{x}) \lvert x - \tilde{x} \rvert }_{\text{vec.}}  \underbrace{-  \int \dd \tilde{x} \varrho (\tilde{x}) \frac{1}{8}  \lvert x- \tilde{x} \rvert }_{\text{bifund. hyper}}  = - \frac{9}{8}  \int \dd \tilde{x} \varrho (\tilde{x}) \lvert x - \tilde{x} \rvert .
			\end{equation*}
			Besides, we stress that, although the possibility of different gauge couplings introduces an asymmetry between the gauge nodes, the condition $\varrho_{j}= \varrho_{j+1} $ comes from minimizing a term which is higher order than the gauge coupling. Thus, having $g_{j} \ne g_{j+1}$ does not change the equality between eigenvalue densities at different gauge nodes.\par
			As in MSYM, the two pieces compete if 
			\begin{equation*}
				1+2\gamma = 2+\gamma \quad \Longrightarrow \quad \gamma = 1 .
			\end{equation*}
			In particular, it follows that the circular quiver $SU(N)^L$ has the $N^3$ behaviour.

		\subsubsection*{Eigenvalue density}
			
			After the cancellation of the $\mathcal{O}(N^{2+3\gamma})=\mathcal{O}(N^5)$ terms, we are left with \eqref{eq:Seff5dCircNLO}. To reduce clutter, we further assume $g_j=g$ for all $j=1, \dots, L$. The saddle point equation derived from there is 
			\begin{equation}
			\label{eq:SPE5dCircular}
				2 x g = \frac{9}{4} \int \varrho (\tilde{x}) \mathrm{sgn}(x-\tilde{x}) .
			\end{equation}
			Differentiating \eqref{eq:SPE5dCircular} with respect to $x$ we find 
			\begin{equation*}
				\varrho (x) = \frac{4}{9} g .
			\end{equation*}
			The normalization condition is 
			\begin{equation*}
				1= \int_{-B}^{B} \dd x \varrho (x) = \frac{8g}{9} B \quad \Longrightarrow \quad B = \frac{9}{8 g} .
			\end{equation*}
			
		\subsubsection*{Free energy}
			Computing the free energy with the constant eigenvalue density we get 
			\begin{equation*}
				\mathcal{F}_{\cs^5} = - \frac{27 \pi}{64 g} L N^3 .
			\end{equation*}
			Therefore, the physical free energy $-\mathcal{F}_{\cs^5}$ is positive, shows a linear scaling in $L$ and a cubic scaling in $N$, and is monotonically decreasing in $g$. This last point is consistent with the $\mf$-theorem, since $g$ is a relevant deformation and increasing $g$ pushes the system to the IR.\par
			\medskip
			{\small\textbf{Exercise:} Compute the free energy for arbitrary couplings $g_1, \dots, g_L$.}

		\subsection{M-theory limit in five dimensions: Higher rank \texorpdfstring{$E_n$}{E} theories}
		\label{sec:MJafferisPufu}
			Let us consider the five-dimensional theory with gauge group $USp(2N)$ coupled to $\mathsf{F} <8$ hypermultiplets in the fundamental representation and one hypermultiplet in the antisymmetric representation. We also define $\mathsf{F}+1 =n$. We will refer to this theory as the rank-$N$ $E_{n}$ theory.\par
			This choice of gauge group does not admit a Chern--Simons term in $d=5$. Moreover, we decide to work directly in the strong coupling regime $g \to 0$, approaching the superconformal point.\par
			
			\subsubsection*{Effective action}
			The large $N$ limit of this gauge theory in the M-theoretical regime was obtained in \cite{Jafferis:2012iv}. The sphere partition function is 
			\begin{equation*}
				\mz_{\cs^5} [E_{n}] = \frac{1}{N!} \int \dd \phi ~ e^{- S_{\mathrm{eff}} ^{E_{n}} (\phi )} \left( 1 + \cdots \right )
			\end{equation*}
			with effective action 
			\begin{equation}
			\begin{aligned}
				S_{\mathrm{eff}} ^{E_{n}} (\phi ) & = \underbrace{ \sum_{1 \le a \ne b \le N } \left[ v_{5} (\phi_{a} - \phi_b)  + v_{5} (\phi_{a} + \phi_b) \right] + \sum_{a=1}^{N} \left[ v_{5} (2 \phi_a ) + v_{5} (-2 \phi_a ) \right] }_{\text{vec.}} \\
					& + \underbrace{ \sum_{1 \le a \ne b \le N } \left[ h_{5} (\phi_{a} - \phi_b)  + h_{5} (\phi_{a} + \phi_b) \right] }_{\text{antisym. hyper.}} + (n-1) \underbrace{ \sum_{a=1}^{N}  \left[ h_{5} (\phi_{a} )  + h_{5} (- \phi_{a} ) \right] }_{\text{fund. hyper.}}  .
			\end{aligned}
			\label{eq:EnJafferisPufuSeff}
			\end{equation}
			The functions $v_d (\phi)$ and $h_d (\phi)$ have been defined in Section \ref{sec:SphereZ}, formula \eqref{eq:defvdhd}.\par
			We now send $N \to \infty$. Since there are no couplings in \eqref{eq:EnJafferisPufuSeff}, we consider the M-theory limit 
			\begin{equation*}
				\phi = N^{\gamma} x 
			\end{equation*}
			for $\gamma >0$. A crucial difference, compared with $d=3$, is that the effective action is real-valued. Therefore, the eigenvalues remain on the real line and we do not need to introduce a function $y(x)$ to retain the imaginary part of $\phi$.\par
			In this limit, the arguments of the functions $v_5 (\phi)$ and $h_5 (\phi)$ are large and we use the asymptotic form \eqref{eq:asymptvdhdlargex}, also given in \eqref{eq:asymptv5dh5dMth} for $d=5$. Plugging these expressions in the large $N$ limit of \eqref{eq:EnJafferisPufuSeff} yields 
			\begin{align*}
				S_{\mathrm{eff}} ^{E_{n}} & = \underbrace{ - \pi N^{2 + \gamma} \int \dd x \varrho (x) \int \dd \tilde{x} \varrho (\tilde{x}) \left[ \lvert x - \tilde{x} \rvert + \lvert x + \tilde{x} \rvert   \right]  + \frac{\pi}{6} N^{1+3 \gamma} \int \dd x \varrho (x) \left[ \lvert 2 x \rvert^3 +  \lvert -2 x \rvert^3 \right]  }_{\text{vec.}} \\
					&  \underbrace{ - \frac{\pi}{8} N^{2 + \gamma} \int \dd x \varrho (x) \int \dd \tilde{x} \varrho (\tilde{x}) \left[ \lvert x - \tilde{x} \rvert + \lvert x + \tilde{x} \rvert   \right]  }_{\text{antisym. hyper.}}  -  (n-1) \underbrace{ \frac{\pi}{6} N^{1+3 \gamma} \int \dd x \varrho (x) \left[ \lvert x \rvert^3 +  \lvert - x \rvert^3 \right] }_{\text{fund. hyper.}} .
			\end{align*}
			In this expression, the leading terms $\mathcal{O}(N^{2+3\gamma})$ from the vector multiplet and the antisymmetric hypermultiplet, which contain the cubic terms $\lvert x - \tilde{x} \rvert^3$ and $\lvert x + \tilde{x} \rvert^3$, have canceled among each other. This cancellation is identical to \eqref{eq:Mlimcancelx3}.\par 
			Rearranging and combining the various surviving terms in $S_{\mathrm{eff}} ^{E_{n}}$ we get 
			\begin{equation}
			\label{eq:SeffMthLargeNJaffPufu}
				S_{\mathrm{eff}} ^{E_{n}}  =  \int \dd x \varrho (x) \left\{ \frac{(9-n)\pi }{3} N^{1+3 \gamma} \lvert x \rvert^3  - \frac{9\pi }{8}  N^{2 + \gamma} \int \dd \tilde{x} \varrho (\tilde{x}) \left[ \lvert x - \tilde{x} \rvert + \lvert x + \tilde{x} \rvert   \right]  \right\} .
			\end{equation}
			The two contributions compete in determining a non-trivial saddle point if 
			\begin{equation*}
				2 + \gamma = 1 + 3 \gamma \quad \Longrightarrow \quad \gamma = \frac{1}{2} .
			\end{equation*}
			This produces the $N^{\frac{5}{2}}$ behaviour of the free energy, in agreement with the M-theory predictions.\par
			
			\subsubsection*{Eigenvalue density}
			The SPE is 
			\begin{equation*}
				\frac{9\pi }{4} \int \dd \tilde{x} \varrho (\tilde{x}) \left[  \text{sgn}(x- \tilde{x}) +  \text{sgn}(x+ \tilde{x})   \right] = \pi (9-n) x^2   \text{sgn}(x).
			\end{equation*}
			Differentiating this SPE we obtain 
			\begin{equation*}
				\frac{9 }{2} \varrho (x) = 2  (9-n) \lvert x \rvert \quad \Longrightarrow \quad \varrho (x) = \frac{4}{9} (9-n)  \lvert x \rvert ,
			\end{equation*}
			supported on $[0,B]$. To understand why the support is on the positive semi-axis, recall that $USp(2N)$ has $2N$ eigenvalues that come in pairs of opposite signs. Therefore, it is possible to choose the $N$ positive ones to be the independent ones, without loss of generality. In this convention, the density of eigenvalues accounts for the positive eigenvalues.\par 
			To determine $B$, we impose the normalization condition 
			\begin{equation*}
				1 = \int_{0} ^{B} \dd x \varrho (x) = \frac{4}{9} (9-n) B^2  \quad \Longrightarrow \quad B = \frac{3}{\sqrt{2(9-n)}} .
			\end{equation*}\par
			
			\subsubsection*{Free energy}
			With this solution at hand, we can insert $\varrho (x)$ back into \eqref{eq:SeffMthLargeNJaffPufu} and find the free energy 
			\begin{align*}
				\mf_{\cs^5} [E_n] & = - N^{\frac{5}{2}} \cdot 2 \cdot  \int_{0} ^{B} \dd x \varrho (x) \left( \frac{(9-n) \pi}{3} \lvert x \rvert^3 \right) \\ 
					& = - N^{\frac{5}{2}} \frac{2 (9-n) \pi}{3}  \int_{0 } ^{\frac{3}{\sqrt{2(9-n)}} } \dd x  \left( \frac{4 (9-n)}{9} \right) \lvert x \rvert^4 \\
					& = - N^{\frac{5}{2}} \pi \frac{2^3 (9-n)^2 }{3^3}  \cdot \frac{1}{5} \left( \frac{3}{\sqrt{2(9-n)}}  \right)^5 = - N^{\frac{5}{2}} \pi \frac{9  \sqrt{2}}{ 5 \sqrt{9-n} } ,
			\end{align*}
			where we recall that $n$ is the number of fundamental hypermultiplets plus one.

	\subsection{(Aside) Phase transitions in five dimensions}
			
			In this subsection we investigate a different aspect of supersymmetric theories in the M-theory limit: the appearance of phase transitions \cite{Nedelin:2015mta,Santilli:2021qyt}. Differently from what we have done so far, we move away from the conformal point by turning on mass deformations.\par
			Let us illustrate the mechanism in a concrete and simple example: five-dimensional $SU(N)$ gauge theory with $\mathsf{F}$ fundamental hypermultiplets. We assign to half of them mass $m$ and to the other half mass $-m$, and set $k=0=g$ in the classical potential. More general choices of masses, gauge couplings, and gauge groups can be made \cite{Santilli:2021qyt}.\par
			The effective action is 
			\begin{equation*}
				S_{\mathrm{eff}} ^{\text{SQCD}_5} = \sum_{a=1}^{N} \left[ \frac{\mathsf{F}}{2} h_{d=5} \left( \phi_a +m \right) + \frac{\mathsf{F}}{2} h_{d=5} \left( \phi_a -m \right) +\sum_{b \ne a} v_{d=5} \left( \phi_a -\phi_b \right)  \right] .
			\end{equation*}
			In the absence of a gauge coupling, we make the M-theory scaling ansatz $\phi = N^{\gamma} x$ with $\gamma>0$, from which we recover the M-theory scaling sending $m \to 0$. We treat real mass parameters and scalar fields equally, so we also impose the scaling $m=N^{\gamma} \mu $.\par
			The SPE in this limit is 
			\begin{equation}
			\label{eq:SPEPT5d}
				\zeta \left[ \left(  x+\mu\right)^2 \text{sgn}\left(  x+\mu\right) +   \left(  x-\mu\right)^2 \text{sgn}\left( x-\mu\right) \right] = 2 \int \dd \tilde{x} \varrho (\tilde{x})  \left( x-\tilde{x} \right)^2 \text{sgn}\left(  x-\tilde{x}\right) ,
			\end{equation}
			where $\zeta = \frac{\mathsf{F}}{2N}$ is the Veneziano parameter for each family of $\frac{\mathsf{F}}{2}$ hypermultiplets of equal mass. Observe that $0 \le \zeta \le 1$.\par
			Taking three derivatives of \eqref{eq:SPEPT5d} we find the solution 
			\begin{equation}
			\label{eq:rhogensol5dPT}
				\varrho (x) = c_A \delta \left( x-A \right) + c_B \delta \left( x-B \right) + c_{m} \left[ \delta \left( x+\mu \right) + \delta \left( x-\mu \right)  \right] ,
			\end{equation}
			supported on $x \in [A,B]$. In \eqref{eq:rhogensol5dPT}, 
			\begin{equation*}
				c_m = \begin{cases} \frac{\zeta}{2} & \text{ if } \mu \in [A,B] \\ 0 & \text{ otherwise} \end{cases}
			\end{equation*}
			is obtained directly from \eqref{eq:rhogensol5dPT}. In turn, the endpoints $A,B$ and their coefficients $c_A, c_B$ are to be determined. Plugging \eqref{eq:rhogensol5dPT} back into the SPE \eqref{eq:SPEPT5d} gives three additional constraints (recall that we have differentiated \eqref{eq:SPEPT5d} three times), 
			\begin{subequations}
			\begin{align}
				- c_A + c_B & =0 \label{eq:PTSPEsys1} \\
				c_A A -c_B B &= ( 2 c_{m} - \zeta )\mu \label{eq:PTSPEsys2} \\
				c_A A^2 + c_B B^2 & = 0 . \label{eq:PTSPEsys3}
			\end{align}
			\end{subequations}
			These three equations are complemented by the normalization 
			\begin{equation*}
				\int_A ^B \dd x \varrho (x) = 1 \quad \Longrightarrow \quad c_A + c_B +2 c_m =1.
			\end{equation*}
			We thus have four equations to solve for the quadruple $(A,B,c_A,c_B)$. The normalization equation together with \eqref{eq:PTSPEsys1} give 
			\begin{equation*}
			\begin{aligned}
				c_A = c_B = \frac{1}{2} - c_m .
			\end{aligned}
			\end{equation*}
			Using \eqref{eq:PTSPEsys1} into \eqref{eq:PTSPEsys3} gives $A=-B$. Substituting everything in \eqref{eq:PTSPEsys2} leads to 
			\begin{equation*}
				-A =B = \mu \frac{\zeta - 2 c_m}{1-2 c_m} .
			\end{equation*}
			For large mass $\mu>B$ (or $\mu < -B$) we find
			\begin{equation}
				\rho (x) = \frac{1}{2} \delta \left( x+\mu \zeta \right) + \frac{1}{2} \delta \left( x-\mu \zeta \right) . \label{eq:symrhoPT5dsol}
			\end{equation}
			This solution would break down at $\mu= \pm B$. However, in this phase, $B=\mu \zeta$ with $0\le \zeta \le 1$; therefore $\lvert \mu\rvert  \ge B$ all along, with $\mu=\pm B$ only at $\mu=0$. At this value, the support has shrunk to a point. In conclusion, there is no phase with $ \pm \mu \in [-B,B]$.\par
			\medskip
			We utilize the eigenvalue density \eqref{eq:symrhoPT5dsol} to compute the free energy in this regime. It reads 
			\begin{align*}
				\mf_{\cs^5} [\text{SQCD}_5] & = N^{2+3\gamma} \frac{\pi}{6} \int_A ^B \dd x \varrho ( x ) \left\{ - \zeta \left[ \lvert x +\mu \rvert^3  + \lvert x -\mu \rvert^3  \right] + \int_{A} ^B \dd \tilde{x} \varrho ( \tilde{x} ) \lvert x - \tilde{x} \rvert^3 \right\} \\
					&= N^{2+3\gamma} \frac{\pi}{3} \left\{  - \frac{\zeta}{2} \left[ c_{\scriptscriptstyle B} \lvert B+\mu \rvert^3 + c_{\scriptscriptstyle A} \lvert A+\mu \rvert^3  + c_{\scriptscriptstyle B} \lvert B-\mu \rvert^3 + c_{\scriptscriptstyle A} \lvert A-\mu \rvert^3 \right] + c_{\scriptscriptstyle A} c_{\scriptscriptstyle B} \lvert B-A \rvert^3  \right\} \\
					&= N^{2+3\gamma} \frac{\pi}{3} \left\{ - \frac{\zeta}{2} \left[ \lvert \mu \zeta + \mu \rvert^3 + \lvert \mu \zeta - \mu \rvert^3  \right] + 2 \lvert \mu \zeta \rvert^3 \right\} \\
					& = N^{2+3\gamma} \frac{\pi}{6} \zeta \lvert \mu \rvert^3 \left\{ 4 \zeta^2 - (1+ \zeta)^3 - (1-\zeta)^3  \right\} ,
			\end{align*}
			where we have used the explicit form \eqref{eq:symrhoPT5dsol} and $0 \le \zeta \le 1 $. Simplifying, we get 
			\begin{equation*}
				\mf_{\cs^5} [\text{SQCD}_5]  = - N^{2+3\gamma} \frac{\pi}{6} \zeta (1+\zeta^2) \lvert \mu \rvert^3 .
			\end{equation*}
			We underline the properties of $- \mf_{\cs^5} [\text{SQCD}_5]$:
			\begin{enumerate}[(i)]
				\item It is non-negative;
				\item It is a monotonically increasing function of $\zeta$;
				\item\label{item:PT3rd} It has a discontinuity in its third derivative at $m=0$, due to the absolute value.
			\end{enumerate}
			The first two points support a strong version of the $\mf$-theorem, holding away from the fixed points: Remember that $\zeta$ is proportional to the number of flavours, thus decreasing $\zeta$ corresponds to move towards the IR.\par
			Point \eqref{item:PT3rd} means that there is a third order phase transition at $m=0$. The reason why it takes place precisely at the conformal point is clear: there are no other mass scales in the problem, thus the distinction between heavy and light fields is whether $m =0$ or not.\par
			The discussion can be extended almost identically to the other classical gauge groups and, for $SU(N)$, also adding a Chern--Simons term \cite{Santilli:2021qyt}.

\section{Long quiver limit}
\label{sec:LargeNLongQ}

	The long quiver limit has been first studied by C. Uhlemann in five-dimensional SCFTs \cite{Uhlemann:2019ypp}. As the name suggests, the setup consists of taking quivers of length $L$ and analyzing them in the regime $L \to \infty $. The validity of the derivation in the present section relies on the following assumptions:
	\begin{itemize}
		\item[(LQ1)]\label{hyp:LQ1} The theory $\Th$ is encoded in a linear or circular quiver of $L$ unitary or special unitary gauge nodes, with $L \to \infty $;
		\item[(LQ2)]\label{hyp:LQ2} The theory is balanced, i.e. $2 N_j - N_{j-1} - N_{j+1} = \mathsf{F}_j$ $\forall j=1, \dots , L$;
		\item[(LQ3)]\label{hyp:LQ3} The theory is at the strong coupling point.
	\end{itemize}\par
	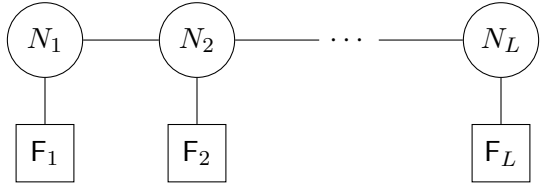
\begin{figure}[t]
\centering
\begin{tikzpicture}[auto,square/.style={regular polygon,regular polygon sides=4}]
	\node[circle,draw] (gauge1) at (3,0) { \hspace{20pt} };
	\node (a1) at (3,0) {$N_{L}$};
	\node[draw=none] (gaugemid) at (1,0) {$\cdots$};
	\node[circle,draw] (gauge3) at (-1,0) { \hspace{20pt} };
	\node[circle,draw] (gauge4) at (-3,0) { \hspace{20pt} };
	\node (a2) at (-3,0) {$N_{1}$};
	\node (a3) at (-1,0) {$N_{2}$};
	\node[square,draw] (fl1) at (3,-1.5) { \hspace{8pt} };
	\node[square,draw] (fl2) at (-3,-1.5) { \hspace{8pt} };
	\node[square,draw] (fl3) at (-1,-1.5) { \hspace{8pt} };
	\node[draw=none] (aux1) at (3,-1.5) {$\mathsf{F}_{L}$};
	\node[draw=none] (aux2) at (-3,-1.5) {$\mathsf{F}_1$};
	\node[draw=none] (aux3) at (-1,-1.5) {$\mathsf{F}_2$};
	\draw[-](gauge1)--(gaugemid);
	\draw[-](gaugemid)--(gauge3);
	\draw[-](gauge4)--(gauge3);
	\draw[-](gauge1)--(fl1);
	\draw[-](gauge4)--(fl2);
	\draw[-](gauge3)--(fl3);
\end{tikzpicture}
\caption{Linear quiver of length $L$. Circular nodes indicate gauge groups, square nodes indicate fundamental hypermultiplets.}
\label{fig:linearquiver}
\end{figure}\par
	The first assumption (\hyperref[hyp:LQ1]{LQ1}) requires that the quiver is of the form shown in Figure \ref{fig:linearquiver}, possibly closed by an additional edge connecting the last with the first. Assumption (\hyperref[hyp:LQ2]{LQ2}) will produce cancellations and simplifications that render the limit explicitly solvable. The third assumption (\hyperref[hyp:LQ3]{LQ3}) means we set the gauge couplings to infinity. This last assumption is unnecessary, but we impose it to streamline the derivation. Besides, note that (\hyperref[hyp:LQ3]{LQ3}) only affects the classical action in $d=5$, while in $d=3$ sphere partition function is independent of the gauge coupling.\par
	The idea behind the long quiver limit is close in spirit to the M-theory limit: the eigenvalues may develop a non-trivial scaling dependence on $L$, self-consistently determined by the requirement that a non-trivial saddle point exists.\par

	\subsection{Long quivers setup}
		To keep the discussion as neat and simple as possible, we consider a $d$-dimensional gauge theory $\Th_{\vec{\nu}} $ described by a long linear quiver as in Figure \ref{fig:linearquiver}, which moreover is balanced by (\hyperref[hyp:LQ2]{LQ2}). Circular quivers are dealt with in a similar way, imposing the periodic identification $L+1 \sim 1$.\par
		The ranks of the gauge groups are written as 
		\begin{equation*}
			N_j = \nu_j N , \qquad \forall j=1, \dots , L ,
		\end{equation*}
		for rational numbers $\nu_j \in \frac{1}{N} \Z$ and $N \in \mathbb{N}$. We will send $N \to \infty$ keeping the fractions $\left\{ \nu_j \right\}_{j=1, \dots, L}$ fixed. The theory is thus essentially specified by giving the list of pairs $\left\{ (\nu_j , \mathsf{F}_j)\right\}_{j=1 , \dots, L}$. In addition, the number of fundamental hypermultiplets $\mathsf{F}_j$ are uniquely determined by the collection $\left\{ \nu_j \right\}_{j=1 , \dots, L}$ through (\hyperref[hyp:LQ2]{LQ2}), whence the notation $\Th_{\vec{\nu}}$.\par
		
		\subsubsection*{Large $N$ limit}
		With the assumptions  (\hyperref[hyp:LQ1]{LQ1})-(\hyperref[hyp:LQ2]{LQ2}) above, but without imposing (\hyperref[hyp:LQ3]{LQ3}) for now, the generic effective action takes the form 
		\begin{equation}
		\label{eq:LQSeff}
			S_{\mathrm{eff}} ^{\Th_{\vec{\nu}}}  = \sum_{j=1} ^{L} \left\{ \underbrace{  \frac{N}{\lambda_{d,j}} \sum_{a=1} ^{N_j} \phi_{j,a} ^2  }_{\text{class.}}  + \underbrace{ \sum_{1 \le a \ne b \le N_j} v_{d} (\phi_{j,a} - \phi_{j,b} )  }_{\text{vec.}} + \mathsf{F}_j \underbrace{ \sum_{a=1} ^{N_j} h_d (\phi_{j,a}) }_{\text{fund. hyper.}} + \underbrace{  \sum_{a=1}^{N_j} \sum_{b=1}^{N_{j+1}} h_d (\phi_{j,a} - \phi_{j+1,b} ) }_{\text{bifund. hyper.}}  \right\} .
		\end{equation}
		The last term, accounting for the contribution of the bi-fundamental hypermultiplet, should be set to zero if $j=L$ in linear quivers, or should use the periodic identification $L+1 \sim 1$ for circular quivers. For linear quivers, it is clear that the over-counting term is sub-leading in the limit $L \to \infty$, so we henceforth abuse notation and omit put it to zero at $j=L$, since it will eventually drop out of our analysis anyway.\par
		We now send $N \to \infty $, introducing the eigenvalue densities $\rho_j (\phi)$, normalized according to 
		\begin{equation*}
			\int \dd \phi \rho_j (\phi) = \nu_j .
		\end{equation*}
		It is also natural to consider the 't Hooft (and Veneziano) limit in this regime, with $\lambda_d$ and $\zeta_j = \frac{\mathsf{F}_j}{N}$ fixed. In this way, \eqref{eq:LQSeff} is rewritten as 
		\begin{equation}
		\label{eq:SeffLargeNFixP}
		\begin{aligned}
			S_{\mathrm{eff}} ^{\Th_{\vec{\nu}}}  = N^2 \sum_{j=1} ^{L} \int \dd \phi \rho_j (\phi) & \left\{  \underbrace{ \frac{\phi^2}{\lambda_{d,j}} }_{\text{class.}}  + \underbrace{ \int_{\sigma \ne \phi} \dd \sigma \rho_j (\sigma) v_d (\phi - \sigma) }_{\text{vec.}} + \zeta_j \underbrace{ h_d (\phi)  }_{\text{fund. hyper.}} \right. \\
			& \ + \left. \underbrace{ \int \dd \sigma \rho_{j+1} (\sigma) h_d (\phi - \sigma)  }_{\text{bifund. hyper.}} \right\} .
		\end{aligned}
		\end{equation}
		
		\subsubsection*{Balancing condition}
			Recall our hypothesis (\hyperref[hyp:LQ2]{LQ2}) that the quiver is balanced, 
			\begin{equation*}
				2 N_j - N_{j-1} - N_{j+1} = \mathsf{F}_j .
			\end{equation*}
			With the substitutions outlined above, this balancing expression becomes 
			\begin{align}
				- N \left[ \left( \nu_{j+1}- \nu_j \right) - \left( \nu_j - \nu_{j-1} \right) \right] & = N \zeta_j \notag \\
				\Longrightarrow \quad \left( \nu_{j+1}- \nu_j \right) - \left( \nu_j - \nu_{j-1} \right)  & = - \zeta_j . \label{eq:nujandzetaj}
			\end{align}
			
		\subsubsection*{Balanced effective action}
			Following \cite{Uhlemann:2019ypp}, a preliminary convenient rewriting consists of using the sum over $j$ and a symmetrization over it to replace
			\begin{align}
				\rho_{j+1} (\phi) \mapsto & \frac{1}{2} \left[ \rho_{j+1} (\phi) + \rho_{j-1} (\phi) \right] \notag \\
				= & \underbrace{\rho_j (\phi)}_{\text{bifund-}j} + \underbrace{ \frac{1}{2} \left[ \left( \rho_{j+1} (\phi) - \rho_j (\phi) \right) - \left( \rho_j (\phi)  - \rho_{j-1} (\phi)\right) \right] }_{\text{bifund-}(j\pm 1) } \label{eq:rewriterhojderiv}
			\end{align}
			in the $j^{\text{th}}$ summand. The effective action is invariant under this rewriting: it simply amounts to count the bifundamental hypermultiplets as contributing half to the left node and half to the right.\par
			\medskip
			{\small\textbf{Exercise:} Plug \eqref{eq:rewriterhojderiv} into \eqref{eq:SeffLargeNFixP}. Write the sum over $j$ explicitly and check that it corresponds to a trivial rewriting.}\par
			\medskip
			Plugging in the replacement \eqref{eq:rewriterhojderiv} together with \eqref{eq:nujandzetaj}, the effective action \eqref{eq:SeffLargeNFixP} becomes  
			\begin{equation*}
			\begin{aligned}
				S_{\mathrm{eff}} ^{\Th_{\vec{\nu}}}  = N^2 \sum_{j=1} ^{L} \int \dd \phi \rho_j (\phi) & \left\{  \underbrace{ \frac{\phi^2}{\lambda_{d,j}} }_{\text{class.}} + \zeta_j \underbrace{ h_d (\phi) }_{\text{fund. hyper.}} \right. \\
					& + \int \dd \sigma \rho_j (\sigma) \left[ \underbrace{ v_d (\phi - \sigma) \cdot 1_{\sigma \ne \phi} }_{\text{vec.}}+ \underbrace{h_d (\phi - \sigma) }_{\text{bifund-}j} \right]  \\
					& \left. + \int \dd \sigma \underbrace{ \frac{1}{2} \left[ \left( \rho_{j+1} (\sigma) - \rho_j (\sigma) \right) - \left( \rho_j (\sigma)  - \rho_{j-1} (\sigma)\right) \right] h_d (\phi - \sigma) }_{\text{bifund-}(j\pm 1) } \right\} .
			\end{aligned}
			\end{equation*}
			To reduce clutter, in the second line we have used the shorthand $1_{\sigma \ne \phi}$ for the indicator function 
			\begin{equation*}
				1_{\sigma \ne \phi} := \begin{cases} 1 & \text{ if } \sigma \ne \phi \\ 0 & \text{ if } \sigma=\phi .\end{cases}
			\end{equation*}\par

		\subsubsection*{Large $L$ limit}
		So far, the procedure is nothing but the 't Hooft limit of a linear (or circular) quiver, with some rewriting based on the balanced hypothesis (\hyperref[hyp:LQ2]{LQ2}). At this point, the stage is set to take the limit $L \to \infty$. We introduce 
		\begin{equation*}
			z= \frac{j}{L+1}, \quad 0 < z < 1 ,
		\end{equation*}
		possibly with periodic identification $z+1 \sim z $ for circular quivers. In this way, all quantities labeled by the discrete index $j$ become functions of the continuous variable $z$, and the sums are replaced by integrals:
		\begin{equation*}
			\frac{1}{L} \sum_{j=1} ^{L} F_j \mapsto  \int_0 ^{1} \dd z F (z) .
		\end{equation*}
		In particular, 
		\begin{equation*}
			\nu(z) = \nu_{\lfloor z L \rfloor }  , \quad 0 < z < 1 ,
		\end{equation*}
		and likewise
		\begin{equation*}
			\lambda_{d,j} \mapsto \lambda_{d} (z) , \qquad \qquad \zeta_{j} \mapsto \zeta (z) . 
		\end{equation*}
		We also gather the $L$ eigenvalue densities $\left\{ \rho_j (\phi ) \right\}_{j=1, \dots, L}$ into the function 
		\begin{equation*}
			\rho (z, \phi) , \quad 0 \le z < 1 .
		\end{equation*}
		In this limit, the second term in \eqref{eq:rewriterhojderiv} becomes:
		\begin{equation*}
		\begin{aligned}
				 &\left[ \left( \rho_{j+1} (\sigma) - \rho_j (\sigma) \right) - \left( \rho_j (\sigma)  - \rho_{j-1} (\sigma)\right) \right]= \left[ \left( \frac{\rho_{j+1} (\sigma) - \rho_j (\sigma)}{ (j+1) -j} \right) - \left( \frac{\rho_j (\sigma)  - \rho_{j-1} (\sigma)}{j-(j-1)}\right) \right]  \\
				 &\stackrel{\delta \equiv \frac{1}{L}}{=} \frac{1}{L\delta}\left[ \left( \frac{\rho(z + \delta, \sigma) - \rho(z,\sigma)}{L (z + \delta) - Lz} \right) - \left(  \frac{\rho(z, \sigma) - \rho(z-\delta,\sigma)}{Lz - L(z -\delta)}\right) \right] \xrightarrow{ \ L \to \infty \ } \frac{1}{L^2}  \frac{\partial^2 \ }{\partial z^2} \rho (z, \sigma) .
		\end{aligned}
		\end{equation*}\par
		The effective action as $L \to \infty$ reads 
		\begin{equation}
			\begin{aligned}
				S_{\mathrm{eff}} ^{\Th_{\vec{\nu}}}  = N^2 L \int_0^1 \dd z \int \dd \phi \rho (z,\phi) & \left\{  \underbrace{ \frac{\phi^2}{\lambda_{d} (z)} }_{\text{class.}} + \zeta (z) \underbrace{ h_d (\phi) }_{\text{fund. hyper.}} \right. \\
					& +  \underbrace{  \int \dd \sigma \rho (z,\sigma) \left[ v_d (\phi - \sigma) \cdot 1_{\sigma \ne \phi} + h_d (\phi - \sigma) \right]  }_{\text{vec. + bifund.}} \\
					& \left. + \underbrace{ \int \dd \sigma  \frac{1}{2 L^2} \left( \frac{\partial^2 \ }{\partial z^2} \rho (z, \phi) \right) h_d (\phi - \sigma) }_{\text{residual bifund.} } \right\} .
			\end{aligned}
			\label{eq:SeffLQlargePnotscale}
			\end{equation}
			
		\subsubsection*{Rank function and kinks}
			
			Sending $L \to \infty$, the balancing condition in the form \eqref{eq:nujandzetaj} yields 
			\begin{equation}
			\label{eq:alancezetad2nu}
				\zeta (z) = - \frac{1}{L^2} \frac{\partial^2 \nu (z)}{\partial z^2} .
			\end{equation}
			If the ranks vary smoothly, the balancing condition (\hyperref[hyp:LQ2]{LQ2}) rules out the addition of flavours in the large $L$ limit. They can only appear at singularities of $\nu (z)$ where it is not of class $C^2 ([0,1])$. In other words, hypermultiplets in the fundamental representation are only attached at those gauge nodes that correspond to a kink for the rank function $\nu (z)$.\par
			We may impose regularity conditions on the rank functions $\nu (z)$, such as being twice differentiable everywhere on $[0,1]$ except a finite number of isolated points, namely 
			\begin{equation*}
				\nu \in C^2 \left( [0,1] \setminus \left\{ z_{\alpha} \right\}_{\alpha \in \mathscr{F}}  \right) .
			\end{equation*}
			In what follows, we consider the set $\mathscr{F}$ to be the collection of a finite number of indices, fixed from the onset, so that the cardinality $\lvert \mathscr{F} \rvert $ remains fixed as $L \to \infty$.\par
			This choice of regularity condition on the rank function $\nu (z)$ means that the flavours are ``sparse'', i.e. they are only attached to a finite number of gauge nodes, whose labels are encoded in $\mathscr{F}$. The Veneziano parameters then become 
			\begin{equation}
			\label{eq:LargePVeneziano}
				\zeta (z) = \sum_{\alpha \in \mathscr{F}} \zeta_{\alpha} \delta \left( z - z_{\alpha} \right) .
			\end{equation}

		\subsubsection*{Long quiver scaling}
		This is not the end of the story, because we must allow for the scaling of the eigenvalues with $L$. The origin and motivation for this scaling is precisely as in the M-theory limit: we do not let the couplings vary with $L$, which, combined with the requirement of a non-trivial saddle point configuration, leads to the ansatz 
		\begin{equation}
		\label{eq:phiscalingLargeP}
			\phi = L^{\alpha} (d-2) x , \quad x= \mathcal{O}(1) .
		\end{equation}
		Here the choice of coefficient $(d-2)$ is merely for convenience, while the power $\alpha >0$ will be determined momentarily by the self-consistency of the problem. Let us define the functions
		\begin{subequations}
		\begin{align}
			F^{(v)} _d (x ) & = \begin{cases} - \pi L^{\alpha} \lvert x \rvert & d=3 ,  \\   \frac{9\pi}{2} L^{3 \alpha} \lvert  x \rvert^3  & d= 5  ,   \end{cases}  \label{eq:LargePFV} \\
			F^{(h)} _d (x ) & = \begin{cases}    \pi L^{\alpha} \lvert x \rvert & d=3 ,  \\ -\frac{9\pi}{2} L^{3 \alpha} \lvert  x \rvert^3  & d= 5 , \end{cases}  \label{eq:LargePFH} \\
			F^{(0)} _d (x ) & = \begin{cases}  \frac{\pi}{4} L^{- \alpha} \delta ( x ) & d= 3,  \\ - \frac{27 \pi}{8} L^{\alpha} \lvert x \rvert & d=5, \end{cases} \label{eq:LargePF0}
		\end{align}
		\label{eq:largePscalingfunctions}
		\end{subequations}
		which are the leading large $L$ contributions to 
		\begin{equation*}
			v_d (L^{\alpha} (d-2) x), \quad h_d (L^{\alpha} (d-2) x), \quad  \left[ v_d (L^{\alpha} (d-2) x)+h_d (L^{\alpha} (d-2) x) \right]  , 
		\end{equation*}
		respectively. The function \eqref{eq:LargePF0} encapsulates the statements:
		\begin{itemize} 
		\item In $d=3$, the vector multiplet contributes $-\log \sinh \pi (\pi - \sigma) 1_{\sigma \ne \phi}$, while the bifundamental hypermultiplet contributes $+\log \cosh \pi (\pi - \sigma) $. Asymptotically for large argument, these two contributions cancel everywhere outside $\phi = \sigma $. The leftover is a peak at $\phi=\sigma$, namely $\delta (\phi-\sigma) = \delta (L^{\alpha} (x-y))$ (recall Figure \ref{fig:LogTanhCancel}). To write the first line of \eqref{eq:LargePF0} we have used $\delta (L^{\alpha} x )= L^{- \alpha} \delta (x)$.
		\item In $d=5$, the vector multiplet contributes $( \frac{\pi}{6} \lvert \phi - \sigma \rvert^3 - \pi  \lvert \phi - \sigma \rvert + \cdots) 1_{\sigma \ne \phi}$, while the bifundamental hypermultiplet contributes $- \frac{\pi}{6} \lvert \phi - \sigma \rvert^3 - \frac{\pi}{8}  \lvert \phi - \sigma \rvert + \cdots$. Asymptotically for large argument, the leading (cubic) contributions cancel everywhere outside $\phi = \sigma $, but the resulting term $\propto \lvert \phi - \sigma \rvert^3$ vanishes at that point. What remains is the sum of their sub-leading contributions, which add up to $\left(-1 - \frac{1}{8} \right)\pi \lvert \phi- \sigma \rvert$. After substituting the argument for $3 L^{\alpha} x$, this yields $ - \frac{9}{8} \pi \cdot 3 L^{\alpha}  \cdot \lvert x \rvert$, reproducing the second line of \eqref{eq:LargePF0}.
		\end{itemize}
		The case $d=4$ should be treated separately.\par
		We introduce the one-parameter family of eigenvalue densities $\varrho (z, x) $ for the scaled variable $x$, which satisfies 
		\begin{equation*}
			\varrho (z, x) \dd x = \rho_{\lfloor z L \rfloor } ( \phi ) \dd \phi  \qquad \forall 0 < z <1 ,
		\end{equation*}
		with $\phi$ and $x$ related as in \eqref{eq:phiscalingLargeP}. The function $\varrho (z, x)$ is also subject to the normalization:
		\begin{align*}
			\int \dd x \varrho (z, x) = \nu (z) ,
		\end{align*}
		plus the periodicity condition $\varrho (z+1, x) = \varrho (z, x)$ if one considers circular quivers.

		\subsubsection*{Effective action in the long quiver limit}
			The next step is to write \eqref{eq:SeffLQlargePnotscale} in the long quiver limit.  We send $L \to \infty $, utilize the ansatz \eqref{eq:phiscalingLargeP}, and make the substitutions of all the quantities as functions of $0 < z <1$. A key observation is that the rewriting \eqref{eq:rewriterhojderiv} allows to combine $v_d (L^{\alpha}x) + h_d (L^{\alpha}x)$, whose leading term is $F^{(0)} _d (x)$ in \eqref{eq:LargePF0}.\par
			Putting all the pieces together we find
			\begin{equation}
			\label{eq:SeffLargeP}
			\begin{aligned}
				S_{\mathrm{eff}} ^{\Th_{\vec{\nu}}}  = N^2 L \int_0 ^{1} \dd z & \int \dd x \varrho (z,x) \left\{  \zeta (z) \underbrace{ F^{(h)} _d (x) }_{\text{fund. hyper.}}\right. \\
				+ & \left.  \int \dd \tilde{x}  \left[ \underbrace{  \varrho (z,\tilde{x}) F^{(0)} _d (x - \tilde{x})  }_{\text{vec. + bifund.}}  +   \underbrace{ \frac{1}{2 L^2} \frac{\partial^2 \ }{\partial z^2} \varrho (z, \tilde{x}) F_d ^{(h)} (x - \tilde{x})  }_{\text{residual bifund.} } \right] \right\} ,
			\end{aligned}
			\end{equation}
			with $F_d ^{(0)}$ and $F_d ^{(h)}$ given in \eqref{eq:largePscalingfunctions}. We are assuming $d=3,5$, thus the classical piece vanishes.\par
			We are then led to a minimization problem for \eqref{eq:SeffLargeP}. In the next two subsections, we solve it in $d=3$ and $d=5$.
			
		\subsubsection*{Determining \texorpdfstring{$\alpha$}{alpha}}
			The definitions \eqref{eq:LargePFH}-\eqref{eq:LargePF0} are such that 
			\begin{equation*}
				F_d ^{(h)} \propto L^{\alpha (d-2)} , \qquad F^{(0)} _d \propto L^{\alpha (d-4)} ,
			\end{equation*}
			and $F_d ^{(h)}$ in the second line of \eqref{eq:SeffLargeP} appears multiplied by $L^{-2}$. The two terms in the second line of \eqref{eq:SeffLargeP} are of the same order in $L$ in the long quiver limit only if 
			\begin{equation*}
				\alpha (d-4) = \alpha (d-2) -2 \quad \Longrightarrow \quad \alpha = 1 .
			\end{equation*}
			We therefore have that, by \eqref{eq:largePscalingfunctions}, the partition function is dominated by eigenvalues that grow linearly with $L$ in both $d=3$ and $d=5$.

		\subsubsection*{Long quivers in diverse dimensions}
			So far, we have only introduced and discussed the long quiver limit in odd dimensions $d \in \left\{3,5\right\}$. The procedure to solve the long quiver limit is completely analogous in these dimensions \cite{Akhond:2022oaf}, the only difference residing in the explicit form of the leading terms in \eqref{eq:largePscalingfunctions}. The long quiver limit applies more generally, but the derivation in $d=4$ requires more effort. We omit it from this overview and refer to the original literature \cite{Nunez:2023loo}.\par 
			The upshot is that the long quiver limit shows a universal behaviour across dimensions.

		\subsection{Long quiver SCFTs in three dimensions}
		\label{sec:3dLQ}
			Let us first address three-dimensional long quivers. We need to plug into \eqref{eq:SeffLargeP} the expressions from \eqref{eq:largePscalingfunctions} corresponding to $d=3$.\par
			We get 
			\begin{equation}
			\begin{aligned}
				\left. S_{\mathrm{eff}} ^{\Th_{\vec{\nu}}} \right\vert_{d=3} = N^2 L \int_0 ^{1} \dd z & \int \dd x \varrho (z,x) \left\{  L^{\alpha} \zeta (z) \pi \lvert x \rvert  \right. \\
					& \left. \ + \int \dd \tilde{x}  \left[ L^{-\alpha} \varrho (z,\tilde{x}) \frac{\pi}{4} \delta (x-\tilde{x}) +  \frac{L^{-2 + \alpha}}{2} \frac{\partial^2 \ }{\partial z^2} \varrho (z, \tilde{x}) \pi \lvert x - \tilde{x} \rvert \right] \right\} 
			\end{aligned}
			\label{eq:SeffLargeP3dLQ}
			\end{equation}
			The two terms in square brackets compete if 
			\begin{equation*}
				- \alpha = -2 + \alpha \quad \Longrightarrow \quad \alpha = 1 .
			\end{equation*}
			As a consequence of the balancing condition \eqref{eq:alancezetad2nu}, the term in the first line $L^{\alpha} \zeta (z)$ is of order $L^{-2+\alpha}$, the same as the bifundamental hypermultiplet.\par
			The SPE for the action \eqref{eq:SeffLargeP3dLQ} is 
			\begin{equation}
			\label{eq:SPELQ3dbeforeD}
				\frac{1}{2} \frac{ \partial \ }{\partial x}  \varrho (z,x) + \int \dd \tilde{x} \frac{\partial^2 \ }{\partial z^2} \varrho (z, \tilde{x}) \text{sgn} (x-\tilde{x})  + L^2 \zeta (z) \text{sgn} (x) = 0 ,
			\end{equation}
			which we now differentiate with respect to $x$. We use 
			\begin{align*}
				 \frac{ \partial \ }{\partial x} \text{sgn} (x-\tilde{x}) &= \lim_{\varepsilon \to 0^{+}} \begin{cases} \frac{+1 - (+1)}{\varepsilon} & x >\tilde{x} + \varepsilon \\ \frac{+1 - (-1)}{\varepsilon} & \tilde{x} - \varepsilon <x  <\tilde{x} + \varepsilon \\ \frac{-1 - (-1)}{\varepsilon} & x <\tilde{x} - \varepsilon \end{cases} \\
					&= 2 \delta (x-\tilde{x})
			\end{align*}
			and, dividing both sides by $2$, \eqref{eq:SPELQ3dbeforeD} implies 
			\begin{equation}
			\label{eq:LQSPE3d}
				\left[ \frac{\partial^2 \ }{\partial z^2}  + \frac{1}{4} \frac{ \partial^2 \ }{\partial x^2 }  \right]  \varrho (z,x) = - L^2 \zeta (z) \delta (x) .
			\end{equation}
			The SPE \eqref{eq:LQSPE3d} is a Laplace equation, thus effectively reducing the problem of finding the eigenvalue density in the large $L$ limit to an electrostatic problem. Inserting flavours into the three-dimensional $\mN=4$ quiver introduces sources (or kinks) for the charges in the electrostatic problem.\par
			
			\subsubsection*{Direct solution to the Laplace equation}
			
			The Laplace equation \eqref{eq:LQSPE3d} has the solution 
			\begin{equation}
			\label{eq:SolVarrhozxLQ3d}
				\varrho (z,x) = - \frac{L^2}{2 \pi} \int_0 ^{1} \dd \tilde{z} \zeta (\tilde{z}) \log \left[  \frac{ \cosh (2 \pi x) - \cos (\pi (z-\tilde{z}))  }{ \cosh (2 \pi x) - \cos (\pi (z+\tilde{z})) } \right] .
			\end{equation}
			With the assumption that the flavour nodes are ``sparse'', we have (cf. \eqref{eq:LargePVeneziano})
			\begin{equation*}
				\zeta (\tilde{z}) = \sum_{\alpha \in \mathscr{F}} \zeta_{\alpha} \delta (\tilde{z} - z_{\alpha}) 
			\end{equation*}
			for only finitely many nodes at the position $z_{\alpha} = \alpha /L$, with $\mathscr{F}$ the collection of indices at which $\zeta_j \ne 0$. With this in mind, and recalling the substitution $\sum_{j=1} ^{L} \mapsto L \int_0 ^1 \dd \tilde{z} $, \eqref{eq:SolVarrhozxLQ3d} is recast in the form 
			\begin{equation*}
				\varrho (z,x) = - \frac{L}{2 \pi} \sum_{\alpha  \in \mathscr{F} } \zeta_{\alpha} \log \left[  \frac{ \cosh (2 \pi x) - \cos (\pi (z-z_{\alpha}))  }{ \cosh (2 \pi x) - \cos (\pi (z+z_{\alpha})) } \right]  .
			\end{equation*}\par

			\subsubsection*{Free energy of three-dimensional long quiver SCFTs}
			
				The leading contribution to the sphere free energy $\mf_{\cs^3} [\Th_{\vec{\nu}}]$ comes from the flavour term in \eqref{eq:SeffLargeP3dLQ}, integrated against the eigenvalue density \eqref{eq:SolVarrhozxLQ3d}. Direct integration gives \cite{Coccia:2020wtk}:
				\begin{align*}
					\mf_{\cs^3} [\Th_{\vec{\nu}}] & = - \frac{N^2 L^2}{2}  \int_0 ^1 \dd z \zeta (z) \int_{0}^{\infty} \dd x \varrho (z,x) \lvert x \rvert \\
						& =  - \frac{N^2 L^4}{4 }  \int_0 ^1 \dd z \zeta (z)  \int_0 ^1 \dd \tilde{z} \zeta (\tilde{z})  \int_{0}^{\infty} \dd x \lvert x \rvert \log \left[  \frac{ \cosh (2 \pi x) - \cos (\pi (z-\tilde{z}))  }{ \cosh (2 \pi x) - \cos (\pi (z+\tilde{z})) } \right] \\
						& \underbrace{=}_{\text{ \eqref{eq:LargePVeneziano}}}  - \frac{N^2 L^2}{4 }   \sum_{\alpha, \beta \in \mathscr{F}} \zeta_{\alpha} \zeta_{\beta} \int_{0}^{\infty} \dd x \lvert x \rvert \log \left[  \frac{ \cosh (2 \pi x) - \cos (\pi (z_{\alpha}-z_{\beta}))  }{ \cosh (2 \pi x) - \cos (\pi (z_{\alpha}+z_{\beta}))) } \right] \\
						& =  -  \frac{ N^2 L}{4 \pi^2 }  \sum_{\alpha, \beta \in \mathscr{F}} \zeta_{\alpha} \zeta_{\beta} \Re \left[ \mathrm{Li}_3 (e^{i \pi (z_{\alpha}+z_{\beta})}) - \mathrm{Li}_3 (e^{i \pi (z_{\alpha}-z_{\beta})})    \right] \\
						& =  N^2 \left(\frac{ L}{2 \pi }  \right)^2 \sum_{\alpha, \beta \in \mathscr{F}} \zeta_{\alpha} \zeta_{\beta} \Re \left[ \mathrm{Li}_3 (e^{i \pi (\alpha - \beta)/L}) - \mathrm{Li}_3 (e^{i \pi (\alpha + \beta)/L})    \right] .
				\end{align*}
				Notice that the term in the sum is only reliable at leading order in $L$, and should be approximated after the collection $\mathscr{F}$ is fixed. Using the large $L$ approximation  
				\begin{equation*}
					\pm \Re \left[ \mathrm{Li}_3 (e^{i \pi (\alpha \mp \beta)/L}) \right] = \pm 2 \zeta (3) \mp \frac{3}{2 L^2} \left( \alpha \mp \beta \right)^2  \mp \frac{1}{L^2} \left( \alpha \mp \beta \right)^2  \log \left\lvert  \frac{\alpha \mp \beta}{L} \right\rvert + \cdots ,
				\end{equation*}
				the constant terms cancel out, and the leading $\mathcal{O} (L^{-2})$ dependence combines with the overall factor of $L^2$. In particular, $\Re \left[ \cdots \right]$ gives a non-negative real number at leading order in $L$, thus confirming $\mf_{\cs^3} [\Th_{\vec{\nu}} ] \ge 0$.

			\subsubsection*{Fourier series solution to the Laplace equation}
				An alternative way to solve \eqref{eq:LQSPE3d} is based on expanding $\varrho (z, x)$ in Fourier modes in the compact coordinate $z$. This approach was taken in \cite{Akhond:2021ffz,Akhond:2022oaf}.\par
				The input that specifies the linear quiver is the array of ranks $\vec{\nu}$, which is collected in $\nu (z)$ for $0<z<1$. The same information can be encapsulated in the Fourier expansion of $\nu (z)$ on $[0,1]$, namely 
				\begin{equation*}
					\nu (z) = \sum_{k=1}^{\infty} \hat{\nu}_k \sin (\pi k z ) .
				\end{equation*}
				We are denoting $\left\{ \hat{\nu}_k \right\}_{k \ge 1}$ the collection of Fourier coefficients of the rank function $\nu (z)$. Note that we are expanding with boundary conditions $\nu (0)=0=\nu (1)$, since there is no quiver there, thus the rank function vanishes. Using the balancing condition in the form \eqref{eq:alancezetad2nu} we also have 
				\begin{equation}
				\label{eq:zetadnuFourier}
					\zeta (z) = \frac{\pi^2}{L^2} \sum_{k=1}^{\infty} k^2 \hat{\nu}_k \sin (\pi k z ) .
				\end{equation}\par
				Expanding  \eqref{eq:LQSPE3d} in Fourier modes and denoting $\hat{\varrho}_k (x)$ the $k^{\text{th}}$ Fourier mode of $\varrho (z,x)$ with respect to the compact coordinate $z$, we find 
				\begin{equation*}
					 \sum_{k=1}^{\infty} \left[  \frac{1}{4} \partial_x ^2 \hat{\varrho}_k (x)  - k^2 \pi^2  \hat{\varrho}_k (x) + k^2 \pi^2 \hat{\nu}_k \delta (x) \right] \sin (\pi k z ) = 0 .
				\end{equation*}
				The SPE is thus reduced to a second order ordinary differential equation in $x$ with sources: 
				\begin{equation*}
					\frac{1}{4} \partial_x ^2 \hat{\varrho}_k (x)  - k^2 \pi^2  \hat{\varrho}_k (x) + k^2 \pi^2 \hat{\nu}_k \delta (x) =0,
				\end{equation*}
				which we rewrite in the more convenient form
				\begin{equation*}
					- \frac{1}{(2 \pi k)^2} \partial_x ^2 \hat{\varrho}_k (x)  +  \hat{\varrho}_k (x) = \hat{\nu}_k \delta (x) .
				\end{equation*}
				Its solution is 
				\begin{equation}
				\label{eq:solrhoLQFourier}
					\varrho (z,x) = C \sum_{k=1}^{\infty} k \hat{\nu}_k \sin (\pi k z ) e^{- 2 \pi k \lvert x \rvert } ,
				\end{equation}
				with $C>0$ an overall constant fixed by normalization:
				\begin{align*}
					\nu (z) &= \int \dd x \varrho (z,x) \\
						&= C \sum_{k=1}^{\infty} k \hat{\nu}_k \sin (\pi k z ) \underbrace{ \int e^{- 2 \pi k \lvert x \rvert } \dd x }_{2 \cdot \frac{1}{2 \pi k}} \\
						&= \frac{C}{\pi}  \sum_{k=1}^{\infty} \hat{\nu}_k \sin (\pi k z ) .
				\end{align*}
				Comparing with the Fourier expansion of $\nu (z)$ fixes $C=\pi$.

			\subsubsection*{Free energy of three-dimensional long quiver SCFTs via Fourier series}
				Let us now compute the free energy using the Fourier series technique. The general formula \eqref{eq:FishalfS1} dictates that 
				\begin{align*}
					\mf_{\cs^3} [\Th_{\vec{\nu}}] & = \left. \frac{1}{2} N^2 L^2 \int_0 ^{1} \dd z \int_{0}^{\infty} \dd x \varrho (z,x) \zeta (z) \pi \lvert x \rvert \right\rvert_{\text{on-shell}} \\
						&=  \frac{\pi}{2} N^2  \int_0 ^{1} \dd z  \underbrace{\left( \pi^2 \sum_{k=1}^{\infty} k^2 \hat{\nu}_k \sin (\pi k z )\right)}_{\text{\eqref{eq:zetadnuFourier}}} \int_{0}^{\infty} \dd x \varrho (z,x) \lvert x \rvert \\
						&=  \frac{\pi}{2} N^2  \int_0 ^{1} \dd z  \left( \pi^2 \sum_{k=1}^{\infty} k^2 \hat{\nu}_k \sin (\pi k z )\right) \underbrace{ \left( \pi \sum_{\ell=1}^{\infty} \ell \hat{\nu}_{\ell} \sin (\pi \ell z )   \right) }_{\text{from \eqref{eq:solrhoLQFourier}}} \underbrace{ \int_{0}^{\infty} e^{- 2 \pi \ell \lvert x \rvert } \lvert x \rvert \dd x }_{\frac{1}{2 \pi^2 \ell^2} } \\
						&= \frac{\pi^2}{4} N^2 \sum_{k=1}^{\infty} k^2 \hat{\nu}_k \sum_{\ell=1}^{\infty} \frac{1}{\ell} \hat{\nu}_{\ell} \underbrace{\int_0 ^{1} \dd z \sin (\pi k z ) \sin (\pi \ell z ) }_{\frac{1}{2} \delta_{k \ell}} \\
						&= \frac{\pi^2}{8} N^2 \sum_{k=1}^{\infty} k \hat{\nu}_k ^2 .
				\end{align*}
				A long linear quiver which moreover is balanced is entirely determined by $\vec{\nu}$. Translating this information into the Fourier coefficients $\left\{\hat{\nu}_k \right\}_{k \ge 1}$, we obtain the free energy 
				\begin{equation*}
					\mf_{\cs^3} [\Th_{\vec{\nu}}] = \frac{\pi^2}{8} N^2 \sum_{k=1}^{\infty} k \hat{\nu}_k ^2 .
				\end{equation*}

		\subsection{Long quiver SCFTs in five dimensions}
		\label{sec:5dLQ}
		
			Let us now go back to \eqref{eq:SeffLargeP} and study it at $d=5$. We simply have to read off the corresponding functions $F_d ^{(h)}, F_d ^{(0)} $ from \eqref{eq:largePscalingfunctions}.\par
			The effective action of a five-dimensional SCFT in the long quiver limit is 
			\begin{equation}
			\begin{aligned}
				S_{\mathrm{eff}} ^{\Th_{\vec{\nu}}} \vert_{d=5} = N^2 L \int_0 ^{1} \dd z & \int \dd x \varrho (z,x) \left\{  - L^{3\alpha} \zeta (z) \frac{\pi}{6} \lvert 3 x \rvert^3  \right. \\
					& \left. \ - \int \dd \tilde{x}  \left[ L^{\alpha} \varrho (z,\tilde{x}) 3^2 \frac{\pi}{8} \lvert 3 (x-\tilde{x}) \rvert +  \frac{L^{-2 + 3 \alpha}}{2} \frac{\partial \ }{\partial z^2} \varrho (z, \tilde{x}) \frac{\pi}{6} \lvert 3 (x - \tilde{x}) \rvert^3 \right] \right\} .
			\end{aligned}
			\label{eq:SeffLargeP5dLQ}
			\end{equation}
			The two terms in the square bracket contribute to the same order in $L$ if 
			\begin{equation*}
				\alpha = -2 + 3 \alpha \quad \Longrightarrow \quad \alpha = 1 .
			\end{equation*}
			Once again, a consequence of the balancing condition \eqref{eq:alancezetad2nu} is that the contribution $L^{3 \alpha} \zeta (z)$ from the fundamental hypermultiplets to the first line in \eqref{eq:SeffLargeP5dLQ} is of order $L^{-2 + 3 \alpha}$, thus of the same order as the rest if $\alpha=1$.\par
			The SPE for \eqref{eq:SeffLargeP5dLQ} is 
			\begin{equation*}
				\int \dd \tilde{x} \left[  - \frac{27 \pi}{4} \varrho (z,\tilde{x}) - \frac{27 \pi}{2}  \left( \frac{\partial^2 \ }{\partial z^2} \varrho (z, \tilde{x}) \right) \lvert x - \tilde{x} \rvert^2 \right]  \text{sgn} (x-\tilde{x}) - L^2 \frac{27 \pi}{2} \zeta (z) x^2 \text{sgn}(x) = 0 ,
			\end{equation*}
			which, after simplifications, reads 
			\begin{equation*}
				\int \dd \tilde{x} \left[ \frac{1}{2}\varrho (z,\tilde{x})  +  \left( \frac{\partial^2 \ }{\partial z^2} \varrho (z, \tilde{x}) \right) \lvert x - \tilde{x} \rvert^2 \right]  \text{sgn} (x-\tilde{x}) + L^2 \zeta (z) x^2 \text{sgn}(x) = 0 .
			\end{equation*}
			The next step is to differentiate this expression three times with respect to $x$. Using:
			\begin{align*}
				\frac{\partial^3 \ }{\partial x^3} \ \int \dd \tilde{x} \frac{1}{2}\varrho (z,\tilde{x}) \text{sgn} (x-\tilde{x}) &= \frac{\partial^2 \ }{\partial x^2} \  \int \dd \tilde{x} \frac{1}{2}\varrho (z,\tilde{x}) \cdot 2 \delta (x-\tilde{x}) \\
				&= \frac{\partial^2 \ }{\partial x^2}  \varrho (z,x) 
			\end{align*}
			and 
			\begin{align*}
				\frac{\partial^3 \ }{\partial x^3} \ \int \dd \tilde{x} \left( \frac{\partial^2 \ }{\partial z^2} \varrho (z, \tilde{x}) \right) \lvert x - \tilde{x} \rvert^2  \text{sgn} (x-\tilde{x}) &=  \frac{\partial^2 \ }{\partial z^2}  \frac{\partial^3 \ }{\partial x^3} \ \int \dd \tilde{x} \varrho (z,\tilde{x}) \lvert x - \tilde{x} \rvert^2  \text{sgn} (x-\tilde{x}) \\
				&= \frac{\partial^2 \ }{\partial z^2} \ \int \dd \tilde{x} \varrho (z,\tilde{x}) \cdot 2! \cdot 2 \delta (x-\tilde{x}) \\
				& = 4 \frac{\partial^2 \ }{\partial z^2} \varrho (z,x)
			\end{align*}
			and dividing both sides by $4$ gives 
			\begin{equation*}
				\left[  \frac{1}{4} \frac{ \partial^2 \ }{\partial x^2}  +  \frac{\partial^2 \ }{\partial z^2}  \right] \varrho (z,x) = - L^2 \zeta (z) \delta (x) ,
			\end{equation*}
			which reproduces the Laplace equation \eqref{eq:LQSPE3d} already found in $d=3$. The solution is therefore \eqref{eq:SolVarrhozxLQ3d}.\par
			\medskip
			{\small\textbf{Exercise:} Repeat the derivation allowing for a finite gauge coupling.}

			\subsubsection*{Free energy of five-dimensional long quiver SCFTs}
				Thanks to the analogy between solutions in three and five dimensions, the computation of the sphere free energy is essentially identical to the $d=3$ setting, the unique difference being in the explicit form of $F_{d=5} ^{(h)}$. We obtain 
				\begin{align*}
					\mf_{\cs^5} [\Th_{\vec{\nu}}] & =  N^2 L^2 \frac{27 \pi}{12}  \int_0 ^1 \dd z \zeta (z) \int_{0}^{\infty} \dd x \varrho (z,x)  \lvert x \rvert^3 \\
						& =  27 N^2 \left(\frac{ L}{2 \pi }  \right)^4 \sum_{\alpha, \beta \in \mathscr{F}} \zeta_{\alpha} \zeta_{\beta}  \Re \left[ \mathrm{Li}_5 (e^{i \pi (\alpha + \beta)/L}) - \mathrm{Li}_5 (e^{i \pi (\alpha - \beta)/L})    \right]    .
				\end{align*}
				This solution was obtained in \cite{Uhlemann:2019ypp}. Again, the derivation is only reliable at leading order in $L$ and, approximating the term $\Re \left[ \cdots \right]$, gives a negative real number, in agreement with the F-theorem.

				\subsubsection*{Free energy of five-dimensional long quiver SCFTs via Fourier series}
				Let us now compute the free energy using the Fourier series technique. The SPE in $d=5$ is identical to $d=3$, thus its Fourier series solution is again \eqref{eq:solrhoLQFourier}. One then takes the general formula \eqref{eq:FishalfS1} and computes exactly as we did in $d=3$, except that now the $d=5$ expression in \eqref{eq:LargePFH} must be used. We get 
				\begin{align*}
					\mf_{\cs^5 } [ \Th_{\vec{\nu}} ] & = \frac{1}{2} N^2 L^4 \int_0 ^{1} \dd z \int_{0}^{\infty} \dd x \varrho (z,x) \zeta (z) \frac{9\pi}{2} \lvert x \rvert^3 \rvert_{\text{on-shell}} \\
						&=  \frac{9 \pi}{4} N^2 L^2 \int_0 ^{1} \dd z  \underbrace{ \left( \pi^2 \sum_{k=1}^{\infty} k^2 \hat{\nu}_k \sin (\pi k z )\right) }_{\text{\eqref{eq:zetadnuFourier}}} \underbrace{ \left( \pi \sum_{\ell=1}^{\infty} \ell \hat{\nu}_{\ell} \sin (\pi \ell z )   \right) }_{\text{from \eqref{eq:solrhoLQFourier}}} \underbrace{ \int_{0}^{\infty} e^{- 2 \pi \ell \lvert x \rvert } \lvert x \rvert^3 \dd x }_{ \frac{3}{4 \pi^4 \ell^4} } \\
						&= \frac{9 }{4} \cdot \frac{3}{4} \cdot N^2 L^2 \sum_{k=1}^{\infty} k^2 \hat{\nu}_k \sum_{\ell=1}^{\infty} \frac{1}{\ell^3 } \hat{\nu}_{\ell} \underbrace{\int_0 ^{1} \dd z \sin (\pi k z ) \sin (\pi \ell z ) }_{\frac{1}{2} \delta_{k \ell} } \\
						&= \frac{27}{32} N^2 L^2 \sum_{k=1}^{\infty} k^{-1} \hat{\nu}_k ^2 .
				\end{align*}
				A long linear quiver which moreover is balanced is entirely determined by $\vec{\nu}$. Translating this information into the Fourier coefficients $\left\{ \hat{\nu}_k \right\}_{k \ge 1}$ we obtain the free energy 
				\begin{equation*}
					\mf_{\cs^5} [\Th_{\vec{\nu}} ] = \frac{27}{32} N^2 L^2 \sum_{k=1}^{\infty} k^{-1} \hat{\nu}_k ^2 .
				\end{equation*}

\clearpage
\begin{appendix}

\section{Further reading}
\label{app:reading}

\subsubsection*{Large \texorpdfstring{$N$}{N} reviews}
	Several excellent previous reviews exist, discussing large $N$ limits. Our hope is to provide a more pedagogical cut to the presentation of existing results in this overview. The interested reader may refer to the following list for more advanced reviews that cover one of the topics introduced here.
	\begin{itemize}
		\item The reviews \cite{Marino:2011nm,Marino:2016new} are focused on supersymmetric Chern--Simons-matter theories, especially ABJM.
		\item A review of large $N$ results from localization in the $d=4$ $\mN=2^{\ast}$ theory is \cite{Russo:2013sba}, while an overview of large $N$ $\cs^4$ partition functions and their role in AdS$_5$/CFT$_4$ is \cite{Zarembo:2016bbk}. For an early review of the large $N$ limit of $\mN=4$ and its string theory dual, consult \cite{Russo:2000ak}, dating back to the pre-localization era.
		\item A useful review of matrix models for gauge theories in $d=5$ is J. Minahan's \cite{Minahan:2016xwk}.
	\end{itemize}

\subsubsection*{Large \texorpdfstring{$N$}{N} without supersymmetry}
	While not directly pertinent to the main theme of these notes, it is worthwhile to mention the large $N$ limit of vector models. A detailed review is \cite{Moshe:2003xn}, and applications to phase transitions and critical phenomena can be found in the textbook \cite{ZinnJustin:2007zz}.

\subsubsection*{Localization}
	The format of supersymmetric localization on spheres that we adopt in this work is based on V. Pestun's pioneering work \cite{Pestun:2007rz}. The idea of localization in quantum field theory, however, has its roots in earlier works by E. Witten, N. Nekrasov, and others.\par
	The approach to localization needed to read these notes is summarized in \cite{Cremonesi:2013twh}. Additionally, the following list includes references specialized to different dimensions.
	\begin{itemize}
		\item A localization formula for the partition function on $\cs^2$ was derived in \cite{Benini:2012ui,Doroud:2012xw}.
		\item In $d=3$, localization was first worked out by A. Kapustin, B. Willet and I. Yaakov \cite{Kapustin:2009kz}. A canonical reference for a detailed review is \cite{Willett:2016adv}, see also the more recent \cite{Closset:2019hyt}.
		\item Localization on $\cs^4$ \cite{Pestun:2007rz} is reviewed in \cite{Pestun:2014mja,Hosomichi:2016flq}. A localization formula for $\mN=2$ topologically twisted theories on (weighted) projective spaces is in \cite{Bershtein:2015xfa,Lundin:2023tzw,Mauch:2024uyt}.
		\item Localization on $\cs^5$ was derived in \cite{Kallen:2012cs,Kallen:2012va,Kim:2012qf}, see also \cite{Alday:2015lta,Lundin:2021zeb,Closset:2022vjj} and the review \cite{Qiu:2016dyj}.
	\end{itemize}
	The maximally supersymmetric theory on $\cs^d$ in arbitrary $d$ has been considered in \cite{Minahan:2015jta}.

\subsubsection*{Sphere free energies in three dimensions}
	In the main text we have focused on a class of examples amenable to showcase the approach and procedures that constitute the main topic of this review. Other approaches exist, especially in three dimensions, where much progress has been made in matching large $N$ sphere free energies with their holographic duals beyond leading order.
	\begin{itemize}
		\item More sophisticated examples of large $N$ supersymmetric theories were considered in \cite{Gulotta:2011vp,Gulotta:2012yd,Crichigno:2012sk,Assel:2012cp,Mezei:2013gqa,Amariti:2019pky}.
		\item A different approach, based on the Fermi gas formalism, was put forward by M. Mari\~{n}o and P. Putrov in \cite{Marino:2011eh}. Related further developments in ABJM theory include \cite{Drukker:2010nc,Fuji:2011km,Moriyama:2014gxa,Nosaka:2024gle,Kubo:2024qhq}.
		\item Large $N$ expansions of partition functions of three-dimensional quiver theories with holographic dual in Type IIA or M-theory have been given in \cite{Hong:2021bsb,Bobev:2022jte,Bobev:2022eus,Bobev:2023lkx,Geukens:2024zmt}.
	\end{itemize}
\subsubsection*{Sphere free energies in four dimensions}
	In four dimensions, only the maximally supersymmetric $\mN=4$ super-Yang--Mills theory has been discussed in the main text. Besides the plethora of references studying all facets of such theory, there are several other theories that can be studied at large $N$. Here we mention the representative example of the $\mN=2^{\ast}$ theory, or adjoint QCD with eight supercharges. It can be derived by a deformation of $\mN=4$ theory in which the adjoint hypermultiplet acquires a mass. Aspects of this theory at large $N$ have been considered in \cite{Russo:2012ay,Bobev:2013cja,Bobev:2018hbq,Russo:2019lgq}, among others.
\subsubsection*{Sphere free energies in five dimensions}
	Additionally, in $d=5$, a thorough analysis of supersymmetric field theories on $\cs^5$ can be found in the series \cite{Chang:2017cdx,Chang:2017mxc,Chang:2019uag}. A systematic study of large $N$ free energies and RG flows in $d=5$ is \cite{Fluder:2020pym}.

\subsubsection*{Long quivers}
	The long quiver limit has been reviewed in Section \ref{sec:LargeNLongQ} and, as mentioned therein, it was first solved by C. Uhlemann \cite{Uhlemann:2019ypp}. It has soon found further applications in five-dimensional \cite{Uhlemann:2019lge,Legramandi:2021uds,Akhond:2022awd,Akhond:2022oaf,Santilli:2023fuh} and three-dimensional \cite{Coccia:2020cku,Coccia:2020wtk,Akhond:2021ffz} quivers. Four-dimensional long quivers are more subtle, and have been solved in \cite{Nunez:2023loo}. Holography approaches to the long quiver SCFTs include \cite{Gutperle:2018vdd,Bergman:2018hin,Fluder:2018chf,Nunez:2019gbg,Fluder:2019szh,Gutperle:2020rty,Apruzzi:2022nax,Macpherson:2024frt,Chatzis:2024top,Chatzis:2024kdu}.

\subsubsection*{Wilson loops}
	The consideration of supersymmetric Wilson loops in supersymmetric field theories on $\cs^d$ using matrix models techniques started with the foundational papers of localization of V. Pestun \cite{Pestun:2007rz} and A. Kapustin, B. Willett and I. Yaakov \cite{Kapustin:2009kz}. There is a vast literature addressing the large $N$ limit of Wilson loops in the fundamental representation, and here we limit ourselves to mention references that consider less standard techniques.
	\begin{itemize}
		\item Wilson loops in higher representations have been studied in SQCD in three \cite{Santilli:2018byi}, four \cite{Hartnoll:2006is,Chen-Lin:2015dfa,Russo:2017ngf} and five \cite{Assel:2012nf,Santilli:2021qyt} dimensions.
		\item Wilson loops in the long quiver limit have been studied in \cite{Uhlemann:2020bek,Fatemiabhari:2022kpv}.
	\end{itemize}

\subsubsection*{Product manifolds}
	In the main text we have focused on supersymmetric field theories on the round sphere $\cs^d$. Of course, there are many other topologies of interest, which have been studied at large $N$ and compared with predictions from AdS/CFT. 
	\begin{itemize}
		\item The other main topology is $\cs^{d-1} \times \cs^1$. The literature on this topic is too broad to be exhaustively reviewed here, thus we limit ourselves to list a small sample of recent works in AdS$_4$/CFT$_3$ \cite{Nian:2019pxj,Bobev:2022wem,BenettiGenolini:2023rkq,Amariti:2023ygn,Bobev:2024mqw}, in AdS$_5$/CFT$_4$ \cite{Cabo-Bizet:2018ehj,Benini:2018ywd,ArabiArdehali:2019tdm,Cabo-Bizet:2019osg,Cabo-Bizet:2019eaf,ArabiArdehali:2019orz,Amariti:2020jyx,Amariti:2021ubd,Choi:2021rxi,Cabo-Bizet:2021jar,ArabiArdehali:2023bpq,Choi:2023tiq}, and in AdS$_6$/CFT$_5$ \cite{Bergman:2013koa,Choi:2019miv,Crichigno:2020ouj}. An EFT approach to the study of these partition functions was put forward by A. Arabi-Ardehali and J. Hong \cite{Ardehali:2021irq}, building on \cite{Cassani:2021fyv}.
		\item It is possible to compactify a $d$-dimensional SCFT on a Riemann surface and study the theory on $\cs^{d-2} \times \Sigma$. Large $N$ limits of various five-dimensional gauge theories on these product manifolds were addressed in \cite{Crichigno:2018adf,Jain:2021sdp,Jain:2022avc}, see also \cite{Santilli:2020uht} for an outlier example and \cite{Bah:2018lyv,Legramandi:2021aqv} for the holographic side of the story.
	\end{itemize}

\section{Planar matrix models}
\label{app:MMHU}

\subsection{'t Hooft limit of Hermitian matrix models}
\label{app:HMM}

The setup is as in Section \ref{sec:MMplanarlimit}, and we complement the discussion therein by showcasing two concrete instances of computations.

\subsubsection*{Eigenvalue density for even potentials}
		To gain familiarity with the type of computations, let us consider a polynomial potential of degree $\ell+1$ containing only even powers:
		\begin{equation*}
			W (\phi) = \sum_{p =1}^{(\ell +1)/2} \frac{c_{2p}}{2p} \phi^{2p} .
		\end{equation*}
		 The derivative $W^{\prime} (z)$ has $\ell$ zeros, which will be all distinct for generic coefficients $c_{2p}$. We thus look for an $\ell$-cut solution. We observe that
		 \begin{itemize}
		 	\item[(i)] By assumption $\ell +1 \in 2 \N$, thus the number of cuts $\ell$ is odd.
		 	\item[(ii)] By assumption the potential is even, and so is the square of the Vandermonde determinant, thus the matrix model is an even function of the eigenvalues. The support of the eigenvalue density will be invariant under $\phi \mapsto -\phi$, thus the $\ell$ cuts will be symmetric with respect to the origin.
		 	\item[(iii)] Convergence of the matrix model requires $c_{\ell +1}>0$ and we can always reabsorb this coefficient into a change of variables $\tilde{\phi}=( c_{\ell +1})^{\frac{1}{\ell +1}} \phi$ together with a redefinition of the coefficients, $\tilde{c}_{2p} := c_{2p} (c_{\ell +1})^{\frac{2p}{\ell +1}}$. We henceforth drop the tilde and set $c_{\ell +1}=1$ without loss of generality.
		 \end{itemize}\par
		 We use formula \eqref{eq:omegaMMpoly} to evaluate the resolvent $\omega (z)$. The polynomial $W^{\prime}$ has no poles, thus we are left with 
		 \begin{equation*}
		 	\omega (z) = \frac{1}{2} \sum_{p =1}^{(\ell +1)/2} c_{2p} z^{2p-1} - \frac{1}{2} \sqrt{ \prod_{s=1} ^{\ell} (z-A_s)(z-B_s) } .
		 \end{equation*}
		 The eigenvalue density is then extracted using \eqref{eq:disceqrhoomega}. Let us assume that $\phi \in [A_{s_{\star}}, B_{s_{\star}}]$ for a fixed $s_{\star} \in \left\{ 1, \dots, \ell \right\}$, and evaluate $\omega (z)$ at the points $z = \phi \pm i \varepsilon$. We find 
		 \begin{align*}
		 	\rho (\phi) &= - \frac{1}{2\pi i} \lim_{\varepsilon \to 0^{+}} \left[ \omega (\phi + i \varepsilon ) - \omega (\phi - i \varepsilon )  \right] \\
		 	&= - \frac{1}{2\pi i} \left( - \frac{1}{2} \right) \lim_{\varepsilon \to 0^{+}} \left[ \prod_{s=1} ^{\ell} \sqrt{ (\phi+i \varepsilon-A_s)(\phi+i \varepsilon-B_s) } -  \prod_{s=1} ^{\ell} \sqrt{ (\phi-i \varepsilon-A_s)(\phi-i \varepsilon-B_s) } \right] \\
		 	&= \frac{1}{4 \pi i} \lim_{\varepsilon \to 0^{+}} \left[  \sqrt{ (\phi+i \varepsilon-A_{s_{\star}})(\phi+i \varepsilon-B_{s_{\star}}) } -  \sqrt{ (\phi-i \varepsilon-A_{s_{\star}})(\phi-i \varepsilon-B_{s_{\star}}) } \right] \\
		 		& \qquad \qquad \times \underbrace{\prod_{s\ne s_{\star}} \sqrt{ (\phi-A_s)(\phi-B_s) } }_{\text{no disc. if }s \ne s_{\star}} .
		 \end{align*}
		 In the latter line we have used that the square roots have a definite sign (either both terms under the square root are positive, or both are negative) if $s \ne s_{\star}$. Rearranging some minus signs, but being careful not to pull them out of the square root, we easily get
		 \begin{align*}
		 	\rho (\phi) &= \frac{1}{4 \pi i} \lim_{\varepsilon \to 0^{+}} \left[  \sqrt{ -(\phi+i \varepsilon-A_{s_{\star}})(B_{s_{\star}} -\phi-i \varepsilon) } -  \sqrt{ -(\phi-i \varepsilon-A_{s_{\star}})( B_{s_{\star}} - \phi+i \varepsilon) } \right] \\
		 	& \qquad \qquad \times\prod_{s\ne s_{\star}} \sqrt{ (\phi-A_s)(\phi-B_s) } .
		 \end{align*}
		 To make the computation more transparent, we now approximate the phase of $B_{s_{\star}} -\phi - i \varepsilon $ by a small negative phase $e^{- i \epsilon}$; likewise the phase of $B_{s_{\star}} -\phi + i \varepsilon $ is approximated by a small and positive phase $e^{i \epsilon}$. Given that $\varepsilon \to 0$ is taken eventually, the phase $\epsilon \to 0^{+}$ as $\varepsilon \to 0^{+}$. Then, 
		 \begin{align*}
		 	\rho (\phi) &= \frac{1}{4 \pi i} \lim_{\varepsilon \to 0^{+}} \left[  \sqrt{ e^{i (\pi - \epsilon)}(\phi-A_{s_{\star}})(B_{s_{\star}} -\phi) } -  \sqrt{ e^{i (-\pi + \epsilon)}(\phi-A_{s_{\star}})( B_{s_{\star}} - \phi) } \right] \\
		 	& \qquad \qquad \times \prod_{s\ne s_{\star}} \sqrt{ (\phi-A_s)(\phi-B_s) } \\
		 	&= \frac{1}{4 \pi i} \left( e^{i \frac{\pi}{2}} - e^{-i \frac{\pi}{2}} \right) \sqrt{(\phi-A_{s_{\star}})(B_{s_{\star}} -\phi) } \prod_{s\ne s_{\star}} \sqrt{ (\phi-A_s)(\phi-B_s) } \\
		 	&= \frac{1}{2\pi} \sqrt{(\phi-A_{s_{\star}})(B_{s_{\star}} -\phi) } \prod_{s\ne s_{\star}} \sqrt{ (\phi-A_s)(\phi-B_s) } .
		 \end{align*}
		In the first line we have been consistent with the choice of branch cut: all phases must lie in $(-\pi, \pi)$, whereby we write $-e^{i \epsilon} = e^{i (-\pi + \epsilon)}$.\par 
		The crux of the matter here is that the two square roots approach the branch cut from two different sides, namely $e^{\pm i(\pi - \epsilon)}$ as $\epsilon \to 0^{+}$. They sit on two distinct Riemann sheets, producing different phases. We thus find
		 \begin{equation*}
		 	\rho (\phi) = \frac{1}{2\pi} \sqrt{(\phi-A_{s_{\star}})(B_{s_{\star}} -\phi) } \prod_{s\ne s_{\star}} \sqrt{ (\phi-A_s)(\phi-B_s) } \qquad \text{ if }\phi \in [A_{s_{\star}}, B_{s_{\star}}] , \quad s_{\star}= 1, \dots, \ell .
		 \end{equation*}
		 The dependence of the eigenvalue density on the couplings $\left\{ c_2, c_4, \dots, c_{\ell} \right\}$ is hidden in the $\ell$ pairs of endpoints $\left\{ (A_s, B_s) \right\}_{s=1, \dots, \ell}$, which depend on the couplings through the equations imposed by the normalization condition.

\subsubsection*{One-cut solution to the quartic matrix model}
		 In general, the mechanism that reduces from an $\ell$-cut to an $(\ell-1)$-cut solution corresponds to the merger of two of the endpoints of the intervals forming $\surho$. Beyond that value of the couplings, the equations fixing the two endpoints do not admit a real solution.\par
		 Let us exemplify the one-cut solution in the quartic matrix model, with potential 
		 \begin{equation*}
			W (\phi) = \frac{1}{2} \phi^2 + \frac{\lambda}{4} \phi^4.
		\end{equation*}
		We change variables in the matrix model according to $\tilde{\phi}= \lambda^{1/2} \phi$, so that (omitting the tilde) the potential becomes 
		\begin{equation*}
			W (\phi) = \frac{1}{\lambda}\left( \frac{\phi^2}{2}  + \frac{\phi^4}{4} \right) \qquad \Longrightarrow \qquad W^{\prime} (\phi) = \frac{1}{\lambda}\left( \phi + \phi^3 \right) .
		\end{equation*}\par
		We now consider the one-cut phase. That is to say, we make an ansatz $\ell=1$ for the saddle point eigenvalue density. The derivative $W^{\prime} (z)$ is holomorphic, thus our general formula \eqref{eq:omegaMMpoly} for the planar resolvent $\omega (z)$ gives:
		\begin{equation*}
			\omega_{\text{quartic}} (z) = \frac{1}{2\lambda}\left( z+z^3 \right) - \frac{1}{2\lambda} (z+iC)(z-iC) \sqrt{(z-A)(z-B)} .
		\end{equation*}
		The term $(z+iC)(z-iC)=z^2+C^2$ should be computed directly from the integral, under the ansatz of a one-cut solution. Here instead we take a more pedestrian approach: we use this ansatz, consistent with the general formula and the one-cut hypothesis, and fix the value of $C$ by self-consistency. Notice that we should in principle allow a more general ansatz $(z+iC) (z+i D)$ but, relying on the $\phi \mapsto -\phi$ symmetry of the matrix model, we already know that, in this case, $D=-C$.\par
		Expanding $\omega (z)$ in $1/z$ we find 
		\begin{align*}
		2 \lambda \omega_{\text{quartic}} (z) &= \frac{1}{2} z^2 (A+B) \\
		&+ \frac{1}{8} z \left(A^2-2 A B+B^2-8 C^2+8\right)\\
		&+\frac{1}{16} \left(A^3-A^2 B-A B^2+8 A C^2+B^3+8 B C^2\right)\\
		&+\frac{1}{128 z}\left(5 A^4-4 A^3 B-2 A^2 B^2+16 A^2 C^2-4 A B^3-32 A B C^2+5B^4+16 B^2 C^2\right) \\
		&+ \mathcal{O} (z^{-2}) ,
	\end{align*}
	to be equated to $1/z + \mathcal{O} (z^{-2}) $. Imposing the vanishing of the term $\mathcal{O}(z^2)$ is simply telling us that $A=-B$, as we already knew it should be the case from the $\Z_2$ symmetry $\phi \mapsto -\phi$. Substituting $A=-B$ the resolvent simplifies:
	\begin{equation*}
		2 \lambda \omega_{\text{quartic}}(z) = z \left(\frac{B^2}{2}-C^2+1\right)+\frac{1}{8 z} \left(B^4+4 B^2 C^2\right) + \mathcal{O} (z^{-2}).
	\end{equation*}
	Setting the coefficient of $z$ to zero, as required by the large $z$ behaviour of $\omega (z)$, fixes the constant $C$ to 
	\begin{equation*}
		C^2= 1+ \frac{B^2}{2} .
	\end{equation*}
	The resolvent simply becomes 
	\begin{equation*}
		\omega_{\text{quartic}} (z) = \frac{1}{2 \lambda}  \left[ z+z^3 - \left(z^2 +1 + \frac{B^2}{2} \right)\sqrt{z^2-B^2} \right] ,
	\end{equation*}
	with $B$ determined by the normalization condition $\lim_{\lvert z \rvert \to \infty} z \omega (z) =1$. This condition reads
	\begin{equation*}
		\frac{1}{2\lambda} \frac{B^2 (3B^2 +4)}{8} = 1 ,
	\end{equation*}
	which we understand as a quadratic equation in the variable $B^2$. Out of the two solutions for $B^2$, only the one that satisfies $B^2 >0 $ when $\lambda>0$ is consistent with the matrix model we started with. We thus get 
	\begin{equation}
	\label{eq:BquarticMM}
		B^2 =\frac{2}{3} \left( -1 + \sqrt{1+12 \lambda} \right) .
	\end{equation}
	We readily obtain the density of eigenvalues from the discontinuity equation \eqref{eq:disceqrhoomega}, finding 
	\begin{equation}
	\label{eq:RhoquarticMM}
		\rho_{\text{quartic}} (\phi) = \frac{1}{2\pi \lambda}  \left(\phi^2 +1 + \frac{B^2}{2} \right)\sqrt{B^2-\phi^2} .
	\end{equation}\par
	\medskip
			{\small\textbf{Exercise:} Derive the equation for $B^2$ and solve it. Plot the two branches and convince yourself of \eqref{eq:BquarticMM}.}

	\subsubsection*{Free energy of the quartic matrix model}
		Let us now evaluate the free energy of the quartic matrix model in the one-cut phase. We start by computing its derivative:
		\begin{align*}
		\frac{1}{N^2}\frac{\partial \ }{\partial (\lambda^{-1})} \mf [\text{\footnotesize quartic}] &= -\frac{1}{N^2}\left\langle - N \sum_{a=1}^{N}\left( \frac{\phi_a^2}{2} + \frac{\phi_a^4}{4} \right) \right\rangle \\
			&= \int_{-B} ^{B} \dd \phi \rho_{\text{quartic}} (\phi) \left( \frac{\phi^2}{2} + \frac{\phi^4}{4} \right) \\
			&= \int_{-B} ^{B} \dd \phi  \underbrace{ \frac{1}{2 \pi \lambda} \left(\phi^2 +1 + \frac{B^2}{2} \right)\sqrt{B^2-\phi^2} }_{\text{from \eqref{eq:RhoquarticMM}}} \left( \frac{\phi^2}{2} + \frac{\phi^4}{4} \right)  \\
			&= \frac{1}{2 \pi \lambda} \cdot \frac{\pi }{512}  B^4 \left(9 B^4+40 B^2+32\right) \\
			&= \frac{1}{2^{10} \lambda} B^4 \left(9 B^4+40 B^2+32\right) .
	\end{align*}
	Using $\frac{\partial \ }{\partial (\lambda^{-1})} \mf= - \frac{1}{\lambda^2} \frac{\partial \ }{\partial \lambda} \mf$ we get 
	\begin{equation}
	\label{eq:appquarticFofB}
		\frac{1}{N^2}\frac{\partial \ }{\partial \lambda} \mf[\text{\footnotesize quartic}] = - \frac{\lambda}{2^{10} } B^4 \left(9 B^4+40 B^2+32\right) ,
	\end{equation}
	with $B^2$ an explicit function of $\lambda$ given in \eqref{eq:BquarticMM}. Expanding in perturbation series and integrating order by order in $\lambda$ is an easy task at this point, and we find:
	\begin{equation*}
		\mf[\text{\footnotesize quartic}] = N^2 \left( -\frac{\lambda ^4}{8}+\frac{\lambda ^5}{10}-\frac{3 \lambda ^6}{8}+\frac{27 \lambda^7}{14}-\frac{189 \lambda ^8}{16}+\mathcal{O}\left(\lambda ^9\right) \right).
	\end{equation*}
	Actually, plugging \eqref{eq:BquarticMM} into \eqref{eq:appquarticFofB} and integrating, we obtain the free energy in closed form:
	\begin{equation*}
		\frac{\mf[\text{\footnotesize quartic}]}{N^2} = -\frac{(12 \lambda +1) \left(181440 \lambda ^3-95760 \lambda ^2+1260 \lambda +320 (12\lambda +1)^{\frac{5}{2}}-448 (12 \lambda +1)^{\frac{3}{2}}-105\right)}{34836480}
	\end{equation*}

\subsection{'t Hooft limit of unitary matrix models}
\label{app:UMM}

The goal of this appendix is to review the solution of unitary matrix models at large $N$. It is analogous to the discussion in Section \ref{sec:MMplanarlimit}, but replacing the integration contour $\R^N$ with $(\mathbb{S}^1)^N$. We follow closely the presentations in \cite{Jain:2013py,Santilli:2019wvq}.\par
Unitary matrix models with single-trace potential are of the form 
\begin{subequations}
\begin{align}
	\mz_{\text{unitary}} & = \int_{U(N)} \dd U  ~\exp\left[  N \mathrm{Tr} W (U) \right] \label{eq:UnitaryMMHaartot} \\
		&= \frac{1}{N!} \oint_{(\cs^1)^N} \prod_{1 \le a < b \le N} \lvert z_a - z_b \rvert^{2} \prod_{a=1}^{N} e^{ N W(z_a)} \frac{ \dd z_a}{2 \pi i z_a } \label{eq:UnitaryMMz} \\
		&= \frac{1}{N!} \int_{[-\pi, \pi]^N }  \prod_{1 \le a \ne b \le N} \left\lvert 2 \sin \left( \frac{ \theta_a - \theta_b }{2} \right)  \right\rvert  \prod_{a=1}^{N} e^{ N \widehat{W}(\theta_a) } \frac{ \dd \theta_a}{2 \pi } . \label{eq:UnitaryMMtheta}
\end{align}
\end{subequations}
In the first line, $\dd U$ is the normalized invariant Haar measure on $U(N)$ and in \eqref{eq:UnitaryMMz} we have diagonalized $U$ and integrated out the off-diagonal terms, producing the usual Vandermonde factor. We are left with integration over the eigenvalues $\left\{ z_a \in \cs^1 \right\}_{a=1, \dots , N}$ of the random matrix $U$. There is also an overall $\frac{1}{N!}$ from the residual Weyl group $S_N$ shuffling the $N$ eigenvalues. In the last step we have changed variables $z_a = e^{i \theta_a}$, $\forall a=1, \dots, N$ and used
\begin{align*}
	\prod_{1 \le a < b \le N} \lvert z_a - z_b \rvert^{2} &=  \prod_{1 \le a \ne b \le N} \lvert e^{i \theta_a} - e^{i \theta_b} \rvert \\
		&=  \prod_{1 \le a \ne b \le N} \underbrace{\lvert e^{i \theta_a /2 } \rvert}_{1} \cdot  \underbrace{\lvert e^{i \theta_b /2 } \rvert}_{1} \cdot \lvert e^{i \frac{ \theta_a - \theta_b}{2} } - e^{i \frac{ \theta_b - \theta_a}{2} }  \rvert \\ 
		&= \prod_{1 \le a \ne b \le N} \left\lvert 2 \sin \left( \frac{ \theta_a - \theta_b }{2} \right)  \right\rvert
\end{align*}
to arrive at \eqref{eq:UnitaryMMtheta}. Besides, in \eqref{eq:UnitaryMMtheta} we have defined 
\begin{equation*}
	\widehat{W} (\theta) \equiv W (e^{i \theta} ) .
\end{equation*}
Notice also the normalization by $N$ in front of the classical potential $W(z)$, in agreement with the argument of Section \ref{sec:MMplanarlimit}.\par
Before delving into the large $N$ analysis, a few remarks are in order.
\begin{enumerate}[(i)]
	\item It is possible to replace the integration over $U(N)$ or $SU(N)$ with integration over the other classical groups $SO(N)$ and $USp(2N)$. The difference at large $N$ is just a factor $\frac{1}{2}$ in front of $\log \mz$.
	\item\label{item:assumptVcos} We henceforth assume that the potential $W(z)$ is symmetric under $z \mapsto z^{-1}$. Under the change of variables $z= e^{i \theta}$, $W$ becomes a function of $\cos \theta$ only.
	\item We restrict ourselves to single-trace potentials. Multi-trace potentials are more relevant in the study of Hagedorn transitions and gauge theories on compact spaces. They can nevertheless be reduced to single-trace unitary models after multiple Hubbard--Stratonovich transformations. In that case, one first performs the large $N$ limit of the model as we show in the following, and then extremizes with respect to the auxiliary Hubbard--Stratonovich fields.
\end{enumerate}
From assumption \eqref{item:assumptVcos}, we write 
\begin{equation*}
	\widehat{W} (\theta)  = \sum_{j \ge 1} \frac{t_j}{j} \cos ( j \theta ) .
\end{equation*}
For later convenience observe that, if we set $t_j=0$ $\forall j \ge 1$, $\widehat{W} (\theta)  =0$ and the unitary matrix model trivializes:
\begin{equation}
\label{eq:appZunitarynorm}
	\mz_{\text{unitary}} (\vec{t}=0) \underbrace{=}_{\text{from \eqref{eq:UnitaryMMHaartot}}} \int_{U(N)} \dd U \underbrace{=}_{\text{Haar measure normalized}} 1 . 
\end{equation}
We then have 
\begin{equation*}
	\mz_{\text{unitary}} = \frac{1}{(2 \pi)^N N!} \int_{[-\pi, \pi]^N }  \dd \theta ~ e^{ S_{\text{eff}} (\theta)} ,
\end{equation*}
with 
\begin{equation*}
	S_{\text{eff}} (\theta) = N \sum_{a=1}^{N}  \underbrace{  \sum_{j \ge 1} \frac{t_j}{j} \cos ( j \theta_a ) }_{\text{potential }\widehat{W}} +  \underbrace{ \sum_{1 \le a \ne b \le N} \log  \left\lvert 2 \sin \left( \frac{ \theta_a - \theta_b }{2} \right)  \right\rvert  }_{\text{Vandermonde}} .
\end{equation*}
Following the general discussion in Section \ref{sec:LargeNtypes} and Section \ref{sec:MMplanarlimit}, in the large $N$ limit we have 
\begin{equation*}
	S_{\text{eff}} (\theta) = N^2 \int_{- \pi} ^{\pi} \dd \theta \rho (\theta) \left\{   \sum_{j \ge 1} \frac{t_j}{j} \cos ( j \theta )  + \int_{- \pi} ^{\pi} \dd \varphi \rho (\varphi) \log  \left\lvert 2 \sin \left( \frac{ \theta - \varphi }{2} \right)  \right\rvert  \right\} .
\end{equation*}
The eigenvalue density $\rho (\theta) $ is normalized, 
\begin{equation*}
	\int_{- \pi} ^{\pi} \dd \theta \rho (\theta) =1 ,
\end{equation*}
and is supported on $\mathrm{supp} \rho \subseteq [-\pi, \pi ]$.\par
A major difference compared to the setup of Section \ref{sec:MMplanarlimit} is the compactness of the original integration cycle. The distance between any two eigenvalues cannot exceed $2 \pi$, which, taking the large $N$ limit, is translated into the constraint 
\begin{equation*}
	\rho (\theta) \ge 0 .
\end{equation*}
As a consequence, whenever $\rho (\theta) $ would become negative in some region of the parameter space, a phase transition takes place, such that the eigenvalue density in the new phase is supported on a collection of arcs $ \surho \subset [-\pi , \pi]$ and is still normalized and nonnegative.\par
\medskip
	{\small\textbf{Exercise:} Show that $\lvert \theta_a - \theta_b\rvert \le 2\pi$ for all $a,b$ implies $\rho (\theta) \ge 0$ in the large $N$ limit. (Hint: use the Weyl invariance to order the eigenvalues in decreasing order and study $\lvert \theta_{a} - \theta_{a-1} \rvert = \frac{\theta_{a} - \theta_{a-1}}{a-(a-1)}$.)}\par
\medskip
In the large $N$ limit, the leading order contribution comes from $\rho (\theta)$ that extremizes the effective action $S_{\text{eff}} (\theta)$. We are led to solve the SPE 
\begin{equation}
\label{eq:SPEUnitary}
	\mathrm{P}\!\int \dd \varphi  \rho (\varphi) \cot \left( \frac{\theta - \varphi}{2}  \right) =  \widehat{W}^{\prime} (\theta) ,
\end{equation} 
where we recall 
\begin{equation*}
	\widehat{W}^{\prime} (\theta) = \sum_{j \ge 1} t_j \sin (j \theta) .
\end{equation*}
Notice also the lack of a factor $\frac{1}{2}$ on the right-hand side, compared to \eqref{eq:SPEMM}.\par

\subsubsection*{Ungapped phase}
\begin{figure}[ht]
\centering
\includegraphics[width=0.35\textwidth]{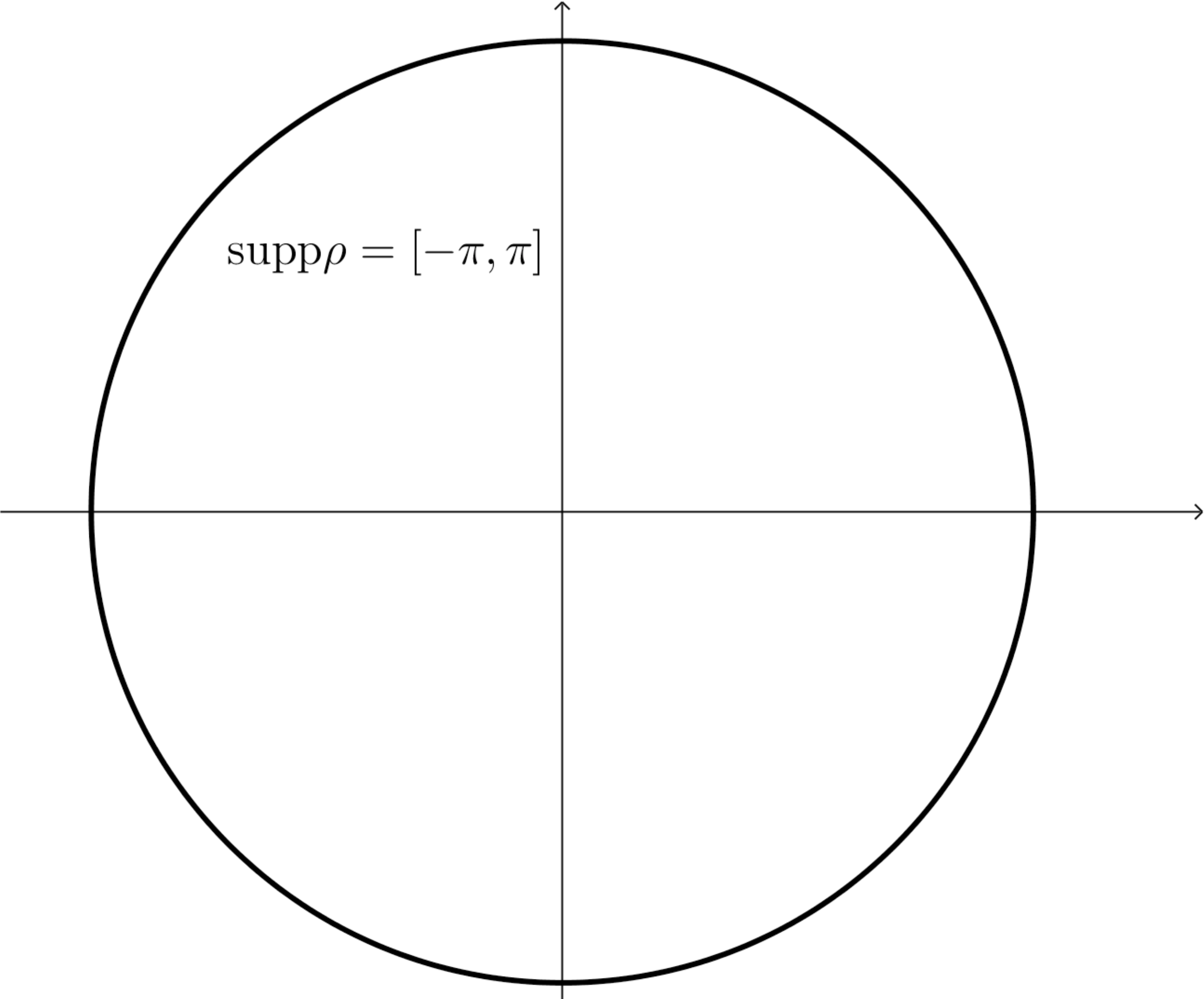}\hspace{0.05\textwidth}
\includegraphics[width=0.35\textwidth]{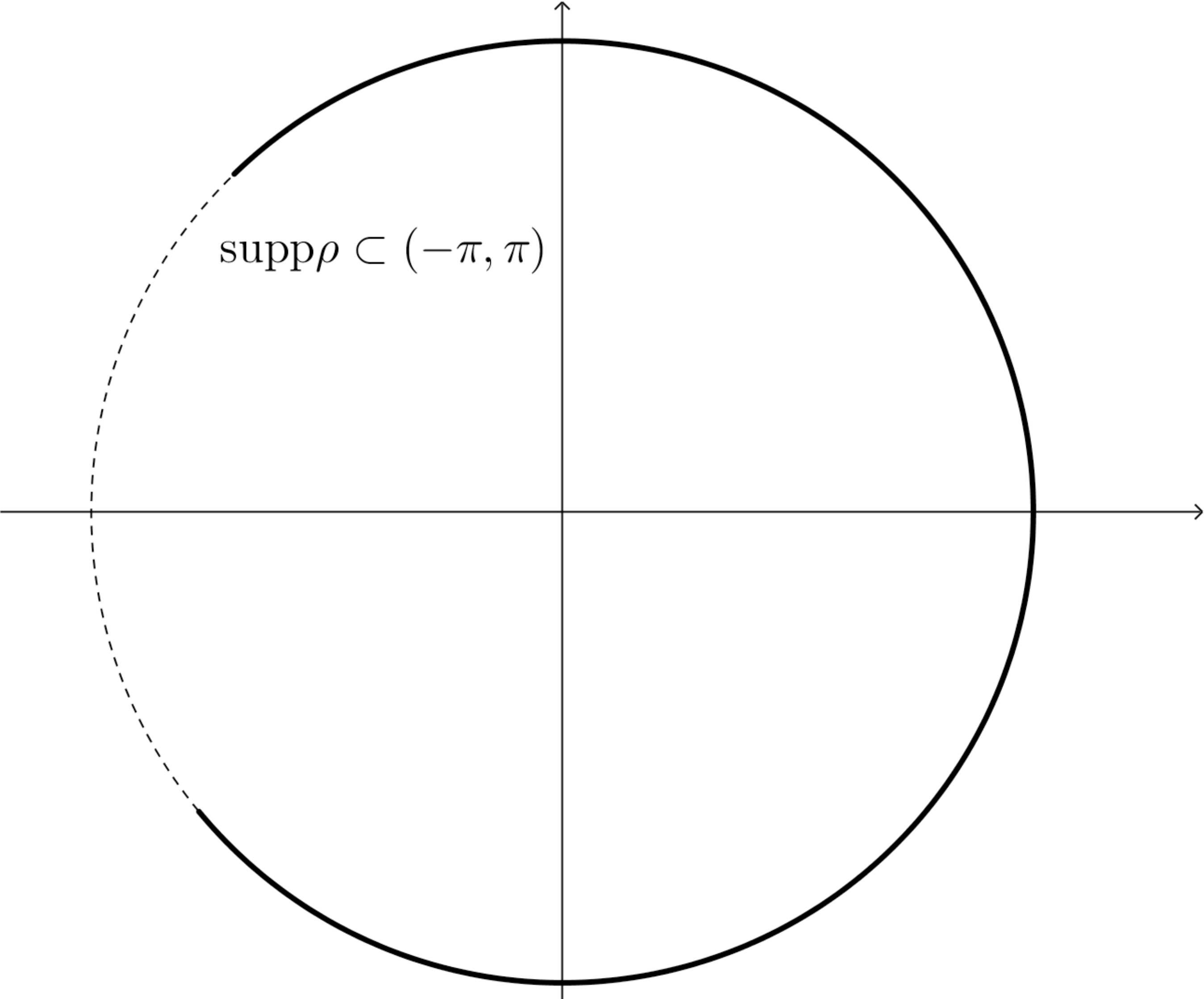}
\caption{Left: Ungapped phase. The eigenvalue density is supported on the whole circle. Right: gapped one-cut phase. The eigenvalue density is supported on a simply-connected arc.}
\label{fig:gapvsungap}
\end{figure}

The problem we want to face in the rest of this appendix is to solve \eqref{eq:SPEUnitary} for $\rho$. To begin with, we make the ansatz 
\begin{equation*}
	\mathrm{supp} \rho = [-\pi, \pi ] ,
\end{equation*}
that is, the eigenvalue density is supported on the whole circle. Solutions of this type go under the name of ``ungapped'', as opposed to the ``gapped'' solutions, for which $\mathrm{supp} \rho$ develops a gap in $[- \pi, \pi]$, see Figure \ref{fig:gapvsungap}. We emphasize that, generically, the ``gapped'' solution merely refers to the shape of $\mathrm{supp} \rho$ and is meant to have no implication on the energy spectrum of the model under consideration.\par
\par
To solve \eqref{eq:SPEUnitary} with the ungapped ansatz, we use the (integral) identity 
\begin{equation}
\label{eq:intidcot}
	\mathrm{P}\!\int \dd \varphi  \rho (\varphi) \cot \left( \frac{\theta - \varphi}{2}  \right) =  2 \sum_{j=1} ^{\infty} \int \dd \varphi  \rho (\varphi) \left[ \cos (j \varphi) \sin (j \theta) -\cos (j \theta ) \sin (j \varphi )  \right]  .
\end{equation}
Denoting by $\rho ^{\cos} _j$ and $\rho^{\sin} _j $ the corresponding Fourier coefficients of the eigenvalue density $\rho (\theta)$, we use \eqref{eq:intidcot} into \eqref{eq:SPEUnitary} and integrate over $-\pi \le \varphi \le \pi$. We get 
\begin{equation*}
	2 \pi \sum_{j=1} ^{\infty} \left[ \rho ^{\cos} _j \sin (j \theta) - \rho ^{\sin} _j \cos (j \theta) \right] = \sum_{j \ge 1} t_j \sin (j \theta) ,
\end{equation*}
which, using the orthogonality of the Fourier basis, yields 
\begin{equation*}
	2 \pi \rho ^{\cos} _j = t_j , \quad 2 \pi \rho ^{\sin} _j = 0, \quad \forall j \ge 1 .
\end{equation*}
In this way, all the Fourier coefficients of $\rho (\theta)$ are obtained except $\rho ^{\cos} _0$, which however is fixed by normalization:
\begin{equation*}
	\int_{-\pi}^{\pi} \dd \theta \rho (\theta) = 2 \pi \rho ^{\cos} _0 \quad \Longrightarrow \quad 2 \pi \rho ^{\cos} _0 = 1 .
\end{equation*}
We thus find 
\begin{equation}
\label{eq:ungapsolrhoUMM}
	\rho (\theta) = \frac{1}{2\pi} \sum_{j \ge 0} t_j \cos (j \theta ) , \quad \theta \in [-\pi, \pi ] ,
\end{equation}
where for shortness we have defined $t_0 = 1$.

\subsubsection*{Free energy in the ungapped phase}
To compute $\mf_{\text{unitary}} = - \log \mz_{\text{unitary}} $, it is convenient to evaluate $\frac{\partial \ }{\partial t_{\alpha}}  \mf_{\text{unitary}}$, $\forall \alpha \ge 1$. We get 
\begin{align*}
	\frac{\partial \ }{\partial t_{\alpha}}  \mf_{\text{unitary}} & = - \frac{1}{\mz_{\text{unitary}} }\frac{\partial \ }{\partial t_{\alpha}} \mz_{\text{unitary}}  \\
		&= - \left\langle \frac{1}{\alpha} \left( -N \sum_{a=1}^{N} \cos (\alpha \theta_a) \right) \right\rangle \\
		&= \frac{N^2}{\alpha} \int_{- \pi } ^{\pi} \dd \theta \rho (\theta) \cos (\alpha \theta) \\
		& = \frac{N^2}{\alpha} \int_{- \pi } ^{\pi} \frac{ \dd \theta }{2 \pi } \cos ( \alpha \theta ) \sum_{j \ge 0} t_j \cos (j \theta) =  \frac{N^2}{\alpha} \sum_{j\ge 0} t_j \delta_{j, \alpha} = \frac{N^2}{\alpha} t_{\alpha} .
\end{align*}
Integrating all these equations gives 
\begin{equation*}
	 \mf_{\text{unitary}} = \frac{N^2}{2} \sum_{j \ge 1} \frac{t_{j}^2}{j} ,
\end{equation*}
up to a constant term independent of all couplings. This constant term is seen to vanish imposing the boundary condition \eqref{eq:appZunitarynorm}, namely $\mz_{\text{unitary}} (\vec{t} = 0) =1 $, for the integration over each $t_{\alpha}$.

\subsubsection*{Gapped solution}
In regions in the $\vec{t}$-parameter space in which $\rho (\theta_0) <0$ for some $\theta_0 \in [-\pi, \pi]$, the ungapped solution \eqref{eq:ungapsolrhoUMM} breaks down and we have to look for a different ansatz.\par
\begin{figure}[ht]
\centering
\includegraphics[width=0.65\textwidth]{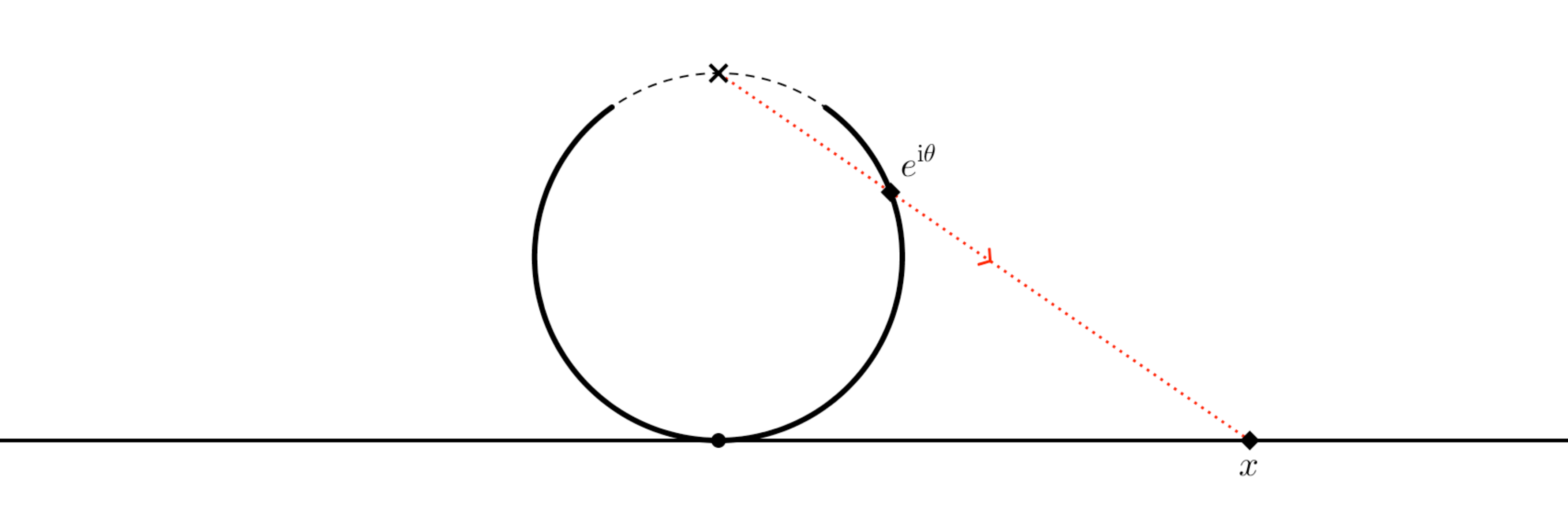}
\caption{The stereographic projection maps the gapped phase of a unitary matrix model to a Hermitian matrix model.}
\label{fig:stereoproj}
\end{figure}

The support of the new solution develops a gap around the point $\theta_0 \in [-\pi, \pi]$ at which the ungapped solution breaks down (see the right panel of Figure \ref{fig:gapvsungap}). The solution in the gapped phase will be much more along the lines of Section \ref{sec:MMplanarlimit} and, in fact, solving \eqref{eq:SPEUnitary} in any gapped phase could be mapped to a problem on $\R$ via the stereographic projection, as in Figure \ref{fig:stereoproj}. Let us reiterate that the adjective ``gapped'' refers only to the form of the solution for $\rho (\theta)$.\par
\par
Let us solve \eqref{eq:SPEUnitary} for 
\begin{equation*}
	\surho := \text{supp} \rho = \bigcup_{s=1} ^{\ell} [\theta_s ^{(-)}, \theta_s ^{(+)} ] , \quad  \theta_s ^{(+)} < \theta_{s+1} ^{(-)}.
\end{equation*}
The collection of endpoints inherits a reflection symmetry $\theta \mapsto - \theta$ from assumption \eqref{item:assumptVcos} on the $\theta \mapsto -\theta$ symmetry of the classical potential $W$.\par
\begin{figure}[ht]
\centering
\includegraphics[width=0.35\textwidth]{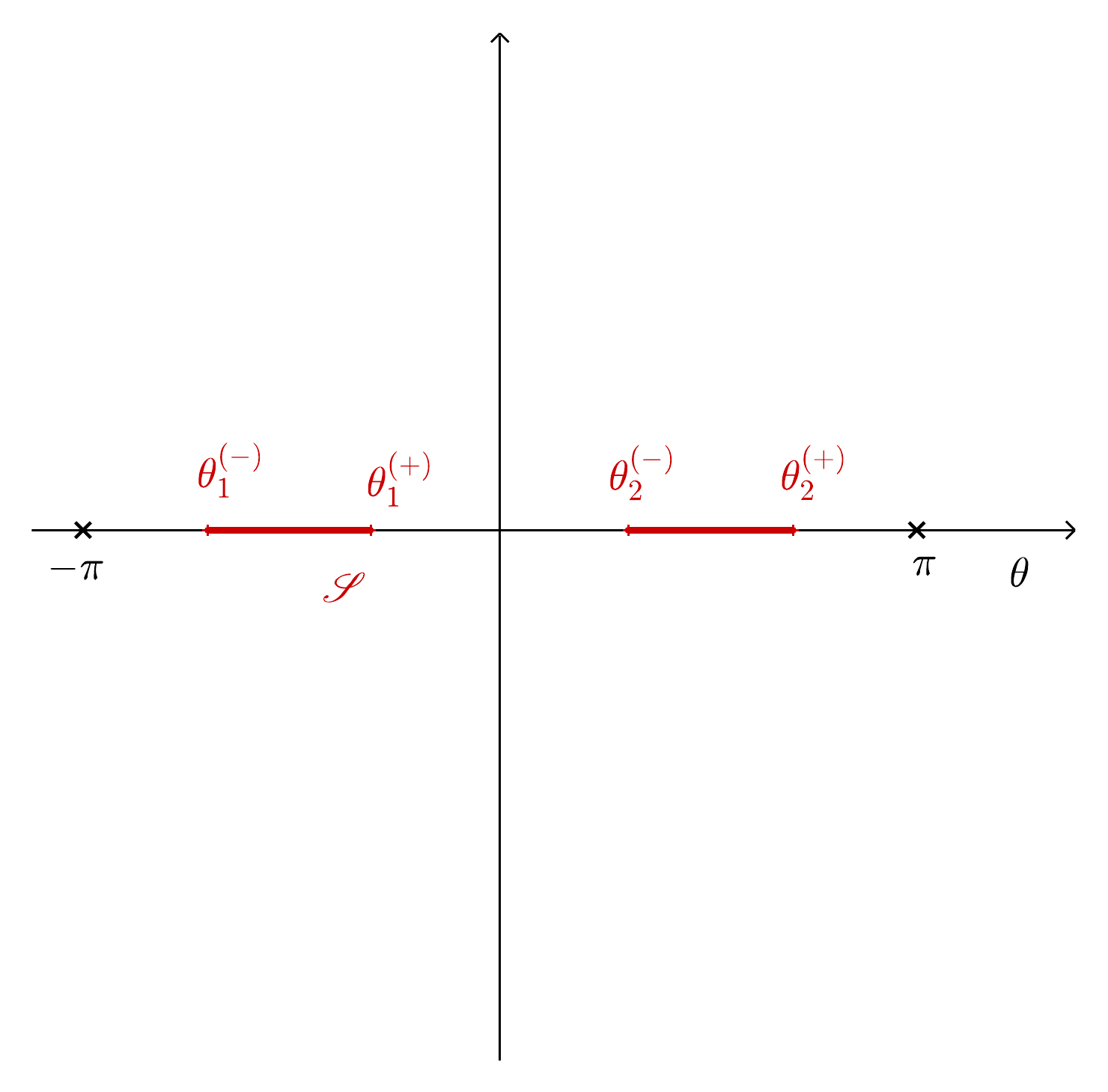}\hspace{0.05\textwidth}
\includegraphics[width=0.35\textwidth]{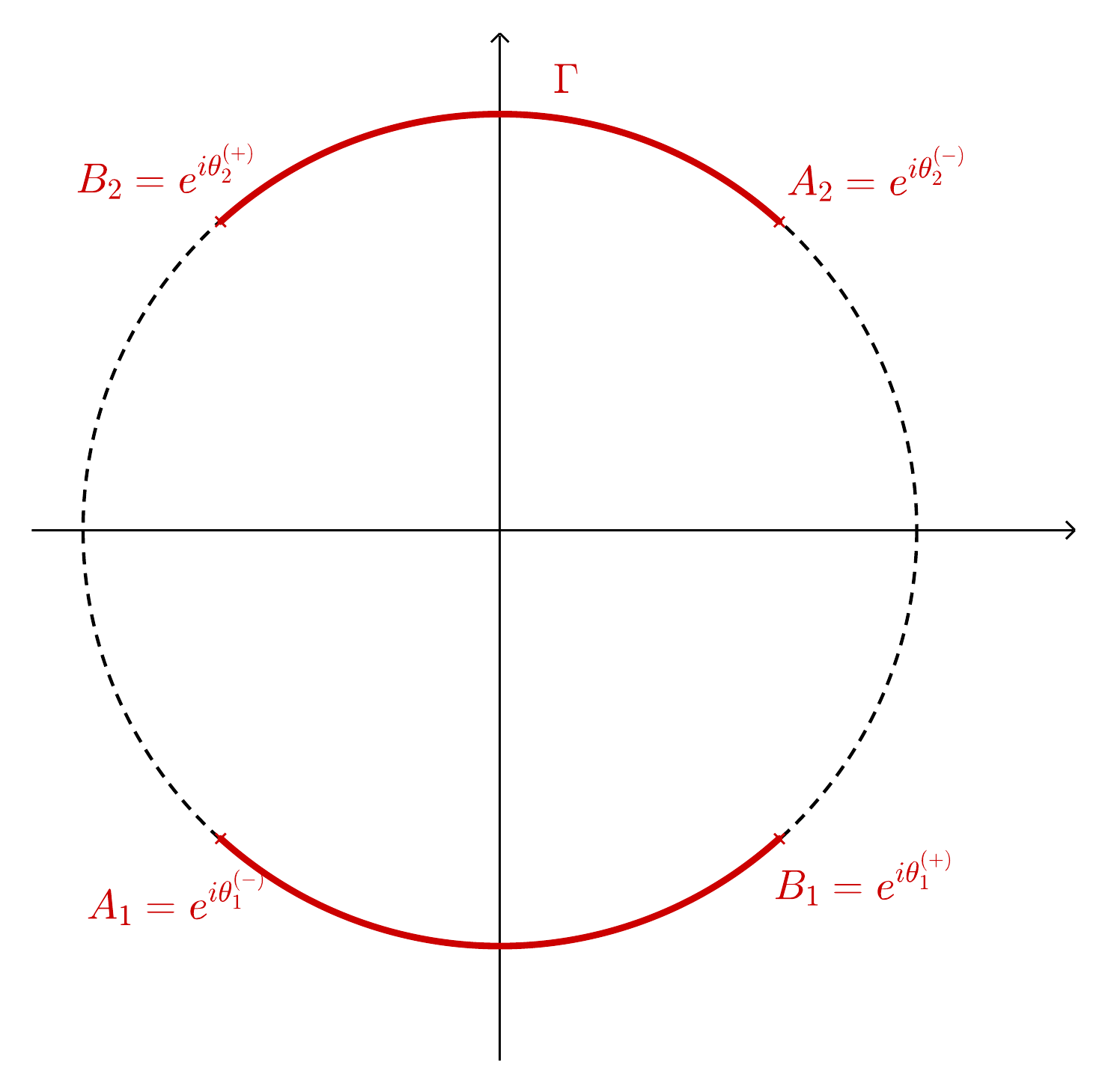}
\caption{Left: Support $\surho$, a union of intervals in $[-\pi, \pi]$. Right: The change of variables $z=e^{i \theta}$ sends $\surho$ to $\Gamma$, a union of arcs in $\C$.}
\label{fig:maptoGamma}
\end{figure}

We make the change of variables $z=e^{i \theta}$ and $u= e^{i \varphi}$. We denote $\hat{\rho} (u)$ the real-valued function of complex variable such that $\hat{\rho}(e^{i \theta}) = \rho (\theta)$, and also define $\Gamma \subset \cs^1$ to be the union of circular arcs in $\C$ running from $e^{i \theta_s ^{(-)}}$ to $e^{i \theta_s ^{(+)} }$. In other words, $\Gamma$ is the image of the embedding $\mathscr{S} \to \C$ under the map $\theta \mapsto e^{i \theta}$, see Figure \ref{fig:maptoGamma} for an illustration. For consistency with the notation of Section \ref{sec:MMplanarlimit}, let us also define $A_s := e^{i \theta_s ^{(-)}}$ and $B_s := e^{i \theta_s ^{(+)}}$.\par

\par

The SPE \eqref{eq:SPEUnitary} is rewritten as 
\begin{equation}
\label{eq:SPEUnitaryzvar}
	\mathrm{P}\!\int_{\Gamma } \frac{ \dd u}{i u} \hat{\rho} (u) \left( i \frac{z+u}{z-u} \right) = \frac{i}{2} \sum_{j \ge 1} \lambda_j \left( z^{j} - \frac{1}{z^j} \right) .
\end{equation}
It is important to underline that we are studying the matrix model in the form \eqref{eq:UnitaryMMtheta} and the saddle point equation is for the angular variables $\theta_a$. The change of variables at this stage is just a convenient rewriting. In particular, we (i) first obtain \eqref{eq:SPEUnitary}, and then (ii) change variables. Hence, the right-hand side of \eqref{eq:SPEUnitaryzvar} is not the derivative of $W(z)$ with respect to $z$; instead, it is $\widehat{W}^{\prime}$ after writing $e^{i \theta} =z$.\par
We introduce the resolvent 
\begin{equation*}
	\omega (z) = \mathrm{P}\!\int_{\Gamma } \frac{ \dd u}{i u} \hat{\rho} (u) \frac{z+u}{z-u} , \quad z \in \C \setminus \Gamma ,
\end{equation*}
where we recall that $\Gamma$ is the union of simply-connected arcs. $\omega (z)$ mimics the resolvent introduced in Section \ref{sec:MMplanarlimit}, but with two differences. First, carefully checking all the factors, one gets 
\begin{equation}
\label{eq:omegajumpunitary}
		\lim_{\varepsilon \to 0^{+}} \left[ \omega (e^{i \theta  + \varepsilon} ) - \omega (e^{i \theta  - \varepsilon}  )  \right] = 4 \pi \hat{\rho} (e^{i \theta}) = 4 \pi \rho (\theta) , \quad e^{i \theta} \in \Gamma ,
\end{equation}
as opposed to \eqref{eq:disceqrhoomega}. Also, because of the term $\frac{z+u}{z-u}$ in the definition, as opposed to $\frac{1}{z-\sigma}$, the resolvent $\omega (z)$ in a unitary matrix model is normalized as 
\begin{equation*}
	\lim_{\lvert z \rvert \to \infty } \omega (z) = 1 .
\end{equation*}\par
\medskip
{\small\textbf{Exercise:} Derive the factor $4 \pi$ in \eqref{eq:omegajumpunitary}.}\par
\begin{figure}[ht]
\centering
\includegraphics[width=0.5\textwidth]{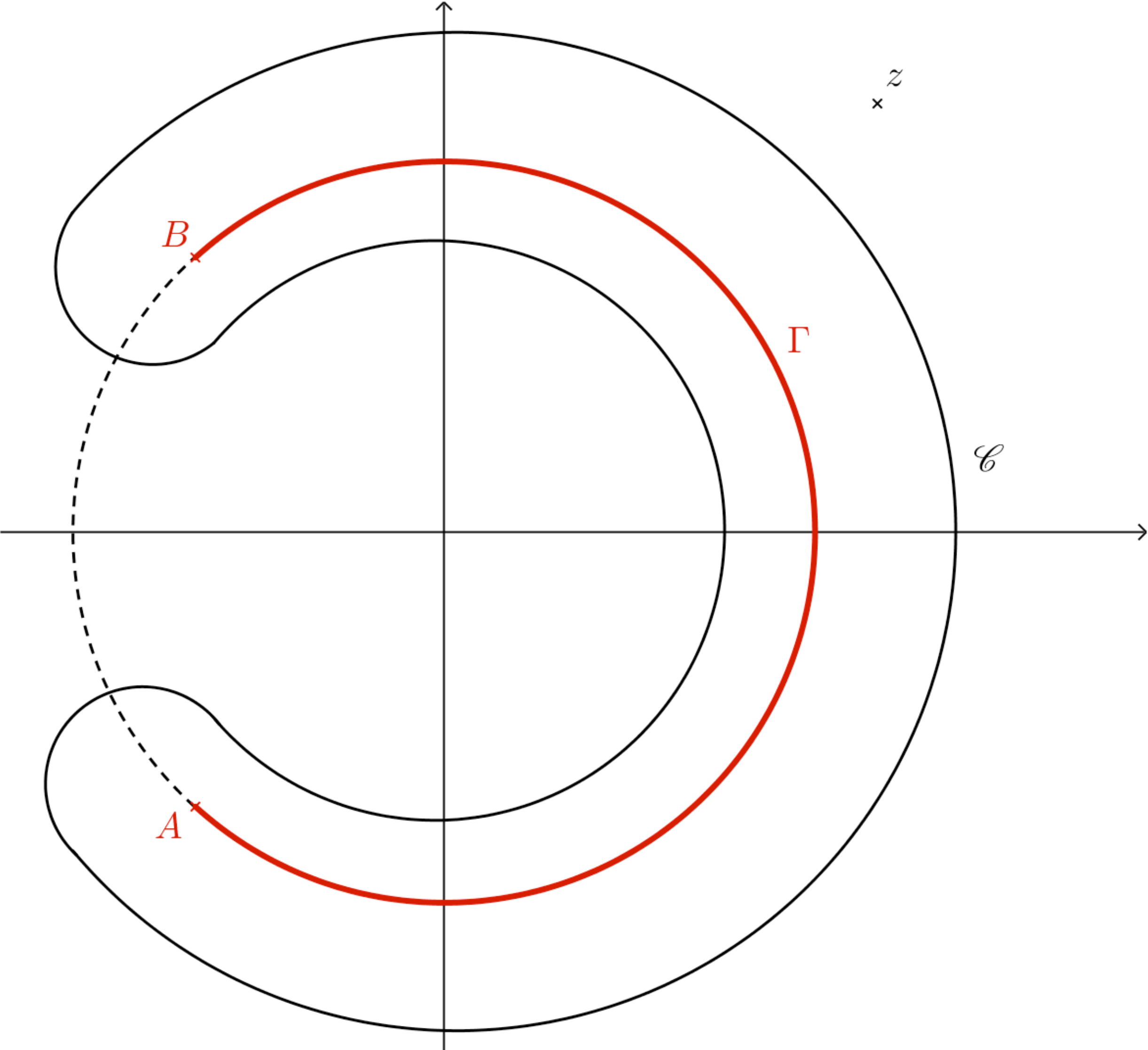}
\caption{Integration contour $\mathscr{C}$ in the one-cut case. It encircles $\Gamma$ (red) leaving $z$ outside.}
\label{fig:ContourUnitary}
\end{figure}

\medskip
At this stage, we take an integration contour $\mathscr{C}$ that encircles the cut $\Gamma $ but leaves outside $z \in \mathbb{C}$, as in Figure \ref{fig:ContourUnitary}.\par
\par
Then, \eqref{eq:SPEUnitaryzvar} is solved by 
\begin{equation}
\label{eq:omegaunitaryres}
	\omega (z) = \sqrt{  \prod_{s=1} ^{\ell} \left(A_s -z \right)  \left( B_s -z \right)} \oint_{\mathscr{C}} \frac{\dd u}{2 \pi } \frac{  \left( - \frac{i}{2} \right) \sum_{j \ge 1} \lambda_j \left( u^{j} - \frac{1}{u^j} \right)  }{ (z-u) \sqrt{   \prod_{s=1} ^{\ell} \left(A_s -u \right)  \left( B_s -u \right) } } 
\end{equation}
(notice the extra minus sign in the numerator). At this point, we deform the integration contour to infinity, picking the poles at $u=z$, at $u=0$ and possibly the poles coming from $\widehat{W}^{\prime}$. Notice that, for specific choices of the coefficients $t_j$, $\widehat{W}^{\prime}$ may sum up to a meromorphic function, whose poles contribute to the integral and must be carefully considered. The solution is essentially analogous to \eqref{eq:omegadefcontour}.\par

\subsubsection*{Example: The Gross--Witten--Wadia model}
As a concrete example, let us apply the machinery to the Gross--Witten--Wadia model, characterized by 
\begin{equation*}
	\widehat{W} (\theta) = t \cos (\theta) .
\end{equation*}
In its field theory interpretation, $t= \lambda^{-1}$ is the inverse of the 't Hooft gauge coupling.\par
In the ungapped phase, \eqref{eq:ungapsolrhoUMM} is specialized to 
\begin{equation*}
	\rho_{\text{GWW}} (\theta) = \frac{1}{2\pi} \left[ 1 + t \cos (\theta) \right] .
\end{equation*}
This eigenvalue density ceases to be a solution at $t=1$, because beyond that point one would get a negative density around $\theta=\pi$. We thus look for a one-cut solution supported on $\theta \in [\theta^{(-)}, \theta^{(+)}]$. The support inherits the reflection symmetry of $\widehat{W} (\theta)$, and we write $\theta^{(\pm)}=\pm \theta_0$.\par
\begin{figure}[ht]
\centering
\includegraphics[width=0.5\textwidth]{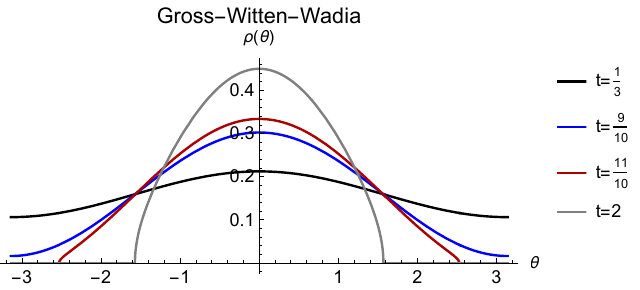}
\caption{Eigenvalue density of the Gross--Witten--Wadia model as $t$ varies.}
\label{fig:GWWplot}
\end{figure}

In this model, \eqref{eq:omegaunitaryres} receives a contribution from the pole at $u=z$, which only contributes to the regular part of $\omega (z)$. Additionally, it has a pole at $u=0$, contributing 
\begin{equation*}
	 \sqrt{ \left( e^{- i \theta_0} - z \right) \left( e^{i \theta_0} - z \right) } \mathrm{Res}_{u=0} \frac{t}{2u } \frac{1}{ (z-u) \sqrt{ \left( e^{- i \theta_0} - u \right) \left( e^{i \theta_0} - u \right) } } =  \sqrt{ \left( e^{- i \theta_0} - z \right) \left( e^{i \theta_0} - z \right) } \frac{t}{2z } ,
\end{equation*}
plus the residual integration from the circle at infinity, which, in the limit of large radius, is 
\begin{equation*}
	 \sqrt{ \left( e^{- i \theta_0} - z \right) \left( e^{i \theta_0} - z \right) } \oint_{\cs^1 _{\infty}} \frac{\dd u}{2 \pi  } \left( - \frac{i t}{2 u} \right) = \frac{t}{2}  \sqrt{ \left( e^{- i \theta_0} - z \right) \left( e^{i \theta_0} - z \right) } .
\end{equation*}
Being careful with the sign coming from the square roots, according to the direction from where the cut $\Gamma$ is approached, we find 
\begin{equation*}
	\lim_{\varepsilon \to 0^{+}}  \omega (e^{i \theta  + \varepsilon} ) - \lim_{\varepsilon \to 0^{+}}  \omega (e^{i \theta  - \varepsilon}  ) = t \left( 1+e^{-i \theta} \right)  \lim_{\varepsilon \to 0^{+}}  \sqrt{ \left( e^{- i \theta_0} - e^{i \theta + \varepsilon } \right) \left( e^{i \theta_0} - e^{i \theta + \varepsilon } \right) } ,
\end{equation*}
which, from \eqref{eq:omegajumpunitary}, gives 
\begin{equation*}
		\rho_{\text{GWW}} (\theta) = \frac{t}{ \pi} \cos \frac{\theta}{2} \sqrt{ \left( \sin \frac{\theta_0}{2} \right)^2 - \left( \sin \frac{\theta}{2} \right)^2 } , \quad \theta \in [- \theta_0, \theta_0 ] .
\end{equation*}
The endpoint $\theta_0$ of the support is fixed by normalization, and we find 
\begin{equation*}
	\int_{- \theta_0} ^{\theta_0} \dd \theta \rho_{\text{GWW}} (\theta) = 1 \quad \Longrightarrow \quad \left( \sin \frac{\theta_0}{2} \right)^2 = \frac{1}{t} .
\end{equation*}
The solution is plotted in Figure \ref{fig:GWWplot}.

\end{appendix}

\clearpage
{\small
\bibliography{LargeNSusy}

\providecommand{\href}[2]{#2}\begingroup\raggedright\begin{thebibliography}{100}

\bibitem{Aharony:1999ti}
O.~Aharony, S.~S. Gubser, J.~M. Maldacena, H.~Ooguri and Y.~Oz, \emph{{Large N
  field theories, string theory and gravity}},
  \href{https://doi.org/10.1016/S0370-1573(99)00083-6}{\emph{Phys. Rept.}
  {\bfseries 323} (2000) 183}
  [\href{https://arxiv.org/abs/hep-th/9905111}{{\ttfamily hep-th/9905111}}].

\bibitem{Maldacena:2000mw}
J.~M. Maldacena and C.~Nunez, \emph{{Supergravity description of field theories
  on curved manifolds and a no go theorem}},
  \href{https://doi.org/10.1142/S0217751X01003937}{\emph{Int. J. Mod. Phys. A}
  {\bfseries 16} (2001) 822}
  [\href{https://arxiv.org/abs/hep-th/0007018}{{\ttfamily hep-th/0007018}}].

\bibitem{Gaiotto:2009gz}
D.~Gaiotto and J.~Maldacena, \emph{{The Gravity duals of N=2 superconformal
  field theories}}, \href{https://doi.org/10.1007/JHEP10(2012)189}{\emph{JHEP}
  {\bfseries 10} (2012) 189} [\href{https://arxiv.org/abs/0904.4466}{{\ttfamily
  0904.4466}}].

\bibitem{Martelli:2005tp}
D.~Martelli, J.~Sparks and S.-T. Yau, \emph{{The Geometric dual of
  a-maximisation for Toric Sasaki-Einstein manifolds}},
  \href{https://doi.org/10.1007/s00220-006-0087-0}{\emph{Commun. Math. Phys.}
  {\bfseries 268} (2006) 39}
  [\href{https://arxiv.org/abs/hep-th/0503183}{{\ttfamily hep-th/0503183}}].

\bibitem{Martelli:2006yb}
D.~Martelli, J.~Sparks and S.-T. Yau, \emph{{Sasaki-Einstein manifolds and
  volume minimisation}},
  \href{https://doi.org/10.1007/s00220-008-0479-4}{\emph{Commun. Math. Phys.}
  {\bfseries 280} (2008) 611}
  [\href{https://arxiv.org/abs/hep-th/0603021}{{\ttfamily hep-th/0603021}}].

\bibitem{Benini:2015eyy}
F.~Benini, K.~Hristov and A.~Zaffaroni, \emph{{Black hole microstates in
  AdS$_{4}$ from supersymmetric localization}},
  \href{https://doi.org/10.1007/JHEP05(2016)054}{\emph{JHEP} {\bfseries 05}
  (2016) 054} [\href{https://arxiv.org/abs/1511.04085}{{\ttfamily
  1511.04085}}].

\bibitem{Cabo-Bizet:2018ehj}
A.~Cabo-Bizet, D.~Cassani, D.~Martelli and S.~Murthy, \emph{{Microscopic origin
  of the Bekenstein-Hawking entropy of supersymmetric AdS$_{5}$ black holes}},
  \href{https://doi.org/10.1007/JHEP10(2019)062}{\emph{JHEP} {\bfseries 10}
  (2019) 062} [\href{https://arxiv.org/abs/1810.11442}{{\ttfamily
  1810.11442}}].

\bibitem{Choi:2018hmj}
S.~Choi, J.~Kim, S.~Kim and J.~Nahmgoong, \emph{{Large AdS black holes from
  QFT}},  \href{https://arxiv.org/abs/1810.12067}{{\ttfamily 1810.12067}}.

\bibitem{Benini:2018ywd}
F.~Benini and E.~Milan, \emph{{Black Holes in 4D $\mathcal{N}$=4
  Super-Yang-Mills Field Theory}},
  \href{https://doi.org/10.1103/PhysRevX.10.021037}{\emph{Phys. Rev. X}
  {\bfseries 10} (2020) 021037}
  [\href{https://arxiv.org/abs/1812.09613}{{\ttfamily 1812.09613}}].

\bibitem{tHooft:1973alw}
G.~'t~Hooft, \emph{{A Planar Diagram Theory for Strong Interactions}},
  \href{https://doi.org/10.1016/0550-3213(74)90154-0}{\emph{Nucl. Phys. B}
  {\bfseries 72} (1974) 461}.

\bibitem{Brezin:1977sv}
E.~Brezin, C.~Itzykson, G.~Parisi and J.~B. Zuber, \emph{{Planar Diagrams}},
  \href{https://doi.org/10.1007/BF01614153}{\emph{Commun. Math. Phys.}
  {\bfseries 59} (1978) 35}.

\bibitem{Witten:1979kh}
E.~Witten, \emph{{Baryons in the 1/n Expansion}},
  \href{https://doi.org/10.1016/0550-3213(79)90232-3}{\emph{Nucl. Phys. B}
  {\bfseries 160} (1979) 57}.

\bibitem{Billo:1996pu}
M.~Bill\`{o}, M.~Caselle, A.~D'Adda and S.~Panzeri, \emph{{Finite temperature
  lattice QCD in the large N limit}},
  \href{https://doi.org/10.1142/S0217751X97001158}{\emph{Int. J. Mod. Phys. A}
  {\bfseries 12} (1997) 1783}
  [\href{https://arxiv.org/abs/hep-th/9610144}{{\ttfamily hep-th/9610144}}].

\bibitem{ZinnJustin:2007zz}
J.~Zinn-Justin, \emph{{Phase transitions and renormalization group}}, Oxford
  graduate texts. {Oxford University Press}, Oxford, UK, 2007.

\bibitem{BenettiGenolini:2020doj}
P.~Benetti~Genolini and L.~Tizzano, \emph{{Instantons, symmetries and anomalies
  in five dimensions}},
  \href{https://doi.org/10.1007/JHEP04(2021)188}{\emph{JHEP} {\bfseries 04}
  (2021) 188} [\href{https://arxiv.org/abs/2009.07873}{{\ttfamily
  2009.07873}}].

\bibitem{Santilli:2021qyt}
L.~Santilli, \emph{{Phases of five-dimensional supersymmetric gauge theories}},
  \href{https://doi.org/10.1007/JHEP07(2021)088}{\emph{JHEP} {\bfseries 07}
  (2021) 088} [\href{https://arxiv.org/abs/2103.14049}{{\ttfamily
  2103.14049}}].

\bibitem{Chen:2023lzq}
Y.~Chen, M.~Heydeman, Y.~Wang and M.~Zhang, \emph{{Probing supersymmetric black
  holes with surface defects}},
  \href{https://doi.org/10.1007/JHEP10(2023)136}{\emph{JHEP} {\bfseries 10}
  (2023) 136} [\href{https://arxiv.org/abs/2306.05463}{{\ttfamily
  2306.05463}}].

\bibitem{Amariti:2024bsr}
A.~Amariti, P.~Glorioso, D.~Morgante and A.~Zanetti, \emph{{Cardy matches Bethe
  on the surface: A tale of a brane and a black hole}},
  \href{https://doi.org/10.1016/j.nuclphysb.2024.116773}{\emph{Nucl. Phys. B}
  {\bfseries 1010} (2025) 116773}
  [\href{https://arxiv.org/abs/2403.17190}{{\ttfamily 2403.17190}}].

\bibitem{Hofman:2017vwr}
D.~M. Hofman and N.~Iqbal, \emph{{Generalized global symmetries and
  holography}},
  \href{https://doi.org/10.21468/SciPostPhys.4.1.005}{\emph{SciPost Phys.}
  {\bfseries 4} (2018) 005} [\href{https://arxiv.org/abs/1707.08577}{{\ttfamily
  1707.08577}}].

\bibitem{Pestun:2007rz}
V.~Pestun, \emph{{Localization of gauge theory on a four-sphere and
  supersymmetric Wilson loops}},
  \href{https://doi.org/10.1007/s00220-012-1485-0}{\emph{Commun. Math. Phys.}
  {\bfseries 313} (2012) 71} [\href{https://arxiv.org/abs/0712.2824}{{\ttfamily
  0712.2824}}].

\bibitem{DiFrancesco:1993cyw}
P.~Di~Francesco, P.~H. Ginsparg and J.~Zinn-Justin, \emph{{2-D Gravity and
  random matrices}},
  \href{https://doi.org/10.1016/0370-1573(94)00084-G}{\emph{Phys. Rept.}
  {\bfseries 254} (1995) 1}
  [\href{https://arxiv.org/abs/hep-th/9306153}{{\ttfamily hep-th/9306153}}].

\bibitem{Marino:2004eq}
M.~Mari\~{n}o, \emph{{Les Houches lectures on matrix models and topological
  strings}},  \href{https://arxiv.org/abs/hep-th/0410165}{{\ttfamily
  hep-th/0410165}}.

\bibitem{Marino:2011nm}
M.~Mari\~{n}o, \emph{{Lectures on localization and matrix models in
  supersymmetric Chern-Simons-matter theories}},
  \href{https://doi.org/10.1088/1751-8113/44/46/463001}{\emph{J. Phys. A}
  {\bfseries 44} (2011) 463001}
  [\href{https://arxiv.org/abs/1104.0783}{{\ttfamily 1104.0783}}].

\bibitem{Minahan:2016xwk}
J.~A. Minahan, \emph{{Matrix models for 5d super Yang\textendash{}Mills}},
  \href{https://doi.org/10.1088/1751-8121/aa5cbe}{\emph{J. Phys. A} {\bfseries
  50} (2017) 443015} [\href{https://arxiv.org/abs/1608.02967}{{\ttfamily
  1608.02967}}].

\bibitem{Coleman:1985rnk}
S.~Coleman, \emph{{Aspects of Symmetry}: {Selected Erice Lectures}}. Cambridge
  University Press, Cambridge, U.K., 1985,
  \href{https://doi.org/10.1017/CBO9780511565045}{10.1017/CBO9780511565045}.

\bibitem{Klebanov:2018fzb}
I.~R. Klebanov, F.~Popov and G.~Tarnopolsky, \emph{{TASI Lectures on Large $N$
  Tensor Models}}, \href{https://doi.org/10.22323/1.305.0004}{\emph{PoS}
  {\bfseries TASI2017} (2018) 004}
  [\href{https://arxiv.org/abs/1808.09434}{{\ttfamily 1808.09434}}].

\bibitem{Cremonesi:2013twh}
S.~Cremonesi, \emph{{An Introduction to Localisation and Supersymmetry in
  Curved Space}}, \href{https://doi.org/10.22323/1.201.0002}{\emph{PoS}
  {\bfseries Modave2013} (2013) 002}.

\bibitem{Pestun:2016qko}
V.~Pestun, \emph{{Review of localization in geometry}},
  \href{https://doi.org/10.1088/1751-8121/aa6161}{\emph{J. Phys. A} {\bfseries
  50} (2017) 443002} [\href{https://arxiv.org/abs/1608.02954}{{\ttfamily
  1608.02954}}].

\bibitem{Szabo:1996md}
R.~J. Szabo, \emph{{Equivariant localization of path integrals}}. Springer,
  2000, \href{https://doi.org/10.1007/3-540-46550-2}{10.1007/3-540-46550-2},
  [\href{https://arxiv.org/abs/hep-th/9608068}{{\ttfamily hep-th/9608068}}].

\bibitem{Duistermaat:1982vw}
J.~J. Duistermaat and G.~J. Heckman, \emph{{On the Variation in the cohomology
  of the symplectic form of the reduced phase space}},
  \href{https://doi.org/10.1007/BF01399506}{\emph{Invent. Math.} {\bfseries 69}
  (1982) 259}.

\bibitem{Atiyah:1984px}
M.~F. Atiyah and R.~Bott, \emph{{The Moment map and equivariant cohomology}},
  \href{https://doi.org/10.1016/0040-9383(84)90021-1}{\emph{Topology}
  {\bfseries 23} (1984) 1}.

\bibitem{Closset:2012vp}
C.~Closset, T.~T. Dumitrescu, G.~Festuccia, Z.~Komargodski and N.~Seiberg,
  \emph{{Comments on Chern-Simons Contact Terms in Three Dimensions}},
  \href{https://doi.org/10.1007/JHEP09(2012)091}{\emph{JHEP} {\bfseries 09}
  (2012) 091} [\href{https://arxiv.org/abs/1206.5218}{{\ttfamily 1206.5218}}].

\bibitem{Kallen:2012cs}
J.~K\"all\'en and M.~Zabzine, \emph{{Twisted supersymmetric 5D Yang-Mills
  theory and contact geometry}},
  \href{https://doi.org/10.1007/JHEP05(2012)125}{\emph{JHEP} {\bfseries 05}
  (2012) 125} [\href{https://arxiv.org/abs/1202.1956}{{\ttfamily 1202.1956}}].

\bibitem{Gaiotto:2008ak}
D.~Gaiotto and E.~Witten, \emph{{S-Duality of Boundary Conditions In N=4 Super
  Yang-Mills Theory}},
  \href{https://doi.org/10.4310/ATMP.2009.v13.n3.a5}{\emph{Adv. Theor. Math.
  Phys.} {\bfseries 13} (2009) 721}
  [\href{https://arxiv.org/abs/0807.3720}{{\ttfamily 0807.3720}}].

\bibitem{ABJM}
O.~Aharony, O.~Bergman, D.~L. Jafferis and J.~Maldacena, \emph{{N=6
  superconformal Chern-Simons-matter theories, M2-branes and their gravity
  duals}}, \href{https://doi.org/10.1088/1126-6708/2008/10/091}{\emph{JHEP}
  {\bfseries 10} (2008) 091} [\href{https://arxiv.org/abs/0806.1218}{{\ttfamily
  0806.1218}}].

\bibitem{Gross:1989vs}
D.~J. Gross and A.~A. Migdal, \emph{{Nonperturbative Two-Dimensional Quantum
  Gravity}}, \href{https://doi.org/10.1103/PhysRevLett.64.127}{\emph{Phys. Rev.
  Lett.} {\bfseries 64} (1990) 127}.

\bibitem{Brezin:1990rb}
E.~Brezin and V.~A. Kazakov, \emph{{Exactly Solvable Field Theories of Closed
  Strings}}, \href{https://doi.org/10.1016/0370-2693(90)90818-Q}{\emph{Phys.
  Lett. B} {\bfseries 236} (1990) 144}.

\bibitem{Douglas:1989ve}
M.~R. Douglas and S.~H. Shenker, \emph{{Strings in Less Than One-Dimension}},
  \href{https://doi.org/10.1016/0550-3213(90)90522-F}{\emph{Nucl. Phys. B}
  {\bfseries 335} (1990) 635}.

\bibitem{Muskhe:1977}
N.~I. Muskhelishvili, \emph{Singular Integral Equations}. Noordhoff, 1977.

\bibitem{Pipkin:1991}
A.~C. Pipkin, \emph{A Course on Integral Equations}, no.~9 in Texts in Applied
  Mathematics. Springer-Verlag, New York, 1991.

\bibitem{Santilli:2022tjt}
L.~Santilli, \emph{{Matrix models and phase transitions in gauge theories}},
  Ph.D. thesis, \href{http://hdl.handle.net/10451/53689}{University of Lisbon},
  2022.

\bibitem{Shimizu:2018pnd}
K.~Shimizu, \emph{{Aspects of Massive Gauge Theories on Three Sphere in
  Infinite Mass Limit}},
  \href{https://doi.org/10.1007/JHEP01(2019)090}{\emph{JHEP} {\bfseries 01}
  (2019) 090} [\href{https://arxiv.org/abs/1809.03679}{{\ttfamily
  1809.03679}}].

\bibitem{Tierz:2002jj}
M.~Tierz, \emph{{Soft matrix models and Chern-Simons partition functions}},
  \href{https://doi.org/10.1142/S0217732304014100}{\emph{Mod. Phys. Lett. A}
  {\bfseries 19} (2004) 1365}
  [\href{https://arxiv.org/abs/hep-th/0212128}{{\ttfamily hep-th/0212128}}].

\bibitem{Barranco:2014tla}
A.~Barranco and J.~G. Russo, \emph{{Large N phase transitions in supersymmetric
  Chern-Simons theory with massive matter}},
  \href{https://doi.org/10.1007/JHEP03(2014)012}{\emph{JHEP} {\bfseries 03}
  (2014) 012} [\href{https://arxiv.org/abs/1401.3672}{{\ttfamily 1401.3672}}].

\bibitem{Tierz:2018fsn}
M.~Tierz, \emph{{Wilson loops and free energies in $3d$ $\mathcal{N}=4$ SYM:
  exact results, exponential asymptotics and duality}},
  \href{https://doi.org/10.1093/ptep/ptz036}{\emph{PTEP} {\bfseries 2019}
  (2019) 053B01} [\href{https://arxiv.org/abs/1804.10845}{{\ttfamily
  1804.10845}}].

\bibitem{Santamaria:2010dm}
R.~C. Santamaria, M.~Mari\~{n}o and P.~Putrov, \emph{{Unquenched flavor and
  tropical geometry in strongly coupled Chern-Simons-matter theories}},
  \href{https://doi.org/10.1007/JHEP10(2011)139}{\emph{JHEP} {\bfseries 10}
  (2011) 139} [\href{https://arxiv.org/abs/1011.6281}{{\ttfamily 1011.6281}}].

\bibitem{Azeyanagi:2012xj}
T.~Azeyanagi, M.~Fujita and M.~Hanada, \emph{{From the planar limit to
  M-theory}}, \href{https://doi.org/10.1103/PhysRevLett.110.121601}{\emph{Phys.
  Rev. Lett.} {\bfseries 110} (2013) 121601}
  [\href{https://arxiv.org/abs/1210.3601}{{\ttfamily 1210.3601}}].

\bibitem{Grassi:2014vwa}
A.~Grassi and M.~Mari\~{n}o, \emph{{M-theoretic matrix models}},
  \href{https://doi.org/10.1007/JHEP02(2015)115}{\emph{JHEP} {\bfseries 02}
  (2015) 115} [\href{https://arxiv.org/abs/1403.4276}{{\ttfamily 1403.4276}}].

\bibitem{Hong:2021bsb}
J.~Hong and J.~T. Liu, \emph{{Subleading corrections to the S$^{3}$ free energy
  of necklace quiver theories dual to massive IIA}},
  \href{https://doi.org/10.1007/JHEP11(2021)183}{\emph{JHEP} {\bfseries 11}
  (2021) 183} [\href{https://arxiv.org/abs/2103.17033}{{\ttfamily
  2103.17033}}].

\bibitem{Beccaria:2023hhi}
M.~Beccaria and A.~A. Tseytlin, \emph{{Comments on ABJM free energy on $S^3$ at
  large $N$ and perturbative expansions in M-theory and string theory}},
  \href{https://doi.org/10.1016/j.nuclphysb.2023.116286}{\emph{Nucl. Phys. B}
  {\bfseries 994} (2023) 116286}
  [\href{https://arxiv.org/abs/2306.02862}{{\ttfamily 2306.02862}}].

\bibitem{Herzog:2010hf}
C.~P. Herzog, I.~R. Klebanov, S.~S. Pufu and T.~Tesileanu, \emph{{Multi-Matrix
  Models and Tri-Sasaki Einstein Spaces}},
  \href{https://doi.org/10.1103/PhysRevD.83.046001}{\emph{Phys. Rev. D}
  {\bfseries 83} (2011) 046001}
  [\href{https://arxiv.org/abs/1011.5487}{{\ttfamily 1011.5487}}].

\bibitem{Drukker:2010nc}
N.~Drukker, M.~Mari\~{n}o and P.~Putrov, \emph{{From weak to strong coupling in
  ABJM theory}}, \href{https://doi.org/10.1007/s00220-011-1253-6}{\emph{Commun.
  Math. Phys.} {\bfseries 306} (2011) 511}
  [\href{https://arxiv.org/abs/1007.3837}{{\ttfamily 1007.3837}}].

\bibitem{Boido:2023ojv}
A.~Boido, A.~L\"uscher and J.~Sparks, \emph{{Matrix models from black hole
  geometries}}, \href{https://doi.org/10.1007/JHEP05(2024)226}{\emph{JHEP}
  {\bfseries 05} (2024) 226}
  [\href{https://arxiv.org/abs/2312.11640}{{\ttfamily 2312.11640}}].

\bibitem{Imamura:2008nn}
Y.~Imamura and K.~Kimura, \emph{{On the moduli space of elliptic
  Maxwell-Chern-Simons theories}},
  \href{https://doi.org/10.1143/PTP.120.509}{\emph{Prog. Theor. Phys.}
  {\bfseries 120} (2008) 509}
  [\href{https://arxiv.org/abs/0806.3727}{{\ttfamily 0806.3727}}].

\bibitem{Jafferis:2008qz}
D.~L. Jafferis and A.~Tomasiello, \emph{{A Simple class of N=3 gauge/gravity
  duals}}, \href{https://doi.org/10.1088/1126-6708/2008/10/101}{\emph{JHEP}
  {\bfseries 10} (2008) 101} [\href{https://arxiv.org/abs/0808.0864}{{\ttfamily
  0808.0864}}].

\bibitem{Jafferis:2011zi}
D.~L. Jafferis, I.~R. Klebanov, S.~S. Pufu and B.~R. Safdi, \emph{{Towards the
  F-Theorem: N=2 Field Theories on the Three-Sphere}},
  \href{https://doi.org/10.1007/JHEP06(2011)102}{\emph{JHEP} {\bfseries 06}
  (2011) 102} [\href{https://arxiv.org/abs/1103.1181}{{\ttfamily 1103.1181}}].

\bibitem{Gulotta:2011si}
D.~R. Gulotta, C.~P. Herzog and S.~S. Pufu, \emph{{From Necklace Quivers to the
  F-theorem, Operator Counting, and T(U(N))}},
  \href{https://doi.org/10.1007/JHEP12(2011)077}{\emph{JHEP} {\bfseries 12}
  (2011) 077} [\href{https://arxiv.org/abs/1105.2817}{{\ttfamily 1105.2817}}].

\bibitem{Kallen:2012zn}
J.~Kallen, J.~A. Minahan, A.~Nedelin and M.~Zabzine, \emph{{$N^3$-behavior from
  5D Yang-Mills theory}},
  \href{https://doi.org/10.1007/JHEP10(2012)184}{\emph{JHEP} {\bfseries 10}
  (2012) 184} [\href{https://arxiv.org/abs/1207.3763}{{\ttfamily 1207.3763}}].

\bibitem{Minahan:2013jwa}
J.~A. Minahan, A.~Nedelin and M.~Zabzine, \emph{{5D super Yang-Mills theory and
  the correspondence to AdS$_7$/CFT$_6$}},
  \href{https://doi.org/10.1088/1751-8113/46/35/355401}{\emph{J. Phys. A}
  {\bfseries 46} (2013) 355401}
  [\href{https://arxiv.org/abs/1304.1016}{{\ttfamily 1304.1016}}].

\bibitem{Jafferis:2012iv}
D.~L. Jafferis and S.~S. Pufu, \emph{{Exact results for five-dimensional
  superconformal field theories with gravity duals}},
  \href{https://doi.org/10.1007/JHEP05(2014)032}{\emph{JHEP} {\bfseries 05}
  (2014) 032} [\href{https://arxiv.org/abs/1207.4359}{{\ttfamily 1207.4359}}].

\bibitem{Nedelin:2015mta}
A.~Nedelin, \emph{{Phase transitions in 5D super Yang-Mills theory}},
  \href{https://doi.org/10.1007/JHEP07(2015)004}{\emph{JHEP} {\bfseries 07}
  (2015) 004} [\href{https://arxiv.org/abs/1502.07275}{{\ttfamily
  1502.07275}}].

\bibitem{Uhlemann:2019ypp}
C.~F. Uhlemann, \emph{{Exact results for 5d SCFTs of long quiver type}},
  \href{https://doi.org/10.1007/JHEP11(2019)072}{\emph{JHEP} {\bfseries 11}
  (2019) 072} [\href{https://arxiv.org/abs/1909.01369}{{\ttfamily
  1909.01369}}].

\bibitem{Akhond:2022oaf}
M.~Akhond, A.~Legramandi, C.~Nunez, L.~Santilli and L.~Schepers, \emph{{Matrix
  Models and Holography: Mass Deformations of Long Quiver Theories in 5d and
  3d}}, \href{https://doi.org/10.21468/SciPostPhys.15.3.086}{\emph{SciPost
  Phys.} {\bfseries 15} (2023) 086}
  [\href{https://arxiv.org/abs/2211.13240}{{\ttfamily 2211.13240}}].

\bibitem{Nunez:2023loo}
C.~Nunez, L.~Santilli and K.~Zarembo, \emph{{Linear Quivers at Large-N}},
  \href{https://doi.org/10.1007/s00220-024-05186-1}{\emph{Commun. Math. Phys.}
  {\bfseries 406} (2025) 6} [\href{https://arxiv.org/abs/2311.00024}{{\ttfamily
  2311.00024}}].

\bibitem{Coccia:2020wtk}
L.~Coccia and C.~F. Uhlemann, \emph{{On the planar limit of 3d $
  {\mathrm{T}}_{\rho}^{\sigma}\left[\mathrm{SU}\left(\mathrm{N}\right)\right]
  $}}, \href{https://doi.org/10.1007/JHEP06(2021)038}{\emph{JHEP} {\bfseries
  06} (2021) 038} [\href{https://arxiv.org/abs/2011.10050}{{\ttfamily
  2011.10050}}].

\bibitem{Akhond:2021ffz}
M.~Akhond, A.~Legramandi and C.~Nunez, \emph{{Electrostatic description of 3d $
  \mathcal{N} $ = 4 linear quivers}},
  \href{https://doi.org/10.1007/JHEP11(2021)205}{\emph{JHEP} {\bfseries 11}
  (2021) 205} [\href{https://arxiv.org/abs/2109.06193}{{\ttfamily
  2109.06193}}].

\bibitem{Marino:2016new}
M.~Mari\~{n}o, \emph{{Localization at large N in
  Chern\textendash{}Simons-matter theories}},
  \href{https://doi.org/10.1088/1751-8121/aa5f69}{\emph{J. Phys. A} {\bfseries
  50} (2017) 443007} [\href{https://arxiv.org/abs/1608.02959}{{\ttfamily
  1608.02959}}].

\bibitem{Russo:2013sba}
J.~G. Russo and K.~Zarembo, \emph{{Localization at Large N}},  in \emph{{100th
  anniversary of the birth of I.Ya. Pomeranchuk}}, pp.~287--311, 2014,
  \href{https://arxiv.org/abs/1312.1214}{{\ttfamily 1312.1214}},
  \href{https://doi.org/10.1142/9789814616850_0015}{DOI}.

\bibitem{Zarembo:2016bbk}
K.~Zarembo, \emph{{Localization and AdS/CFT Correspondence}},
  \href{https://doi.org/10.1088/1751-8121/aa585b}{\emph{J. Phys. A} {\bfseries
  50} (2017) 443011} [\href{https://arxiv.org/abs/1608.02963}{{\ttfamily
  1608.02963}}].

\bibitem{Russo:2000ak}
J.~G. Russo, \emph{{Large $N$ field theories from superstrings}},
  \href{https://doi.org/10.22323/1.005.0044}{\emph{PoS} {\bfseries silafae-III}
  (2000) 044} [\href{https://arxiv.org/abs/hep-th/0008212}{{\ttfamily
  hep-th/0008212}}].

\bibitem{Moshe:2003xn}
M.~Moshe and J.~Zinn-Justin, \emph{{Quantum field theory in the large N limit:
  A Review}}, \href{https://doi.org/10.1016/S0370-1573(03)00263-1}{\emph{Phys.
  Rept.} {\bfseries 385} (2003) 69}
  [\href{https://arxiv.org/abs/hep-th/0306133}{{\ttfamily hep-th/0306133}}].

\bibitem{Benini:2012ui}
F.~Benini and S.~Cremonesi, \emph{{Partition Functions of ${\mathcal{N}=(2,2)}$
  Gauge Theories on S$^{2}$ and Vortices}},
  \href{https://doi.org/10.1007/s00220-014-2112-z}{\emph{Commun. Math. Phys.}
  {\bfseries 334} (2015) 1483}
  [\href{https://arxiv.org/abs/1206.2356}{{\ttfamily 1206.2356}}].

\bibitem{Doroud:2012xw}
N.~Doroud, J.~Gomis, B.~Le~Floch and S.~Lee, \emph{{Exact Results in D=2
  Supersymmetric Gauge Theories}},
  \href{https://doi.org/10.1007/JHEP05(2013)093}{\emph{JHEP} {\bfseries 05}
  (2013) 093} [\href{https://arxiv.org/abs/1206.2606}{{\ttfamily 1206.2606}}].

\bibitem{Kapustin:2009kz}
A.~Kapustin, B.~Willett and I.~Yaakov, \emph{{Exact Results for Wilson Loops in
  Superconformal Chern-Simons Theories with Matter}},
  \href{https://doi.org/10.1007/JHEP03(2010)089}{\emph{JHEP} {\bfseries 03}
  (2010) 089} [\href{https://arxiv.org/abs/0909.4559}{{\ttfamily 0909.4559}}].

\bibitem{Willett:2016adv}
B.~Willett, \emph{{Localization on three-dimensional manifolds}},
  \href{https://doi.org/10.1088/1751-8121/aa612f}{\emph{J. Phys. A} {\bfseries
  50} (2017) 443006} [\href{https://arxiv.org/abs/1608.02958}{{\ttfamily
  1608.02958}}].

\bibitem{Closset:2019hyt}
C.~Closset and H.~Kim, \emph{{Three-dimensional \ensuremath{\mathscr{N}} = 2
  supersymmetric gauge theories and partition functions on Seifert manifolds: A
  review}}, \href{https://doi.org/10.1142/S0217751X19300114}{\emph{Int. J. Mod.
  Phys. A} {\bfseries 34} (2019) 1930011}
  [\href{https://arxiv.org/abs/1908.08875}{{\ttfamily 1908.08875}}].

\bibitem{Pestun:2014mja}
V.~Pestun, \emph{{Localization for $\mathcal {N}=$ 2 Supersymmetric Gauge
  Theories in Four Dimensions}},  in \emph{{New Dualities of Supersymmetric
  Gauge Theories}} (J.~Teschner, ed.), Mathematical Physics Studies,
  pp.~159--194.
\newblock Springer, 2016.
\newblock \href{https://arxiv.org/abs/1412.7134}{{\ttfamily 1412.7134}}.
\newblock \href{https://doi.org/10.1007/978-3-319-18769-3_6}{DOI}.

\bibitem{Hosomichi:2016flq}
K.~Hosomichi, \emph{{${{{\mathcal N}}=2}$ SUSY gauge theories on S$^4$}},
  \href{https://doi.org/10.1088/1751-8121/aa7775}{\emph{J. Phys. A} {\bfseries
  50} (2017) 443010} [\href{https://arxiv.org/abs/1608.02962}{{\ttfamily
  1608.02962}}].

\bibitem{Bershtein:2015xfa}
M.~Bershtein, G.~Bonelli, M.~Ronzani and A.~Tanzini, \emph{{Exact results for $
  \mathcal{N} $ = 2 supersymmetric gauge theories on compact toric manifolds
  and equivariant Donaldson invariants}},
  \href{https://doi.org/10.1007/JHEP07(2016)023}{\emph{JHEP} {\bfseries 07}
  (2016) 023} [\href{https://arxiv.org/abs/1509.00267}{{\ttfamily
  1509.00267}}].

\bibitem{Lundin:2023tzw}
J.~Lundin, R.~Mauch and L.~Ruggeri, \emph{{From 5d flat connections to 4d
  fluxes (the art of slicing the cone)}},
  \href{https://doi.org/10.1007/JHEP10(2023)155}{\emph{JHEP} {\bfseries 10}
  (2023) 155} [\href{https://arxiv.org/abs/2305.02313}{{\ttfamily
  2305.02313}}].

\bibitem{Mauch:2024uyt}
R.~Mauch and L.~Ruggeri, \emph{{Super Yang-Mills on branched covers and
  weighted projective spaces}},
  \href{https://doi.org/10.1007/JHEP08(2024)106}{\emph{JHEP} {\bfseries 08}
  (2024) 106} [\href{https://arxiv.org/abs/2404.11600}{{\ttfamily
  2404.11600}}].

\bibitem{Kallen:2012va}
J.~K\"all\'en, J.~Qiu and M.~Zabzine, \emph{{The perturbative partition
  function of supersymmetric 5D Yang-Mills theory with matter on the
  five-sphere}}, \href{https://doi.org/10.1007/JHEP08(2012)157}{\emph{JHEP}
  {\bfseries 08} (2012) 157} [\href{https://arxiv.org/abs/1206.6008}{{\ttfamily
  1206.6008}}].

\bibitem{Kim:2012qf}
H.-C. Kim, J.~Kim and S.~Kim, \emph{{Instantons on the 5-sphere and
  M5-branes}},  \href{https://arxiv.org/abs/1211.0144}{{\ttfamily 1211.0144}}.

\bibitem{Alday:2015lta}
L.~F. Alday, P.~Benetti~Genolini, M.~Fluder, P.~Richmond and J.~Sparks,
  \emph{{Supersymmetric gauge theories on five-manifolds}},
  \href{https://doi.org/10.1007/JHEP08(2015)007}{\emph{JHEP} {\bfseries 08}
  (2015) 007} [\href{https://arxiv.org/abs/1503.09090}{{\ttfamily
  1503.09090}}].

\bibitem{Lundin:2021zeb}
J.~Lundin and L.~Ruggeri, \emph{{SYM on quotients of spheres and complex
  projective spaces}},
  \href{https://doi.org/10.1007/JHEP03(2022)204}{\emph{JHEP} {\bfseries 03}
  (2022) 204} [\href{https://arxiv.org/abs/2110.13065}{{\ttfamily
  2110.13065}}].

\bibitem{Closset:2022vjj}
C.~Closset and H.~Magureanu, \emph{{Partition functions and fibering operators
  on the Coulomb branch of 5d SCFTs}},
  \href{https://doi.org/10.1007/JHEP01(2023)035}{\emph{JHEP} {\bfseries 01}
  (2023) 035} [\href{https://arxiv.org/abs/2209.13564}{{\ttfamily
  2209.13564}}].

\bibitem{Qiu:2016dyj}
J.~Qiu and M.~Zabzine, \emph{{Review of localization for 5d supersymmetric
  gauge theories}}, \href{https://doi.org/10.1088/1751-8121/aa5ef0}{\emph{J.
  Phys. A} {\bfseries 50} (2017) 443014}
  [\href{https://arxiv.org/abs/1608.02966}{{\ttfamily 1608.02966}}].

\bibitem{Minahan:2015jta}
J.~A. Minahan and M.~Zabzine, \emph{{Gauge theories with 16 supersymmetries on
  spheres}}, \href{https://doi.org/10.1007/JHEP03(2015)155}{\emph{JHEP}
  {\bfseries 03} (2015) 155}
  [\href{https://arxiv.org/abs/1502.07154}{{\ttfamily 1502.07154}}].

\bibitem{Gulotta:2011vp}
D.~R. Gulotta, J.~P. Ang and C.~P. Herzog, \emph{{Matrix Models for
  Supersymmetric Chern-Simons Theories with an ADE Classification}},
  \href{https://doi.org/10.1007/JHEP01(2012)132}{\emph{JHEP} {\bfseries 01}
  (2012) 132} [\href{https://arxiv.org/abs/1111.1744}{{\ttfamily 1111.1744}}].

\bibitem{Gulotta:2012yd}
D.~R. Gulotta, C.~P. Herzog and T.~Nishioka, \emph{{The ABCDEF's of Matrix
  Models for Supersymmetric Chern-Simons Theories}},
  \href{https://doi.org/10.1007/JHEP04(2012)138}{\emph{JHEP} {\bfseries 04}
  (2012) 138} [\href{https://arxiv.org/abs/1201.6360}{{\ttfamily 1201.6360}}].

\bibitem{Crichigno:2012sk}
P.~M. Crichigno, C.~P. Herzog and D.~Jain, \emph{{Free Energy of $\hat{D}_n$
  Quiver Chern-Simons Theories}},
  \href{https://doi.org/10.1007/JHEP03(2013)039}{\emph{JHEP} {\bfseries 03}
  (2013) 039} [\href{https://arxiv.org/abs/1211.1388}{{\ttfamily 1211.1388}}].

\bibitem{Assel:2012cp}
B.~Assel, J.~Estes and M.~Yamazaki, \emph{{Large N Free Energy of 3d N=4 SCFTs
  and $AdS_4/CFT_3$}},
  \href{https://doi.org/10.1007/JHEP09(2012)074}{\emph{JHEP} {\bfseries 09}
  (2012) 074} [\href{https://arxiv.org/abs/1206.2920}{{\ttfamily 1206.2920}}].

\bibitem{Mezei:2013gqa}
M.~Mezei and S.~S. Pufu, \emph{{Three-sphere free energy for classical gauge
  groups}}, \href{https://doi.org/10.1007/JHEP02(2014)037}{\emph{JHEP}
  {\bfseries 02} (2014) 037} [\href{https://arxiv.org/abs/1312.0920}{{\ttfamily
  1312.0920}}].

\bibitem{Amariti:2019pky}
A.~Amariti, M.~Fazzi, N.~Mekareeya and A.~Nedelin, \emph{{New 3d
  $\mathcal{N}=2$ SCFT's with $N^{3/2}$ scaling}},
  \href{https://doi.org/10.1007/JHEP12(2019)111}{\emph{JHEP} {\bfseries 12}
  (2019) 111} [\href{https://arxiv.org/abs/1903.02586}{{\ttfamily
  1903.02586}}].

\bibitem{Marino:2011eh}
M.~Mari\~{n}o and P.~Putrov, \emph{{ABJM theory as a Fermi gas}},
  \href{https://doi.org/10.1088/1742-5468/2012/03/P03001}{\emph{J. Stat. Mech.}
  {\bfseries 1203} (2012) P03001}
  [\href{https://arxiv.org/abs/1110.4066}{{\ttfamily 1110.4066}}].

\bibitem{Fuji:2011km}
H.~Fuji, S.~Hirano and S.~Moriyama, \emph{{Summing Up All Genus Free Energy of
  ABJM Matrix Model}},
  \href{https://doi.org/10.1007/JHEP08(2011)001}{\emph{JHEP} {\bfseries 08}
  (2011) 001} [\href{https://arxiv.org/abs/1106.4631}{{\ttfamily 1106.4631}}].

\bibitem{Moriyama:2014gxa}
S.~Moriyama and T.~Nosaka, \emph{{Partition Functions of Superconformal
  Chern-Simons Theories from Fermi Gas Approach}},
  \href{https://doi.org/10.1007/JHEP11(2014)164}{\emph{JHEP} {\bfseries 11}
  (2014) 164} [\href{https://arxiv.org/abs/1407.4268}{{\ttfamily 1407.4268}}].

\bibitem{Nosaka:2024gle}
T.~Nosaka, \emph{{Large N expansion of mass deformed ABJM matrix model:
  M2-instanton condensation and beyond}},
  \href{https://doi.org/10.1007/JHEP03(2024)087}{\emph{JHEP} {\bfseries 03}
  (2024) 087} [\href{https://arxiv.org/abs/2401.11484}{{\ttfamily
  2401.11484}}].

\bibitem{Kubo:2024qhq}
N.~Kubo, T.~Nosaka and Y.~Pang, \emph{{Exact large $N$ expansion of mass
  deformed ABJM theory on squashed sphere}},
  \href{https://arxiv.org/abs/2411.07334}{{\ttfamily 2411.07334}}.

\bibitem{Bobev:2022jte}
N.~Bobev, J.~Hong and V.~Reys, \emph{{Large N Partition Functions, Holography,
  and Black Holes}},
  \href{https://doi.org/10.1103/PhysRevLett.129.041602}{\emph{Phys. Rev. Lett.}
  {\bfseries 129} (2022) 041602}
  [\href{https://arxiv.org/abs/2203.14981}{{\ttfamily 2203.14981}}].

\bibitem{Bobev:2022eus}
N.~Bobev, J.~Hong and V.~Reys, \emph{{Large N partition functions of the ABJM
  theory}}, \href{https://doi.org/10.1007/JHEP02(2023)020}{\emph{JHEP}
  {\bfseries 02} (2023) 020}
  [\href{https://arxiv.org/abs/2210.09318}{{\ttfamily 2210.09318}}].

\bibitem{Bobev:2023lkx}
N.~Bobev, J.~Hong and V.~Reys, \emph{{Large N partition functions of 3d
  holographic SCFTs}},
  \href{https://doi.org/10.1007/JHEP08(2023)119}{\emph{JHEP} {\bfseries 08}
  (2023) 119} [\href{https://arxiv.org/abs/2304.01734}{{\ttfamily
  2304.01734}}].

\bibitem{Geukens:2024zmt}
S.~Geukens and J.~Hong, \emph{{Subleading analysis for S$^{3}$ partition
  functions of $ \mathcal{N} $ = 2 holographic SCFTs}},
  \href{https://doi.org/10.1007/JHEP06(2024)190}{\emph{JHEP} {\bfseries 06}
  (2024) 190} [\href{https://arxiv.org/abs/2405.00845}{{\ttfamily
  2405.00845}}].

\bibitem{Russo:2012ay}
J.~G. Russo and K.~Zarembo, \emph{{Large N Limit of N=2 SU(N) Gauge Theories
  from Localization}},
  \href{https://doi.org/10.1007/JHEP10(2012)082}{\emph{JHEP} {\bfseries 10}
  (2012) 082} [\href{https://arxiv.org/abs/1207.3806}{{\ttfamily 1207.3806}}].

\bibitem{Bobev:2013cja}
N.~Bobev, H.~Elvang, D.~Z. Freedman and S.~S. Pufu, \emph{{Holography for $N =
  2^*$ on $S^4$}}, \href{https://doi.org/10.1007/JHEP07(2014)001}{\emph{JHEP}
  {\bfseries 07} (2014) 001} [\href{https://arxiv.org/abs/1311.1508}{{\ttfamily
  1311.1508}}].

\bibitem{Bobev:2018hbq}
N.~Bobev, F.~F. Gautason and J.~Van~Muiden, \emph{{Precision Holography for
  $\mathcal{N}=2^{*}$ on $S^4$ from type IIB Supergravity}},
  \href{https://doi.org/10.1007/JHEP04(2018)148}{\emph{JHEP} {\bfseries 04}
  (2018) 148} [\href{https://arxiv.org/abs/1802.09539}{{\ttfamily
  1802.09539}}].

\bibitem{Russo:2019lgq}
J.~G. Russo, E.~Wid\'en and K.~Zarembo, \emph{{$N = 2^{\ast}$ phase transitions
  and holography}}, \href{https://doi.org/10.1007/JHEP02(2019)196}{\emph{JHEP}
  {\bfseries 02} (2019) 196}
  [\href{https://arxiv.org/abs/1901.02835}{{\ttfamily 1901.02835}}].

\bibitem{Chang:2017cdx}
C.-M. Chang, M.~Fluder, Y.-H. Lin and Y.~Wang, \emph{{Spheres, Charges,
  Instantons, and Bootstrap: A Five-Dimensional Odyssey}},
  \href{https://doi.org/10.1007/JHEP03(2018)123}{\emph{JHEP} {\bfseries 03}
  (2018) 123} [\href{https://arxiv.org/abs/1710.08418}{{\ttfamily
  1710.08418}}].

\bibitem{Chang:2017mxc}
C.-M. Chang, M.~Fluder, Y.-H. Lin and Y.~Wang, \emph{{Romans Supergravity from
  Five-Dimensional Holograms}},
  \href{https://doi.org/10.1007/JHEP05(2018)039}{\emph{JHEP} {\bfseries 05}
  (2018) 039} [\href{https://arxiv.org/abs/1712.10313}{{\ttfamily
  1712.10313}}].

\bibitem{Chang:2019uag}
C.-M. Chang, M.~Fluder, Y.-H. Lin and Y.~Wang, \emph{{Proving the 6d Cardy
  Formula and Matching Global Gravitational Anomalies}},
  \href{https://doi.org/10.21468/SciPostPhys.11.2.036}{\emph{SciPost Phys.}
  {\bfseries 11} (2021) 036}
  [\href{https://arxiv.org/abs/1910.10151}{{\ttfamily 1910.10151}}].

\bibitem{Fluder:2020pym}
M.~Fluder and C.~F. Uhlemann, \emph{{Evidence for a 5d F-theorem}},
  \href{https://doi.org/10.1007/JHEP02(2021)192}{\emph{JHEP} {\bfseries 02}
  (2021) 192} [\href{https://arxiv.org/abs/2011.00006}{{\ttfamily
  2011.00006}}].

\bibitem{Uhlemann:2019lge}
C.~F. Uhlemann, \emph{{AdS$_6$/CFT$_5$ with O7-planes}},
  \href{https://doi.org/10.1007/JHEP04(2020)113}{\emph{JHEP} {\bfseries 04}
  (2020) 113} [\href{https://arxiv.org/abs/1912.09716}{{\ttfamily
  1912.09716}}].

\bibitem{Legramandi:2021uds}
A.~Legramandi and C.~Nunez, \emph{{Electrostatic description of
  five-dimensional SCFTs}},
  \href{https://doi.org/10.1016/j.nuclphysb.2021.115630}{\emph{Nucl. Phys. B}
  {\bfseries 974} (2022) 115630}
  [\href{https://arxiv.org/abs/2104.11240}{{\ttfamily 2104.11240}}].

\bibitem{Akhond:2022awd}
M.~Akhond, A.~Legramandi, C.~Nunez, L.~Santilli and L.~Schepers, \emph{{Massive
  flows in AdS6/CFT5}},
  \href{https://doi.org/10.1016/j.physletb.2023.137899}{\emph{Phys. Lett. B}
  {\bfseries 840} (2023) 137899}
  [\href{https://arxiv.org/abs/2211.09824}{{\ttfamily 2211.09824}}].

\bibitem{Santilli:2023fuh}
L.~Santilli and C.~F. Uhlemann, \emph{{3d defects in 5d: RG flows and defect
  F-maximization}}, \href{https://doi.org/10.1007/JHEP06(2023)136}{\emph{JHEP}
  {\bfseries 06} (2023) 136}
  [\href{https://arxiv.org/abs/2305.01004}{{\ttfamily 2305.01004}}].

\bibitem{Coccia:2020cku}
L.~Coccia, \emph{{Topologically twisted index of $T[SU(N)]$ at large $N$}},
  \href{https://doi.org/10.1007/JHEP05(2021)264}{\emph{JHEP} {\bfseries 05}
  (2021) 264} [\href{https://arxiv.org/abs/2006.06578}{{\ttfamily
  2006.06578}}].

\bibitem{Gutperle:2018vdd}
M.~Gutperle, A.~Trivella and C.~F. Uhlemann, \emph{{Type IIB 7-branes in warped
  AdS$_{6}$: partition functions, brane webs and probe limit}},
  \href{https://doi.org/10.1007/JHEP04(2018)135}{\emph{JHEP} {\bfseries 04}
  (2018) 135} [\href{https://arxiv.org/abs/1802.07274}{{\ttfamily
  1802.07274}}].

\bibitem{Bergman:2018hin}
O.~Bergman, D.~Rodr\'\i{}guez-G\'omez and C.~F. Uhlemann, \emph{{Testing
  AdS$_{6}$/CFT$_{5}$ in Type IIB with stringy operators}},
  \href{https://doi.org/10.1007/JHEP08(2018)127}{\emph{JHEP} {\bfseries 08}
  (2018) 127} [\href{https://arxiv.org/abs/1806.07898}{{\ttfamily
  1806.07898}}].

\bibitem{Fluder:2018chf}
M.~Fluder and C.~F. Uhlemann, \emph{{Precision Test of AdS$_6$/CFT$_5$ in Type
  IIB String Theory}},
  \href{https://doi.org/10.1103/PhysRevLett.121.171603}{\emph{Phys. Rev. Lett.}
  {\bfseries 121} (2018) 171603}
  [\href{https://arxiv.org/abs/1806.08374}{{\ttfamily 1806.08374}}].

\bibitem{Nunez:2019gbg}
C.~N\'u\~nez, D.~Roychowdhury, S.~Speziali and S.~Zacar\'\i{}as,
  \emph{{Holographic aspects of four dimensional ${\cal N }=2$ SCFTs and their
  marginal deformations}},
  \href{https://doi.org/10.1016/j.nuclphysb.2019.114617}{\emph{Nucl. Phys. B}
  {\bfseries 943} (2019) 114617}
  [\href{https://arxiv.org/abs/1901.02888}{{\ttfamily 1901.02888}}].

\bibitem{Fluder:2019szh}
M.~Fluder, S.~M. Hosseini and C.~F. Uhlemann, \emph{{Black hole microstate
  counting in Type IIB from 5d SCFTs}},
  \href{https://doi.org/10.1007/JHEP05(2019)134}{\emph{JHEP} {\bfseries 05}
  (2019) 134} [\href{https://arxiv.org/abs/1902.05074}{{\ttfamily
  1902.05074}}].

\bibitem{Gutperle:2020rty}
M.~Gutperle and C.~F. Uhlemann, \emph{{Surface defects in holographic 5d
  SCFTs}}, \href{https://doi.org/10.1007/JHEP04(2021)134}{\emph{JHEP}
  {\bfseries 04} (2021) 134}
  [\href{https://arxiv.org/abs/2012.14547}{{\ttfamily 2012.14547}}].

\bibitem{Apruzzi:2022nax}
F.~Apruzzi, O.~Bergman, H.-C. Kim and C.~F. Uhlemann, \emph{{Generalized
  quotients and holographic duals for 5d S-fold SCFTs}},
  \href{https://doi.org/10.1007/JHEP04(2023)027}{\emph{JHEP} {\bfseries 04}
  (2023) 027} [\href{https://arxiv.org/abs/2211.13243}{{\ttfamily
  2211.13243}}].

\bibitem{Macpherson:2024frt}
N.~T. Macpherson, P.~Merrikin and C.~Nunez, \emph{{Marginally deformed
  AdS$_{5}$/CFT$_{4}$ and spindle-like orbifolds}},
  \href{https://doi.org/10.1007/JHEP07(2024)042}{\emph{JHEP} {\bfseries 07}
  (2024) 042} [\href{https://arxiv.org/abs/2403.02380}{{\ttfamily
  2403.02380}}].

\bibitem{Chatzis:2024top}
D.~Chatzis, A.~Fatemiabhari, C.~Nunez and P.~Weck, \emph{{Conformal to
  confining SQFTs from holography}},
  \href{https://doi.org/10.1007/JHEP08(2024)041}{\emph{JHEP} {\bfseries 08}
  (2024) 041} [\href{https://arxiv.org/abs/2405.05563}{{\ttfamily
  2405.05563}}].

\bibitem{Chatzis:2024kdu}
D.~Chatzis, A.~Fatemiabhari, C.~Nunez and P.~Weck, \emph{{SCFT deformations via
  uplifted solitons}},
  \href{https://doi.org/10.1016/j.nuclphysb.2024.116659}{\emph{Nucl. Phys. B}
  {\bfseries 1006} (2024) 116659}
  [\href{https://arxiv.org/abs/2406.01685}{{\ttfamily 2406.01685}}].

\bibitem{Santilli:2018byi}
L.~Santilli and M.~Tierz, \emph{{Phase transitions and Wilson loops in
  antisymmetric representations in Chern-Simons-matter theory}},
  \href{https://doi.org/10.1088/1751-8121/ab335c}{\emph{J. Phys. A} {\bfseries
  52} (2019) 385401} [\href{https://arxiv.org/abs/1808.02855}{{\ttfamily
  1808.02855}}].

\bibitem{Hartnoll:2006is}
S.~A. Hartnoll and S.~P. Kumar, \emph{{Higher rank Wilson loops from a matrix
  model}}, \href{https://doi.org/10.1088/1126-6708/2006/08/026}{\emph{JHEP}
  {\bfseries 08} (2006) 026}
  [\href{https://arxiv.org/abs/hep-th/0605027}{{\ttfamily hep-th/0605027}}].

\bibitem{Chen-Lin:2015dfa}
X.~Chen-Lin and K.~Zarembo, \emph{{Higher Rank Wilson Loops in N = 2*
  Super-Yang-Mills Theory}},
  \href{https://doi.org/10.1007/JHEP03(2015)147}{\emph{JHEP} {\bfseries 03}
  (2015) 147} [\href{https://arxiv.org/abs/1502.01942}{{\ttfamily
  1502.01942}}].

\bibitem{Russo:2017ngf}
J.~G. Russo and K.~Zarembo, \emph{{Wilson loops in antisymmetric
  representations from localization in supersymmetric gauge theories}},
  \href{https://doi.org/10.1142/9789813233867_0021}{\emph{Rev. Math. Phys.}
  {\bfseries 30} (2018) 1840014}
  [\href{https://arxiv.org/abs/1712.07186}{{\ttfamily 1712.07186}}].

\bibitem{Assel:2012nf}
B.~Assel, J.~Estes and M.~Yamazaki, \emph{{Wilson Loops in 5d N=1 SCFTs and
  AdS/CFT}}, \href{https://doi.org/10.1007/s00023-013-0249-5}{\emph{Annales
  Henri Poincare} {\bfseries 15} (2014) 589}
  [\href{https://arxiv.org/abs/1212.1202}{{\ttfamily 1212.1202}}].

\bibitem{Uhlemann:2020bek}
C.~F. Uhlemann, \emph{{Wilson loops in 5d long quiver gauge theories}},
  \href{https://doi.org/10.1007/JHEP09(2020)145}{\emph{JHEP} {\bfseries 09}
  (2020) 145} [\href{https://arxiv.org/abs/2006.01142}{{\ttfamily
  2006.01142}}].

\bibitem{Fatemiabhari:2022kpv}
A.~Fatemiabhari and C.~Nunez, \emph{{Wilson loops for 5d and 3d conformal
  linear quivers}},
  \href{https://doi.org/10.1016/j.nuclphysb.2023.116125}{\emph{Nucl. Phys. B}
  {\bfseries 989} (2023) 116125}
  [\href{https://arxiv.org/abs/2209.07536}{{\ttfamily 2209.07536}}].

\bibitem{Nian:2019pxj}
J.~Nian and L.~A. Pando~Zayas, \emph{{Microscopic entropy of rotating
  electrically charged AdS$_{4}$ black holes from field theory localization}},
  \href{https://doi.org/10.1007/JHEP03(2020)081}{\emph{JHEP} {\bfseries 03}
  (2020) 081} [\href{https://arxiv.org/abs/1909.07943}{{\ttfamily
  1909.07943}}].

\bibitem{Bobev:2022wem}
N.~Bobev, S.~Choi, J.~Hong and V.~Reys, \emph{{Large N superconformal indices
  for 3d holographic SCFTs}},
  \href{https://doi.org/10.1007/JHEP02(2023)027}{\emph{JHEP} {\bfseries 02}
  (2023) 027} [\href{https://arxiv.org/abs/2210.15326}{{\ttfamily
  2210.15326}}].

\bibitem{BenettiGenolini:2023rkq}
P.~Benetti~Genolini, A.~Cabo-Bizet and S.~Murthy, \emph{{Supersymmetric phases
  of AdS$_{4}$/CFT$_{3}$}},
  \href{https://doi.org/10.1007/JHEP06(2023)125}{\emph{JHEP} {\bfseries 06}
  (2023) 125} [\href{https://arxiv.org/abs/2301.00763}{{\ttfamily
  2301.00763}}].

\bibitem{Amariti:2023ygn}
A.~Amariti, J.~Nian, L.~A. Pando~Zayas and A.~Segati, \emph{{Universal
  Cardy-Like Behavior of 3D A-Twisted Partition Functions}},
  \href{https://arxiv.org/abs/2306.05462}{{\ttfamily 2306.05462}}.

\bibitem{Bobev:2024mqw}
N.~Bobev, S.~Choi, J.~Hong and V.~Reys, \emph{{Superconformal indices of 3d $
  \mathcal{N} $ = 2 SCFTs and holography}},
  \href{https://doi.org/10.1007/JHEP10(2024)121}{\emph{JHEP} {\bfseries 10}
  (2024) 121} [\href{https://arxiv.org/abs/2407.13177}{{\ttfamily
  2407.13177}}].

\bibitem{ArabiArdehali:2019tdm}
A.~Arabi~Ardehali, \emph{{Cardy-like asymptotics of the 4d $ \mathcal{N}=4 $
  index and AdS$_{5}$ blackholes}},
  \href{https://doi.org/10.1007/JHEP06(2019)134}{\emph{JHEP} {\bfseries 06}
  (2019) 134} [\href{https://arxiv.org/abs/1902.06619}{{\ttfamily
  1902.06619}}].

\bibitem{Cabo-Bizet:2019osg}
A.~Cabo-Bizet, D.~Cassani, D.~Martelli and S.~Murthy, \emph{{The asymptotic
  growth of states of the 4d $ \mathcal{N}=1 $ superconformal index}},
  \href{https://doi.org/10.1007/JHEP08(2019)120}{\emph{JHEP} {\bfseries 08}
  (2019) 120} [\href{https://arxiv.org/abs/1904.05865}{{\ttfamily
  1904.05865}}].

\bibitem{Cabo-Bizet:2019eaf}
A.~Cabo-Bizet and S.~Murthy, \emph{{Supersymmetric phases of 4d $ \mathcal{N} $
  = 4 SYM at large $N$}},
  \href{https://doi.org/10.1007/JHEP09(2020)184}{\emph{JHEP} {\bfseries 09}
  (2020) 184} [\href{https://arxiv.org/abs/1909.09597}{{\ttfamily
  1909.09597}}].

\bibitem{ArabiArdehali:2019orz}
A.~Arabi~Ardehali, J.~Hong and J.~T. Liu, \emph{{Asymptotic growth of the 4d $
  \mathcal{N} $ = 4 index and partially deconfined phases}},
  \href{https://doi.org/10.1007/JHEP07(2020)073}{\emph{JHEP} {\bfseries 07}
  (2020) 073} [\href{https://arxiv.org/abs/1912.04169}{{\ttfamily
  1912.04169}}].

\bibitem{Amariti:2020jyx}
A.~Amariti, M.~Fazzi and A.~Segati, \emph{{The SCI of $ \mathcal{N} $ = 4
  USp(2N$_{c}$) and SO(N$_{c}$) SYM as a matrix integral}},
  \href{https://doi.org/10.1007/JHEP06(2021)132}{\emph{JHEP} {\bfseries 06}
  (2021) 132} [\href{https://arxiv.org/abs/2012.15208}{{\ttfamily
  2012.15208}}].

\bibitem{Amariti:2021ubd}
A.~Amariti, M.~Fazzi and A.~Segati, \emph{{Expanding on the Cardy-like limit of
  the SCI of 4d $ \mathcal{N} $ = 1 ABCD SCFTs}},
  \href{https://doi.org/10.1007/JHEP07(2021)141}{\emph{JHEP} {\bfseries 07}
  (2021) 141} [\href{https://arxiv.org/abs/2103.15853}{{\ttfamily
  2103.15853}}].

\bibitem{Choi:2021rxi}
S.~Choi, S.~Jeong, S.~Kim and E.~Lee, \emph{{Exact QFT duals of AdS black
  holes}}, \href{https://doi.org/10.1007/JHEP09(2023)138}{\emph{JHEP}
  {\bfseries 09} (2023) 138}
  [\href{https://arxiv.org/abs/2111.10720}{{\ttfamily 2111.10720}}].

\bibitem{Cabo-Bizet:2021jar}
A.~Cabo-Bizet, \emph{{Quantum phases of 4d SU(N) $ \mathcal{N} $ = 4 SYM}},
  \href{https://doi.org/10.1007/JHEP10(2022)052}{\emph{JHEP} {\bfseries 10}
  (2022) 052} [\href{https://arxiv.org/abs/2111.14942}{{\ttfamily
  2111.14942}}].

\bibitem{ArabiArdehali:2023bpq}
A.~Arabi~Ardehali, M.~Martone and M.~Rossell\'o, \emph{{High-temperature
  expansion of the Schur index and modularity}},
  \href{https://arxiv.org/abs/2308.09738}{{\ttfamily 2308.09738}}.

\bibitem{Choi:2023tiq}
S.~Choi, S.~Kim and J.~Song, \emph{{Large N universality of 4d $ \mathcal{N} $
  = 1 superconformal index and AdS black holes}},
  \href{https://doi.org/10.1007/JHEP08(2024)105}{\emph{JHEP} {\bfseries 08}
  (2024) 105} [\href{https://arxiv.org/abs/2309.07614}{{\ttfamily
  2309.07614}}].

\bibitem{Bergman:2013koa}
O.~Bergman, D.~Rodr\'\i{}guez-G\'omez and G.~Zafrir, \emph{{5d superconformal
  indices at large N and holography}},
  \href{https://doi.org/10.1007/JHEP08(2013)081}{\emph{JHEP} {\bfseries 08}
  (2013) 081} [\href{https://arxiv.org/abs/1305.6870}{{\ttfamily 1305.6870}}].

\bibitem{Choi:2019miv}
S.~Choi and S.~Kim, \emph{{Large AdS$_{6}$ black holes from CFT$_{5}$}},
  \href{https://doi.org/10.1007/JHEP08(2024)228}{\emph{JHEP} {\bfseries 08}
  (2024) 228} [\href{https://arxiv.org/abs/1904.01164}{{\ttfamily
  1904.01164}}].

\bibitem{Crichigno:2020ouj}
P.~M. Crichigno and D.~Jain, \emph{{The 5d Superconformal Index at Large $N$
  and Black Holes}}, \href{https://doi.org/10.1007/JHEP09(2020)124}{\emph{JHEP}
  {\bfseries 09} (2020) 124}
  [\href{https://arxiv.org/abs/2005.00550}{{\ttfamily 2005.00550}}].

\bibitem{Ardehali:2021irq}
A.~A. Ardehali and J.~Hong, \emph{{Decomposition of BPS moduli spaces and
  asymptotics of supersymmetric partition functions}},
  \href{https://doi.org/10.1007/JHEP01(2022)062}{\emph{JHEP} {\bfseries 01}
  (2022) 062} [\href{https://arxiv.org/abs/2110.01538}{{\ttfamily
  2110.01538}}].

\bibitem{Cassani:2021fyv}
D.~Cassani and Z.~Komargodski, \emph{{EFT and the SUSY Index on the 2nd
  Sheet}}, \href{https://doi.org/10.21468/SciPostPhys.11.1.004}{\emph{SciPost
  Phys.} {\bfseries 11} (2021) 004}
  [\href{https://arxiv.org/abs/2104.01464}{{\ttfamily 2104.01464}}].

\bibitem{Crichigno:2018adf}
P.~M. Crichigno, D.~Jain and B.~Willett, \emph{{5d Partition Functions with A
  Twist}}, \href{https://doi.org/10.1007/JHEP11(2018)058}{\emph{JHEP}
  {\bfseries 11} (2018) 058}
  [\href{https://arxiv.org/abs/1808.06744}{{\ttfamily 1808.06744}}].

\bibitem{Jain:2021sdp}
D.~Jain, \emph{{Notes on 5d Partition Functions - I}},
  \href{https://arxiv.org/abs/2106.15126}{{\ttfamily 2106.15126}}.

\bibitem{Jain:2022avc}
D.~Jain, \emph{{Notes on 5d Partition Functions -- II}},
  \href{https://arxiv.org/abs/2203.17203}{{\ttfamily 2203.17203}}.

\bibitem{Santilli:2020uht}
L.~Santilli, R.~J. Szabo and M.~Tierz, \emph{{Five-dimensional cohomological
  localization and squashed $q$-deformations of two-dimensional Yang-Mills
  theory}}, \href{https://doi.org/10.1007/JHEP06(2020)036}{\emph{JHEP}
  {\bfseries 06} (2020) 036}
  [\href{https://arxiv.org/abs/2003.09411}{{\ttfamily 2003.09411}}].

\bibitem{Bah:2018lyv}
I.~Bah, A.~Passias and P.~Weck, \emph{{Holographic duals of five-dimensional
  SCFTs on a Riemann surface}},
  \href{https://doi.org/10.1007/JHEP01(2019)058}{\emph{JHEP} {\bfseries 01}
  (2019) 058} [\href{https://arxiv.org/abs/1807.06031}{{\ttfamily
  1807.06031}}].

\bibitem{Legramandi:2021aqv}
A.~Legramandi and C.~Nunez, \emph{{Holographic description of SCFT$_{5}$
  compactifications}},
  \href{https://doi.org/10.1007/JHEP02(2022)010}{\emph{JHEP} {\bfseries 02}
  (2022) 010} [\href{https://arxiv.org/abs/2109.11554}{{\ttfamily
  2109.11554}}].

\bibitem{Jain:2013py}
S.~Jain, S.~Minwalla, T.~Sharma, T.~Takimi, S.~R. Wadia and S.~Yokoyama,
  \emph{{Phases of large $N$ vector Chern-Simons theories on $S^2 \times
  S^1$}}, \href{https://doi.org/10.1007/JHEP09(2013)009}{\emph{JHEP} {\bfseries
  09} (2013) 009} [\href{https://arxiv.org/abs/1301.6169}{{\ttfamily
  1301.6169}}].

\bibitem{Santilli:2019wvq}
L.~Santilli and M.~Tierz, \emph{{Phase transition in complex-time Loschmidt
  echo of short and long range spin chain}},
  \href{https://doi.org/10.1088/1742-5468/ab837b}{\emph{J. Stat. Mech.}
  {\bfseries 2006} (2020) 063102}
  [\href{https://arxiv.org/abs/1902.06649}{{\ttfamily 1902.06649}}].

\end{thebibliography}\endgroup
}
\end{document}